\documentclass[10pt]{iopart}

\usepackage{graphicx}

\usepackage{iopams}
\usepackage[english]{babel}
\usepackage{hyperref}
\usepackage{url}
\hypersetup{
colorlinks,
menucolor=blue,
linkcolor=red,
citecolor=blue,
urlcolor=blue
}

\newcommand{\bm}[1]{\boldsymbol{#1}}
\newcommand{\aver}[1]{\left\langle #1 \right\rangle}

\newcommand{\rr}{\bm{r}}

\newcommand{\rry}{\bm{r}_1}

\newcommand{\p}{\bm{p}}
\newcommand{\q}{\bm{q}}

\newcommand{\kron}[1]{\delta_{#1}}

\def \k{ {\bf{k}} }

\newcommand{\varG}[3]{\frac{\delta #1}{\delta G(#2,#3)}}
\newcommand{\varu}[3]{\frac{\delta #1}{\delta \phi(#2,#3)}}
\newcommand{\unolla}[1]{\left( #1 \right)_{\phi=0}}

\newcommand{\bGj}{\bm{\mathcal{G}}_{\mathrm{Bath},j}}

\newcommand{\text}[1]{\mathrm{#1}}

\newcommand{\beq}{\begin{equation}}
\newcommand{\enq}{\end{equation}}

\newcommand{\eqref}[1]{\eref{#1}}

\begin{document}

\title{The Fulde-Ferrell-Larkin-Ovchinnikov state for ultracold fermions in lattice and harmonic potentials: a review}

\author{Jami J. Kinnunen}
\author{Jildou E. Baarsma}
\author{Jani-Petri Martikainen}
\author{P{\"{a}}ivi T{\"{o}}rm{\"{a}}}
\ead{paivi.torma@aalto.fi}
\address{COMP Center of Excellence, Department of Applied Physics, Aalto University, Fi-00076, Aalto, Finland}

\date{\today}

\begin{abstract}
We review the concepts and the present state of theoretical studies of spin-imbalanced superfluidity, in particular the elusive Fulde-Ferrell-Larkin-Ovchinnikov (FFLO) state, in the context of ultracold quantum gases. The comprehensive presentation of the theoretical basis for the FFLO state that we provide is useful also for research on the interplay between magnetism and superconductivity in other physical systems. We focus on settings that have been predicted to be favourable for the FFLO state, such as optical lattices in various dimensions and spin-orbit coupled systems. These are also the most likely systems for near-future experimental observation of the FFLO state. Theoretical bounds, such as Bloch's and Luttinger's theorems, and experimentally important limitations, such as finite-size effects and trapping potentials, are considered. In addition, we provide a comprehensive review of the various ideas presented for the observation of the FFLO state. We conclude our review with an analysis of the open questions related to the FFLO state, such as its stability, superfluid density, collective modes and extending the FFLO superfluid concept to new types of lattice systems.
\end{abstract}

\maketitle

\ioptwocol

\tableofcontents

\section{Introduction}
\label{sec:introduction}

\subsection{Superconductivity, superfluidity and magnetism}

Superfluidity appears in range of systems from ultracold atomic gases to neutron stars, and from electrons in metals to quark-gluon plasma. The microscopic Bardeen-Cooper-Schrieffer (BCS)~\cite{bcs} theory explains superfluidity in fermionic systems as the formation and condensation of Cooper pairs of indistinguishable fermions, for instance particles with different internal spin states~\footnote{Throughout this review, by "superfluid" and "superfluidity" we refer to superfluidity of paired fermions, unless stated otherwise.}.
Magnetism and superfluidity of charged particles (i.e., superconductivity) are, to a large extent, mutually incompatible: magnetic fields cause eddy currents in superconductors, which either results in expulsion of the magnetic field from the superconductor known as the Meissner effect, or the suppression of superconductivity in the regions of non-zero magnetic field. 
If the magnetic field becomes sufficiently strong, the free energy associated with the superfluid state may exceed the free energy of the normal state and the material loses superconductivity altogether. This is because Cooper pairing favours equal amounts of particles in the two spin states while Pauli paramagnetism implies that the spins of the charge carriers prefer to align with the magnetic field.
Still, it has been theorized that superfluidity and magnetism can coexist, and there has been tremendous interest in trying to realize such systems experimentally. 

Notable theory proposals of ground states that combine superfluidity and magnetism, such as the Fulde-Ferrell-Larkin-Ocvhinnikov (FFLO) state~\cite{FF,LOrus,LOeng}, assume a finite Zeeman energy, but no orbital effects that are typically caused by magnetic fields. The search for such exotic forms of superconductivity and superfluidity thus calls for systems where orbital effects are absent. By considering two-dimensional layered superconductors for example, one can suppress orbital effects against in-plane magnetic fields, as circulating eddy currents around magnetic field lines are blocked by geometric constraints.
This is the approach used in many solid-state experiments to realize magnetized superconductors~\cite{casalbuoni_inhomogeneous_2004,buzdin_fflo_2007,buzdin_non-uniform_2012,beyer_emerging_2013}. Another possibility to avoid orbital effects is to study superfluidity in neutral systems. Experiments on ultracold neutral atoms allow independent knobs for orbital effects (via fluid rotation) and the Zeeman effect (via pseudo-spin species imbalance). This provides a serious advantage of ultracold quantum gases (UQG) over solid state fermionic systems for exploring the competition between paired superfluidity and Zeeman energy. Here, we review the latest advances in the theoretical and numerical study of FFLO superfluids in neutral atomic gases. We focus on lattice systems but cover also recent developments on the effects of trap geometry and spin-orbit coupling.

\subsection{Chandrasekhar-Clogston limit and exotic superfluidity}

In spin-singlet superconductors, electrons are paired with opposite spins.
Switching  the electron spin would thus result in breaking  a Cooper pair, imposing energy cost equal to the binding energy of the pair.
This is known as the Chandrasekhar-Clogston limit~\cite{Chandra1962,clogston_upper_1962}, which tells us that the Zeeman energy due to the magnetic field $h$ must exceed Cooper pair binding energy $\Delta$ before Pauli paramagnetism will play a role. 
A closer inspection yields the bound $h = \Delta/\sqrt{2}$.
If the magnetic field is less than this critical value, the system will remain non-magnetised and a superfluid.
Beyond this limit, the magnetic field provides sufficient energy to switch the spins, break the pairs, and suppress superfluidity.
However this argument is limited to simple BCS-type superfluids. The salient question is whether some more exotic superfluid phases, such as breached pair~\cite{Sarma1963,liu_interior_2003,gubankova_breached_2003}, FFLO~\cite{FF,LOeng} or deformed Fermi surface~\cite{muther_spontaneous_2002,sedrakian_pairing_2005} superfluids can exceed the Chandrasekhar-Clogston limit and realize magnetized superfluidity.
Furthermore, the limit can be broken in spin-orbit coupled
superfluids where spin-orbit coupling can stabilize the superfluid and enable
different topological phases~\cite{kubasiak_topological_2010}.

Breached pair superfluids retain the homogeneous nature of the BCS superfluid but allow the coexistence of spin-imbalance by opening an unoccupied breach in the momentum distribution. This results in gapless excitations, showing that not all atoms are paired.
Deformation of Fermi surfaces, associated with spontaneous breaking of the rotational symmetry, has also been predicted as a possible mechanism for spin-imbalanced superfluidity with zero-momentum pairs~\cite{muther_spontaneous_2002,sedrakian_pairing_2005}.
In the FFLO state, Cooper pairs are condensed in one or more finite-center-of-mass-momentum states.
In a Fulde-Ferrell (FF)~\cite{FF} superfluid, all pairs have the same momentum ${\bf q}$ while in the Larkin-Ovchinnikov (LO)~\cite{LOeng} one pairs involve two opposite momenta $\pm {\bf q}$. Gapless excitations exist in FFLO superfluids like in the breached pair one, but in the FFLO state the order parameters become spatially varying, with the period given by the inverse of the FFLO wave vector ${\bf q}$. 
As condensates typically, the FFLO state is associated with breaking of the $U(1)$ symmetry for the phase of the order parameter $\Delta$, but in addition, other symmetries may be broken. The essential difference between FF- and LO-type states is that the FF states break the time-reversal symmetry but retain the translational one while LO states are time-reversal-symmetric but there is no translational invariance.  The symmetry breaking properties are determined by having a complex (FF) or real (LO) order parameter, and by its spatial dependence and symmetry; the number of FFLO vectors $\mathbf{q}$ involved in the pairing can in general be more than one or two. The Goldstone modes that are related to breaking of continuous symmetries are thus distinct for different types of FFLO states. 

\subsection{FFLO studies with ultracold atomic gases}

In UQG, the pseudospin states of the atoms are determined in the preparation of the sample, and these spin degrees of freedom are effectively frozen. The system is thus somewhat different from the Pauli paramagnetism in metals, where electrons will realign the magnetic moments with the external magnetic field. Instead, magnetism in ultracold gases is itinerant: atoms are mobile and free to rearrange into different magnetic domains. Any exotic magnetized superfluidity would appear as a coexistence region involving both superfluidity and spin imbalance, whereas absence of such exotic phases would be observable as either phase separation into a spin-balanced superfluid and spin-imbalanced normal fluid, or as an absence of superfluidity altogether.

A phase with spin-imbalance and vanishing Cooper pair momentum ${\bf q}=0$ is known to be unstable at low temperatures, with instability in favor of the FFLO state. On the other hand the FFLO state is unstable towards the aforementioned phase separation in a large region of parameter space, in particular in three-dimensional (3D) systems. Indeed early UQG experiments on spin-imbalanced superfluids~\cite{zwierlein_fermionic_2006,partridge_pairing_2006} showed phase separation with a spin-balanced superfluid core surrounded by a spin-imbalanced normal state gas in a harmonic trap.

There are already reviews regarding the FFLO state in continuum 3D UQG~\cite{radzihovsky_imbalanced_2010,chevy_ultra-cold_2010,gubbels_imbalanced_2013}. Reduced dimensionality has been predicted to favour FFLO. This was exemplified by promising experimental results in references~\cite{liao_spin-imbalance_2010,revelle_1d-to-3d_2016}, in which a spin-imbalanced fermionic gas was prepared in an array of one-dimensional (1D) tubes. Density profiles were in good agreement with theories and suggestive of an FFLO state, but clear observation of FFLO-type pairing was not obtained. These experiments were done in continuum and are thus outside the scope of this review, but they have been covered in a review~\cite{guan_fermi_2013}.
In low dimensions, the long-range order of the FFLO state is hampered by quantum and thermal fluctuations. Even in 3D the nature of the FFLO state is quite peculiar, with the FF state having vanishing superfluid stiffness in transverse (to the FFLO momentum) direction and in the LO state the superfluid density being highly anisotropic~\cite{radzihovsky_imbalanced_2010}. The small parameter regime where the FFLO state is predicted in continuum systems, together with vulnerability to phase separation and fluctuations, motivates the search for systems and conditions that could improve the stability of this elusive state of matter.

The focus of this review is on the FFLO state in 2D and 3D optical lattices.
We mention some works of FFLO in 1D lattices but do not provide a comprehensive review of the 1D case, for that see~\cite{Feiguin2012,guan_fermi_2013,yin_quench_2016}. Lattice systems improve the stability of the FFLO state due to the shape of Fermi surfaces that can increase the available phase space for pair formation for finite-momentum pairs. 
This effect is called Fermi surface nesting, and is closely connected with existence of van Hove singularities that provide high density of states. The nesting effects are further enhanced in lattices with a lower dimension. 
Low-dimensional optical lattices are thus promising for experimental realization of the FFLO state. Two-dimensional (2D) lattices are  
particularly attractive with the advent of quantum gas microscopes for fermionic atoms~\cite{edge_imaging_2015,omran_microscopic_2015,cheuk_quantum-gas_2015,haller_single-atom_2015,parsons_site-resolved_2015,Mitra2017}, which allow observation of individual atoms and probing of non-local correlations.
Mott~\cite{Jordens2008,Schneider2008} and short-range antiferromagnetic~\cite{Greif2013,Boll1257,Cheuk1260,Brown2017}
%HERE
 correlations have been observed for repulsively interacting fermions in optical lattices, and recently the use of quantum gas microscopes has enabled observing antiferromagnetic correlations ranging over the whole sample~\cite{mazurenko_cold-atom_2017}. These are closely connected with the FFLO state with attractive interactions through particle-hole mapping: the striped phase in the doped repulsive Hubbard model corresponds to the FFLO state in the attractive Hubbard model. This striped phase may compete with d-wave superfluidity, which would be important for understanding the origin of high temperature superconductivity. Quantum gas microscopes provide a promising platform for studying these phenomena and for the actual observation of the FFLO state.

An interesting new candidate for realizing FFLO superfluids is spin-orbit coupled atomic gases.
UQG systems with spin-orbit coupling have recently been created for bosons~\cite{lin_spin-orbit-coupled_2011,dalibard_textitcolloquium_2011,galitski_spin-orbit_2013} and for fermions~\cite{wang_spin-orbit_2012-1,cheuk_spin-injection_2012}. These systems are realized by utilizing Raman lasers that couple the spin- and spatial degrees of freedom of the atoms, effectively providing a ${\bf L \cdot S}$-type spin-orbit coupling (SOC). This is a neutral atom analogy of the spin-orbit coupling in solid state physics or of the electrons orbiting the nucleus within the atom. In SOC systems, the spatial degrees of freedom become mixed with the spin degrees of freedom, making the concept of magnetization less well defined. Like optical lattices, these systems also exhibit large parameter regions with FFLO-type superfluids, but here finite-momentum pairing can already be realized with spin-balanced systems.
The SOC introduces preferred momenta in the dispersion, moreover, some forms of SOC may break the rotational symmetry already at the Hamiltonian level. Furthermore, unlike in non-spin-orbit coupled systems, even the single-plane wave FF state has been shown to be realizable in the ground state. In sum, the FFLO state in SOC systems is expected to be quite different from its conventional counterpart in continuum and lattices.

\subsection{Structure of this review}

We start this review by providing in section~\ref{BasicFFLOTheoryLattice} a detailed pedagogic description of the mean-field theory applied to spin-imbalanced gases leading to FFLO states and other spin-imbalanced superfluid states. 
We describe the use of thermodynamic potentials and Ginzbug-Landau theory for determining the stability of the various phases and nature of the phase transitions. 
Important theorems, such as the Bloch's theorem for the absence of current in the ground state and the Luttinger's theorem connecting Fermi surfaces with particle densities, are discussed. The mapping to the repulsive Hubbard model and the corresponding connection of the FFLO state to the striped phase and d-wave superconductors is explained. 

We continue with mean-field studies in section~\ref{MFstudies} by providing a review of theoretical mean-field works on the FFLO state in optical lattices and explaining the important role played by Fermi surface nesting and van Hove singularities. Also works on mass imbalanced gases in optical lattices are briefly reviewed in section~\ref{section_mass_imbalance}.

As a step toward a quantum field theoretical treatment, section~\ref{GreenFFLO} describes the FFLO physics using Green's function formalism. 
The formalism is applied in section~\ref{CollectiveSection} for studying collective modes. 
In section~\ref{Beyond-mean-fieldSection} we review the studies done with more advanced numerical methods, namely dynamical mean-field theory (DMFT) and quantum Monte Carlo (QMC) studies. 
These studies are important in determining the actual stability of the FFLO state against fluctuations neglected in mean-field studies.
Especially the role of dimensional crossover is considered, as low dimensional systems provide larger parameter ranges for the FFLO state while higher dimensions give the necessary stability for the long-range order. Despite the focus on lattice systems, we present also a review on trapped continuum systems in section~\ref{section_traps}. We concentrate on advances missing from the previous review~\cite{gubbels_imbalanced_2013}, particularly new studies done in non-spherical traps.

We give an analysis of the literature on the FFLO state in spin-orbit coupled systems in section~\ref{FFLOSOC}. In section~\ref{section_detection}, we provide a comprehensive review of the various detection schemes that have been proposed for observing the FFLO state in continuum systems, optical lattices, various dimensionalities, and spin-orbit coupled systems. 
This is a stand-alone part of the review and can be accessed without first reading the rest of the review. Finally, in section~\ref{conclusions} we conclude with an overview of the salient points of the latest research on FFLO physics in UQG, and of the key open questions.

\section{Mean-field theory of the FFLO state in a lattice} \label{BasicFFLOTheoryLattice}

Here, we present the basic mean-field theory description of the FFLO state using the approach of BCS theory and the Bogoliubov transformation. We start from a generic Hamiltonian for two component fermions, which describes an ultracold Fermi gas well. A lattice potential is included and continuous position variables are used to begin with, discrete spatial variables are introduced later. The reader can easily convert the following lattice description back to free space, if desired. We perform the BCS mean-field approximation, including the possibility of a population imbalance between the two spin components and exotic pairing, such as the FFLO and or breached pair (BP) states. The Hamiltonian is expressed in the (lattice) momentum basis and a Bogoliubov transformation is applied to diagonalize the Hamiltonian. The grand potential and the Helmholtz free energy can then be determined in terms of the eigenenergies. We discuss those quasiparticle eigenenergies in detail since they provide intuitive understanding about the various exotic superfluids. Finally, the fundamental relations between the FFLO state and Bloch's and Luttinger's theorems are described.

\subsection{The Hamiltonian for a two-component Fermi gas in a lattice}
We present here the Hamiltonian for a two-component Fermi system, including a lattice potential, and show how to obtain the lattice Hamiltonian from the continuous one. For a more detailed derivation and information on the approximations used, see for instance \cite{jaksch_cold_2005,bloch_many-body_2008,giorgini_theory_2008,stoof_ultracold_2008,lewenstein_ultracold_2012}.
We consider a system of two types of fermions, which we label spin up ($\uparrow$) and spin down ($\downarrow$). 
The main focus are systems of ultracold gases, where this labeling is a pseudospin, which means atoms in two different internal atomic states, such as hyperfine states, or different atomic species or isotopes. The interaction between the two components of the gas is described by the potential $V_{\uparrow\downarrow}\left(\mathbf{r},\mathbf{r}'\right)$, and the components experience an external lattice potential $V_{\text{ext}, \sigma}(\mathbf{r})$, which could include also trapping potential. 
We consider three dimensional systems, but transferring these results to other dimensions is straightforward.
 The system is described by the Hamiltonian
\begin{eqnarray}
\nonumber\widehat{H}= \int d\mathbf{r}\sum_{\sigma=\uparrow,\downarrow}\hat{\psi}_{\sigma}^{\dagger}\left(\mathbf{r}\right)\left({\hat K_\sigma}({\mathbf r})-\mu^0_{\sigma}\right)\hat{\psi}_{\sigma}\left(\mathbf{r}\right)\\
+\int \int d\mathbf{r} d\mathbf{r'} V_{\uparrow\downarrow}\left(\bf{r,r'}\right)\hat{\psi}_{\uparrow}^{\dagger}\left(\mathbf{r}\right)
\hat{\psi}_{\downarrow}^{\dagger}\left(\mathbf{r'}\right)
\hat{\psi}_{\downarrow}\left(\mathbf{r'}\right)\hat{\psi}_{\uparrow}\left(\mathbf{r}\right),\label{eq:a}
\end{eqnarray}
where ${\hat K_\sigma}({\mathbf r}) = -\frac{\hbar^2 \nabla^2}{2m_\sigma} + V_{\mathrm{ext},\sigma}({\bf r})$, $m_\sigma$ is the mass and $\mu^0_{\sigma}$ the chemical potential for a particle in state $\sigma$. 
A lattice potential for ultracold atoms is created by interfering laser beams, where the periodicity of the lattice is half the laser wavelength and the lattice depth depends on the intensity of the laser \cite{metcalf_laser_2001,bloch_many-body_2008}. 
Nowadays, a lattice can also be created using holographic masks~\cite{Bakr2009} and
such a lattice was used for a single site resolution of individual atoms. For other 
experiments on quantum gas microscopes, see references~\cite{sherson_single-atom-resolved_2010,miranda_site-resolved2015,cheuk_quantum-gas_2015,omran_microscopic_2015,Mitra2017}.
For ultracold dilute gases, where the interactions are short range and temperatures low, the interaction potential can be approximated by a contact interaction, see for instance references~\cite{bloch_many-body_2008,stoof_ultracold_2008,zwerger_bcs-bec_2012,lewenstein_ultracold_2012},
\begin{equation}
V_{\uparrow\downarrow}\left(\mathbf{r},\mathbf{r}'\right)=V_{0}\delta\left(\mathbf{r}-\mathbf{r}'\right),
\label{eq:a-2}
\end{equation}
where $V_0<0$ for attractive interactions. The Hamiltonian then reads
\begin{eqnarray}
\widehat{H}=& \int d\mathbf{r}\sum_{\sigma=\uparrow,\downarrow}\hat{\psi}_{\sigma}^{\dagger}\left(\mathbf{r}\right)
\left({\hat K_\sigma}({\mathbf r})-\mu^0_\sigma\right)\hat{\psi}_{\sigma}\left(\mathbf{r}\right) \nonumber \\
&+ V_{0}\int d\mathbf{r}\,\hat{\psi}_{\uparrow}^{\dagger}\left(\mathbf{r}\right)\hat{\psi}_{\downarrow}^{\dagger}\left(\mathbf{r}\right)\hat{\psi}_{\downarrow}\left(\mathbf{r}\right)\hat{\psi}_{\uparrow}\left(\mathbf{r}\right).\label{eq:a-3}
\end{eqnarray}
In the presence of a lattice potential, it is convenient to expand the Fermi operators in the continuum as
\begin{eqnarray}
\hat{\psi}_\sigma^\dagger({\bf r})=\sum_{\bf n,i}{\hat c}^\dagger_{\sigma{\bf n,i}}w^*_{\bf n}({\bf r-r_i}),
\end{eqnarray}
where ${\hat c}^\dagger_{\sigma{\bf n,i}}$ creates a $\sigma$ particle in the lattice at site ${\bf i}$ in an energy band ${\bf n}$ and where $w^*_{\bf n}({\bf r-r_i})$ are the Wannier functions \cite{kohn_analytic_1959}. In the tight-binding limit, the Wannier functions can be approximated by harmonic oscillator states on each site. At low temperatures and for sufficiently small interaction energies, which are criteria met by ultracold atoms in optical lattices, particles only occupy the lowest band with ${\bf n}=0$ and consequently the band index can be omitted. Note, however, that if one wishes to consider very strong interactions created by Feshbach resonances, a proper description involves higher bands~\cite{buechler_FeshbachHubbard_2010}. Using the above expansion, the continuum Hamiltonian can be rewritten into the standard Hubbard Hamiltonian
\begin{eqnarray}
\widehat{H}=-&\sum_{\sigma}
\sum_{\langle{\bf i,j}\rangle}
t_{\sigma{\bf ij}}{\hat c}_{\sigma{\bf i}}^\dagger{\hat c}_{\sigma{\bf j}}
+\sum_{\sigma,{\bf i}}\left(\varepsilon_\sigma-\mu^0_\sigma\right){\hat c}_{\sigma{\bf i}}^\dagger{\hat c}_{\sigma{\bf i}} \nonumber \\
&+U\sum_{{\bf i}}{\hat c}_{\uparrow{\bf i}}^\dagger{\hat c}_{\downarrow{\bf i}}^\dagger {\hat c}_{\downarrow{\bf i}}{\hat c}_{\uparrow{\bf i}},
\label{eq:FFLO_Hamiltonian_position_space1}
\end{eqnarray}
where the summation indices $\langle{\bf i,j}\rangle$ are over nearest neighbour indices and nearest neighbour tunneling (hopping) energy $t_{\sigma{\bf ij}}$, the on-site energy $\varepsilon_\sigma$, and the on-site interaction energy $U$ are~\cite{jaksch_cold_1998}
\begin{eqnarray}
t_{\sigma{\bf ij}}&=-\int d{\bf r}\,w^*({\bf r-r_i}){\hat K_\sigma}({\mathbf r}) w({\bf r-r_j})\\
\varepsilon_\sigma&=\int d{\bf r}\,w^*({\bf r-r_i}){\hat K_\sigma}({\mathbf r})w({\bf r-r_i})\\
U&=V_0\int d{\bf r}\,|w({\bf r-r_i})|^4. \label{eq:Udefinition}
\end{eqnarray}
In the equations~(\ref{eq:FFLO_Hamiltonian_position_space1})--(\ref{eq:Udefinition}), we have allowed for spin dependent and anisotropic hopping $t_{\sigma{\bf ij}}$. Anisotropic lattices can be created by using different laser intensities and wavelengths for different spatial directions. A spin dependence can result both from different masses for the two spin components or from a spin-dependent lattice. In both cases,  the onsite energies are also spin dependent $\varepsilon_\uparrow\neq\varepsilon_\downarrow$.  A spin dependent lattice can be created by applying a magnetic field gradient modulated in time \cite{jotzu_creating_2015}.

In the rest of this section, we assume the masses of the two spin components to be the same and the lattice potential to be spin independent, so that $t_{\uparrow{\bf ij}}=t_{\downarrow{\bf ij}}\equiv t_{{\bf ij}}$ and $\varepsilon_{\uparrow}=\varepsilon_{\downarrow}\equiv \varepsilon$. The latter is in this case a constant energy shift and can be taken equal to zero.
The Hamiltonian now reads
\begin{eqnarray}
\widehat{H}=-\sum_{\sigma}
\sum_{\langle{\bf r,r'}\rangle}
t_{\bf ij}{\hat c}_{\sigma{\bf r}}^\dagger{\hat c}_{\sigma{\bf r'}}
-\sum_{\sigma{\bf r}}\mu^0_\sigma{\hat c}_{\sigma{\bf r}}^\dagger{\hat c}_{\sigma{\bf r}}\nonumber \\
+U\sum_{{\bf r}}{\hat c}_{\uparrow{\bf r}}^\dagger{\hat c}_{\downarrow{\bf r}}^\dagger {\hat c}_{\downarrow{\bf r}}{\hat c}_{\uparrow{\bf r}},
\label{eq:FFLO_Hamiltonian_position_space2}
\end{eqnarray}
where we use the same notation as in the continuum for the position ${\bf r}$, which is here a discrete variable corresponding to the lattice sites, and here $U<0$ for attractive interactions.

\subsection{The BCS mean-field approximation}
To study the transition from a normal state to a superfluid consisting of paired fermions, the so-called Cooper pairs, we introduce  a field describing the pairs. This is done using the BCS mean-field approximation, which also introduces the Hartree fields and replaces, through the use of  Wick's theorem, some of the operators in the four-operator interaction term by complex numbers. More details on BCS theory in general and in cold gases can be found in numerous places, for instance references~\cite{mahan_many-particle_2000,fetter_quantum_2003,zwerger_bcs-bec_2012,lewenstein_ultracold_2012}.
 
The {\bf order parameter} of the BCS theory is the pairing field $\hat{\Delta}({\bf r})$, which is on average related to the expectation value of 
$\hat{\psi}_\downarrow\hat{\psi}_\uparrow$
\begin{equation}
\langle\hat{\Delta}\left(\mathbf{r}\right)\rangle\equiv\Delta({\bf r})= V_0 \left\langle \hat{\psi}_{\downarrow}\left(\mathbf{r}\right)\hat{\psi}_{\uparrow}\left(\mathbf{r}\right)\right\rangle,
\label{eq:a-10}
\end{equation}
and describes the Cooper pairs. As we see in section \ref{subsec:bogoliubov_transformation}, the pairing field is related to the energy gap in the excitation spectrum of the BCS theory. The usual BCS theory describes transitions from a normal gas of fermionic particles to a homogeneous superfluid consisting of Cooper pairs, which is described by a mean-field ansatz independent of position $\Delta({\bf r})=\Delta$.
However to describe exotic superfluidity, the spatial dependence is a key feature.

In the lattice model, the pairing field is
\begin{equation}
\Delta\left(\mathbf{r}\right) = U\langle {\hat c}_{\downarrow{\bf r}} {\hat c}_{\uparrow{\bf r}}\rangle.
\label{eq:order_parameter}
\end{equation}
We include the Hartree fields by redefining the chemical potentials. Denoting $\left\langle {\hat c}_{\sigma}^{\dagger}{\hat c}_{\sigma}\right\rangle=n_{\sigma}$, we get
\begin{eqnarray}
(-\mu^0_\sigma+Un_{-\sigma}){\hat c}_{\sigma}^{\dagger}{\hat c}_{\sigma}\equiv-\mu_\sigma{\hat c}_{\sigma}^{\dagger}{\hat c}_{\sigma},
\label{eq:a-13}
\end{eqnarray}
where it is assumed that the densities are uniform. In the mean-field approximation, the pairing field is approximated by its average value, $\hat{\Delta}({\bf r})=\Delta({\bf r})$, but we do not yet assume any specific form for this average. The Hamiltonian now reads
\begin{eqnarray}
  \label{eq:a-3-1} 
\widehat{H} = -\sum_{\sigma}\sum_{\langle{\bf r,r'}\rangle}t_{\bf ij}{\hat c}_{\sigma{\bf r}}^\dagger {\hat c}_{\sigma{\bf r'}} 
-\sum_\sigma\sum_{\bf r} \mu_\sigma{\hat c}_{\sigma{\bf r}}^\dagger {\hat c}_{\sigma{\bf r}}\\
+\sum_{{\bf r}}\left({\hat c}_{\uparrow{\bf r}}^\dagger{\hat c}_{\downarrow{\bf r}}^\dagger\Delta ({\bf r}) + \Delta^*({\bf r}){\hat c}_{\downarrow{\bf r}}{\hat c}_{\uparrow{\bf r}}
- \frac{\Delta^*({\bf r})\Delta({\bf r})}{U}\right), \nonumber
\end{eqnarray}
where we also included the Hartree shifts. Note that the assumption of uniform densities is not always strictly correct because a spatial modulation of the order parameter amplitude $|\Delta({\bf r})|$ is typically accompanied by a modulation of density in the FFLO state. To take this into account one should keep the spatial dependence of the Hartree terms and solve the problem self-consistently for instance by a Bogoliubov-de Gennes (BdG) approach. In practice this is quite difficult, and fortunately (for the theorist, not for the experimentalist), the modulations of the density are quite small so one can justify the uniform density approximation for the Hartree shifts in the Hamiltonian (\ref{eq:a-3-1}). One should keep in mind, however, that if such Hamiltonian then leads to modulated density of the ground state, the treatment is not fully self-consistent. 

\subsection{The FF and LO mean-field approximation}
\label{sec:FFLO_meanfield}
The possibility of pairing with unequal chemical potentials and with a spatially dependent order parameter is included in the general pairing order parameter introduced in equation~(\ref{eq:order_parameter}). 
In the FF state, $\uparrow$ and $\downarrow$ particles in different momentum states form a Cooper pair, such that the Cooper pair carries a net momentum ${\bf q}$. This is in contrast to the BCS superfluid, where the Cooper pairs are formed between particles in momentum states ${\bf k}$ and ${\bf -k}$. When describing the FF superfluid state, the Cooper pair momentum ${\bf q}$ is  a parameter determined by minimizing the system's energy, and its absolute value turns out to scale with the population imbalance between the two spin components $|\mathbf{q}|\propto|k_{F\uparrow} -k_{F\downarrow}|$ \footnote{Note that the relation $|\mathbf{q}|\propto|k_{F\uparrow} -k_{F\downarrow}|$ is given by mean-field theory, and some beyond mean-field studies on finite lattices have found FFLO wavevectors deviating from the mean-field results \cite{moreo_cold_2007}.}. The FF order parameter is
\begin{equation}
\Delta({\bf r}) = \Delta e^{i{\bf q}\cdot{\bf r}},
\label{eq:order_parameter_FF}
\end{equation}
where $\Delta\in\mathbb{R}$ is a real number. This plane wave form of the order parameter was first considered by Fulde and
Ferrell~\cite{FF}. Larkin and Ovchinnikov~\cite{LOeng} around the same time considered an ansatz containing both ${\bf q}$ and $-{\bf q}$, so that the order parameter is of the form
\begin{equation}
\Delta({\bf r})= \Delta \cos ({\bf q}\cdot{\bf r}).
\label{eq:order_parameter_LO}
\end{equation}
In the FF phase, the order parameter is complex as equation~(\ref{eq:order_parameter_FF}) shows, and time-reversal symmetry is broken. The term FF phase (or FF state) can be used to denote also time-reversal symmetry broken states with several pairing wave vectors ${\bf q}$. Likewise, the essential feature of the LO state is the preservation of the time-reversal symmetry; two opposite FFLO wave vectors ${\bf q}$ and $-{\bf q}$ is the minimal setting for this, but there could be more. 

According to many studies, both for homogeneous and for lattice systems, the cosine wave LO ansatz gives a lower energy than the FF plane wave, see for instance references~\cite{mora_transition_2005,yoshida_larkin-ovchinnikov_2007,batrouni_exact_2008,baarsma_inhomogeneous_2013,baarsma_larkin-ovchinnikov_2016}. 
Fortunately in many different systems, the use of FF and LO ansatzes predict the same parameter regime for the existence of the density-imbalanced superfluid. This means that one can use the simpler FF ansatz  to explore different systems and parameter regimes, keeping in mind that the true ground state could be the LO state. However even when the parameter regimes of the FF and LO state overlap, the order of the phase transitions predicted by using an FF or LO ansatz can be different, for instance compare references~\cite{koponen_finite-temperature_2007,koponen_fflo_2008} and reference~\cite{baarsma_larkin-ovchinnikov_2016}. 

Since the LO state is in general more stable than the FF state, loosely speaking we can also say that the FFLO state is described by the FF ansatz in equation~(\ref{eq:order_parameter_FF}). However, there are also systems where the above is simply not true, for example for Fermi gases with spin orbit coupling. In these systems, it is actually found that the FF state is energetically more favorable than the LO state, see section \ref{FFLOSOC}.
Throughout this review as well as often in the literature, the term FFLO superfluid is used even when only the FF ansatz is actually applied.
In general, the FFLO order parameter can be composed of any number of Fourier components with different ${\bf q}$s, leading to arbitrarily complicated spatial dependence $\Delta({\bf r})$. This is discussed further in section~\ref{section_traps} where the Bogoliubov-deGennes formalism is introduced. 

Let us now proceed here with the FF ansatz (\ref{eq:order_parameter_FF}), so that the Hamiltonian becomes
\begin{eqnarray}
  \label{eq:FFLO_Hamiltonian_position_space}
  \hat{H} =& -\sum_{\sigma}\sum_{\langle{\bf r,r'}\rangle}t_{\bf ij}{\hat c}_{\sigma{\bf r}}^\dagger {\hat c}_{\sigma{\bf r'}} 
-\sum_{\sigma}\sum_{\bf r}\mu_\sigma{\hat c}_{\sigma{\bf r}}^\dagger {\hat c}_{\sigma{\bf r}} \\
&+ \sum_{{\bf r}}\left(\Delta e^{i{\bf q}\cdot{\bf r}}
{\hat c}_{\uparrow{\bf r}}^\dagger{\hat c}_{\downarrow{\bf r}}^\dagger + \Delta
e^{-i{\bf q}\cdot{\bf r}} {\hat c}_{\downarrow{\bf r}}{\hat c}_{\uparrow{\bf r}}
- \frac{\Delta^2}{U}\right).  \nonumber
\end{eqnarray}
It is convenient to write the Hamiltonian in the (lattice-)momentum representation, that is, represent the operators in the plane wave basis. This means Fourier transforming equation~(\ref{eq:FFLO_Hamiltonian_position_space}) by using
\begin{eqnarray}
\label{eq:fourier}
{\hat c}_{\sigma{\bf r}} &=\frac{1}{\sqrt{M}}\sum_{\bf k}e^{i{\bf k}\cdot{\bf r}}{\hat c}_{\sigma{\bf k}}
\end{eqnarray}
where $M$ is the number of lattice sites and the momentum ${\bf k}$ runs through the reciprocal lattice. The density terms transform as
\begin{eqnarray}
\sum_{{\bf r}}{\hat c}_{\sigma{\bf r}}^\dagger{\hat c}_{\sigma{\bf r}} &=
   \frac{1}{M}\sum_{{\bf r}}\sum_{\bf k,k'}
  e^{i({\bf k}'-{\bf k})\cdot{\bf r}}{\hat c}_{\sigma{\bf k}}^\dagger
  {\hat c}_{\sigma{\bf k}'} \\
  &=  \sum_{\bf k}
  {\hat c}_{\sigma{\bf k}}^\dagger {\hat c}_{\sigma{\bf k}},
  \nonumber
\end{eqnarray}
since $\frac{1}{M} \sum_{{\bf r}} e^{i({\bf k}'-{\bf k})\cdot{\bf r}} = \delta_{{\bf k},{\bf k}'}$ for the reciprocal lattice vectors ${\bf k},{\bf k}'$. Similarly, the pairing terms become 
\begin{eqnarray}
  \label{FFpairingfourier}
  \sum_{{\bf r}}&\left(\Delta e^{i{\bf q\cdot r}}{\hat c}_{\uparrow{\bf r}}^\dagger{\hat c}_{\downarrow{\bf r}}^\dagger+\Delta e^{-i{\bf q\cdot r}}
{\hat c}_{\downarrow{\bf r}}{\hat c}_{\uparrow{\bf r}}\right) = \\
&\sum_{\bf k}\left(\Delta {\hat c}_{\uparrow{\bf k}}^\dagger {\hat c}_{\downarrow{\bf q-k}}^\dagger+\Delta {\hat c}_{\downarrow{\bf q-k}}{\hat c}_{\uparrow{\bf k}}\right).
\nonumber
\end{eqnarray}
We now assume a cubic lattice, where the nearest neighbor hopping term gives rise to a cosine dispersion, for example, the Fourier transform for the $x$-direction is 
\begin{eqnarray}
\nonumber \sum_\sigma&\sum_{\langle{\bf r,r'}\rangle_{\hat{x}}}{\hat c}_{\sigma{\bf r}}^\dagger {\hat c}_{\sigma{\bf r}}
= \sum_\sigma\sum_{\bf r}\left({\hat c}_{\sigma{\bf r}+d_x}^\dagger
+ {\hat c}_{\sigma{\bf r}-d_x}^\dagger\right){\hat c}_{\sigma{\bf r}}\\
\nonumber &=  \frac{1}{M}\sum_\sigma\sum_{\bf r,k,k'}
e^{-i(({\bf k}-{\bf k}')\cdot{\bf r})}\underbrace{\left(e^{ik_x d_x}+e^{-ik_x d_x}
\right)}_{=2\cos (k_x d_x)} {\hat c}_{\sigma{\bf k}}^\dagger
{\hat c}_{\sigma{\bf k}'} \\
&=\sum_\sigma\sum_{\bf k}2\cos(k_x d_x){\hat c}_{\sigma{\bf k}}^\dagger{\hat c}_{\sigma{\bf k}},
\end{eqnarray}
where $d_x$ is the lattice spacing in the $\hat{x}$-direction.

Finally, we arrive at the mean-field FFLO Hubbard Hamiltonian in momentum space,
\begin{eqnarray}
\label{eq:hamiltonian}
\widehat{H} = \sum_{\bf k} &\left(\xi_{\uparrow{\bf k}}{\hat c}_{\uparrow{\bf k}}^\dagger {\hat c}_{\uparrow{\bf k}} + \xi_{\downarrow{\bf k}}{\hat c}_{\downarrow{\bf k}}^\dagger
{\hat c}_{\downarrow{\bf k}}\right.\\
&\left.+\Delta {\hat c}_{\uparrow{\bf k}}^\dagger{\hat c}_{\downarrow{\bf q-k}}^\dagger
 +\Delta{\hat c}_{\downarrow{\bf q-k}}{\hat c}_{\uparrow{\bf k}} -  \frac{\Delta^2}{U}\right),\nonumber
\end{eqnarray}
where the single particle energies are given by
\begin{equation}
\xi_{\sigma {\bf k}} = \epsilon_{\bf k} - \mu_\sigma = \sum_{\alpha\in\{x,y,z\}} 2t_{\alpha} \big[1 -\cos(k_\alpha d_\alpha)\big] - \mu_\sigma,
\label{particledispersions}
\end{equation} 
with $t_\alpha$ the tunneling coefficient for hopping in the direction $\alpha$.
In the case that all particles in the system occupy the lowest energy levels $\xi_{\sigma{\bf k}}$, the highest occupied energy state is the Fermi energy $E_\text{F}$, and the corresponding momentum state is the Fermi momentum $k_\text{F}$. At zero temperature, the chemical potential is equal to the Fermi energy for a non-interacting system.
To get the dispersions to correspond to those of free particles in the limit of small $k$, the following terms have been added to the Hamiltonian:
\begin{equation}
  2\sum_{\sigma,\alpha,{\bf k}}\left(t_{\alpha}{\hat c}_{\sigma {\bf k}}^\dagger
  {\hat c}_{\sigma{\bf k}} \right) =2 \sum_{\sigma,\alpha}t_{\alpha}N_\sigma.
\end{equation}

\subsection{Bogoliubov transformation in the FFLO case}   
\label{subsec:bogoliubov_transformation}
In this subsection, we diagonalize the Hamiltonian in equation~(\ref{eq:hamiltonian}) by a Bogoliubov transformation to be able to determine the number and gap equations, and to calculate the thermodynamic potential. By performing this transformation, from the particle basis to the quasiparticle basis, we find the quasiparticle energies.
Let us start by rewriting the Hamiltonian equation~(\ref{eq:hamiltonian}) using matrix formalism 
\begin{eqnarray}
  \nonumber
\hat{H}=\sum_{\bf k} &\left(\begin{array}{cc}{\hat c}_{\uparrow{\bf k}}^\dagger, & {\hat c}_{\downarrow{\bf q-k}}\end{array}\right)
\left(\begin{array}{cc} \xi_{\uparrow{\bf k}}& \Delta\\\Delta & -\xi_{\downarrow{\bf q-k}} \end{array} \right)
\left(\begin{array}{cc}{\hat c}_{\uparrow{\bf k}}\\{\hat c}_{\downarrow{\bf q-k}}^\dagger\end{array}\right) \\
  &+\sum_{\bf k} \left(\xi_{\downarrow{\bf q-k}}-\frac{\Delta^2}{U}\right),
\label{eq:matrix}
\end{eqnarray}
where we have used that ${\hat c}_{\downarrow{\bf k}}^\dagger{\hat c}_{\downarrow{\bf k}}=1-{\hat c}_{\downarrow{\bf k}}{\hat c}^\dagger_{\downarrow{\bf k}}$.
The essential point in this representation is that the Hamiltonian can be expressed as a sum of independent $2\times2$ matrices for each lattice momentum ${\bf k}$ that can be diagonalized separately. In other words, if the above summation over momentum ${\bf k}$ is written explicitly, the Hamiltonian can be expressed with a (large) block-diagonal matrix, which allows for rewriting the Hamiltonian as a sum of $2\times2$ matrices.

This simplicity results from the plane wave ansatz used for the FF state. If a Cooper pair ansatz is considered including more plane waves, it is no longer possible to rewrite the Hamiltonian using $2\times2$ matrices. This is already the case for the LO ansatz in equation~(\ref{eq:order_parameter_LO}), containing two plane waves. Namely, the pairing part of the Hamiltonian for the LO Cooper pairs reads
\begin{eqnarray}
\sum_{\bf k}\bigg\{{\hat c}_{\uparrow{\bf k}}^\dagger{\hat c}_{\downarrow{\bf q-k}}^\dagger\Delta &+{\hat c}_{\uparrow{\bf k}}^\dagger{\hat c}_{\downarrow{\bf -q-k}}^\dagger\Delta\nonumber\\
&+\Delta{\hat c}_{\downarrow{\bf q-k}}{\hat c}_{\uparrow{\bf k}}+\Delta{\hat c}_{\downarrow{\bf -q-k}}{\hat c}_{\uparrow{\bf k}}\bigg\},
\label{LOcoupling}
\end{eqnarray} 
which contains more terms than the FF pairing Hamiltonian equation~(\ref{FFpairingfourier}), since in the LO case an $\uparrow$ particle in momentum state ${\bf k}$ couples to both a $\downarrow$ fermion with momentum ${\bf q-k}$ as well as to one with ${\bf -q-k}$. For rewriting the complete LO Hamiltonian using matrix multiplication, it means that the resulting matrix is not block-diagonal and cannot be split into smaller independent matrices. Instead, even with the summation over momentum states ${\bf k}$, the LO matrix is limited only by the number of lattice sites and therefore diverges for infinite lattices. We truncate this matrix here to show a small part
\begin{eqnarray}
\label{eq:LOmatrix}
\hat{H}_\text{LO}=&\frac23\sum_{\bf k} \left( \begin{array}{ccc}{\hat c}_{\uparrow{\bf k}}^\dagger, & {\hat c}_{\downarrow{\bf q-k}}, & {\hat c}_{\downarrow{\bf -q-k}} \end{array}\right) \times \\
&\times \left(\begin{array}{ccc} 
\xi_{\uparrow{\bf k}}	& \Delta					& \Delta				\\
\Delta 			& -\xi_{\downarrow{\bf q-k}}	&0					\\
\Delta			&0						&-\xi_{\downarrow{\bf-q-k}}\\
\end{array} \right) \left(\begin{array}{c}{\hat c}_{\uparrow{\bf k}}\\{\hat c}_{\downarrow{\bf q-k}}^\dagger\\{\hat c}_{\downarrow{\bf -q-k}}^\dagger\end{array}\right)\nonumber\\ 
  &+\sum_{\bf k} \bigg\{\xi_{\downarrow{\bf q-k}}+\xi_{\downarrow{\bf -q-k}}-\frac{\Delta^2}{U} \bigg\}, \nonumber 
\end{eqnarray}
where the factor 2/3 is added to compensate for the fact that two spin components are here described by a $3\times3$ matrix. In the above matrix (\ref{eq:LOmatrix}), all couplings from the ${\hat c}_{\uparrow{\bf k}}$ field to the $\downarrow$ fields are included, but not all couplings are in turn included for the $\downarrow$ fermions. For instance according to equation~(\ref{LOcoupling}), ${\hat c}_{\downarrow{\bf q-k}}$ couples to both ${\hat c}^\dagger_{\uparrow{\bf k}}$ and ${\hat c}^\dagger_{\uparrow{\bf k-2q}}$. The latter coupling is not included in the above matrix. 

To determine thermodynamic quantities for the LO state, one needs to diagonalize the Hamiltonian using a Bogoliubov transformation. This is done in the same way as explained in the following for the FF state, except that it can only be done numerically. In practice, thermodynamic quantities are calculated for a certain matrix dimension, and consequently the matrix is enlarged by including more coupling terms until convergence is reached. When describing a continuous, second order phase transition from the normal state to the superfluid phase the order parameter $\Delta$ is small, and a $6\times6$ or $10\times10$ matrix could be large enough (for LO phase breaking translational symmetry in one out direction only). However when describing the superfluid phase or a first order phase transition, much bigger matrix sizes are needed. For more details, see references~\cite{baarsma_inhomogeneous_2013,baarsma_larkin-ovchinnikov_2016}.

Let us now turn back to the FF Hamiltonian and apply a Bogoliubov transformation for the matrices in \eqref{eq:matrix}, by introducing 
the new basis operators 
$\hat{\gamma}_+$ and $\hat{\gamma}_-$
\begin{eqnarray}
\left(\begin{array}{c}
\hat{\gamma}_{+,{\bf k,q}}\\
\hat{\gamma}_{-,{\bf k,q}}^\dagger
\end{array}\right)
= \left(\begin{array}{cc} u_{\bf k,q}&-v_{\bf k,q}\\v_{\bf k,q}&u_{\bf k,q} \end{array} \right)\left(\begin{array}{c} {\hat c}_{\uparrow{\bf k}} \\ {\hat c}_{\downarrow{\bf q-k}}^\dagger \end{array}\right)
\label{Bogoliubov1}
\end{eqnarray}
and we require $u_{\bf k,q}^2+v_{\bf k,q}^2=1$ to ensure the fermionic anticommutation relations for the $\hat{\gamma}$ operators. The inverse of the above transformation is
\begin{eqnarray}
\left(\begin{array}{c}
{\hat c}_{\uparrow{\bf k}}\\
{\hat c}_{\downarrow{\bf q-k}}^\dagger
\end{array}\right)
= \left(\begin{array}{cc} u_{\bf k,q}&v_{\bf k,q}\\-v_{\bf k,q}&u_{\bf k,q} \end{array}\right)\left(\begin{array}{c} \hat{\gamma}_{+,{\bf k,q}} \\ \hat{\gamma}_{-,{\bf k,q}}^\dagger \end{array}\right).
\label{inverseBogoliubov}
\end{eqnarray}
By inserting this transformation in equation~(\ref{eq:matrix}) and demanding that the resulting Hamiltonian is diagonal, we arrive at the quasiparticle energies
 \begin{eqnarray}
  E_{\pm,{\bf k,q}} &= \sqrt{\left(\frac{\xi_{\uparrow{\bf k}}+\xi_{\downarrow{\bf q-k}}}{2}\right)^2
    + \Delta^2} \pm \frac{\xi_{\uparrow{\bf k}}-\xi_{\downarrow{\bf q-k}}}{2} \nonumber\\
  &\equiv E_{\bf k,q} \pm \frac{\delta\xi_{\bf k,q}}{2}.
  \label{eq:qp_energies}
\end{eqnarray}
For comparison, the quasiparticle energies for the standard BCS theory with ${\bf q}=0$ read
\begin{eqnarray}
\label{eq:qp_energies_PT}
E_{\pm,{\bf k}}=\sqrt{\left(\epsilon_{\bf k} - \frac{\mu_\uparrow + \mu_\downarrow}{2}\right)^2+ \Delta^2} \pm \frac{\mu_\downarrow - \mu_\uparrow}{2}.
\end{eqnarray}
When the chemical potentials are equal for the two spin components, $\mu_\uparrow = \mu_\downarrow$, then $E_+=E_-$ and thier minimum value is $\Delta$. This shows that the quasiparticle excitations have a minimum energy, that is, the energy gap, the magnitude of which is $\Delta$.

The Bogoliubov coefficients are found by diagonalizing the Hamiltonian:
\begin{eqnarray}
\label{Bogoliubovsimpleuv}
u_{\bf k,q}^2 &= 1-v_{\bf k,q}^2=\frac{1}{2}\left(1 + \frac{\xi_{\uparrow{\bf k}}+\xi_{\downarrow{\bf q-k}}}{2E_{\bf k,q}} \right)
\end{eqnarray}
\begin{eqnarray}
u_{\bf k,q}v_{\bf k,q} &= -\frac{\Delta}{2E_{\bf k,q}}
\label{Boguv}
\end{eqnarray}
and we arrive at
\begin{eqnarray}
  \nonumber\widehat{H}& = \sum_{\bf k} \left(\begin{array}{cc}\hat{\gamma}_{+,{\bf k,q}}^\dagger, &\hat{\gamma}_{-,{\bf k,q}}\end{array}\right) \times \\
\nonumber &\quad\quad\times \left(\begin{array}{cc} E_{+,{\bf q,k}} & 0 \\
  0 & -E_{-,{\bf k,q}}\end{array}\right) \left(\begin{array}{c}\hat{\gamma}_{+,{\bf k,q}}\\ \hat{\gamma}_{-,{\bf k,q}}^\dagger\end{array}\right) \\
\nonumber  &\quad\quad+ \xi_{\downarrow{\bf q-k}}-\frac{\Delta^2}{U} \\
\nonumber&=\sum_{\bf k} \left( E_{+,{\bf q,k}}\hat{\gamma}_{+,{\bf k,q}}^\dagger\hat{\gamma}_{+,{\bf k,q}}+E_{-,{\bf k,q}}\hat{\gamma}_{-,{\bf k,q}}^\dagger\hat{\gamma}_{-,{\bf k,q}} \right.\\
\nonumber&\quad\quad+ \left.\xi_{\downarrow{\bf q-k}}-E_{-,{\bf k,q}}-\frac{\Delta^2}{U}  \right),
\label{simpleBogoliubovHamiltonian}
\end{eqnarray}
where in the last step we used the anticommutation relation $\gamma_{-{\bf k}}\gamma_{-{\bf k}}^\dagger=1-\gamma_{-{\bf k}}^\dagger\gamma_{-{\bf k}}$.

The operators $\hat{\gamma}_\pm$ are the quasiparticle operators that obey fermionic anticommutation relations, like the original particle operators. Within the BCS theory, quasiparticles are non-interacting. To derive some basic properties of the superfluids such as collective modes, quasiparticle interactions have to be introduced via the generalized random phase approximation (GRPA), for instance, as discussed in section~\ref{CollectiveSection}. For the moment, we consider non-interacting quasiparticles, which means that the system is described as an ideal Fermi gas with dispersions given by $E_+$ and $E_-$ in equation~(\ref{eq:qp_energies}).

\subsection{The quasiparticle dispersions}

From the diagonalized Hamiltonian, equation~(\ref{simpleBogoliubovHamiltonian}), the mean-field thermodynamic potential can be calculated. Consequently the phase diagram of the two-component Fermi gas can be determined from the thermodynamic potential by minimizing it with respect to the gap $\Delta$ and the FFLO wavevector ${\bf q}$. The thermodynamic potential depends on the quasiparticle dispersions $E_{\pm}$, and before turning to the calculation of the thermodynamic potential, we therefore take a closer look at the quasiparticle dispersions.

The quasiparticles describe the creation of excitations in the BCS state. Explicitly, the operator $\hat{\gamma}_{+}$ is a superposition of creating an up ($\uparrow$) particle in some momentum state and at the opposite momentum creating a hole for the down ($\downarrow$) component. Similarly, the quasiparticle $\hat{\gamma}_{-}$ creates a $\downarrow$ particle and an $\uparrow$ hole.  The BCS state has either both $\uparrow$ and $\downarrow$ components occupying a certain $({\bf k},-{\bf k})$ pair of states, or no particles occupying these states. Therefore an excitation, that is a single $\uparrow$ particle at state $\vec{k}$ but no $\downarrow$ particle at $-{\bf k}$, can be created either by destroying the $\downarrow$ particle in the case where initially there were particles in the $({\bf k},-{\bf k})$ states, or by creating an $\uparrow$ particle in the case where the states $({\bf k},-{\bf k})$ were empty. This is why the quasiparticle operators are superpositions of particles and holes. 

\begin{figure}
\centering
\includegraphics[width=0.45\textwidth]{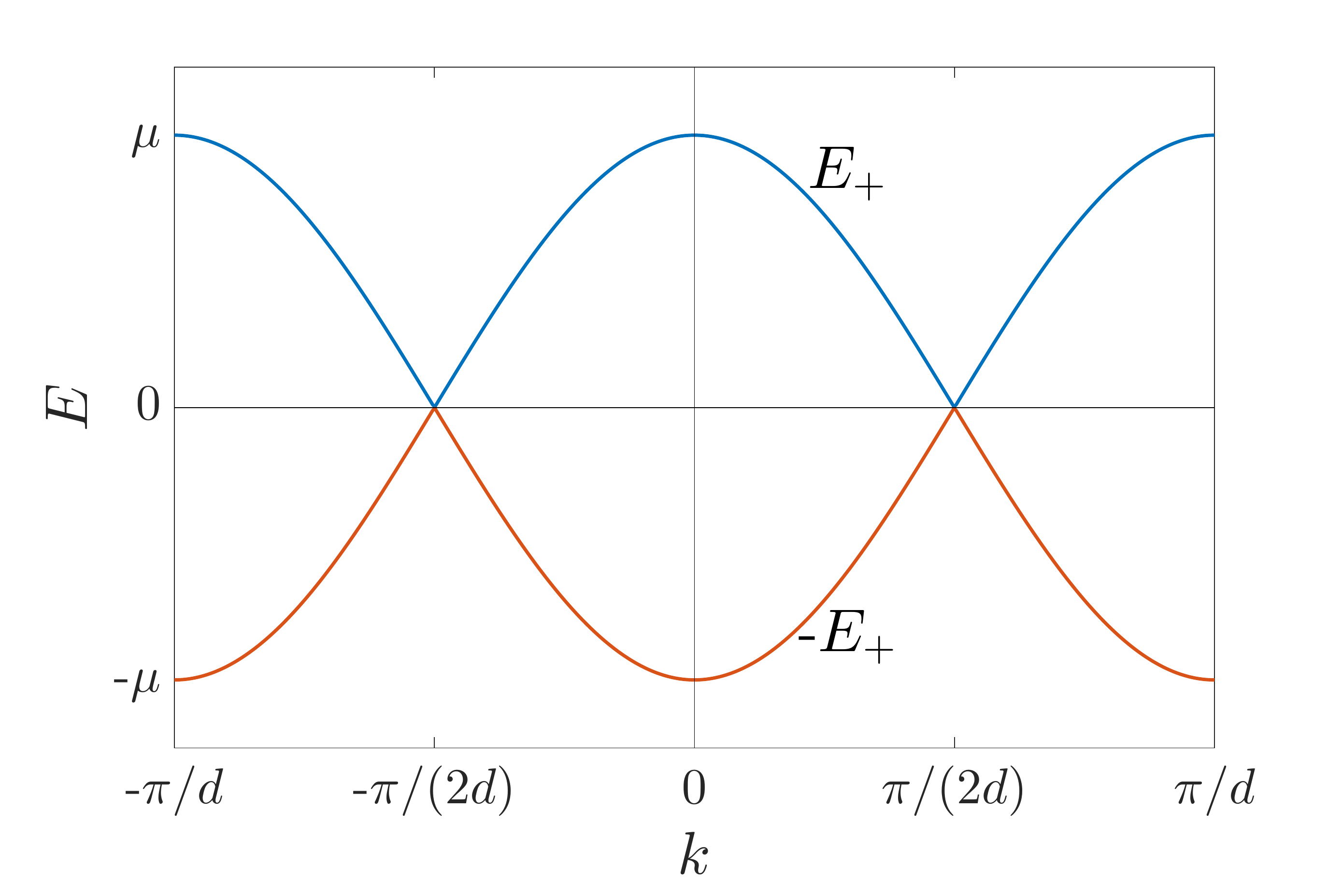}
\includegraphics[width=0.45\textwidth]{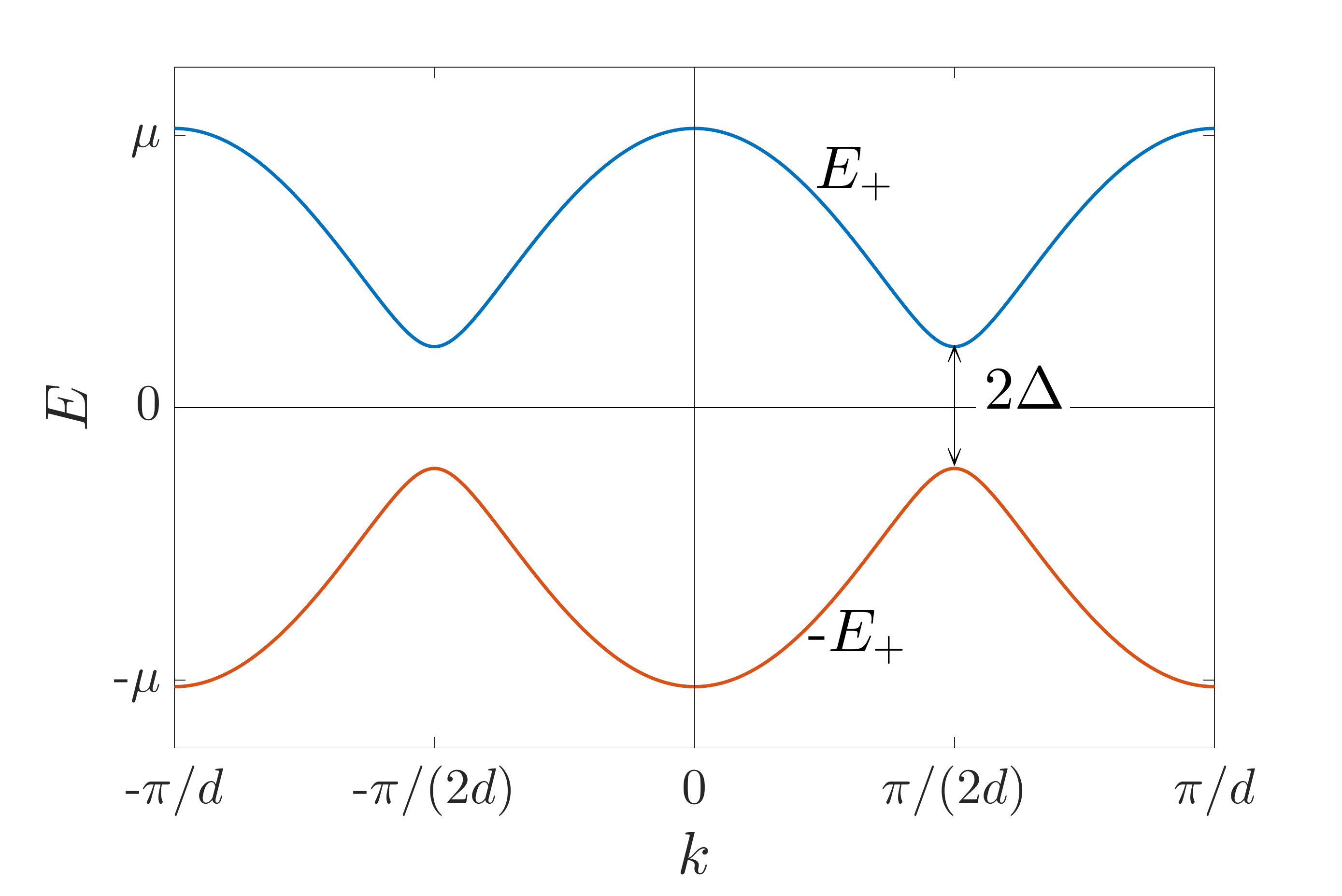}
\caption{Examples of dispersions of  eigenenergies for the non-interacting (left) and the BCS states (right) as functions of $k$ in one dimension. The Fermi momentum here is  at $k=\pi /(2d)$. In the non-interacting case, at the Fermi momentum the quasiparticle energy is zero. Here $E_+=E_-$ and also the hole branch is shown, which is $-E_{\pm}$}
\label{fig:normalBCS}
\end{figure}

When the system is in the normal state, there are no Cooper pairs, $\Delta=0$. The quasiparticle dispersions are then equal to the particle dispersions, $ E_{+,{\bf k}} = \xi_{\uparrow{\bf k}}$ and $ E_{-,{\bf k,q}} = \xi_{\downarrow{\bf q-k}}$, for momenta above the Fermi levels. For $k>k_\text{F}$, one can then associate $ E_{+,{\bf k}}$ with an $\uparrow$ particle and $ E_{-,{\bf k}}$ with a $\downarrow$ one, see the left panel in figure~\ref{fig:normalBCS}. There also the hole branch is shown, which is $-E_{\pm}$.
As soon as the Fermi levels for the two spin components are unequal, the association of the quasiparticles with the particles becomes more complicated due to the square root term in equation~(\ref{eq:qp_energies}).

In the two-component Fermi mixture, a phase transition can occur to a superfluid, consisting of condensed Cooper pairs. 
For an equal number of particles, $\mu_\uparrow=\mu_\downarrow$, this transition in the square lattice is toward a BCS superfluid with ${\bf q}=0$. 
The quasiparticles for this case are presented in the right panel in figure~\ref{fig:normalBCS}, where $\Delta \neq 0$. 
In the paired case, an excitation gap is formed. Here the two eigenenergies are degenerate, meaning that $E_+ = E_-$.  

\begin{figure}
\centering
\includegraphics[width=0.45\textwidth]{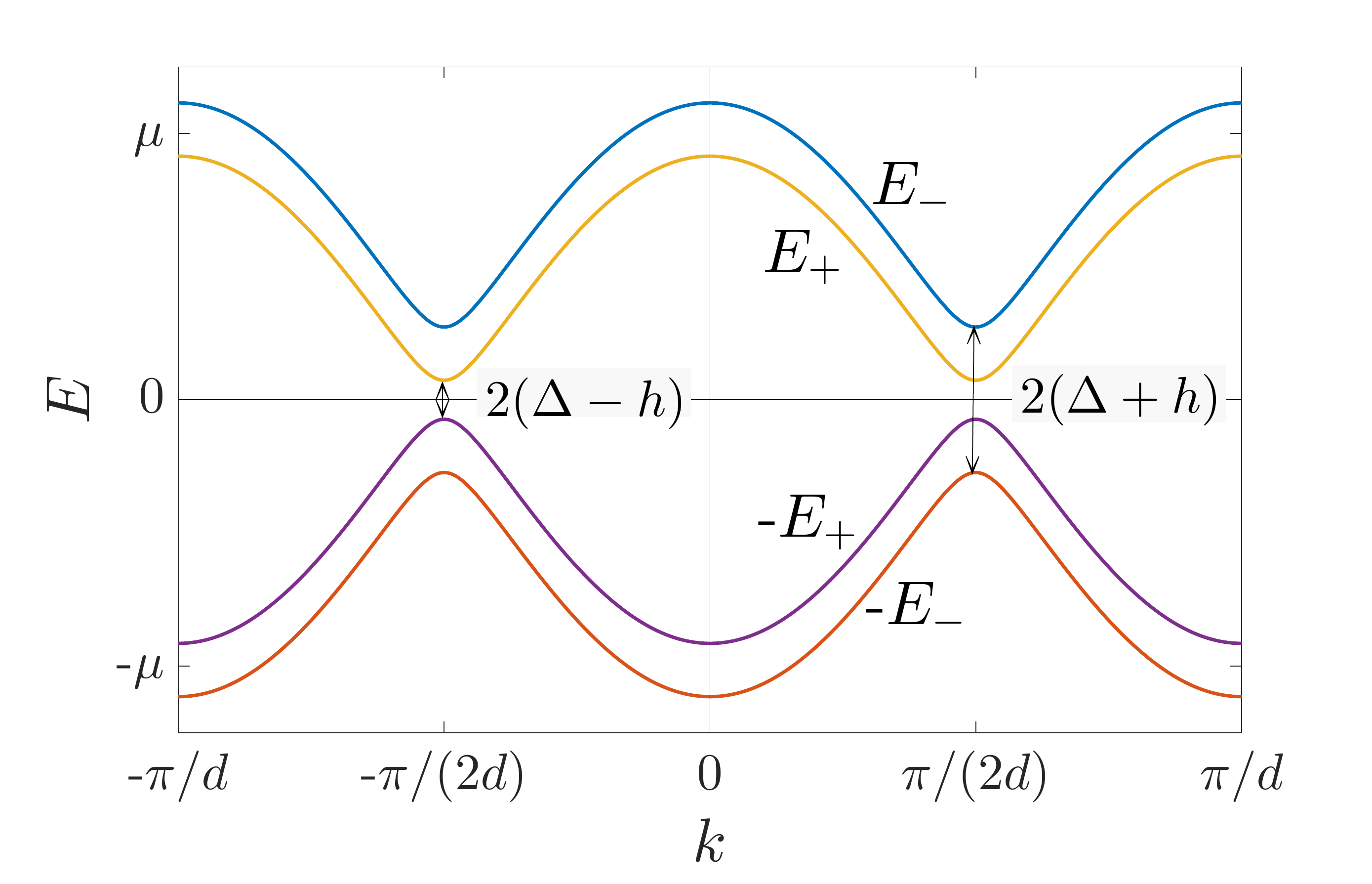}
\includegraphics[width=0.45\textwidth]{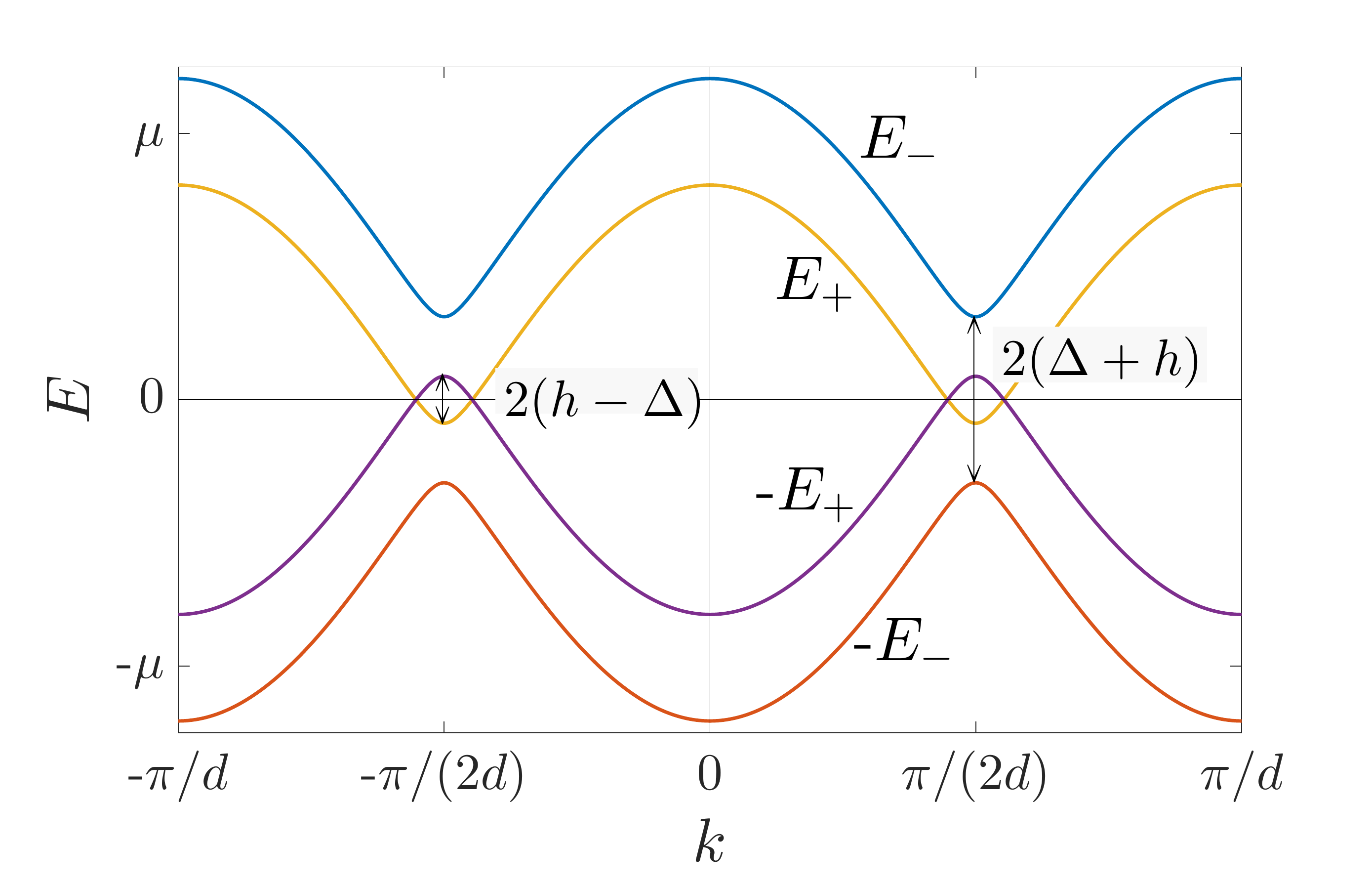}
\caption{Examples of the quasiparticle dispersions in the superfluid phase in the presence of a population imbalance, as a function of $k$. The momentum corresponding to the average chemical potential $(\mu_\uparrow + \mu_\downarrow)/2$ is at $k=\pi /(2d)$, which is where the minima of the quasiparticle dispersions appear. Here $E_+ \neq E_-$, giving two distinct dispersions, where the upper one is $E_{-}$ and the line below $E_{+}$. Hole branches are marked as well with purple and red. In the left panel, the parameters are such that the system is gapped and in the BCS state, although there are two distinct excitation energies now. In the example in the right panel, gapless excitations appear when one of the dispersions crosses zero and the system is in the Sarma state.}
\label{fig:BPenergies}
\end{figure}

If there is a chemical potential imbalance in the Fermi mixture, $\mu_\uparrow\neq\mu_\downarrow$,  a phase transition to a superfluid phase with ${\bf q}=0$ could in principle occur, see figure~\ref{fig:BPenergies}. With respect to the degenerate case in figure~\ref{fig:normalBCS}, the quasiparticle dispersions are shifted by the difference in chemical potentials, $h=(\mu_\uparrow-\mu_\downarrow)/2$. In the left figure, $h<\Delta$ and both excitations are gapped, corresponding to the BCS superfluid. In the right figure, the chemical potential imbalance is even bigger and $h>\Delta$. Now $E_+$ crosses zero, and there is a range of momenta for which gapless excitations exist in $E_+$. This solution of the gap equation is has the peculiar feature that even when most particles (momentum states) are associated with the pairing gap $\Delta$, there are momentum values where gapless excitations are possible. In contrast, in the BCS state all eigenvalues are positive, and there are no quasiparticles at zero temperature. 
It is important to note that, for fixed chemical potentials, weak interactions, continuum systems or simple square lattices, this non-BCS solution of the gap equation is not a state of matter since it is a maximum, not minimum of the grand potential. Sarma explained this in his original work~\cite{Sarma1963}. This solution of the gap equation was later suggested as "breached pair (BP) state"~\cite{liu_interior_2003}, although the authors soon noted that one needs to have fixed particle numbers, not fixed chemical potentials, in order for it to be a state of matter (minimum of energy)~\cite{gubankova_breached_2003}. Later it was shown that the same argument about fixed particle numbers versus chemical potentials holds in square lattices~\cite{koponen_fermion_2006}. The terms Sarma state and BP state are nowadays used for referring to this type of solution of the gap equation, which under specific conditions (e.g., fixed particle numbers or complex lattice geometry) can also correspond to a stable state of matter. In the following, we mostly use the terms Sarma phase or Sarma state.  

\begin{figure}
\centering
\includegraphics[width=0.45\textwidth]{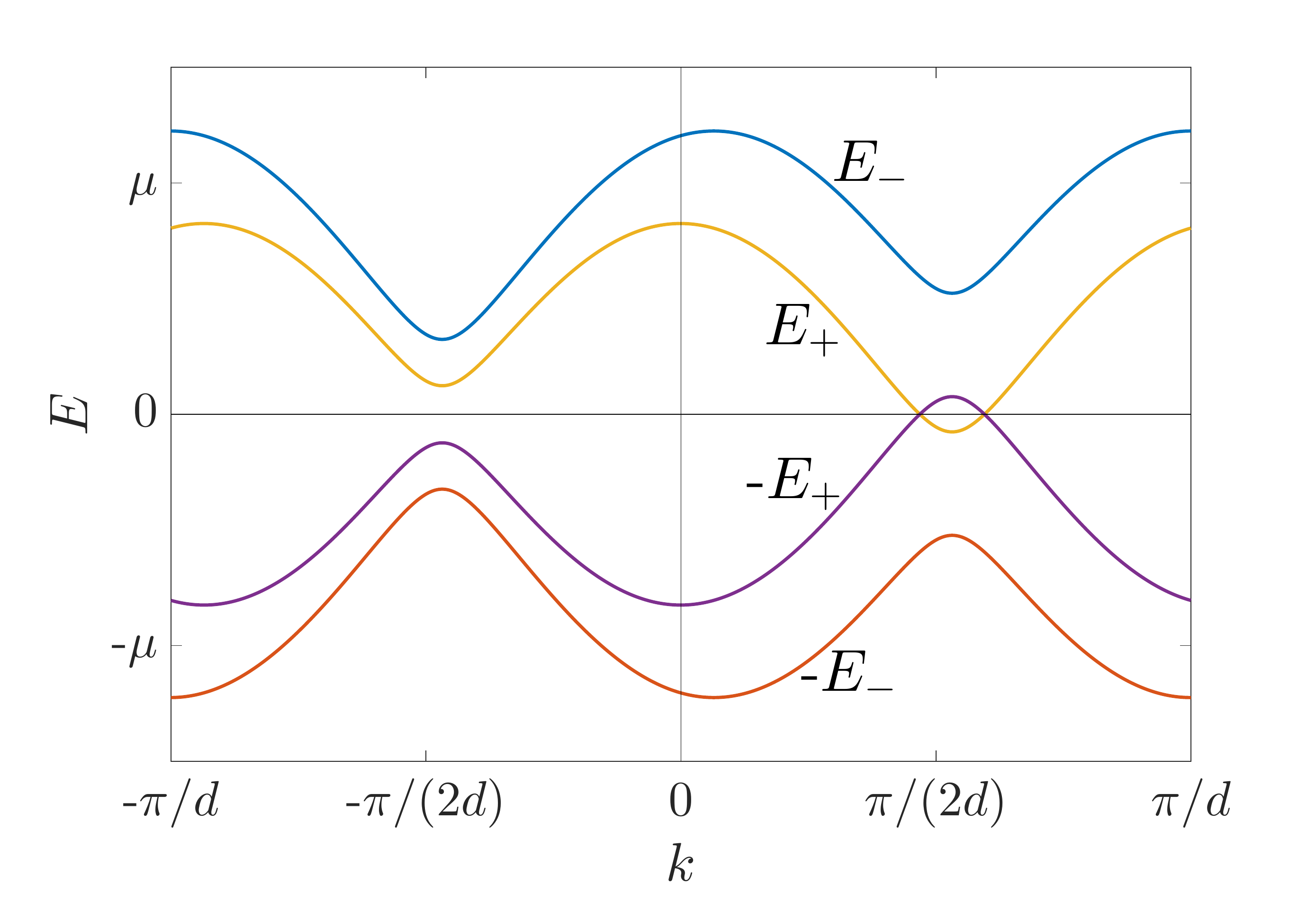}
\includegraphics[width=0.45\textwidth]{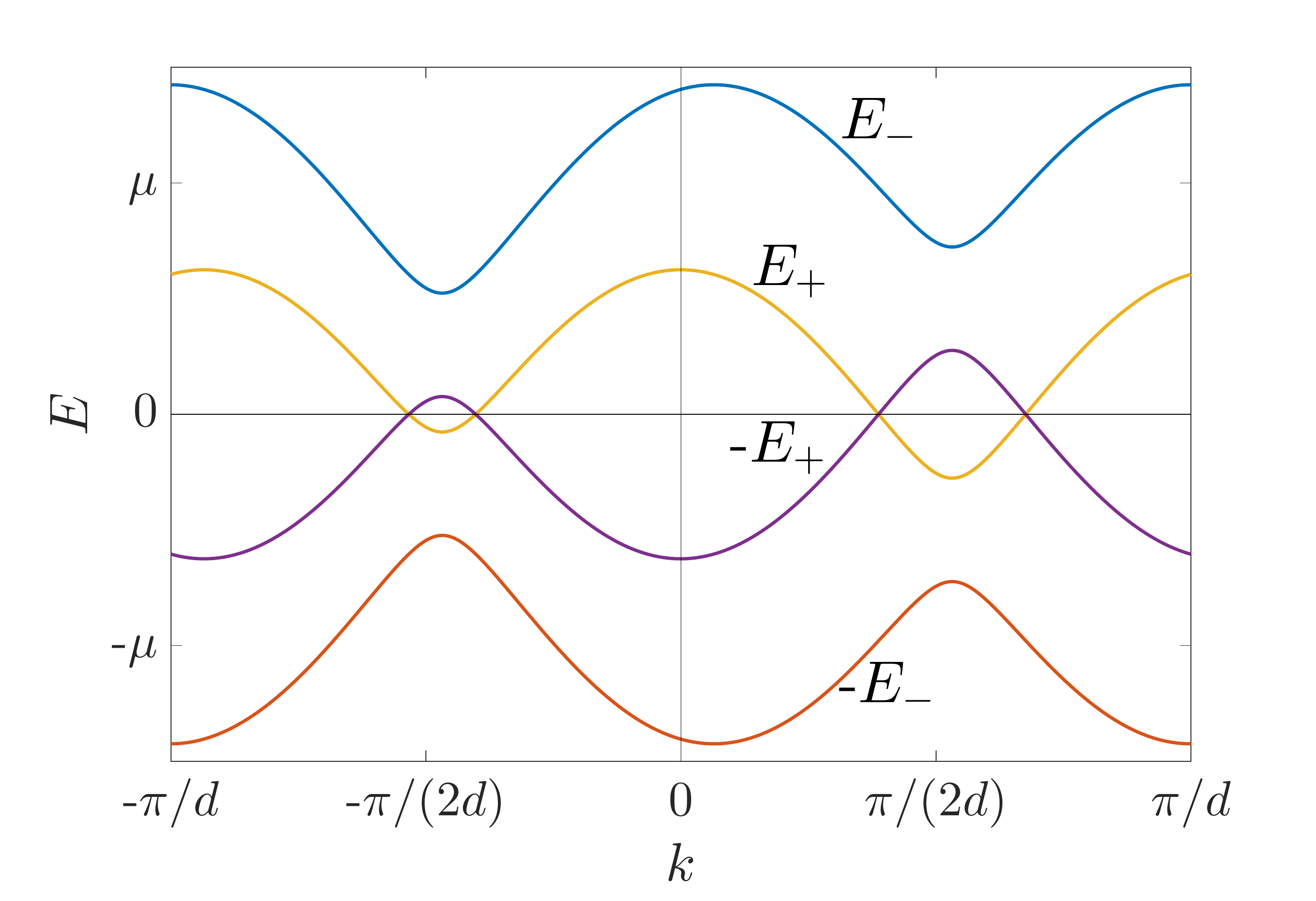}
\caption{Examples of dispersions of the eigenenergies for two cases of an FF state as function of $k$. The momentum corresponding to the average
chemical potential $(\mu_\uparrow + \mu_\downarrow)/2$ is at $k=\pi /(2d)$. Here $E_+ \neq E_-$ giving the distinct dispersions marked in blue and orange. Hole branches are marked in red and purple. Depending on the parameters, gapless excitations may appear for one sign of momentum in one quasiparticle branch (left) or for both negative and positive momenta (right) when both branches cross zero.}
\label{fig:FFLO12}
\end{figure}

The phase transition from the normal state is can also be toward an FFLO superfluid phase, where the Cooper pairs carry a net momentum ${\bf q}$. The quasiparticle dispersions for that case are shown in figure~\ref{fig:FFLO12}. It turns out that the wave vector ${\bf q}$ found by minimizing the thermodynamic potential is proportional to the population imbalance $|{\bf q}| \simeq |k_{F\uparrow} - k_{F\downarrow}|$. In other words, the bigger the difference between the chemical potentials or particle numbers, the bigger $|{\bf q}|$. Also in the FFLO case, there are gapless excitations, and 
quasiparticles can exist in the superfluid state. 
The difference with the Sarma state is that the two quasiparticle branches $E_+$ and $E_-$ are not only shifted with respect to each other as in figure~\ref{fig:BPenergies}, but they also become asymmetric with respect to $k$ and $-k$ due to the nonzero ${\bf q}$, see figure~\ref{fig:FFLO12}. 

\subsection{Self-consistent gap and number equations}
It is now possible to derive a set of coupled equations consisting of the gap and number equations, from which the gap parameter $\Delta$ and the chemical potentials, $\mu_\uparrow$ and $\mu_\downarrow$, can be solved. 
Let us start with deriving the number equations. The total number of particles in one spin component is given by
\begin{equation}
N_\sigma = \sum_{\bf k}
\left\langle{\hat c}_{\sigma,{\bf k}}^\dagger
{\hat c}_{\sigma,{\bf k}}\right\rangle .
\end{equation}
Using the inverse of the Bogoliubov transformation equation~(\ref{inverseBogoliubov}), we obtain for the $\uparrow$ component
\begin{eqnarray}
\label{eq:mu_up}
\nonumber N_\uparrow &= \sum_{\bf k} \left\langle{\hat c}_{\uparrow{\bf k}}^\dagger
{\hat c}_{\uparrow{\bf k}}\right\rangle\\
\nonumber &= \sum_{{\bf k}}\left\{u_{\bf k,q}^2\left\langle\hat{\gamma}_{+,{\bf k,q}}^\dagger \hat{\gamma}_{+,{\bf k,q}} \right\rangle + v_{\bf k,q}^2\left\langle
\hat{\gamma}_{-,{\bf k,q}}\hat{\gamma}_{-,{\bf k,q}}^\dagger\right\rangle \right\}\\
&=\sum_{{\bf k}}\left\{ u_{\bf k,q}^2 n_F(E_{+,{\bf k,q}}) + v_{{\bf k,q}}^2 n_F(-E_{-,{\bf k,q}})\right\},
\end{eqnarray}
where the fact that the expectation values of the form $\langle \hat{\gamma}_+ \hat{\gamma}_-\rangle$ are zero for non-interacting (quasi)particles \cite{mahan_many-particle_2000,stoof_ultracold_2008,lewenstein_ultracold_2012} was used. The chemical potential of the $\uparrow$ component depends on the number of $\downarrow$ particles via the Hartree shifts, see equation~(\ref{eq:a-13}). A similar equation holds for the number of down ($\downarrow$)
particles:
\begin{eqnarray}
\label{eq:mu_down}
N_\downarrow = \sum_{{\bf k}}\left\{ u_{\bf k,q}^2 n_F(E_{-,{\bf k,q}}) + v_{\bf k,q}^2 n_F(-E_{+,{\bf k,q}})\right\},
\end{eqnarray}
where the chemical potentials contain the Hartree shifts.

Using the original definition of the order parameter, $U\langle{\hat c}_{\downarrow{\bf r}}{\hat c}_{\uparrow{\bf r}}\rangle =\Delta e^{i{\bf q\cdot r}}$, it is possible to derive the so-called \emph{gap equation}. The BCS gap equation, where ${\bf q}=0$, is
\begin{eqnarray}
\nonumber \frac{\Delta}{U} &= \left\langle{\hat c}_{\downarrow {\bf r}}{\hat c}_{\uparrow {\bf r}}\right\rangle  \\
\nonumber &= \sum_{\bf k,k'} e^{-i{\bf (k+k')\cdot n}}\left(-u_{\bf k} v_{\bf -k'}
\left\langle\hat{\gamma}_{+,{\bf -k}}^\dagger \hat{\gamma}_{+,{\bf k}'}\right\rangle+ \right.\\
\nonumber &\quad\quad\left.+u_{-{\bf k}'}v_{{\bf k}}\left\langle\hat{\gamma}_{-,-{\bf k}}\hat{\gamma}_{-,{\bf k}'}^\dagger\right\rangle\right)\\
&= \sum_{{\bf k}} u_{{\bf k,q}}v_{{\bf k,q}} \left(1 -n_F(E_{+,{\bf k,q}}) - n_F(E_{-,{\bf k,q}}) \right).
\end{eqnarray}
After dividing the above equation by $\Delta$ and subsequently using equation~(\ref{Boguv}), 
the gap equation reads
\begin{equation}
\label{eq:gap}
1 = -U\sum_{{\bf k}} \frac{1 - n_F(E_{+,{\bf k,q}}) -n_F(E_{-,{\bf k,q}})}{2E_{{\bf k,q}}},
\end{equation}
where the prefactor on the right side is positive, since $U<0$ for attractive interactions.

The summation on the right side of the above gap equation equation~(\ref{eq:gap}) diverges at high momenta ${\bf k}$. This divergence directly results from the approximation of the interatomic interaction by the unphysical contact interaction and is absent if the true interatomic potential is used. In the case we consider here, the single band lattice model, this divergence does not cause any issues because the single band approximation provides a natural cutoff for the above summation, namely the band width (assuming that interactions are so weak that the single band approximation is valid). In the homogeneous Fermi gas, without a lattice, the ultraviolet divergence can be removed, for example, by introducing a cutoff and renormalizing the coupling constant, see for instance references~\cite{bloch_many-body_2008,stoof_ultracold_2008,lewenstein_ultracold_2012}. 

This coupled set of equations, consisting of the number equations \eqref{eq:mu_up} and \eqref{eq:mu_down}, and the
gap equation \eqref{eq:gap}, is equivalent to the standard BCS-Leggett theory in a lattice (although there often the chemical potentials of the components are assumed equal). In the population balanced case, where $N_\uparrow = N_\downarrow$, it is possible to choose $\mu_\uparrow = \mu_\downarrow$ by
hand and eliminate one of the number equations. If the interactions
are weak, that is $|U|$ is small, the chemical potential can be
approximated with the Fermi energy. However when the interaction
strength increases, the chemical potential has to be solved from the
number equation to get the correct results. The interactions become large when
exploring the BCS-BEC crossover \cite{gurarie_BCSBECreview_2007,zwerger_bcs-bec_2012,torma_bcsbec_2014} and approaching the BEC regime from the BCS one, then the use of
the number equation becomes mandatory. Therefore the combined gap and number equations are sometimes called crossover equations. It is essential to remember that the gap equations have also solutions that are not minima of energy and are thus unphysical. It is always essential to examine the physical nature of the solutions, which can be done for instance by considering the thermodynamic potential.

\subsection{Thermodynamic potential and Helmholtz free energy} \label{EnergiesAll}

To determine the ground state of a two-component Fermi mixture in a lattice for certain system parameters, such as densities and temperatures, one needs to minimize the total energy of the system. Depending on the ensemble considered, this total energy is either the grand thermodynamic potential or the (Helmholtz) free energy. The minimum of the grand thermodynamic potential $\Omega$ corresponds to thermodynamic equilibrium  in the grand-canonical ensemble, where the chemical potentials are fixed and the particle numbers are fluctuating. The thermodynamic potential is the Legendre transform of the free energy $F$. In the canonical ensemble, where the particle numbers are fixed, thermodynamic equilibrium is reached by minimizing the free energy. Here we show a short derivation of the thermodynamic potential $\Omega$. For a more elaborate derivation and more background on ensembles and Legendre transforms, see for instance references~\cite{mahan_many-particle_2000,fetter_quantum_2003,gurarie_BCSBECreview_2007,sheehy_becbcs_2007,stoof_ultracold_2008}.

To derive the thermodynamic potential, from which subsequently other thermodynamic quantities can be obtained, we first need the
partition function for the grand-canonical ensemble, which can be calculated from the diagonalized Hamiltonian,
\begin{equation}
 \label{eq:partition_function}
  Z_{\text{G}} = \text{Tr} e^{-\beta \widehat{H}} = \sum_{\gamma} \langle
  \gamma |e^{-\beta \widehat{H}}| \gamma \rangle,
\end{equation}
where the sum goes through the quasiparticle basis (i.e., the basis
where the quasiparticle occupation numbers are good quantum numbers) of
the Hilbert space and $\beta = 1/k_{\text{B}}T$, the inverse thermal energy with $k_\text{B}$ Boltzmann's constant. The Hamiltonian in equation~(\ref{simpleBogoliubovHamiltonian}) is of the form
\begin{equation}
\widehat{H} = \sum_{{\bf k}} \left(E_{+,{\bf k,q}} \hat{\gamma}_{+,{\bf k}}^\dagger\hat{\gamma}_{+,{\bf k}} 
+ E_{-,{\bf k,q}} \hat{\gamma}_{-,{\bf k}}^\dagger\hat{\gamma}_{-,{\bf k}}\right) + C,
\end{equation}
where $C = \sum_{{\bf k}} \left(\xi_{\downarrow -{\bf k+q}} -E_{-,{\bf k,q}} - \Delta^2/U\right)$. As a constant, $C$ is factored out of the expectation value and $Z$ reads 
\begin{equation}
  Z_{\text{G}} = e^{-\beta C}\sum_{\gamma} \langle \gamma|e^{-\beta\sum_{{\bf k}} \left(E_{+,{\bf k,q}} \hat{n}_{+,{\bf k}} + E_{-,{\bf k,q}} \hat{n}_{-,{\bf k}} \right)}| \gamma \rangle,
\end{equation}
where $\hat{n}_{\sigma,{\bf k}} = \hat{\gamma}_{\sigma,{\bf k}}^\dagger
\hat{\gamma}_{\sigma,{\bf k}}$, with $\sigma=\pm$. Since all commutators 
of the type $[\hat{n}_{+,{\bf k}},\hat{n}_{\pm,{\bf k'}}]$ 
are zero for all ${\bf k,k'}$, the exponential of the sum can be written as the product of exponentials,
\begin{equation}
 Z_{\text{G}} = e^{-\beta C}\sum_{\gamma} \langle \gamma| \prod_{{\bf k}}
 e^{-\beta E_{+,{\bf k,q}}\hat{n}_{+,{\bf k}}} e^{-\beta E_{-,{\bf k,q}}\hat{n}_{-,{\bf k}}}
|\gamma \rangle.
\end{equation}
For fermionic quasiparticles, the possible occupation numbers
for a given state are $0$ and $1$, leading to
\begin{equation}
Z_{\text{G}} = e^{-\beta C} \prod_{{\bf k}} \left(1 + e^{-\beta
  E_{+,{\bf k,q}}} \right) \left(1 + e^{-\beta
  E_{-,{\bf k,q}}} \right).
\end{equation}
The thermodynamic grand potential can now be obtained via
\begin{equation}
\Omega = -\frac{1}{\beta} \ln Z_{\text{G}} 
\end{equation}
and reads in this case 
\begin{eqnarray}
\Omega = \sum_{{\bf k}}\bigg\{&\xi_{\downarrow{\bf -k+q}}-
E_{-,{\bf k,q}}- \frac{1}{\beta}\left[\ln\left(1 + e^{-\beta  E_{+,{\bf k,q}}} \right)\right. \nonumber\\
  &\left.+\ln \left(1 + e^{-\beta E_{-,{\bf k,q}}} \right)\right] - \frac{\Delta^2}{U} \bigg\}.
\label{MFthermpot}
\end{eqnarray}

In thermodynamic equilibrium, the order parameter takes the value that minimizes the thermodynamic potential. Here the order parameter is 4-dimensional because $\Delta$ is the amplitude of the order parameter and ${\bf q}$ gives the magnitude and direction of the wave vector in the plane wave form. To find the equilibrium state of the system and determine the phase diagram, one thus has to  minimize the thermodynamic potential. When the global minimum of the thermodynamic potential is located at $\Delta=0$ and ${\bf q}=0$, it means there are no Cooper pairs and the system is in the normal phase. The system is in the BCS state when the particle densities are equal $n_\uparrow=n_\downarrow$ and the global minimum of $\Omega$ is attained at $\Delta\neq0$ and ${\bf q}=0$. The Sarma phase would form the ground state of the system if the energy was minimized with a nonzero number of Cooper pairs carrying no momentum, $\Delta\neq0$ and ${\bf q}=0$, in the presence of a population imbalance $n_\uparrow \neq n_\downarrow$. Finally when the global minimum of the thermodynamic potential is located at $\Delta\neq0$ and ${\bf q}\neq0$, the system is in the FFLO state.

At zero temperature, the BCS state can exist as long as the chemical potential difference between components does not exceed the so-called
Chandrasekhar-Clogston limit. At weak coupling, this implies that the chemical potential difference is smaller than $\Delta/\sqrt{2}$~\cite{recati_role_2008}. At finite temperature, a sufficiently small density difference may coexist with the superfluid since thermal quasiparticles can accommodate some imbalance even in the BCS state, see figure~\ref{fig:BPenergies}a) where one of the quasiparticle branches has a smaller gap and can thus have larger thermal population. In the Sarma state, imbalance can be accommodated both at zero and finite temperature because some of the quasiparticle energies take negative values as seen from figure ~\ref{fig:BPenergies}b). Equation~\ref{eq:qp_energies_PT} shows that existence of negative quasiparticle energies requires $\Delta$ to be smaller than $(\mu_\uparrow - \mu_\downarrow)/2$; this condition should be valid for the Sarma/BP state. The finite temperature BCS state with imbalance and the Sarma/BP state at finite temperature both display uniform superfluid order parameter and spin-density imbalance, therefore these terms are sometimes used rather loosely and interchangeably in the literature. However, the absence and presence of a Fermi surface is a fundamental difference between these states.

Alternatively, the above mean-field thermodynamic potential can be calculated using a path-integral formalism and subsequently making a saddle-point approximation. The criteria obtained for phase transitions are the same, see references~\cite{devreese_competition_2011,devreese_fuldeferrelllarkinovchinnikov_2012} and the references therein.

\subsubsection{Traps and Ensembles}
It is important to choose correctly whether to minimize the thermodynamic potential or the free energy, based on which system parameters are fixed. In our cases of interest, the fixed parameters are temperature, and depending on the situation, either the chemical potentials $\mu_\uparrow$ and $\mu_\downarrow$ or the total particle numbers, $N_\uparrow$ and $N_\downarrow$, as explained below. 

In ultracold gas experiments, there is typically no particle bath. Rather the numbers of particles in the total system, $N_\uparrow$ and $N_\downarrow$, are constant in the experiment. Furthermore since the time scale of spin relaxation is much larger than the time scale of the experiment, the atoms are prevented from converting their spin. However in a  harmonic trap, the particles can distribute themselves non-uniformly, and therefore 
the local particle number is not fixed by external constraints. In this situation, the atom distribution assumes a form where the chemical potentials are constant throughout the trap and the grand potential $\Omega(\Delta,{\bf q},\mu_\uparrow,\mu_\downarrow)$ is the relevant energy to be minimized. This is usually the case in studies of trapped ultracold gases, also in the case of optical lattices since the lattice potential is typically superimposed by a harmonic trap. 

In principle it is possible to create potentials that are nearly flat, such as higher order traps approaching box potentials or new types of traps created by masks or holographic techniques~\cite{meyrath_bose-einstein_2005,es_box_2010,gaunt_bose-einstein_2013,mukherjee_homogeneous_2016}. In these cases, the density of the system is homogeneous, and now the relevant thermodynamic quantity to minimize is the Helmholtz free energy $F$, which can be calculated from the grand potential with
\begin{equation}
F = \Omega + \mu_\uparrow f_\uparrow + \mu_\downarrow f_\downarrow,
\label{eq:F}
\end{equation}
where in the lattice case $f_\uparrow$ and $f_\downarrow$ are the numbers of atoms per lattice site, that is the filling fractions or filling factors. Because of the single band model, it is not possible to have more than one atom of one species at a single lattice site, thus the filling factors are bounded between $0$ and $1$. When minimizing the free energy $F$, it is important to take care that the chemical potentials satisfy the number equations \eqref{eq:mu_up} and \eqref{eq:mu_down}. Numerically this means that for every $(\Delta,{\bf q})$ -pair, before the value of $F$ can be calculated, the number equations must be solved. Since in the interacting case $\mu_\uparrow$ and $\mu_\downarrow$ are not independent, and the value of $\mu_\uparrow$ affects the value of $\mu_\downarrow$ and vice versa, these equations have to be solved iteratively until both chemical potentials have converged. 

\subsection{Ginzburg-Landau expansion}
As explained in section \ref{EnergiesAll}, the phase diagram of a two-component Fermi mixture can be determined from the thermodynamic potential $\Omega$ (or free energy $F$) by minimizing it with respect to the order parameter. However if one is interested in the phase transitions of the system like the onset of superfluidity, it can be enough to study the change in behavior of the thermodynamic potential, which is explained in this subsection by doing a Ginzburg-Landau expansion of the thermodynamic potential. For further reading, see for instance references~\cite{mahan_many-particle_2000,fetter_quantum_2003,gubbels_imbalanced_2013}.

\begin{figure}
\begin{center}
\includegraphics[width=0.95\columnwidth]{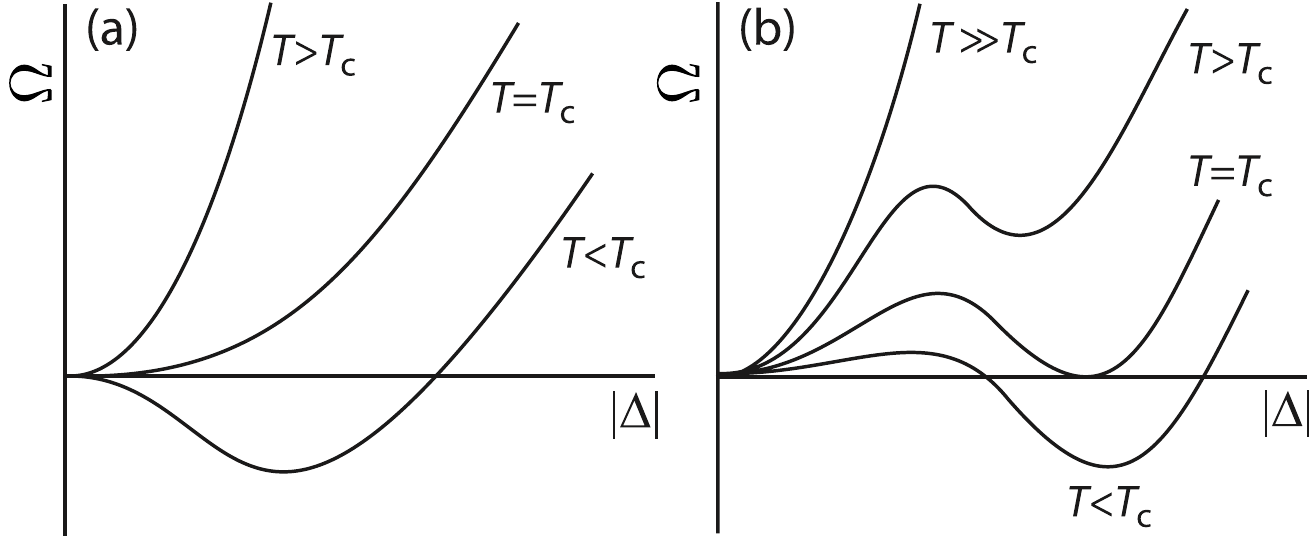}
\end{center}
\caption{The thermodynamic potential $\Omega(|\Delta|)$ as a function of the order parameter $\Delta$. Panel (a) shows the behavior of $\Omega(|\Delta|)$ for different temperatures when a second-order phase transition occurs and panel (b) when a first-order phase transition occurs.} \label{fases}
\end{figure}

Near a continuous phase transition, the BCS order parameter $\Delta({\bf r})$ is small, and the thermodynamic potential in equation~(\ref{MFthermpot}) can be expanded as
\begin{equation}
\Omega(\Delta;\mu_{\sigma},T)=\gamma|\nabla\Delta|^{2}+\alpha|\Delta|^{2}+\frac{\beta}{2}|\Delta|^{4}+\ldots,
\label{omexpansion}
\end{equation}
where the dots denote the higher orders in $|\Delta|^{2}$ and in the gradients $|\nabla\Delta|^{2}$. The expansion contains only even terms in $\Delta$, because the thermodynamic potential is even in $\Delta$, see equation~(\ref{MFthermpot}) and figure~\ref{fases}. We only include the smallest possible gradient term that is even in $\Delta$, because here we are only interested in the onset of possible inhomogeneous FFLO phases.

The Landau coefficients in the  equation~(\ref{omexpansion}) all depend on the system parameters, such as chemical potentials and temperature. If all these Landau coefficients are positive, the minimum of the thermodynamic potential is located at $\Delta({\bf r})$ equal to zero, and the system is in the normal state. A phase transition to a superfluid state has occurred when the position of the global minimum is located at a nonzero order parameter $\Delta({\bf r})$, which describes a condensate of Cooper pairs.

If the coefficient $\gamma$ is positive, it costs energy to have a spatially varying superfluid, such as an FFLO state. A homogeneous phase is then energetically more favorable, and we can therefore restrict ourselves to a pairing field $\Delta$ independent of position. In the opposite case, where $\gamma<0$, the system can gain energy when the order parameter varies in space. The plane wave FF and standing wave LO state are the two simplest examples of the infinitely many possibilities of position-dependent order parameters.

Now in the case where $\gamma$ is positive and the system is homogeneous, both continuous and discontinuous phase transitions from the normal to the superfluid phase can occur. If in the above Landau thermodynamic potential $\alpha$ is negative and all other coefficients are positive, the minimum of the thermodynamic potential is attained at some nonzero $\Delta$ and the Fermi gas is in the superfluid state. Thus as $\alpha$ changes sign, a continuous phase transition takes place, which can be determined by solving $\alpha=0$.

It is also possible to have a first-order phase transition, when in the thermodynamic potential density all coefficients but the fourth-order coefficient $\beta$ are positive. The thermodynamic potential now has two minima, one located at $\Delta$ equal to zero and the other  at a nonzero value of the order parameter. For temperatures above $T_c$, the global minimum of the thermodynamics potential is located at $\Delta$ equal to zero and the Fermi gas is in the normal state. At the critical temperature, the two minima are equal, and for even lower temperatures, the minimum located at a nonzero order parameter $\Delta$ is the global minimum and the system is in the superfluid state. The criterion for a first-order phase transition thus is $\Omega(0)=\Omega(\Delta)$. In contrast to a second-order phase transition, the location of the global minimum of the thermodynamic potential now changes discontinuously, see figure~\ref{fases}b, which is associated with phase separation. In contrast to the continuous transition, for determining a first-order phase transition, the full thermodynamic potential is needed.

\begin{figure}
\begin{center}
\includegraphics[width=0.95\columnwidth]{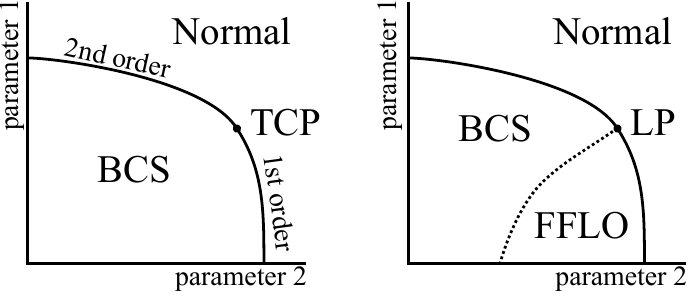}
\end{center}
\caption{Different scenarios of phase diagrams that can occur in the two-component Fermi gas. 
On the left, the phase transition from the normal state to the BCS phase is continuous (second order) or discontinuous (first order), depending on the system parameters. The multicritical point in between is  a tricritical point. On the right, there is also an FFLO region in the phase diagram, and the multicritical point where the normal, the BCS and the FFLO region meet is a Lifshitz point (LP).} 
\label{PDExamples}
\end{figure}

For population imbalanced Fermi gases in optical lattices, both continuous and discontinuous phase transitions occur, as we  discuss in section~\ref{MFstudies}. Therefore there must exist a point in the phase diagram where the character of the phase transition changes, that is, a multicritical point, and a sketch of this scenario is given in the left panel in figure~\ref{PDExamples}.  If the phase transition changes from being second order to a first order transition, the multicritical point is a tricritical point, which can be found by setting 
\begin{equation}
\alpha=\beta=0. \label{tricrit}
\end{equation}

Also inhomogeneous phases can occur for the population imbalanced Fermi gas in optical lattices. Therefore there must exist a point in the phase diagram where the phase transition changes from a transition from a normal to a homogeneous superfluid phase into a transition from the normal  to an inhomogeneous superfluid phase, which is called a Lifshitz point, see figure~\ref{PDExamples} right panel. The criterion for the Lifshitz point is 
\begin{equation}
\alpha=\gamma=0.\label{lifshitz}
\end{equation}
Superfluidity at nonzero momentum can be established in many ways, of which FF and LO are only two examples, and due to this variety of possibilities it is difficult to predict which kind of inhomogeneous superfluidity will actually occur in the system. However they all have to emerge from the Lifshitz point, making it worthwhile to know about its existence and position.
Depending on the strength of the interaction, the Lifshitz point and tricritical point can overlap, $\alpha=\beta=\gamma=0$. In that case, it is necessary to study the full thermodynamic potential to determine what the ground state of the system below this multicritical point actually is.

For the two-component Fermi gas in continuum, many studies exist that investigate the vicinity of the Lifshitz point, for instance references~\cite{combescot_transition_2002,casalbuoni_inhomogeneous_2004,mora_transition_2005,radzihovsky_fluctuations_2011}. For Fermi gases in optical lattices, we are not aware of any such studies.

\begin{figure}
\begin{center}
\includegraphics[width=0.95\columnwidth]{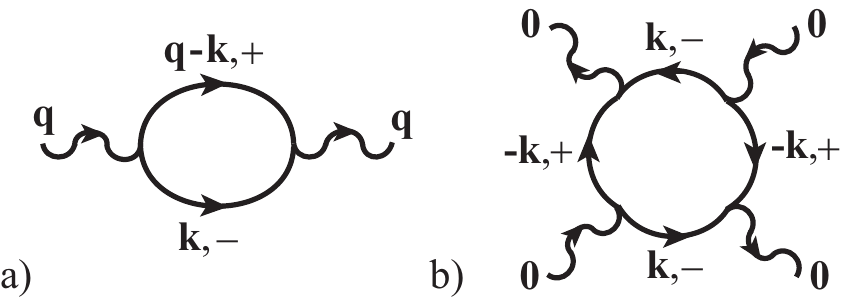}
\end{center}
\caption{a) The ladder diagram with external momentum $\mathbf{q}$. The wiggly lines denote the Cooper pairs, which can break up into two fermions of different spin, denoted by the normal lines. Here $\mathbf{q}$ is the wavevector of the Cooper pairs, while $\mathbf{k}$ and $\mathbf{q}-\mathbf{k}$ are the wavevectors of the fermions. When $\mathbf{q}$ is equal to zero, the amplitude of this diagram corresponds to the quadratic Landau coefficient $\alpha$ in the thermodynamic potential. b) The Feynman diagram corresponding to the fourth-order Landau coefficient $\beta$.} \label{ladderq}
\end{figure}

Now that we have some understanding about the different coefficients in equation~(\ref{omexpansion}), we can start to calculate them. In mean-field theory, the quadratic Landau coefficient $\alpha$ is explicitly given by
\begin{eqnarray}
  \nonumber \alpha&=\left.\frac{\partial\Omega}{\partial|\Delta|^{2}}\right|_{|\Delta|^{2}=0,{\bf q}=0} \\
  &=-\frac{1}{U}
+\sum_{\bf k}\left(\frac{n_{\text F}(\xi_{\uparrow,{\bf k}})+n_{\text F}(\xi_{\downarrow,{\bf k}})-1}{\xi_{\uparrow,{\bf k}}+\xi_{\downarrow,{\bf k}}}\right),
\label{alpha}
\end{eqnarray}
and can straightforwardly be obtained by taking a derivative of the full thermodynamic potential in equation~(\ref{MFthermpot}). Note that when $\alpha$ is set to zero, the above equation is exactly the gap equation in equation~(\ref{eq:gap}) with ${\bf q}=0$. An alternative way to determine the above coefficient, which is rather simple and elegant, is by using Feynman diagrams \cite{kleinert_h._collective_1978}. Also the other coefficients $\gamma$ and $\beta$ can be determined using a diagrammatic language. Namely the quadratic coefficient $\alpha$ corresponds to the amplitude of the so-called ladder diagram where the incoming and outgoing bosonic pairing fields $\hat{\Delta}$ carry zero momentum, see figure~\ref{ladderq}a. The external momentum of the diagram is here taken equal to zero, since the Cooper pairs in the BCS state do not have a net momentum.
Physically, $\alpha$ can be interpreted as being proportional to the chemical potential of the Cooper pairs. The fourth-order coefficient $\beta$ has a
diagrammatic representation with four external bosonic fields carrying zero momentum, see figure~\ref{ladderq}b, or can be calculated by taking derivatives from the full thermodynamic potential. It is given by
\begin{eqnarray}
\label{beta}\beta&=\left.\frac{\partial^{2}\Omega}{(\partial|\Delta|^{2})^{2}}\right|_{|\Delta|^{2}=0,{\bf q}=0}\\
  \nonumber &=\sum_{\bf k}\frac{-1}{4(\epsilon_{\bf k}-\mu)^{2}}
  \Bigg[\beta n_{\text F}(\xi_{\uparrow,{\bf k}})n_{\text F}(-\xi_{\uparrow,{\bf k}})+\\
    &+\beta n_{\text F}(\xi_{\downarrow,{\bf k}})n_{\text F}(-\xi_{\downarrow,{\bf k}})
    +\frac{n_{\text F}(\xi_{\uparrow,{\bf k}})-n_{\text F}(-\xi_{\downarrow,{\bf k}})}{\epsilon_{\bf k}-\mu}\Bigg].
\nonumber
\end{eqnarray}

\begin{figure}
\begin{center}
\includegraphics[width=0.95\columnwidth]{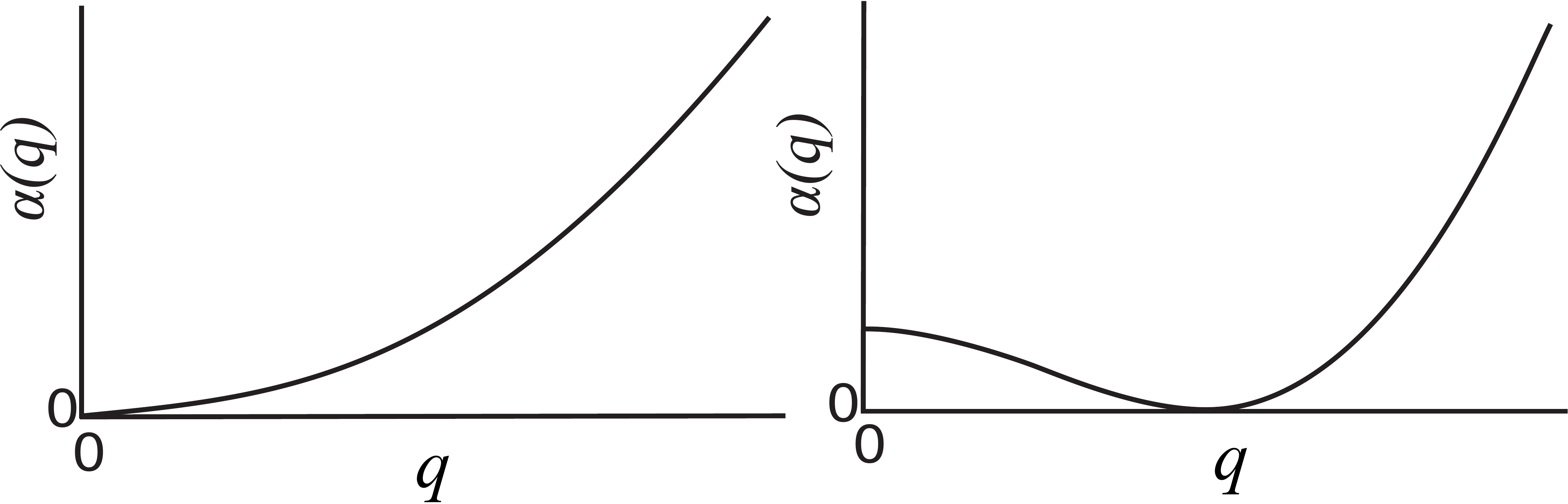}
\end{center}
\caption{Depending on the system parameters, the minimum of $\alpha(\mathbf{q})$ is located either at zero or at nonzero external momentum $|{\bf q}|=q$.} \label{alphaq}
\end{figure}

In the FFLO phase, the Cooper pairs carry a net momentum. So for the transition to the inhomogeneous superfluid phase, we consider the ladder diagram where the bosonic pairing fields carry nonzero momentum $\mathbf{q}$. The expression for the amplitude of this diagram reads
\begin{eqnarray}
\nonumber&\alpha(\mathbf{q})=-\frac{1}{U}+\sum_{\bf k}\left(\frac{n_{\text F}(\xi_{\uparrow,{\bf q-k}})+n_{\text F}(\xi_{\downarrow,{\bf k}})-1}{\xi_{\uparrow,{\bf q-k}}+\xi_{\downarrow,{\bf k}}}\right).
\label{extern}
\end{eqnarray}
For the onset of an FFLO phase, it is enough to study the dependence of this quadratic coefficient on system parameters. To find the location of the Lifshitz point, the first point where FFLO occurs, $\alpha({\bf q})$ is expanded in powers of ${\bf q}$
\begin{equation}
\alpha(\mathbf{q})=a_{0}+a_{1}\mathbf{q}^{2}+a_{2}\mathbf{q}^{4}+\ldots,
\label{external}
\end{equation}
which contains only even powers, because $\alpha({\bf q})$ is even in ${\bf q}$, see figure~\ref{alphaq}, and where the dots denote higher order powers in $\mathbf{q}^{2}$. If all coefficients $a_{i}$ are positive, $\alpha(\mathbf{q})$ has its global minimum at ${\bf q}=0$, and the corresponding phase transition is from the normal to a homogeneous superfluid phase, see the left panel in figure~\ref{alphaq}. Whereas, if $a_{1}$ is negative, $\alpha(\mathbf{q})$ has its global minimum at a nonzero momentum ${\bf q}$, right panel in figure~\ref{alphaq}. In the latter case, the phase transition is from a normal to an inhomogeneous superfluid phase. Comparing this with the Landau expansion of the thermodynamic potential in equation~(\ref{omexpansion}), we see that in the expansion of $\alpha(\mathbf{q})$, $a_{0}$ can be identified with the quadratic coefficient $\alpha$ and $a_{1}$ can be identified with the gradient coefficient $\gamma$. 
Thus from the ladder diagram with external momentum, an expression for the $\gamma$ coefficient can be found, namely
\begin{equation}\label{gamma}
\gamma=\left.\frac{\partial\alpha(\mathbf{q})}{\partial\mathbf{q}^{2}}\right|_{\mathbf{q}=0}.
\end{equation}
Physically, $\gamma$ can be interpreted as being proportional to the inverse effective mass of the Cooper pairs.

The above expression for $\alpha({\bf q})$ can actually be obtained from the FF thermodynamic potential, by differentiating it with respect to $|\Delta|^2$ and consequently taking $\Delta=0$, but not taking ${\bf q}$ to zero, and is equal to the gap equation in equation~(\ref{eq:gap}). But importantly (and as seen from the above explanation using Feynman diagrams) the expression for $\alpha({\bf q})$ is an expression for a position dependent order parameter containing a single wave vector ${\bf q}$. Namely in the above calculation, the only assumption made was that the bosonic field carries a net momentum, whereas no assumption was made on the dependence on position of the order parameter.

Also importantly, the Ginzburg-Landau expansion can only be applied close to a continuous phase transition. In practice, one thus always needs the full thermodynamic potential to check if this is the case. The Ginzburg-Landau expansion is therefore expected to apply at the transition from normal Fermi liquid to the FFLO state, but not for the transition from FFLO to BCS which happens for smaller values of imbalance (chemical potential difference). 

\subsection{The FFLO state and the Bloch's theorem}
\label{subsec:bloch_theorem}

\begin{figure*}
	\centering
	\includegraphics[width=0.95\textwidth]{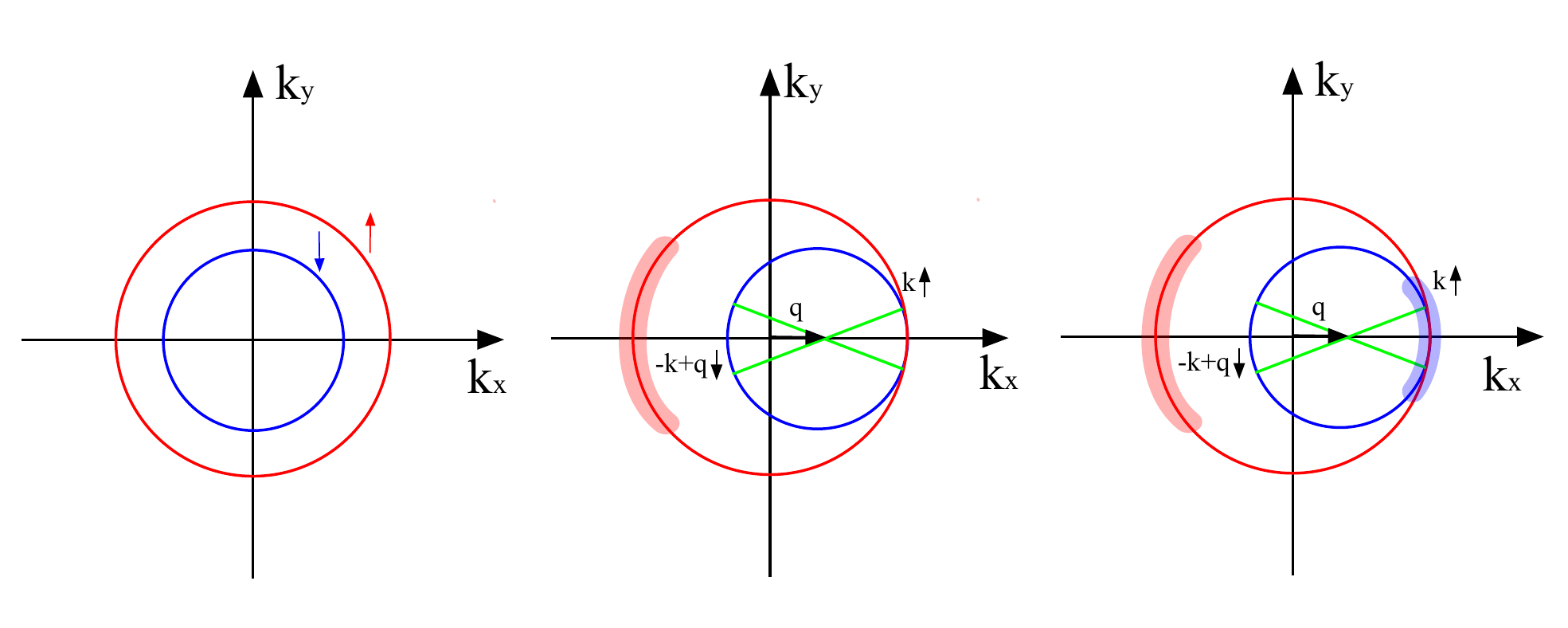}
	\caption{Left: A non-interacting density-imbalanced Fermi gas has different sizes of Fermi surfaces for the two components. Middle: In the FFLO state, pairing occurs in such a way that the minority Fermi surface can be thought of as shifted by the FFLO wave vector $\mathbf{q}$. The green lines show examples of up and down spin particles that pair. The pairs now carry the momentum $\mathbf{q}$. This is compensated by the momenta of the unpaired majority particles at negative momenta, the red region. Right: For some parameters, unpaired minority particles may also appear, see the smaller blue area. Bloch's theorem is still fulfilled, and the total net momentum remains zero.}
	\label{fig:BlochsTheorem}
\end{figure*}

The Bloch's theorem states that there is no net current in the ground state. Since the FFLO state is characterized by a non-zero momentum of the Cooper pairs, it is important to confirm whether it fulfils the Bloch's theorem. We use the following definition for the net momentum contained in the FFLO state:
\begin{eqnarray}
\mathbf{P} &=& \sum_{\bf k} \mathbf{k}
\left\langle\hat{c}_{\uparrow,{\bf k}}^\dagger
\hat{c}_{\uparrow,{\bf k}}\right\rangle + \sum_{\bf k} \mathbf{k}
\left\langle\hat{c}_{\downarrow,{\bf k}}^\dagger
\hat{c}_{\downarrow,{\bf k}}\right\rangle  \nonumber \\
&=& \sum_{\bf k} \mathbf{k}
\left\langle\hat{c}_{\uparrow,{\bf k}}^\dagger
\hat{c}_{\uparrow,{\bf k}}\right\rangle + \sum_{\bf k} (\mathbf{q} - \mathbf{k})
\left\langle\hat{c}_{\downarrow,{\bf q}-{\bf k}}^\dagger
\hat{c}_{\downarrow,{\bf q}-{\bf k}}\right\rangle . \label{definition_net_momentum}
\end{eqnarray}
Periodic boundary conditions were assumed. Now one can apply the inverse of the Bogoliubov transformation equation~\eqref{inverseBogoliubov} to obtain
\begin{eqnarray}
  \mathbf{P} &=&
  	 	\label{eq:net_momentum}
	\sum_{{\bf k}}\mathbf{k} \left\{ u_{\bf k,q}^2 n_F(E_{+,{\bf k,q}}) + v_{{\bf k,q}}^2 n_F(-E_{-,{\bf k,q}})\right\}\\
	 &+& \sum_{{\bf k}}(\mathbf{q}-\mathbf{k})\left\{ u_{\bf k,q}^2 n_F(E_{-,{\bf k,q}}) + v_{\bf k,q}^2 n_F(-E_{+,{\bf k,q}})\right\}\nonumber .
\end{eqnarray}
At zero temperature this becomes
\begin{eqnarray}
  \mathbf{P} =
  \label{eq:net_momentum_zeroT}
	\sum_{{\bf k} (E_{+,{\bf k,q}},E_{-,{\bf k,q}}>0)} \mathbf{q} v_{{\bf k,q}}^2  
	+ \\\sum_{{\bf k}(E_{+,{\bf k,q}}>0,E_{-,{\bf k,q}}<0)} \mathbf{q}-\mathbf{k} 	
	+ \sum_{{\bf k}(E_{+,{\bf k,q}}<0,E_{-,{\bf k,q}}>0)} \mathbf{k} .\nonumber 
\end{eqnarray}
The first term corresponds to paired particles and the momentum $\mathbf{q}$ they carry. The terms $\mathbf{k}$ and $-\mathbf{k}$ cancel in the paired case. Note that the case where both eigenenergies would be negative cannot occur, as seen from the definition~\eqref{eq:qp_energies}. The other two terms in equation~\eqref{eq:net_momentum_zeroT} describe the momenta of the unpaired particles. In general, the three contributions cancel each other to fulfill  Bloch's theorem. The last term corresponds to the unpaired majority particles, thus one can immediately see that an area of unpaired particles should form to momenta opposite  the direction of the FFLO wave vector $\mathbf{q}$ to cancel the first term. The case where only majority particles are unpaired corresponds to the left panel of figure~\ref{fig:FFLO12} where the dispersions cross zero (gapless excitations, i.e., free particles) only for one sign of momenta. But for some parameter regimes, there can be also free (gapless) minority particles, corresponding to the second term in the above equation~\eqref{eq:net_momentum_zeroT}. These are visible in the eigenvalues of figure~\ref{fig:FFLO12}, the right panel, as crossings of zero at both positive and negative momenta. The formation of unpaired regions is illustrated schematically in figure~\ref{fig:BlochsTheorem}, and via numerical data in figure~\ref{Koponen_FFmomentum}. 

One can easily calculate the net momentum equation~\eqref{eq:net_momentum} or equation~\eqref{eq:net_momentum_zeroT} from a numerical solution of the problem and check whether it is zero. The zero temperature case can be approached analytically. Bloch's theorem was central in the original work of Fulde and Ferrell \cite{FF}: it was used to select the ground state from a larger class of possible solutions. A further analytical treatment was done in early classic work by Takada and Izuyama \cite{Takada1969}. They derived analytical formulas for the areas in momentum space where the unpaired particles exist, and they expressed the system energy in terms of these formulas. In the calculation, an approximation was made by neglecting terms of the order $q^2/k_F$, that is, it was assumed that the FFLO wave vector's absolute value is small compared to that of the Fermi wave vector. The authors then showed that the net current defined analogously to equation~\ref{definition_net_momentum} becomes proportional, up to constants, to the derivative of the total energy with respect to the FFLO wave vector:
\begin{equation}
\mathbf{P} \propto \frac{\partial E}{\partial q} .
\end{equation}   
Bloch's theorem is therefore equivalent to the demand that the FFLO solution is an extremum of energy. Fulfilling Bloch's theorem is thus a necessary condition for the existence of an FFLO solution that can minimize the system energy. However it is not sufficient because one further needs to show that the extremum is actually a minimum. The connection between the current being zero and minimizing the energy is also shown in detail in \cite{sheehy_becbcs_2007}.

A similar analysis can be performed in the case where the order parameter ansatz is more complicated than the FF single $\mathbf{q}$ case. When the order parameter ansatz contains two opposing momenta, as with the LO ansatz, it is quite intuitive to assume that there is no net current. In general, one should always confirm whether the solution for a more complex FFLO state fulfills Bloch's theorem.

The FF state breaks time-reversal symmetry (TRS) because it is characterized by a single wavevector $\mathbf{q}$ with a definite direction. It might seem counter intuitive that nevertheless net current is absent. However note that there actually exists spin current in the FF state: the pairs with wavevector $\mathbf{q}$ have the same amount of up and down particles while the unpaired particles that carry momenta that cancel $\mathbf{q}$ are mostly or only majority particles, leading to net spin current.

\subsection{Luttinger's theorem and Fermi surfaces}

Luttinger's theorem states that a volume enclosed by a Fermi surface is directly proportional to the particle density. This is easy to see in the case of non-interacting, zero temperature Fermi gas. In case of the BCS state, the Fermi surface disappears. This is consistent with Luttinger's theorem because in the BCS state, the particle number (density) fluctuates, only its mean value is fixed. This is connected to the breaking of the $U(1)$ symmetry and appearance of a condensate phase. Now in FFLO and Sarma/BP states, there are gapless excitations, that is, Fermi surface(s). One may ask, does this somehow contradict Luttinger's theorem, since these are superfluids and the particle number fluctuates? But remember that it is the number of {\it pairs} that fluctuates in the superfluid: actually the difference $n_\uparrow - n_\downarrow$ remains constant. The volumes enclosed by Fermi surfaces in the FFLO and Sarma states are related to particle density {\it difference}. 

Sachdev and Yang~\cite{sachdev_fermi_2006} considered a model where two fermion species can form a boson (which can be understood as a molecule or a Cooper pair) and rigorously showed  that the particle number difference of the two species is given by
\begin{equation}
n_\uparrow - n_\downarrow = \frac{1}{(2\pi)^d} \left( \Omega_\uparrow - \Omega_\downarrow \right) , 
\end{equation}
where $\Omega_\uparrow$ and $\Omega_\downarrow$ are the volumes of the two Fermi surfaces and $d$ the dimension. This holds also for vanishing Fermi surface volumes, that is, there can be two, one or zero Fermi surfaces depending on the phase of the system. In case of a periodic order parameter, such as in FFLO state, the Fermi surface volumes are 
well-defined modulo the Brillouin zone volume. Quantum phase transitions are associated with the change in the number of Fermi surfaces when the system parameters are varied. In a recent article \cite{Pieri2017}, the validity of the Luttinger's theorem was proven for any approximate theory based on a conserving approximation in case of spin-density imbalance.

\subsection{Connection to the repulsive Hubbard model: the FFLO state and striped phases vs.\ d-wave superconductors} \label{U-Umapping}

In bipartite lattices, it is possible to make simple particle-hole transformations that connect phenomena that occur for attractive and repulsive interactions \cite{emery_theory_1976}. This provides an interesting connection between the FFLO state in lattices and important open questions in high-$T_c$ superconductivity. From experiments, the order parameter in many high-$T_c$ materials is known to have d-wave symmetry. The repulsive Hubbard model is viewed as a minimal model that could perhaps describe high-temperature superconductors, such as  cuprates. Analytical solutions of the Hubbard model are not available, and numerical methods like quantum Monte Carlo are plagued by the numerical sign problem in the regime where superconducting phases would occur, namely away from half filling. Thus the true ground state over the whole parameter space is not known. 

\begin{figure}
	\centering
	\includegraphics[width=0.95\columnwidth]{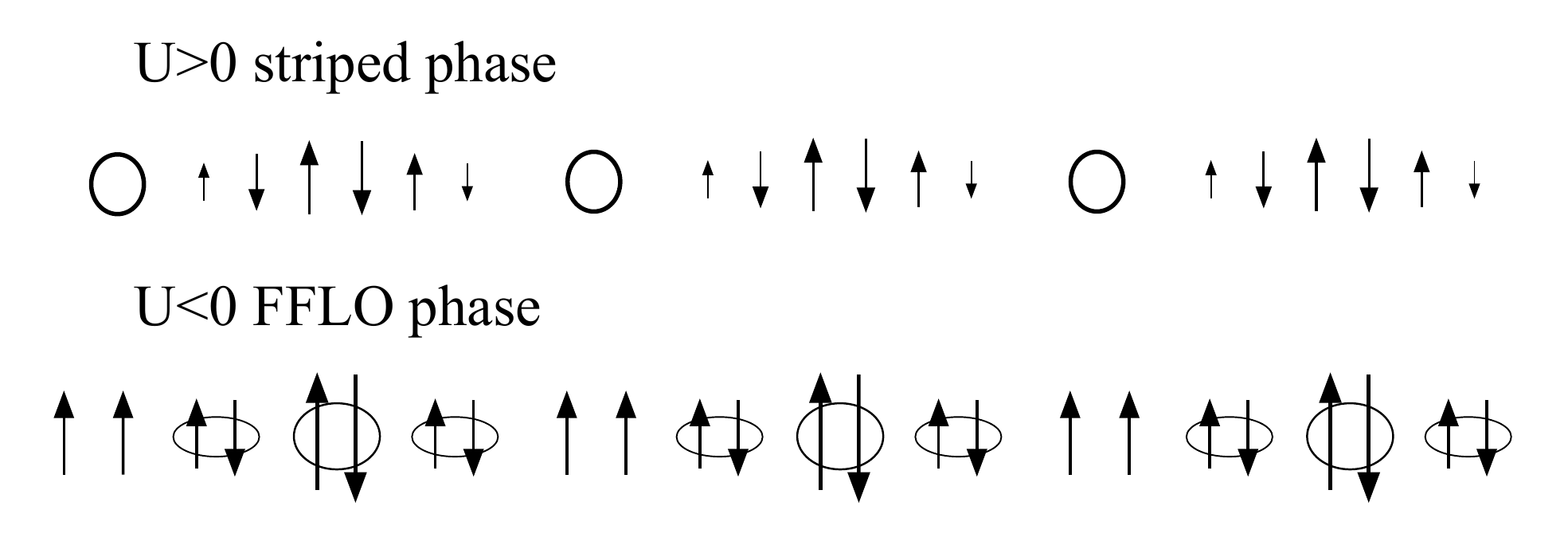}
	\caption{A simplified schematic of the striped phase in the repulsive Hubbard model and the FFLO phase in the attractive one. These phases are equivalent within the particle-hole transformation. In the striped phase, antiferromagnetic ordering alternates with areas of hole doping, while in the FFLO phase, the pairing order parameter oscillates corresponding to the antiferromagnetic order, and hole stripes are replaced by areas of excess majority particles.}
	\label{StripeSchematic}
\end{figure}

At half filling, where quantum Monte Carlo can give reliable results, numerical simulations predict anti-ferromagnetic (AFM) order \cite{Dagotto1994,LeBlanc2015}. Away from half filling, several competing phases have been predicted. For instance, approximative numerical approaches predict a d-wave superfluid where the pairing involves correlations between particles on several neighbouring sites. Such pairing cannot be described by a mean-field theory within the basic Hubbard model that only has on-site interactions. A d-wave superfluid is of course exciting, considering the observed d-wave symmetry of the order parameter in many real materials. The d-wave superfluid may compete or co-exist with the AFM order in the Hubbard model \cite{Dagotto1994,Demler2004}, but not only with that. There are suggestions that also a so-called striped phase could exist \cite{Votja2009}. In the striped phase, the amount of doping varies spatially, on top of AFM order, see figure~\ref{StripeSchematic}. Such a phase can be described within mean-field theory. It is not known which of these phases are true ground states at each area of the parameter space, and whether there are regimes where some or all of them co-exist. Co-existence of AFM and superconducting order has been studied, see for instance \cite{Lichtenstein2000,Capone2006} and references therein, as well as the interplay between stripes and d-wave superconductivity (\cite{Himeda2002,Zheng2016} and references therein). Recently cluster DMFT calculations that included the possibility of AFM order, d-wave superconductivity and the striped phase simultaneously were performed.   
The results show coexistence of the d-wave superconductor and striped phases for a large doping region and a smaller area of a uniform d-wave superfluid~\cite{Vanhala2017}. Existence of the striped phase for one value of doping was also recently shown by a study combining several different numerical approaches~\cite{Zheng2017}, and for a Lieb-lattice geometry~\cite{kumar_temperature_2017}.

As we explain below, the striped phase is connected to the FFLO state via a particle-hole mapping \cite{hirsch_two-dimensional_1985,schulz_domain_1989,scalettar_phase_1989,salkola_inhomogeneous_1998}. Therefore, studies of the FFLO state in optical lattices will provide new knowledge on the coexistence of superfluidity and magnetization, and profoundly important information about the Hubbard model in general and its relevance in explaining high-$T_c$ superconductivity.   

Let us consider the Hubbard model in equation~\eqref{eq:FFLO_Hamiltonian_position_space2}, but with the terms $\sum_{\sigma{\bf r}}\left(\frac{\mu^0_\sigma}{2} + \frac{U}{4} \right) $ and $-\sum_{\sigma{\bf r}}\frac{U}{2}\hat{c}_{\sigma{\bf r}}^\dagger\hat{c}_{\sigma{\bf r}}$ added to the Hamiltonian. The first one is a constant, and the second means a trivial shift of the chemical potentials. This produces the so-called particle-hole symmetric Hubbard Hamiltonian
\begin{eqnarray}
    \label{eq:Hamiltonian_particleholesymm}
  \widehat{H}_=&-\sum_{\sigma}
\sum_{\langle{\bf r,r'}\rangle}
t_{\bf ij}\hat{c}_{\sigma{\bf r}}^\dagger\hat{c}_{\sigma{\bf r'}}
-\sum_{\sigma{\bf r}}\mu^0_\sigma \left( \hat{c}_{\sigma{\bf r}}^\dagger\hat{c}_{\sigma{\bf r}} - \frac{1}{2} \right)\\
\nonumber &+U\sum_{{\bf r}} \left(\hat{c}_{\uparrow{\bf r}}^\dagger\hat{c}_{\uparrow{\bf r}} - \frac{1}{2}\right) \left( \hat{c}_{\downarrow{\bf r}}^\dagger \hat{c}_{\downarrow{\bf r}}  - \frac{1}{2}\right) .
\end{eqnarray}
This Hamiltonian possesses certain symmetries with respect to the repulsive and attractive interactions in bipartite lattices. We first consider the case of having the same chemical potential and particle numbers for $\uparrow$ and $\downarrow$ particles. There exists a unitary transformation $U_1$ (the label one is just to make a difference to the
interaction energy $U$ of the Hubbard model) that connects the two at half filling, namely
\begin{eqnarray}
	U_1^\dagger c_{i\uparrow} U_1 = \epsilon (i)d^\dagger_{i\uparrow} \\
	U_1^\dagger c_{i\downarrow} U_1 = d_{i\downarrow}  ,
\end{eqnarray}
where $\epsilon (i) = 1$ for one sublattice of the bipartite lattice and $\epsilon (i) = -1$ for the other. With this transformation
\begin{equation}
	U_1^\dagger \hat{H}_\mathrm{half-filled} (U) U_1 = \hat{H}_\mathrm{half-filled} (-U) ,
\end{equation}
which means that the Hamiltonians for both signs of $U$ must have the same energy spectrum and the same form of the ground state. 
Therefore at half filling, if one knows the ground state for $U$, the ground state for $-U$ can be straightforwardly obtained. It turns out that the usual BCS superfluid for attractive interactions maps to a N{\'e\'e}l-ordered anti-ferromagnet on the repulsive side. If strong attractive interactions are considered,
on-site pairs are formed and they Bose-condense; there exists a BCS-BEC crossover in the lattice. The BEC of on-site pairs corresponds to a Mott antiferromagnet
at strong interactions. The existence of on-site pairs at higher temperatures, before one reaches the superfluid below the critical temperature, corresponds to having a Mott state before reaching a Mott antiferromagnet. The identical behaviour at negative and positive interaction $U$ has been seen, for instance, in expansion experiments of ultracold Fermi gases~\cite{Schneider2012,Kajala2011}. Importantly in an experimental situation where the symmetry of the system may be broken by some unwanted or desired features, the mapping may not be exact.

Away from half filling and for differing chemical potentials and particle densities, there is still a mapping, but it is more complicated. Now it turns out that the negative and positive $U$ Hubbard models map to each other in such a way that spin imbalance, that is, different chemical potentials $\mu_\uparrow \neq \mu_\downarrow$ in one model maps to chemical potential away from half filling in the model with opposite sign of $U$. To be precise, with the notation $\mu\equiv (\mu_\uparrow + \mu_\downarrow)/2$, $\Delta \mu \equiv (\mu_\uparrow - \mu_\downarrow)/2$ the mapping is the following: 
\begin{eqnarray}
U \longleftrightarrow - U \\
\mu \longleftrightarrow - \Delta \mu \label{mu_mapping}\\
\Delta \mu \longleftrightarrow - \mu  .  
\end{eqnarray}
Since the total density $\frac{1}{M}\sum_{{\bf r}}\langle\hat{c}_{\uparrow{\bf r}}^\dagger \hat{c}_{\uparrow{\bf r}}\rangle + \langle 
\hat{c}_{\downarrow{\bf r}}^\dagger \hat{c}_{\downarrow{\bf r}} \rangle = n$ ($M$ is the number of the lattice sites) and density difference 
$\frac{1}{M}\sum_{{\bf r}}\langle\hat{c}_{\uparrow{\bf r}}^\dagger \hat{c}_{\uparrow{\bf r}}\rangle - \langle 
\hat{c}_{\downarrow{\bf r}}^\dagger \hat{c}_{\downarrow{\bf r}} \rangle = \Delta n$ transform as $\frac{1}{M}\sum_{{\bf r}}
\langle 1 - \hat{d}_{\uparrow{\bf r}}^\dagger\hat{d}_{\uparrow{\bf r}}\rangle \pm \langle\hat{d}_{\downarrow{\bf r}}^\dagger\hat{d}_{\downarrow{\bf r}}\rangle$ one straightforwardly obtains
relations for the total densities in the mapping:
\begin{eqnarray}
n \longleftrightarrow 1 - \Delta n \\
\Delta n \longleftrightarrow 1 - n  \label{n_mapping}.
\end{eqnarray}  
Likewise the polarization $P=(n_\uparrow - n_\downarrow)/(n_\uparrow + n_\downarrow)=\Delta n/n$ maps as $P \longleftrightarrow (1-\Delta n)/(1-n)$. At half filling, $n_\uparrow + n_\downarrow=n=1$, $\mu=0$ and the relations can be further simplified. For instance assuming an attractive interaction system with finite polarization $P$, the situation where the FFLO state may occur, then at half filling for the attractive system (($\mu_{\uparrow \mathrm{attr}} + \mu_{\downarrow \mathrm{attr}})/2 = 0$, $n_\mathrm{attr}=1$) we have $\Delta \mu_\mathrm{attr} = -\mu_\mathrm{rep}$ and $1-P_\mathrm{attr} = n_\mathrm{rep}$. The essence of the particle-hole transformation is essentially unchanged when using the standard Hubbard Hamiltonian of the form in equation~\eqref{eq:FFLO_Hamiltonian_position_space2}, but then one has to take into account extra chemical potential shifts and constants appearing in the transformation when comparing actual numbers. The particle-hole symmetric Hamiltonian~\eqref{eq:Hamiltonian_particleholesymm} in contrast leads to the simple relations between chemical potentials and total densities as given in equations~\eqref{mu_mapping}--\eqref{n_mapping}. 

This means that the FFLO state in the attractive Hubbard model corresponds to the striped phase that has been predicted for the doped repulsive Hubbard model \cite{moreo_cold_2007}. Therefore, observing the FFLO state would address open questions, discussed above, related to the repulsive Hubbard model as well, with potentially great relevance to high temperature superconductivity.

In ultracold quantum gas (UQG) experiments, however, one has to be cautious with the connection between the repulsive and attractive Hubbard models because the lattice potential is typically superimposed by another trapping potential, for instance a harmonic one. This has important consequencies to the particle-hole mapping and to the possible equivalence of phases predicted for the attractive and repulsive models. This is thoroughly analysed by Ho, Cazalilla and Giamarchi \cite{ho_quantum_2009}. The basic ingredients of the Hubbard Hamiltonian are the hopping and the interaction terms as seen from equation~\eqref{eq:Hamiltonian_particleholesymm}; let us write here a generic form to the rest of the Hamiltonian and call it $\hat{H}_\mathrm{ext}$ (external potential), also including the chemical potentials:
\begin{eqnarray}
  \hat{H}_\mathrm{ext} = &\sum_i (V_i - \mu) (n_{i\uparrow} + n_{i\downarrow} - 1) \\
\nonumber  &- \sum_i h_i (n_{i\uparrow}- n_{i\downarrow}) ,
\end{eqnarray} 
where $V_i$ is the energy shift caused by a spatially varying external trapping potential and $h_i$ is a Zeeman field. If the particle number polarization is determined in the preparation of the gas and is uniform, as is usually the case in quantum gas experiments, $h_i=h$ does not depend on position and thus does not disturb the particle-hole mapping. But the term $\sum_i (V_i - \mu)$ does depend on position and has important consequencies. In the repulsive case, it might be difficult to obtain the desired doping uniformly since it is energetically favorable to have the center of the trap fully filled and the edges with lower density. On the attractive side with population imbalance, phase separation into a BCS-paired trap centre and polarized edges might be favored analogously to trapped continuum gases (see section~\ref{section_traps}). All this depends on parameters though, and one may hope to find parameter regimes where the trapping effects are negligible, or that the trap potential changes so slowly that particular phases can be found in different trap areas (c.f.\ discussion in section~\ref{1D3Dcrossover}). Also the newly achieved box potentials~\cite{meyrath_bose-einstein_2005,gaunt_bose-einstein_2013,Schmidutz_quamtum_joule2014,mukherjee_homogeneous_2016}
 offer interesting solutions for uniform gases and thus also straightforward applicability of the particle-hole mapping.

\section{Mean-field studies on FFLO in 2D and 3D optical lattices}
\label{MFstudies}

\begin{figure}
\centering
\includegraphics[width=0.95\columnwidth]{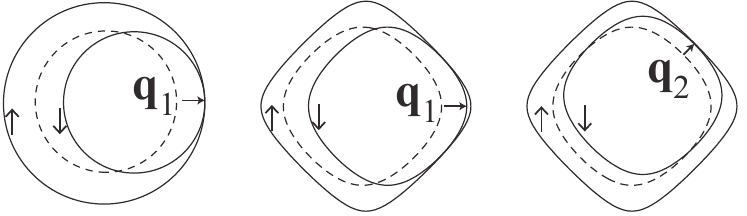}
\caption{Fermi surfaces in momentum space for an imbalanced Fermi mixture, where the $\uparrow$ component forms the majority component. The leftmost figure illustrates
Fermi surfaces in the continuum while others are in a lattice. Here ${\bf q}_1$ and ${\bf q}_2$ are different options for the FFLO wave vector. The match between the Fermi surfaces in the lattice for ${\bf q}_1$  is larger than in the continuum,
and furthermore the match is better with  ${\bf q}_1$ rather than  ${\bf q}_2$.
This nesting effect in the lattice causes FFLO to be more stable in lattices than in continuum~\cite{koponen_finite-temperature_2007}.}
\label{nesting}
\end{figure}

\begin{figure}
\centering
\includegraphics[width=0.45\columnwidth]{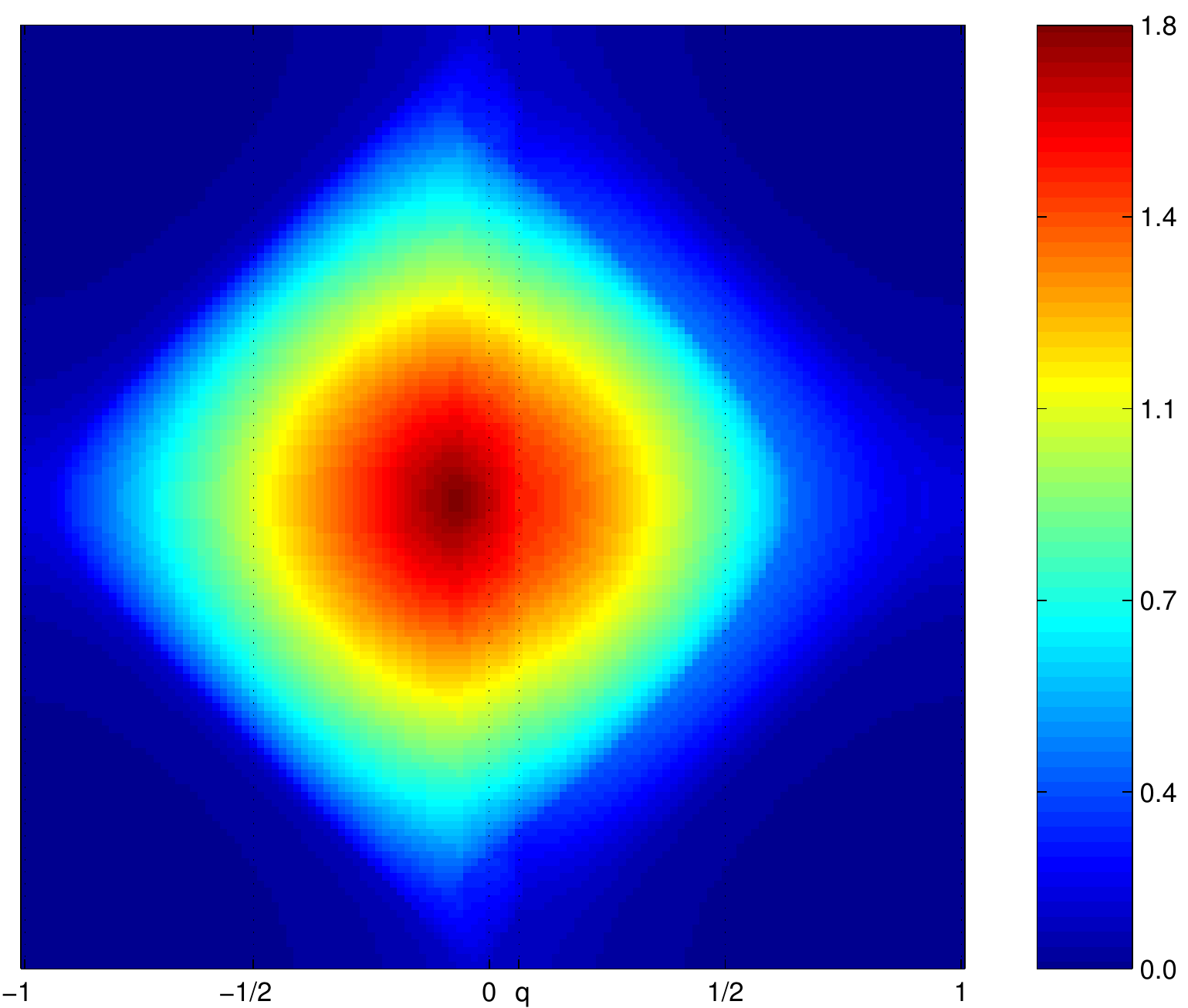}
\includegraphics[width=0.45\columnwidth]{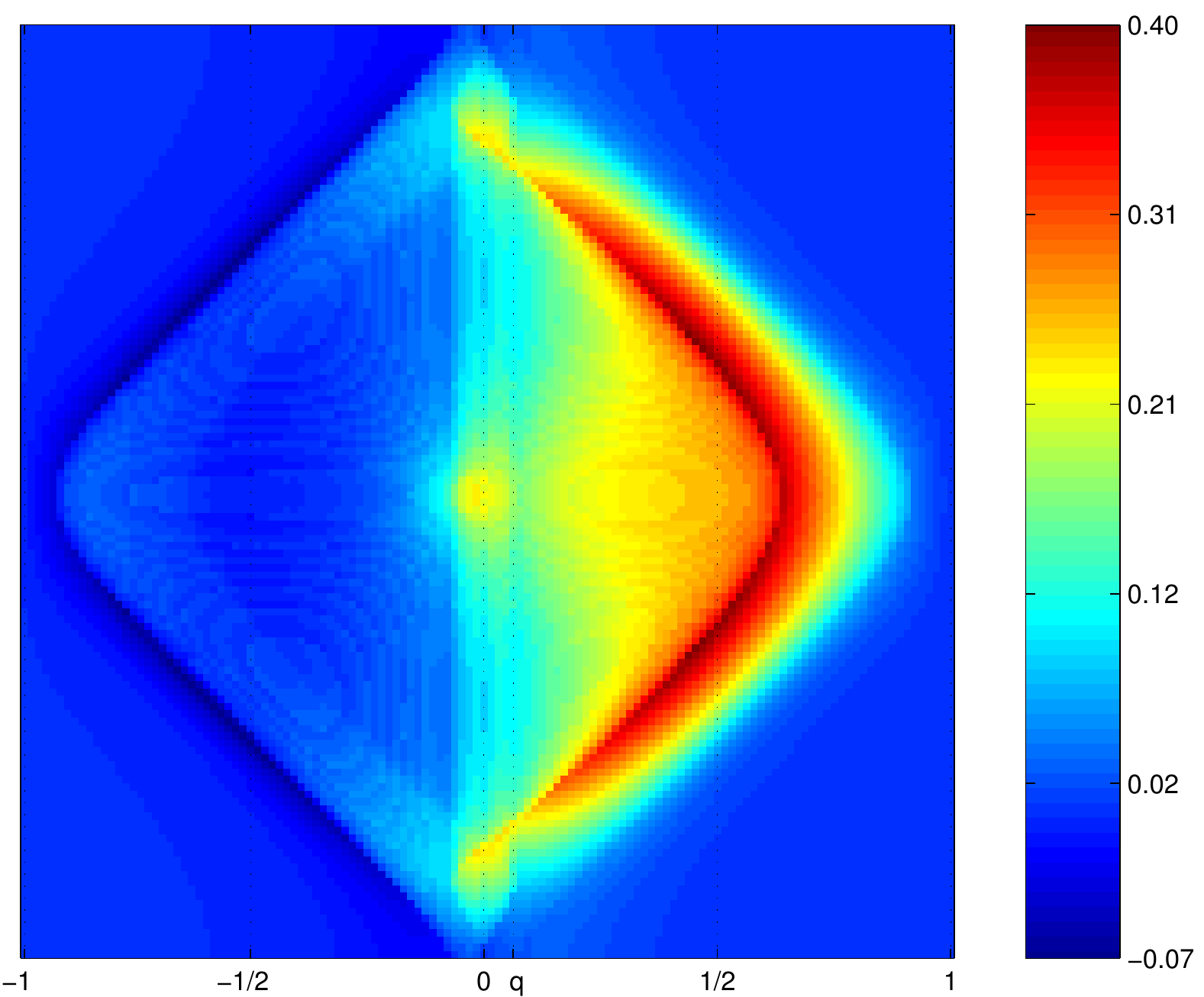}
\caption{The momentum distribution $f_\uparrow+f_\downarrow$ and $f_\uparrow-f_\downarrow$ (filling fraction in $k_x$-$k_y$ momentum space), integrated over the $z$-direction. Here the total filling fraction is $f=0.4$ and $P=0.15$. The state is FFLO with $q=2\pi/(64d)$ in the $x$-direction.
Re-used under Creative Commons Attribution CC BY license 3.0 from~\cite{koponen_fermion_2006}.}
\label{Koponen_FFmomentum}
\end{figure}

The mean-field theory of the FFLO state in two-component Fermi mixtures in optical lattices was considered by Koponen {\it et al.}~\cite{koponen_fermion_2006,KoponenThesis2008}. The authors found that both the Sarma and the FF state can form the ground state of the system. The Sarma state was found to be stable for fixed densities while the FF state is the energy minimum for fixed chemical potentials; this result is similar to the continuum case where fixed densities can stabilize the Sarma state~\cite{gubankova_breached_2003,he_loff-pairing_2006}. The authors of~\cite{koponen_fermion_2006} show that the parameter regime for which the system is in the FF state is relatively large compared to the continuum due to the nesting effect. Namely in the lattice, the overlap of the Fermi levels for the two components can become substantial by displacing the minority Fermi sea with the FF wavevector. This effect is illustrated in figure~\ref{nesting}. Furthermore the authors present the momentum distributions for the FF superfluid phase, see figure~\ref{Koponen_FFmomentum} where this displacement can be seen. It is shown that the system has no total momentum because the momentum distributions of the individual components cancel the effect of the FF wavevector ${\bf q}$.

\begin{figure}
\centering
\includegraphics[width=0.95\columnwidth]{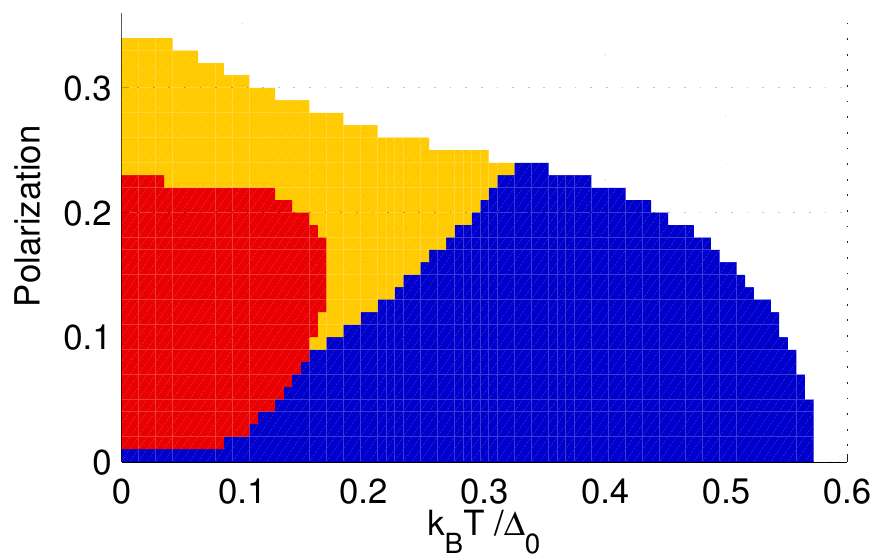}
\caption{The phase diagram of the imbalanced Fermi gas in a lattice. Colors: FFLO=yellow, Sarma/breach pair=blue, phase separation=red, normal=white. The average filling is $f=0.2$ atoms/lattice site in each component, $J=0.07E_R$, and $U=-0.26E_R$, $E_R=\hbar^2\k^2/2m$ is the recoil energy, $k=2\pi/\lambda$ and $\lambda=1030$nm. Here $\Delta_0$ (given in terms of temperature) means the gap at $T=0$ and $P=0$. 
The $T_c$ at $P=0$ is 41 nK. Reproduced with permission from~\cite{koponen_finite-temperature_2007}.}
\label{Koponen_PhaseDiagram}
\end{figure}

In the papers following this first study by the same authors \cite{koponen_finite-temperature_2007,koponen_fflo_2008}, the complete phase diagrams were determined in one-, two- and three-dimensional lattices. In a Letter~\cite{koponen_finite-temperature_2007} from 2007, the three-dimensional case was considered and the phase diagram as a function of temperature and polarization at fixed filling was shown to contain an FF region. Koponen {\it et al.}~\cite{koponen_finite-temperature_2007}
also computed the phase diagram as a function of polarization and filling at zero temperature and showed it has has a  substantial FFLO region, see figure~\ref{Koponen_PhaseDiagram}. Importantly, they found that even after considering
phase separated states (phase separated normal phase and a BCS phase), FFLO states cover a large area of the phase diagram.
 The broad range of parameters for which the FFLO state was the ground state is in striking contrast to three dimensional homogeneous systems, where the FFLO exists only in an exceedingly narrow region of the parameter space~\cite{FF,LOeng,sheehy_bec-bcs_2006,radzihovsky_quantum_2009}.  In reference~\cite{koponen_fflo_2008}, Koponen {\it et al.} extended a mean field exploration of the FFLO state in optical lattices to different dimensions and discussed the role of the Hartree corrections. In all dimensions, nesting and the Van Hove singularity plays an important role, explaining the large regions of the parameter space where FFLO is the ground state. Momentum distributions and density-density correlations were also computed.

By using a more elaborate variational mean-field approach based on the 
Bogoliubov-de Gennes (BdG) method and including Hartree corrections, Loh {\it et al.}~\cite{loh_detecting_2010} 
studied the stability and observable signatures of the LO phase in a cubic lattice.
In their approach, six variational parameters were self-consistently computed at each site: complex valued order parameter, chemical potential, and three Zeeman fields. In a cubic lattice, 
the LO phase is stable in a much broader range of parameters than expected in the continuum. Furthermore, they computed that the LO phase could exist in a lattice if
entropy per particle is below $\sim 0.5\, k_B$. If the lattice was anisotropic, the stability of the LO phase would be enhanced further. While the pairing energy is small compared to the overall energy, compared to the FF phases that break the time-reversal symmetry, the LO phases could have a pairing energy that exceeds the pairing energy of the FF phases even by a factor of $50$.

\begin{figure}
\centering
\includegraphics[width=0.95\columnwidth]{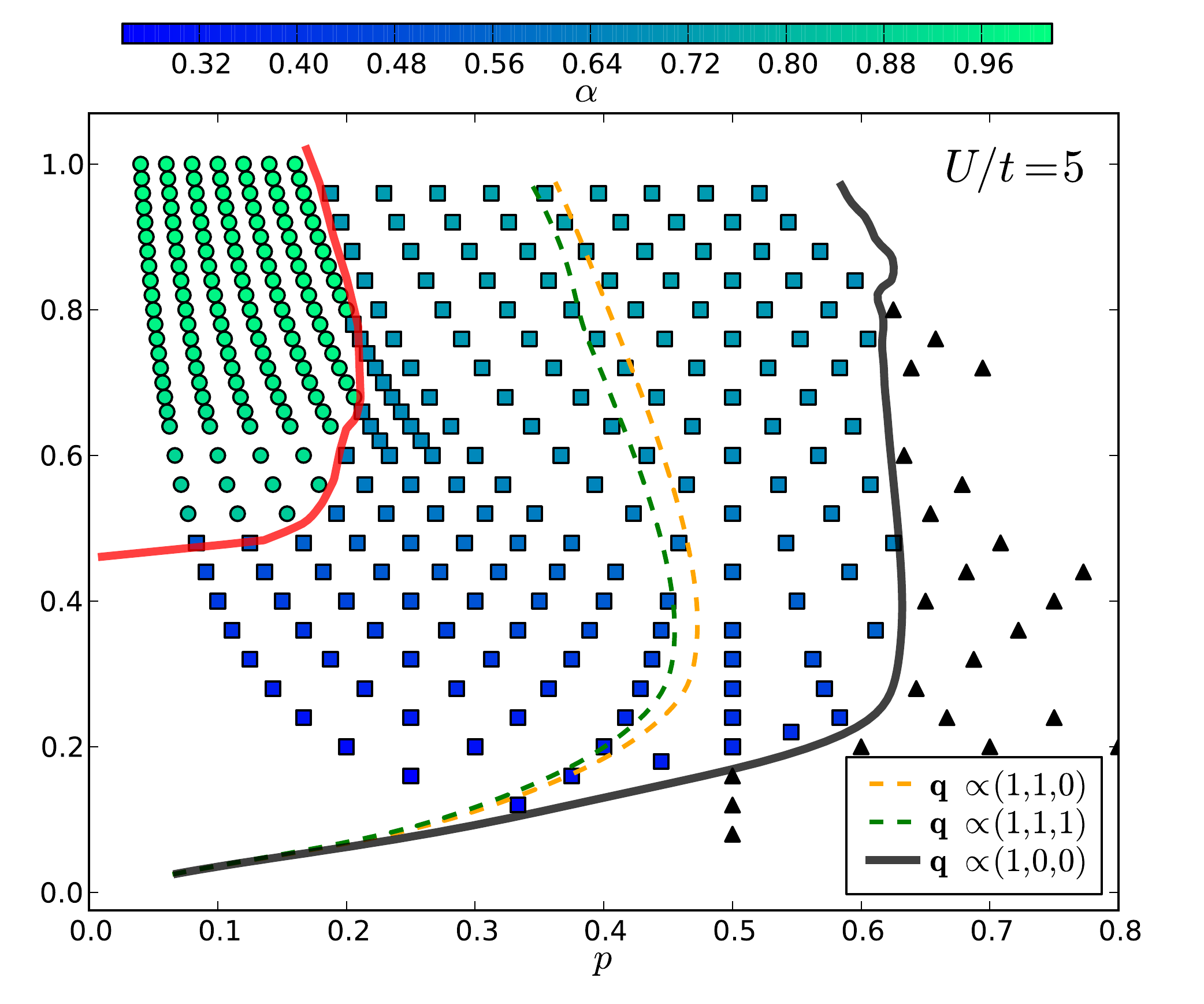}
\caption{The LO phase diagram in a three-dimensional square lattice as a function of
polarization and density.
Circles represent diagonal order with ${\bf q}\propto (1,1,1)$, squares are LO states with
${\bf q}\propto (0,0,1)$, while triangles are regions without order parameter. The color scale
indicates the total density of the excess spin within each nodal region.
Re-used under Creative Commons Attribution CC BY license 3.0 from~\cite{rosenberg_fflo_2015}.}
\label{Rosenberg_q_transitions}
\end{figure}

A mean-field Hartree-Fock-Bogoliubov theory was applied by Chiesa and Zhang
as well as by Rosenberg {\it et al.}~\cite{chiesa_phases_2013,rosenberg_fflo_2015}
to study a polarized Fermi gas in a three-dimensional square lattice. Without targeting states of specific form and via extensive numerical work,
they found that the LO state with a single ${\bf q}$ is energetically favored and intriguingly that the favored direction of the 
LO state ordering can change as the interaction strength is varied. In particular,
a "diagonal" order with ${\bf q}\propto (1,1,1)$ can become favored over
${\bf q}\propto (0,0,1)$- order as interactions become stronger and the system becomes
more strongly filled. An example of the relevant phase diagram is shown in
figure~\ref{Rosenberg_q_transitions} for a relatively large interaction strength. Diagonal
order is indicated by green circles and appears for high densities and relatively small polarizations.

\begin{figure}
\centering
\includegraphics[width=0.95\columnwidth]{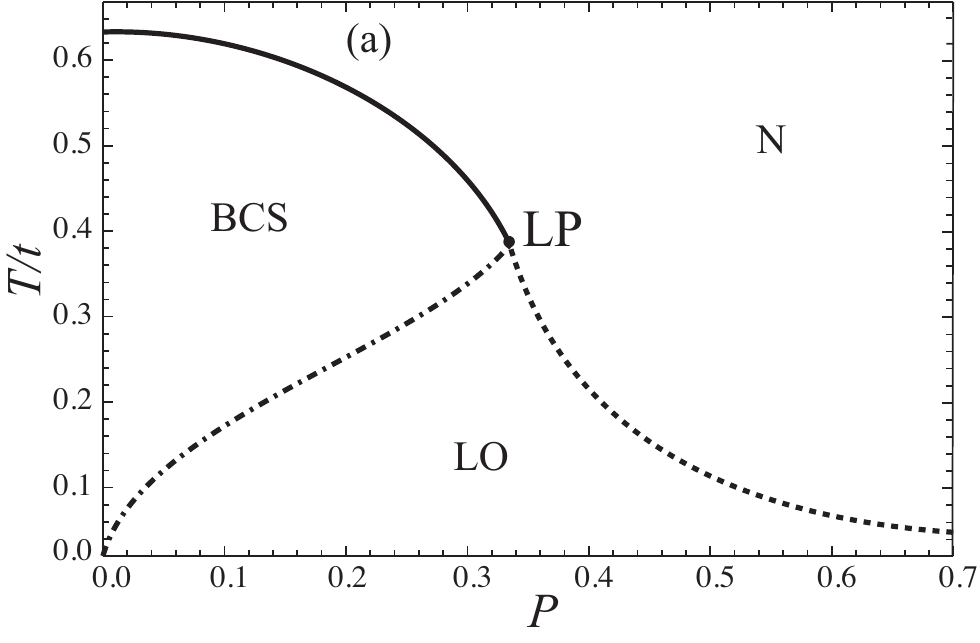}
\includegraphics[width=0.95\columnwidth]{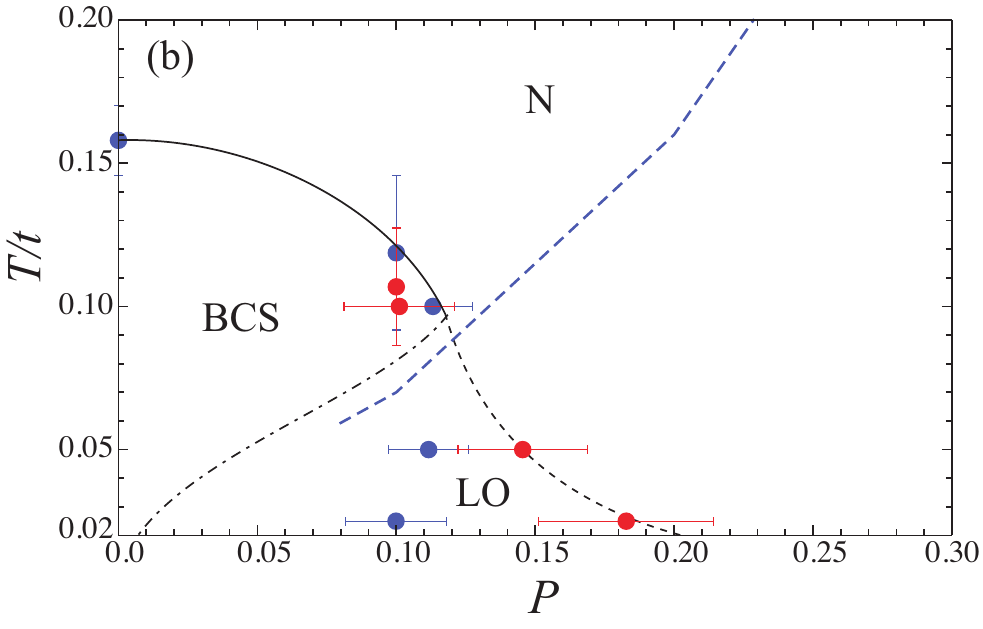}
\caption{(a) Mean-field phase-diagram at quarter filling as a function of polarization and temperature
for a two-component Fermi gas in a two-dimensional square lattice. (b) Same as (a) except that temperature  and polarization of the mean-field results are rescaled and compared with diagrammatic Monte Carlo results from reference~\cite{gukelberger_fulde-ferrell-larkin-ovchinnikov_2015}. Temperature was rescaled by the ratio of the mean field and quantum Monte Carlo critical temperatures at $P=0$ while polarizations were rescaled with the ratio of polarizations at a low temperature of $T=0.02\, t$, where $t$ is the tunneling strength for the nearest neighbor hopping. Reproduced with permission from~\cite{baarsma_larkin-ovchinnikov_2016}.}
\label{Baarsma_TransitionLines}
\end{figure}

In~\cite{baarsma_larkin-ovchinnikov_2016}, an explicit comparison was made between the FF and the LO ansatz within mean-field theory in a two-dimensional square lattice and, in agreement with earlier studies~\cite{sheehy_becbcs_2007,loh_detecting_2010,xu_competing_2014}, the LO state is energetically more favorable than the FF ansatz. Moreover in contrast
to results in a three-dimensional system~\cite{rosenberg_fflo_2015},
it was concluded that it is energetically more favorable to have the LO wave vector ${\bf q}$ along one of the lattice vectors, as compared to a LO ${\bf q}$ that makes a 45${}^o$ angle with the lattice vectors. In the LO state the translational symmetry is broken, which is reflected in the particle densities. An important comparison between between the mean-field phase diagram and a phase diagram obtained by quantum Monte Carlo calculations~\cite{gukelberger_fulde-ferrell-larkin-ovchinnikov_2015} was also made in~\cite{baarsma_larkin-ovchinnikov_2016}. Interestingly the transition lines agree very well if the mean-field temperatures and polarizations are simply scaled by a constant, see figure~\ref{Baarsma_TransitionLines}.

A somewhat different problem, intermediate between continuum and lattice results, was studied by Devreese {\it et al.}~\cite{devreese_resonant_2011,devreese_competition_2011}. They considered a three dimensional system where a one-dimensional optical lattice was imposed
along the $z$-direction. By studying the mean-field theory as a function of interaction strength and lattice parameters, they found that the FFLO state can persist to higher spin imbalance when the lattice period was reduced.

Reduced dimensionality is known to enhance possibilities for FFLO states. Using
a mean-field theory, with a possibility of also pair tunneling, Sun and
Bolech~\cite{sun_pair_2013} mapped the behavior of a system with coupled
one-dimensional tubes. When barriers separating tubes were high, the FFLO state appeared
in the center of the tubes. As barriers were lowered, tunneling rate increased, and
the central FFLO state gave way to a central BCS-like state.
Mean field studies in a 1D lattice found in addition to the FFLO state the possibility of an $\eta$ phase, that is, a phase where the FFLO wave vector is at the first Brillouin zone edge~\cite{ptok_critical_2017}.

\section{Mass-imbalanced Fermi gases in lattices}
\label{section_mass_imbalance}

It is also possible to have mixtures of fermions with different masses, and such mixtures 
have indeed been experimentally realized in ultracold atom experiments~\cite{Taglieber2008,wille_exploring_2008,voigt_ultracold_2009,naik_feshbach_2011,hansen_production_2013}. 
Mass imbalance causes some important
physical changes. First, having equal chemical potentials does not mean equal densities in the mass-imbalanced limit. Second, the dispersions for the two components are different, meaning that already in the population balanced case there is  a Fermi level mismatch.

Qualitatively it is easy to see that mass imbalance and chemical potential
imbalance can give rise to similar effects. For an ideal gas at zero temperature
in a continuum, we have
\beq
\mu_\sigma=\frac{\hbar^2k_{F,\sigma}^2}{2m}
\enq
and consequently $k_{F,\sigma}=\sqrt{2m\mu_\sigma}/\hbar$. Now if the chemical potentials differ but the masses are the same, Fermi surfaces are separated by
\beq
k_{F,\uparrow}-k_{F,\downarrow}=\frac{\sqrt{2m}}{\hbar}\left(\sqrt{\mu_\uparrow}
-\sqrt{\mu_\downarrow}\right).
\enq
On the other hand, if the masses differ but the chemical potentials are the same
\beq
k_{F,\uparrow}-k_{F,\downarrow}=\frac{\sqrt{2\mu}}{\hbar}\left(\sqrt{m_\uparrow}
-\sqrt{m_\downarrow}\right).
\enq
Mismatch in Fermi surfaces can be achieved either by changing
the chemical potentials or by changing the particle masses. Similar arguments also
apply  in the lattice. For example for tight binding dispersion
in a one-dimensional lattice, the connection between chemical potential
and $k_{F,\sigma}$ is given by
$2J_\sigma\cos(k_{F,\sigma})=\mu_\sigma$ and inverse of 
the hopping strength $1/J_\sigma$ plays an equivalent role as a particle mass.
However the analogue in Fermi momentum mismatch does not necessarily
mean that the pairing physics is the same in both cases.

BCS quasiparticle dispersions in a continuum system with unequal masses~\cite{baarsma_population_2010}
are
\begin{equation*}
 E_\pm= \pm \frac{\epsilon_km_-}{2}\mp{\tilde \mu}_- + \sqrt{\left(\frac{\epsilon_k m_+}{2}-\tilde{\mu}_+\right)^2+\Delta^2},
\end{equation*}
where $m_\pm=1 \pm m_\uparrow/m_\downarrow$,     $\tilde{\mu}_\pm=(\mu_\uparrow \pm \mu_\downarrow)/2$, and $\epsilon_{k}=\hbar^2k^2/(2m_\uparrow)$.
If masses are the same $m=m_\downarrow=m_\uparrow$ while chemical potentials are not, we have
\beq
E_{\pm}=\mp {\tilde \mu}_-+\sqrt{(\epsilon-\tilde{\mu}_+)^2+\Delta^2}.
\label{eq_uneqmu}
\enq
On the other hand, we could consider the case with unequal masses, but with equal
chemical potentials $\mu_\uparrow=\mu_\downarrow=\mu$.
In the weak coupling limit, the wavevectors for which
$\epsilon_km_+/2\approx \mu$ are the most relevant. Then quasiparticle dispersions can  be approximated as
\beq
E_\pm\approx \pm\mu\left(\frac{m_-}{m_+}\right)+\sqrt{(\epsilon_r-\mu)^2+\Delta^2},
\label{eq_uneqmass}
\enq
where
\beq
\epsilon_r=\frac{m_+}{2m_\uparrow}\hbar^2k^2=\hbar^2k^2/2M_r\nonumber
\enq
and $M_r=(1/m_\uparrow+1/m_\downarrow)^{-1}$ is the reduced mass.
We then see that equations (\ref{eq_uneqmu}) and (\ref{eq_uneqmass})
are the same as long as we make the mappings
\beq
M_r\leftrightarrow m \nonumber,
\enq
\beq
\mu\leftrightarrow\tilde{\mu} \nonumber,
\enq
and
\beq
\mu\left(\frac{m_-}{m_+}\right)\leftrightarrow \frac{\mu_\downarrow-\mu_\uparrow}{2}.
\enq
This demonstrates that in this limit the phenomena expected in mass imbalanced mixtures
are similar to those in mixtures with unequal chemical potentials. However this argument
is only valid in the weak coupling limit (and in the low filling limit in lattices). In the strong coupling regime, the physics in these two cases is expected to be different. Note that the above argument did not consider the possibility of a FFLO state,
and more complex physics results when both mass and chemical potential imbalance 
are present. 

To physically realize an unequal mass mixture in a lattice, one could put unequal mass fermions in a lattice, but effective mass imbalance can also be introduced via different hopping parameters $J_{\sigma, \alpha}$ for the two different spin components. This is because the hopping parameters are inversely proportional to the (effective) masses of the particles, namely
\begin{equation}
\frac{J_\uparrow}{J_\downarrow}=\frac{m_\downarrow}{m_\uparrow}.
\end{equation}
As an example, a very promising prospect in this regard is the recent experiment presented in \cite{jotzu_creating_2015}, where a versatile method for creating state-dependent optical lattices by applying a magnetic field gradient modulated in time was demonstrated. Thus different hopping parameters can be engineered for different internal spin states.

Mass imbalanced mixtures in optical lattices in the context of FFLO states
were considered by Wang {\it et al.}~\cite{wang_quantum_2009}. 
Using a time-evolving block decimation (TEBD) technique they find that with increasing mass imbalance, the FFLO region in the phase diagram decreases in size and eventually disappears for a large enough mass ratio.
The one-dimensional system was also considered by Dalmonte {\it et al.}~\cite{dalmonte_dimer_2012} by applying density matrix renormalization group and quantum Monte Carlo techniques to
study population and mass imbalanced Fermi mixtures in a trap.  
They found that mass imbalance helps to stabilize paired phases against spin-imbalance
and concluded that unequal mass mixtures are promising systems to observe
FFLO superfluidity in an experimentally realistic system.

Pahl and Koinov~\cite{pahl_phase_2014} considered ${}^6$Li-${}^{40}$K mixtures in square optical lattices using a mean-field theory. They found a normal,  Sarma and  FF phase to be present in the polarization versus a temperature phase diagram. The FF region occurs for non-zero polarization, and the region is larger for a majority of heavier ${}^{40}$K atoms. While the literature on FFLO states in mass imbalanced systems
in lattices is not very broad, based on the results in continuum~\cite{wang_enhancement_2017}, there are reasons to believe that mass imbalance might enhance FFLO type pairing also in a lattice, motivating further studies on this topic.

\section{FFLO state in the Green's function formalism} \label{GreenFFLO}
  
Here we briefly present  how superfluids, such as the FFLO state, can be described using Green's functions. This is convenient for many purposes, such as obtaining the collective modes and determining the ground state by beyond-mean-field approaches. 
We  first define the Green's functions in the Matsubara formalism, which allows consideration of finite temperatures. The possibility of pairing is taken into account by using the Nambu formalism where the Green's function becomes a matrix with the off-diagonal elements accounting to pairing correlations (sometimes called anomalous Green's functions), while the diagonals are the usual Green's functions. Forms of the Green's functions in the Hartree-Fock-Gor'kov mean-field approximations are given and are shown to lead to eigenenergies and coherence factors of the same form as obtained in section~\ref{BasicFFLOTheoryLattice}. After introducing the formalism, we proceed in 
section~\ref{CollectiveSection} to describe interactions between quasiparticles and thereby obtain collective modes of the system. In section~\ref{Beyond-mean-fieldSection} beyond mean-field studies of the FFLO state are discussed.  

The single particle Green's function is defined as follows 
in the finite temperature Matsubara Green's function formalism~\cite{mahan_many-particle_2000,bruus2004,vanLeeuwen2013}
\begin{equation}
  G(1,1^\prime)= - \left \langle T_\tau \left( {\hat c}(1){\hat c}^{\dagger}(1^\prime)\right) \right \rangle.
  \label{SPGreenDef}
\end{equation}
Here, we use the notation 1 for the variables $\rr_1\tau_1\sigma_1$,
denoting the position coordinate, the imaginary time of the Matsubara formalism, and the (pseudo-)spin, respectively, while $T_\tau$ is the time ordering operator
and the brackets denote the thermodynamic average
defined for the operator ${\hat O}$ as
\begin{equation}
\aver{{\hat O}} = \frac{Tr \left\{ e^{-\beta {\hat H}} {\hat O}  \right\} }{Tr \left\{ e^{-\beta {\hat H}} \right\} },
\end{equation}
where $\beta=\frac{1}{k_\text{B} T}$ is the inverse thermal energy 
with $k_\text{B}$ the Boltzmann constant and $T$ the temperature.
We describe the system in Nambu formalism with an extended spin index $\sigma\in\{1,2,3,4\}$ 
where ${\hat c}^{}_1={\hat c}_\uparrow$, ${\hat c}^{}_2={\hat c}_\downarrow$, ${\hat c}^{}_3={\hat c}^\dagger_1={\hat c}^\dagger_\uparrow$ and ${\hat c}^{}_4={\hat c}^\dagger_2={\hat c}^\dagger_\downarrow$.
Compared to equation~\eqref{SPGreenDef}, we moved the spin index from the argument of ${\hat c}$ to its subscript.
Thus the anomalous Green's functions with two creation or two annihilation operators
are also defined by equation~\eqref{SPGreenDef}.
These are essential for describing pairing correlations on the mean field level. The total Green's function can be also written in matrix form by taking the spin indices to correspond to the rows and columns of a matrix, as done explicitly in equation~(\ref{hirmu}) and the discussion following it.

The single particle Green's function has the following equation of motion
\begin{eqnarray}
\int G_0^{-1}(1,\bar{1})G(\bar{1},1^\prime)=
\delta(1,1^\prime)+\int \Sigma(1,\bar{1})G(\bar{1},1^\prime). \label{eom1}
\end{eqnarray}
The integral sign is a shorthand notation for summation over position and spin 
as well as integration over time, while 
variables of summation and integration are indicated by the overbar.
Here $G_0$ is the non-interacting Green's function, which is usually easy to 
determine exactly. The self energy $\Sigma$, for a general two-body interaction, is defined via

\begin{equation}
\Sigma(1,1^\prime)
=-\int V (1,\bar{1})G_2(1,\bar{1}^-,\bar{2},\bar{1}^+)G^{-1}(\bar{2},1^\prime),
\end{equation}
where $V(1,2)$ is the inter-particle interaction potential between two particles and can generally depend on the particle spatial coordinates and spin as well as on time.   
Here the notations $\tau^+$ and $\tau^-$ specify the time ordering
$\tau^+ > \tau > \tau^-$ and imply taking the limit $\tau^{\pm}\rightarrow\tau$,
while the two particle Green's function $G_2$ is defined as
\begin{equation}
G_2(1,2,3,4)=-\aver{T_\tau\left( {\hat c}(1){\hat c}(2){\hat c}^\dagger_{}(4){\hat c}^\dagger_{}(3)\right)} .\label{Green2}
\end{equation}

Here we consider on-site interactions described by the potential $V(1,2)=U_{\sigma_1 \sigma_2}\delta(\rr_1-\rr_2)\delta(\tau_1-\tau_2)$, with $U_{34}=U_{12}=U$ for the extended index. In the Hartree-Fock-Gor'kov mean-field approximation, the self energy -- now written explicitly in matrix form -- is
\begin{eqnarray}
	\Sigma=& 
	-U\delta_{11'}
	\left[\begin{array}{cccc}
		-G_{22}& G_{12} & 0 &  -G_{14}\\
		G_{21} & -G_{11} & -G_{23} & 0\\
		0 & -G_{32} & -G_{44} & G_{34}\\
		-G_{41} & 0 & G_{43} & -G_{33}\\
	\end{array}\right],
	\label{hirmu}
\end{eqnarray}
where $\delta_{11'} = \delta(\rr_1-\rr_1')\delta(\tau_1-\tau_1')$.
Here the single particle Green's functions appearing in $\Sigma$ have the variables $(\rr_1\tau_1,\rr_1\tau_1^+)$, and the spin indices $\sigma$, $\sigma'$ now appear in these Green's functions as $G_{\sigma \sigma'}(\rr_1\tau_1,\rr_1\tau_1^+)=- 
\aver{T_\tau\left({\hat c}_\sigma(\rr_1\tau_1){\hat c}_{\sigma'}^{\dagger}(\rr_1\tau_1^+)\right)}$. 

The Hartree terms can be included in the chemical potentials 
on the diagonal of the self-energy matrix 
in a way similar to equation~\eqref{eq:a-13}.
The Fock-exchange terms $G_{12}$, $G_{21}$, $G_{34}$ and $G_{43}$ 
are zero in a Fermi gas where spin-flips do not occur. 
Finally we introduce the key element of the mean field FFLO theory. We assume that
the pairing fields of the self-energy have an oscillating structure
\begin{eqnarray}
	\Sigma
	=&
	\Delta \delta_{11'}
	\left[\begin{array}{cccc}
	0&0&0&   e^{i\q\cdot\rry}\\
	0&0& - e^{i\q\cdot\rry}&0\\
	0&- e^{-i\q\cdot\rry}&0&0\\
	e^{-i\q\cdot\rry}&0&0&0\\
	\end{array}\right],  \label{hirmu2}
\end{eqnarray}
where the FFLO pairing vector is denoted by $\q$. The choice $\q$ instead of $2\q$, which is sometimes used, corresponds to the ansatz for the FF pairing field as in equation~\eqref{eq:order_parameter_FF}.
 
Although the choice of the spin indexing $\sigma\in\{1,2,3,4\}$ leads to four by four matrices, as used in equations~(\ref{hirmu})--(\ref{hirmu2}), there is redundancy in the labeling because indices $3$ and $4$ simply correspond to Hermitian conjugates of $1$ and $2$. Therefore the essential physics of the system is described by a two by two matrix with usual 
single particle Green's functions for the two spins in the diagonal and the anomalous Green's functions in the off-diagonals; this is the standard form of the Nambu Green's function. After taking a proper Fourier transform and inversion of equation~(\ref{SPGreenDef})~\cite{HeikkinenThesis2014}, one can write the Nambu Green's function as \footnote{Note that in~\cite{HeikkinenThesis2014,heikkinen_collective_2011}, the convention $\Delta\exp(2i{\bf q\cdot x})$ was used instead of $\Delta\exp(i{\bf q\cdot x})$ chosen in this review, another difference in notation is that $v(\p)$ here corresponds to $-v(\p)$ in~\cite{HeikkinenThesis2014}.}
\begin{eqnarray}
&
\left[\begin{array}{cc}
G_{11}(\p_1,\p_2,\omega) & G_{14}(\p_1,\q-\p_2,\omega)\\
G_{41}(\q-\p_1,\p_2,\omega) & G_{44}(\q-\p_1,\q-\p_2,\omega)\\
\end{array}\right]\nonumber\\
=&
\frac{\kron{\p_1,\p_2}}{i\omega-E_+(\p_1)}
\left[\begin{array}{cc}
u(\p_1)^2 & -u(\p_1)v(\p_1) \\
-u(\p_1)v(\p_1) & v(\p_1)^2\\
\end{array}\right]\nonumber\\
+&
\frac{\kron{\p_1,\p_2}}{i\omega+E_-(\p_1)}
\left[\begin{array}{cc}
v(\p_1)^2 & u(\p_1)v(\p_1) \\
u(\p_1)v(\p_1) & u(\p_1)^2\\
\end{array}\right]
\label{fflogreen3}.
\end{eqnarray}

In the spin-block $\sigma_1,\sigma_2\in\{2,3\}$, a similar
calculation yields the expression
\begin{eqnarray}
&
\left[\begin{array}{cc}
G_{22}(\q-\p_1,\q-\p_2,\omega) & G_{23}(\q-\p_1,\p_2,\omega)\\
G_{32}(\p_1,\q-\p_2,\omega) & G_{33}(\p_1,\p_2,\omega)\\
\end{array}\right]\nonumber\\
=&
\frac{\kron{\p_1,\p_2}}{i\omega-E_-(\p_1)}
\left[\begin{array}{cc}
u(\p_1)^2 & u(\p_1)v(\p_1) \\
u(\p_1)v(\p_1) & v(\p_1)^2\\
\end{array}\right]\nonumber\\
+&
\frac{\kron{\p_1,\p_2}}{i\omega+E_+(\p_1)}
\left[\begin{array}{cc}
v(\p_1)^2 & -u(\p_1)v(\p_1) \\
-u(\p_1)v(\p_1) & u(\p_1)^2\\
\end{array}\right]
\label{fflogreen4}.
\end{eqnarray} 
Here appear the quasiparticle energies $E_\pm$ 
\begin{eqnarray}
E_\pm(\p)&=\pm\xi^-_{\p,\q} 
+\sqrt{\left(\xi^+_{\p,\q}\right)^2+\Delta^2},\label{FFquasi}
\end{eqnarray}
where $\xi^\pm_{\p,\q} = \left(\xi_1(\p) \pm \xi_2(\q-\p)\right)/2$, the free particle energies correspond to those defined in equation~\eqref{particledispersions} as $\xi_1=\xi_\uparrow$ and $\xi_2=\xi_\downarrow$, 
and the coherence factors $u$ and $v$ 
\begin{equation}
	u(\p)^2=1-v(\p)^2 = \frac{1}{2} \left( 1 + \frac{\xi^+_{\p,\q}}{\sqrt{\left(\xi^+_{\p,\q}\right)^2 + \Delta^2}} \right).
\end{equation}
The quasiparticle energies and the coherence factors have the same form as when derived within the simple BCS formalism, for comparison of the quasiparticle energies see equations~\eqref{eq:qp_energies} and~\eqref{Bogoliubovsimpleuv} in section~\ref{BasicFFLOTheoryLattice}. 

\section{Collective modes for the FFLO state} \label{CollectiveSection}

Superfluidity can be explained within the Landau symmetry breaking paradigm. The $U(1)$ symmetry that allows freedom to choose the overall phase factor of a quantum state becomes broken: the superfluid wave function obtains a well-defined phase~\cite{mahan_many-particle_2000,fetter_quantum_2003}. In the case of a superfluid of Cooper-paired fermions, this is the phase $\varphi$ of the order parameter $e^{i\varphi}\Delta$, usually chosen to be zero for convenience. As an important consequence, a low-energy collective mode is formed related to this symmetry breaking: it is possible to distort the phase $\varphi$ by a small gradient with low energy cost. Physically this corresponds to a collective movement of all condensed particles, for instance the Cooper pairs. This is an example of the more general Goldstone theorem which states that spontaneous breaking of a continuous symmetry leads to the existence of a Nambu-Goldstone mode, which is a bosonic excitation of the ground state and has a gapless dispersion at the long wavelength limit. 

In the case of superfluids, the Nambu-Goldstone mode is called the Anderson-Bogoliubov phonon and is one of the defining characteristics of superfluidity. However it exists as a low-energy excitation only for neutral superfluids, such as those in ultracold quantum gases (UQG). For charged systems like superconductors, the situation is different. There the order parameter phase (i.e., the movement of the charged particles) becomes coupled to the vector potential. As a result, the collective mode becomes gapped, that is, massive; this is an example of the Higgs mechanism. The Anderson-Bogoliubov phonon naturally affects the system's thermodynamics. BCS theory in its simplest form does not describe such collective excitations, nevertheless it is able to predict some basic quantities like the critical temperature. However one has to be cautious of phases of matter predicted by minimizing the energy within a simple mean-field theory because collective modes, related to order parameter and density fluctuations, may destabilize the state. 
This can be especially critical for the FFLO case where fluctuations may become strong~\cite{radzihovsky_quantum_2009,radzihovsky_fluctuations_2011}; in their original article~\cite{FF}, Fulde and Ferrell already showed that the FF state does not have supercurrent for small values of the vector potential in the direction transverse to the FF wavevector, hinting to instability to fluctuations. Extending mean-field theories, collective excitations can be captured by introducing interactions between quasiparticles, as described below. 

The existence and properties of low-energy collective modes becomes an intriguing question in the case of the FFLO state because not only the $U(1)$ symmetry is broken as the superfluid order parameter forms, but also the translational and rotational symmetries may be broken due to the spatial dependence of the order parameter. The breaking of rotational symmetry is apparent from the FF form of the order parameter $\Delta e^{i\mathbf{q}\cdot \mathbf{r}}$, but since the spatial dependence there is of plane wave type, one needs an LO-type state (e.g., $\Delta cos(\mathbf{q}\cdot \mathbf{r})$) to have the translational symmetry broken as well. In LO-type states also the densities and order parameter amplitudes depend on position. One can ask whether and what kind of collective modes appear as a consequence of such symmetry breaking. Moreover the FFLO state hosts unpaired particles in addition to Cooper pairs, that is, a Fermi surface exists, which could affect the collective modes. 

The collective mode spectrum of the FFLO state has been considered in the context of superconductivity in quantum chromodynamics (see section~IV of the review~\cite{casalbuoni_inhomogeneous_2004}). Effective Lagrangians for the Goldstone bosons were derived, and it was shown that indeed Nambu-Goldstone modes exist. The system is quite complex since quarks have both color and flavor degrees of freedom, in addition to different masses, and the condensates can be of vector form. When two-component, attractively interacting ultracold Fermi gases emerged as a potential (charge neutral) system to experimentally realize the FFLO state, it became relevant to reconsider the question of FFLO states and collective modes in such systems. The purpose of this section is to review theory work on FFLO state collective modes in the ultracold gases context. We first briefly present one method that can be used for calculating the collective modes, namely linear response with the generalized random phase approximation (GRPA). We  proceed to discuss some selected publications where the question of collective modes of the FFLO state in UQG has been approached. 
 
Calculation of the collective mode spectrum can be done within the linear response theory, assuming thermodynamic equilibrium. 
The change in the expected value of a relevant observable ${\hat O}$ 
in response to a time-dependent perturbation described by a interaction-picture operator ${\hat H}'(t)$
is given by the Kubo formula~\cite{mahan_many-particle_2000,bruus2004,vanLeeuwen2013}
\begin{equation}
\delta \aver{{\hat O}(t)}=
-i\int\limits_{-\infty}^\infty dt^\prime \, \Theta(t-t^\prime) \aver{ \left[{\hat O}(t),{\hat H}^\prime(t^\prime)\right]}.
\label{kubo}
\end{equation}
The perturbation ${\hat H}'(t)$ is assumed vanishing for $t<0$, and $\Theta(t)$ is the
Heaviside step function. 
Here both ${{\hat H}}'(t)$ and ${\hat O}(t)$ are single-particle operators
of the generic form
\begin{eqnarray}
{{\hat H}}'(t)&=\int\,\phi(\bar{\mathbf{r}}\bar{\sigma},\bar{\mathbf{r}'}\bar{\sigma'},t){\hat \psi}^\dagger(\bar{\mathbf{r}}\bar{\sigma},t)
{\hat \psi}(\bar{\mathbf{r}'}\bar{\sigma'},t),\\
{\hat O}(t)&=\int\,\tilde{\phi}(\bar{\mathbf{r}}\bar{\sigma},\bar{\mathbf{r}'}\bar{\sigma'},t){\hat \psi}^\dagger(\bar{\mathbf{r}}\bar{\sigma},t)
{\hat \psi}(\bar{\mathbf{r}'}\bar{\sigma'},t), \label{single_particle_operator}
\end{eqnarray}
where $\phi$ now characterizes the strength and  spatial, temporal and spin-dependence of the perturbing field. One should understand $\tilde{\phi}$ as a generic way of defining the observable of interest: it is usually independent of time, selects the spin components of interest, and defines whether the observable is dependent on a spatial coordinate or whether all spatial coordinates are integrated over.

In the case of single particle operators as in equation~(\ref{single_particle_operator}), the integrand of the Kubo formula, that is the integrand of $\delta \aver{{\hat O}(t)}$
is a retarded two-body correlation function, or linear response function, 
multiplied by the strength of the perturbation $\phi$.
The definition of the retarded linear response function is
\begin{eqnarray}
  \label{RetLinearResponsePaivi}
L_{\sigma_1\sigma_2\sigma_3\sigma_4}(\rr,\rr^\prime,t-t^\prime)=\\
\nonumber -i\theta(t-t^\prime)\aver{ \left[ \psi^{\dagger}_{\sigma_3}(\rr,t) \psi_{\sigma_1}(\rr,t),\psi^{\dagger}_{\sigma_4}(\rr^\prime,t^\prime) \psi_{\sigma_2}(\rr^\prime,t^\prime)\right]}.  
\end{eqnarray}
One can now obtain collective modes as a response to, for instance a density perturbation, from
the linear response function. However its exact evaluation is in practice an impossible task for an 
interacting many-body system in dimensions higher than one, for realistic system sizes. Approximations
are needed, but with too crude approximations, the collective modes can be missed. Basically one has to 
include the existence of quasiparticles as well as interactions between them to arrive at the relevant
collective modes. Random phase approximation (RPA)~\cite{Bohm1953} was introduced originally to describe collective effects
in an electron gas and was later generalized (GRPA) to describe collective modes in superconductors~\cite{Anderson1958}. The GRPA
form of the linear response function is usually sufficient to find the low-energy collective modes of superfluids.

The Kadanoff-Baym formalism~\cite{Baym1961,Baym1962} is a method to rigorously
derive the linear response function (for example the GRPA response) so that it is conserving and thermodynamically 
consistent regarding the approximations made in the many-body problem.
A conserving approximation is such that it obeys the conservation 
laws arising from the symmetries of the physical system. Thermodynamical 
consistency means that the partition function is unique:
direct evaluation of observables and their calculation by differentiating
the partition function give the same result. The GRPA linear response function derived from the Kadanoff-Baym formalism
is (compared to equation~\eqref{RetLinearResponsePaivi}, the spin indices are now moved from the subscript to the arguments)
\begin{eqnarray}
L(12,1^\prime 2^\prime) = 
\int G ( 1 , \bar{3} )_0  G(\bar{4},1^\prime)_0 \left({\varu{\phi(\bar{3},\bar{4})}{2^\prime}{2}}\right)_0  
\nonumber\\
+ \int G (1,\bar{3} )_0  G(\bar{4},1^\prime)_0   
\left(\varG{\Sigma(\bar{3},\bar{4})}
	{\bar{5}}{\bar{6}}\right)_0 
L(\bar{5} 2,\bar{6} 2^\prime), \label{Leqq}
\end{eqnarray}
where the subindex $0$ signifies the evaluation in absence of perturbation, $\phi = 0$.
In mean-field approximations of the Hartree-Fock-Bogoliubov type, the functional derivative
$\unolla{\varG{\Sigma(\bar{3},\bar{4})}{{5}}{{6}}}$ is a combination of
delta functions, as one can conclude from equation~(\ref{hirmu}), and does not produce any additional Green's functions to the equation.

\begin{figure}
	\centering
	\includegraphics[width=0.95\columnwidth]{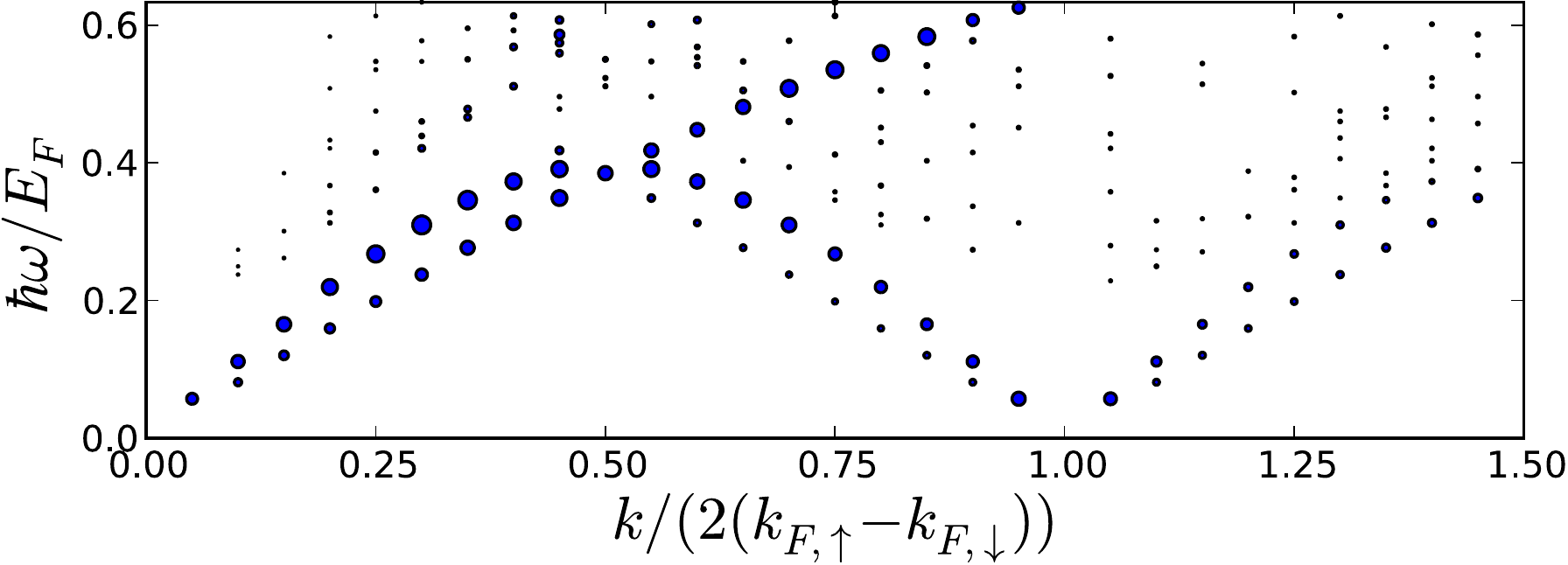}
	\caption{The density response in the FFLO phase in a 1D two-component Fermi gas obtained by mean-field theory, from~\cite{edge_signature_2009}. The amplitude of the response is reflected in the area of a circle. The data is for the polarization $P=0.15$ and for a fixed interaction strength (the interaction energy density divided by the kinetic energy density equals 1.5). Two gapless sound modes emerge around $k=0$ and $k^*=2(k_{F\uparrow}-k_{F\downarrow})$.}
	\label{fig:EdgeCooper}
\end{figure}

Edge and Cooper studied a 1D two-component Fermi gas with attractive contact interactions using a mean-field BdG approach (c.f.\ section~\ref{section_traps} of this review) and the GRPA formalism~\cite{edge_signature_2009,edge_collective_2010}. The BdG approach can describe a spatially oscillating order parameter, for instance of cosine-type, going beyond the simple FF ansatz $\Delta e^{i\mathbf{q}\cdot \mathbf{r}}$. In this case, both translational and rotational symmetries and the densities and absolute value of the order parameter amplitude are spatially varying. Edge and Cooper studied the response of the system to a density perturbation periodic in time and space, thus giving the momentum $k$ and frequency $\omega$ to the system. They found two collective mode branches in the density response at low frequency and wavelength. One of them appears at $k=0$, and the other one at $k^*=2(k_{F\uparrow}-k_{F\downarrow})$, see figure~\ref{fig:EdgeCooper}. Here the factor of two compared with the FFLO wave vector $q=k_{F\uparrow}-k_{F\downarrow}$ appears because the densities oscillate in space with a period only half of the cosine-form order parameter oscillation, because they follow the absolute value. The appearance of two gapless modes is due to the FFLO state breaking both the $U(1)$ and translational symmetries. The period of the density and order parameter absolute value oscillations give $k^*$. The two modes involve spatial oscillations of both density and spin-density. This is a consequence of the breaking of time reversal symmetry by the imbalance. A detailed analysis of how the equivalents of such modes appear in a trapped gas is presented in references~\cite{edge_signature_2009,edge_collective_2010}.
   
While the mean-field analysis of refs.~\cite{edge_signature_2009,edge_collective_2010} can provide qualitative insight to the behaviour of collective modes, it cannot be an accurate description of the 1D system. In 1D no continuous symmetries are spontaneously broken, and only power-law decaying correlations exist. However the qualitative results match well with the knowledge obtained by exact methods in 1D~\cite{Yang2001,zhao_theory_2008,Frahm2008,feiguin_spectral_2009,yin_quench_2016}. A 1D interacting two-component (spin-half) system is known to form a Luttinger liquid whose elementary excitations at zero spin-polarization are collective density fluctuations that carry only charge or spin. When transitioning from the unpolarized 1D superfluid to the 1D FFLO equivalent, the spin gap is closed and a second gapless sound mode emerges. This can be viewed as a Luttinger liquid of the excess fermions with a Fermi wavevector $k_{F\uparrow}-k_{F\downarrow}$. With spin polarization, the excitations become of mixed spin and density type, reflecting the breaking of time-reversal symmetry by the spin-polarization. For more information about the 1D FFLO state, we recommend~\cite{Feiguin2012,guan_fermi_2013,yin_quench_2016}. 

\begin{figure}
	\centerline{
	\includegraphics[width=0.95\columnwidth]{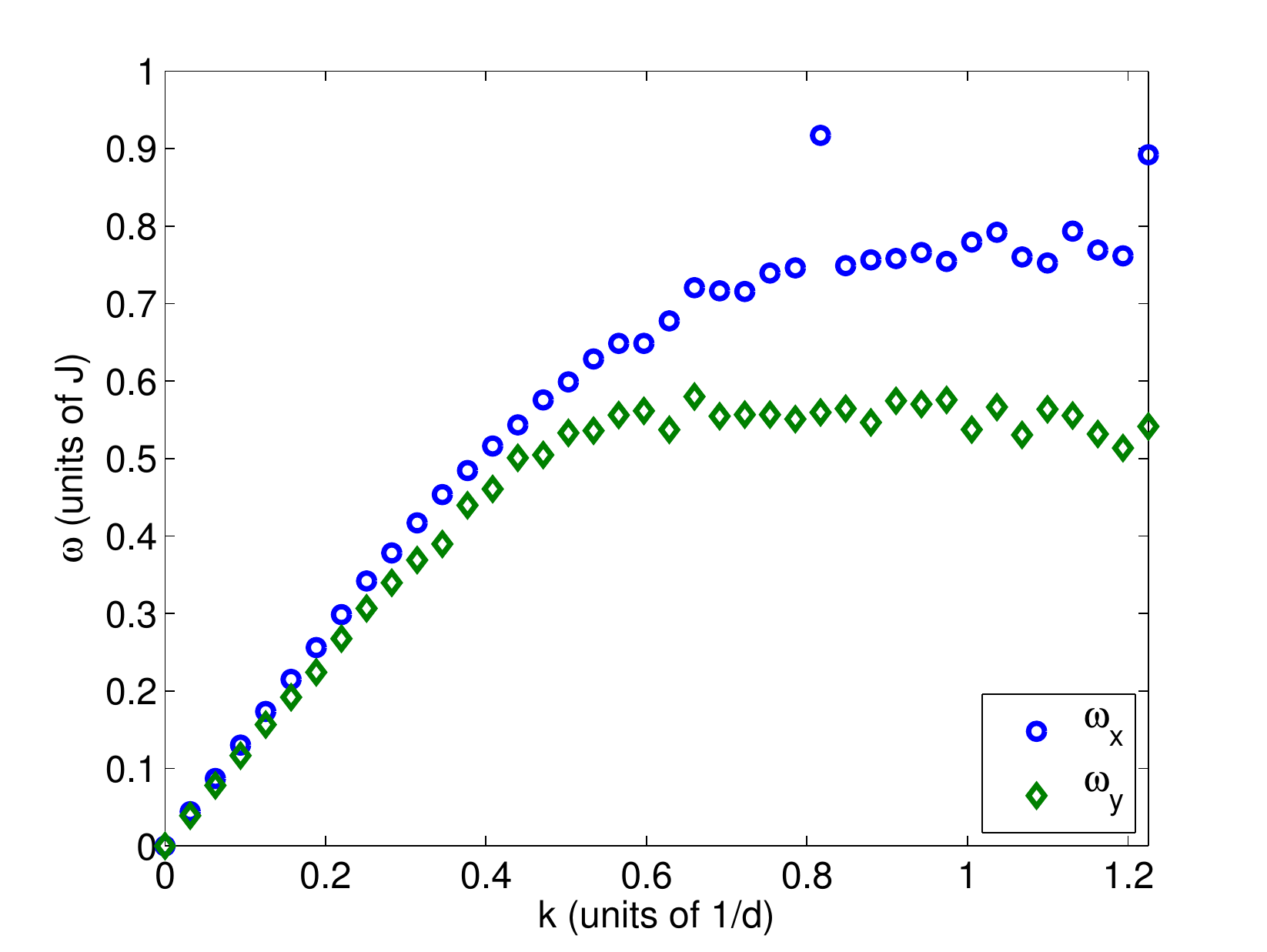}	
	}
	
	\caption{
		The dispersion relation of the collective density modes in the FFLO state in a 2D lattice, from~\cite{heikkinen_collective_2011}. The
		wave vector is parallel to the the FFLO vector $\q$ in the $x$-direction $(\omega_x)$, and perpendicular
		to it in $y$ $(\omega_y)$. The slope of the dispersion at $k=0$ gives the speed of sound.
		The anisotropy of the FFLO state leads to an anisotropy in the speed of sound. 
		\label{MiikkaSound}
	}
\end{figure}

Collective modes in 2D and 3D lattice systems of two-component Fermions were studied by Heikkinen and T\"orm\"a~\cite{heikkinen_collective_2011}, using the FF ansatz. The density response was derived based on the Kadanoff-Baym approach to GRPA, as given by equation~(\ref{Leqq}). The FF ansatz of the form $\Delta e^{i\mathbf{q}\cdot \mathbf{r}}$, with a fixed FF wavevector $\mathbf{q}$, makes the system asymmetric with respect to directions. In a lattice system as discussed in section~\ref{MFstudies}, $\mathbf{q}$ aligned with one of the lattice axes, in most cases, minimizes the energy. It was found in~\cite{heikkinen_collective_2011} that this anisotropy of the order parameter leads to a collective mode dispersion relation that is different in the directions parallel and perpendicular to  $\mathbf{q}$, see figure~\ref{MiikkaSound}. This means that the speed of sound is anisotropic as well. A rigorous derivation is presented in the Ph.D.\ thesis of M.O.J.\ Heikkinen~\cite{HeikkinenThesis2014}, section~4.2.1. which shows that the mode obtained by this approach indeed corresponds to the Nambu-Goldstone mode. Specifically, the response function is shown to diverge at zero frequency and momentum, which proves that the mode is gapless. The proof explicitly shows that the mode is gapless because the BCS gap equation is consistently included in the formalism calculating the collective mode spectrum.

\begin{figure}
	\centerline{
	\includegraphics[width=0.95\columnwidth]{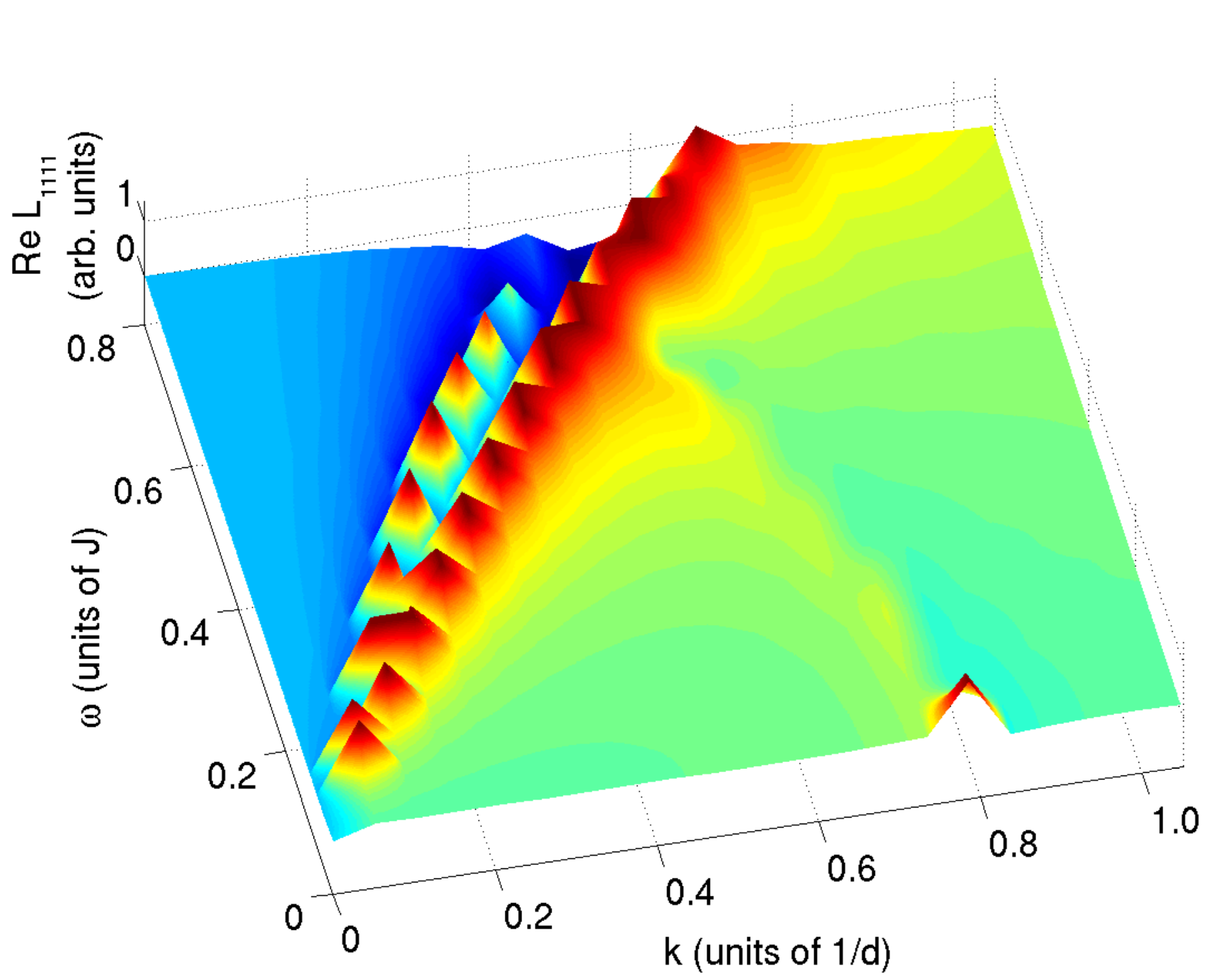}
	}
	\caption{
		Real part of the density response of the FF state in a quasi-1D lattice, from~\cite{heikkinen_collective_2011}, obtained by mean-field theory. The lower branch corresponds to the Anderson-Bogoliubov phonon and the upper to quasiparticle excitations that are gathered into a narrow stripe due to the reduced dimensionality. 
		\label{MiikkaSound1D}
	}
\end{figure}

Heikkinen and T\"orm\"a also considered a quasi-1D lattice to compare with the work of Edge and Cooper~\cite{edge_signature_2009,edge_collective_2010}. They found two branches around $k=0$, similar to~\cite{edge_signature_2009,edge_collective_2010}, see figure~\ref{MiikkaSound1D}. One branch corresponds to the collective mode in the higher dimensional case, while the other comes from the quasi-1D nature: quasiparticle excitations gather into a narrow stripe. The existence of quasiparticles, even at low energy, is due to the gapless nature of the FF state, however in higher dimensions, they are more dispersed and do not form a stripe. The results of~\cite{heikkinen_collective_2011} differ at higher momenta from those in~\cite{edge_signature_2009,edge_collective_2010}: there is no gapless sound mode around $k^*=2(k_{F\uparrow}-k_{F\downarrow})$ because in the FF case, $|\Delta|$ is constant and the excess fermions are uniformely distributed. The backbending of the collective mode dispersion in 2D and 3D lattices at large momenta has also been interpreted as a roton-like behaviour~\cite{koinov_rotonlike_2011,mendoza_superfluidity_2013,mendoza_collective_2014,koinov_collective_2015}.

Radzihovsky~\cite{radzihovsky_quantum_2009,Radzihovsky2011} derived a low-energy Landau theory for the LO state in a two-component Fermi gas and found two Goldstone modes, one corresponding to superfluidity and one that he refers to as a smectic phonon. Drawing analogy to the above discussion, the former one is related to the Anderson-Bogoliubov phonon and the latter to the existence of the excess fermions. However, Radzihovsky emphasizes the liquid-crystalline character and softness of the smectic phonon, which may in some cases lead to vanishing FFLO order parameter, or peculiar topological defects such as composite half-integer vortex-dislocation defects. In a lattice, the rotational symmetry is broken and such features are not expected to appear, except perhaps for special conditions such as low filling where the dispersion felt by the particles is almost rotationally symmetric. 

The Anderson-Bogoliubov phonon can be understood as fluctuation of the {\it phase} of the order parameter. Fluctuation of the order parameter {\it amplitude} is referred to as the pair vibration mode, or the Higgs mode. This has been theoretically considered in the ultracold Fermi gas context in refs.~\cite{Bruun2002,Korolyuk2014,Korolyuk2011,Bruun2014,Bjerlin2016} for trapped gases. The Leggett mode is a collective mode associated with pairing fluctuations of multiple bands, and a harmonic trap analogue of such mode was found in~\cite{Korolyuk2014}. We are not aware of any works on Higgs or Leggett modes in the FFLO state, at least not in the ultracold gases context.

\section{Beyond mean-field studies of the FFLO state in 2D and 3D lattices} \label{Beyond-mean-fieldSection}

The mean-field approximation is known to be typically good in the case of weak interactions and high dimensions when the system is not likely to be strongly correlated. Pairing in the FFLO state is obviously stronger if the interaction strength is increased. Moreover, there is evidence that reduced dimensionality helps to stabilize the FFLO state. Thus both from the conceptual and practical (e.g., higher critical temperatures) point of view, it is of interest to explore the FFLO state in strongly interacting Fermi gases and in reduced dimensions. Correspondingly because quantum and thermal fluctuations are typically stronger in low dimensions and for strong interactions, one has to harness beyond mean-field methods for the theoretical description of the state. 
Even in the weakly interacting limit in 3D, the subtle nature of the FFLO order parameter (c.f.\ the discussion on fluctuations in section~\ref{CollectiveSection}), deserves a closer look with beyond mean-field methods to test its robustness against both quantum and thermal fluctuations. 

There are a handful of beyond mean-field studies of the FFLO state in lattices, in 2D, 3D, and quasi-1D systems~\cite{dukelsky_integrable_2006,kim_exotic_2011,Wolak2012,heikkinen_finite-temperature_2013,heikkinen_nonlocal_2014,gukelberger_fulde-ferrell-larkin-ovchinnikov_2015,Karmakar2016}.\footnote{In 1D, of course, the situation is competely different since exact methods such as Bethe ansatz can be applied; we do not review the works that focus on 1D FFLO state~\cite{guan_fermi_2013}.} Of these, three references~\cite{kim_exotic_2011,heikkinen_finite-temperature_2013,heikkinen_nonlocal_2014} are based on dynamical mean-field theory (DMFT), and the other works~\cite{Wolak2012,gukelberger_fulde-ferrell-larkin-ovchinnikov_2015,Karmakar2016} on quantum Monte Carlo (QMC). The early inspirational work~\cite{dukelsky_integrable_2006} presents an in-principle exact solution of the problem using the algebraic techniques of the Richardson-Gaudin model, but in practice they resort to solving the exact equations for a small (6x6) lattice, also utilizing QMC for the search of candidate ground states. The obtained zero-temperature phase diagram does not easily compare to later mean-field and beyond-mean-field work, possibly due to the small system size. They predict the possibility of a breached FFLO state.   

To be able to describe how the results obtained by DMFT go beyond mean field thery, we first sketch the basic idea of how to apply the DMFT method to study fermion superfluidity and especially FFLO (for a general review on the DMFT method, see~\cite{DMFTreview}). We then review results on the FFLO state in the 1D-3D crossover using the method in section~\ref{1D3Dcrossover}. In section~\ref{QMCstudies}, we discuss results on FFLO physics in 2D lattices obtained utilizing QMC.  

\subsection{Dynamical mean-field theory (DMFT) for studies of the FFLO state}

DMFT~\cite{DMFTreview} is a non-perturbative method where the many-body problem is self-consistently mapped 
to a quantum impurity problem, as described in figure~\ref{fig:dmft_idea}. It is assumed that the self-energy of the
system is local (here $j$ and $l$ label lattice sites, and $\omega_n$ is the Matsubara frequency)
\begin{eqnarray}
\Sigma_{jl}(i\omega_n)\approx\delta_{jl}\Sigma_{j}(i\omega_n).
\end{eqnarray}
This assumption can be relaxed by using a cluster version of DMFT where the self-energy can be non-local within the cluster, as discussed later. The DMFT method as described in figure~\ref{fig:dmft_idea}(a) applies to spatially homogeneous lattice systems, meaning the absence of external potentials, such as a harmonic trap potential typical in ultracold gases. This assumption of homogeneity leads also to a spatically homogeneous order parameter.
To study the FFLO state where the order parameter is non-uniform in space, one needs a real-space version of 
DMFT~\cite{RDMFT,Snoek2008,Koga2011rdmft,kim_exotic_2011}. The idea of real-space DMFT is depicted in figure~\ref{fig:dmft_idea}(b). We briefly 
present how fermionic superfluidity and the FFLO state can be described by the real-space DMFT method.
{\begin{figure*}
	\centerline{
		\includegraphics[width=1.95\columnwidth]{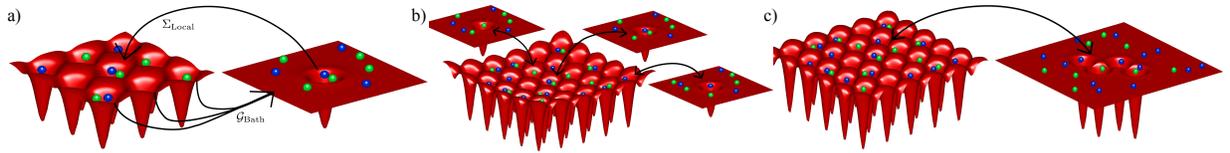}
	}
	\caption{
		(a) In dynamical mean field theory (DMFT), the system is mapped to an interacting impurity problem  at one lattice
		site. The remaining system is represented as a bath
		of non-interacting particles tunneling in and out of the chosen 
		lattice site. The local self-energy is obtained from the impurity problem
		and fed back to the original lattice geometry. Self-consistency is demanded, that is, a
		self-energy which creates its own bath is the physically correct solution. 
		It can be found by
		fixed point iteration. (b) The idea of real-space DMFT: the quantum impurity problem
		is solved for each lattice site which allows describing a spatially non-uniform system. (c) Cluster DMFT: the impurity problem includes more than one interacting lattice site,
		enabling the study of non-local quantum fluctuations. 
		\label{fig:dmft_idea}
	}
\end{figure*}
}

To describe superfluid symmetry breaking in an interacting fermion system,
the anomalous pairing correlations need to be incorporated in to DMFT. This can be done using 
the Nambu formalism. 
The Nambu Green's function is defined as a two by two matrix
\begin{eqnarray}
\bm{G}_{jl}(\tau)
=
\left[\begin{array}{cc}
G_{\uparrow,jl}(\tau) & F_{jl}(\tau) \\
F^\dagger_{jl}(\tau) & -G_{\downarrow,lj}(-\tau)
\end{array}\right] \\
= 
\left[\begin{array}{cc}
-\aver{T_\tau {\hat c}^{}_{j,\uparrow}(\tau){\hat c}^{\dagger}_{l,\uparrow}(0)} & -\aver{T_\tau {\hat c}^{}_{j,\uparrow}(\tau){\hat c}^{}_{l,\downarrow}(0)} \\
-\aver{T_\tau {\hat c}^{\dagger}_{j,\downarrow}(\tau){\hat c}^{\dagger}_{l,\uparrow}(0)} & -\aver{T_\tau {\hat c}^{\dagger}_{j,\downarrow}(\tau){\hat c}^{}_{l,\downarrow}(0)} 
\end{array}\right].
\end{eqnarray}
This is equivalent to the $\sigma=\{1,4\}$ matrix block of Section~\ref{GreenFFLO}. The indices $j,l$ label the lattice sites.
Here for the anomalous Green's function, we have $F^\dagger_{jl}(\tau)= [F_{jl}(\tau)]^\dagger$, and
$[{\hat O}(\tau)]^\dagger={\hat O}^\dagger(-\tau)$ for the Matsubara time evolution.
The lattice dispersion in the non-interacting case enters the non-interacting Green's function, which in the Nambu formalism is of the form
\begin{equation}
\bm{G}_{0,jl}(i\omega_n)=
\left[\begin{array}{cc} G_{\uparrow,jl}^0(i\omega_n) &0\\0& -G_{\downarrow,jl}^0(-i\omega_n)\end{array}\right] ,
\end{equation}
where $G_{\sigma,jl}^0(i\omega_n) = 1/\left[i\omega_n+t_{jl}+\left(\mu_\sigma-V_{j,\sigma}\right)\delta_{jl}\right]$, $t_{jl}$ is the hopping energy and $V_{j,\sigma}$ the single-particle potential that can correspond to the trap potential, for instance.
The order parameter of the superfluid phase at a given site is of the usual form
\begin{equation}
\Delta_j = U \aver{{\hat c}^{\dagger}_{j,\uparrow}{\hat c}^{\dagger}_{j,\downarrow}},
\end{equation}
where $U<0$ for attractive interactions.

The Dyson equation gives the relation between the interacting and non-interacting Green's functions and the self-energy, and is of the form (in general, the Dyson equation can have terms corresponding to spin-flip processes and spin-dependent interactions, but this form is sufficient to describe the two-component ultracold Fermi gases typically considered for the FFLO state):
\begin{eqnarray}
\bm{G}_{jl}(i\omega_n)&=[\bm{G}_{0}^{-1}(i\omega_n)-\bm{\Sigma}(i\omega_n)]^{-1}_{jl}   . \label{dmft1}
\end{eqnarray}
In DMFT, the self-energy is approximated to be site-diagonal,
$\bm{\Sigma}_{jl}(i\omega_n)\approx\delta_{jl}\bm{\Sigma}_{j}(i\omega_n)$.
Direct matrix inversion is used for solving the Dyson equation in real-space DMFT. Now all this defines the 
full lattice problem. In DMFT, one divides the problem into a local impurity problem and a bath that represents the
rest of the lattice. We now proceed to first discuss the impurity problem and then how to connect it to the bath.

The quantum impurity problem of lattice site $j$
can be defined as the problem of solving the full local Green's function $\bm{\mathcal{G}}_j$
for the interaction strength $U$ and a local bath Green's function $\bGj$.
This means 
summation of all connected diagrams of $\bGj$ and $U$. Once the full local Green's function $\bm{\mathcal{G}}_j$
is obtained, one can obtain the local self-energy $\bm{\Sigma}_{j}(i\omega_n)$
from the Dyson equation
\begin{eqnarray}
\bm{\Sigma}_{j}(i\omega_n)=\bGj^{-1}(i\omega_n)-\bm{\mathcal{G}}_j^{-1}(i\omega_n).\label{dmft2}
\end{eqnarray}

Next, the quantum impurity problem~\eqref{dmft2} should be connected to the full lattice problem~\eqref{dmft1}.
This is done by forming the bath of the impurity
problem via removing the self-energy of site $j$ from the local Green's function $\bm{G}_{jj}(i\omega_n)$.
This means that the bath Green's function is given by the equation
\begin{eqnarray}
\bGj^{-1}(i\omega_n)&=[\bm{G}_{jj}(i\omega_n)]^{-1}+\bm{\Sigma}_j(i\omega_n).\label{dmft3}
\end{eqnarray}
The right hand side of this equation specifically contains the inverse of the local
component of $\bm{G}$ and not the $jj$-element of $\bm{G}^{-1}$.
This bath Green's function includes the effect of interactions on all other lattice sites except the site $j$.

The solution of the full problem is found iteratively, for instance by making an initial guess for the self-energy at all sites.
Combined with the fact that the non-interacting lattice Green's function is known, this allows us to calculate the full lattice Green's function and the bath Green's function. 
With the bath Green's function thus determined, one can solve the impurity problems at each site, and thereby obtain improved self-energies. The iteration is repeated until 
convergence is found.

The quantum impurity problem at the heart of the DMFT method, basically obtaining $\bm{\mathcal{G}}_j$, can be solved by various means. For instance,
exact diagonalization (ED) utilizing an Anderson impurity problem, QMC, or density matrix renormalization group (DMRG) methods can be used. Typically ED and DMRG are suited for zero or low temperature descriptions, while QMC is most efficient at intermediate temperatures. The use of a QMC solver in DMFT thus captures, locally, both quantum and thermal fluctuations. 

DMFT goes beyond the mean-field approximation because it solves the local lattice problem exactly. To describe interactions in the lattice system, all local diagrams are included. In some problems, however, non-local quantum correlations may become significant. These correlations can be addressed by a cluster version of DMFT, described in figure~\ref{fig:dmft_idea}(c). There are many variants of cluster DMFT, the most widespread being
cellular dynamical mean field theory (CDMFT) and dynamical cluster approximation (DCA).
In CDMFT, the idea is to define the cluster on the real-space lattice,
while DCA can be characterized as a coarse graining in the momentum space. The cluster selection of the CDMFT method breaks the translation
invariance of the original lattice, whereas DCA enforces the translation invariance.

We now discuss the results obtained in beyond mean-field studies of the FFLO state using the DMFT and QMC methods. The first work using DMFT to study the FFLO state by Kim et al.~\cite{kim_exotic_2011} is discussed in section~\ref{section_traps} since it specifically addresses issues related to the trapping geometry. Another FFLO study in the context of ultracold gases going beyond mean field was done by Wolak {\it et al.}~\cite{Wolak2012} using determinant QMC for 2D square lattice. This work, as discussed in section~\ref{QMCstudies} with a recent related work by Gukelberger {\it et al.}~\cite{gukelberger_fulde-ferrell-larkin-ovchinnikov_2015}. Next, we discuss DMFT studies that consider the FFLO state, particularly in the context of 1D-3D crossover.

\subsection{The FFLO state in the 1D-3D crossover}
\label{1D3Dcrossover}

In 1D, exact methods can be used to show~\cite{guan_fermi_2013} that at zero temperature, the 1D equivalent of the FFLO state exists for any interaction and any polarization $P=(N_\uparrow-N_\downarrow)/(N_\uparrow+N_\downarrow)=(n_\uparrow-n_\downarrow)/(n_\uparrow+n_\downarrow)$, where $N_\sigma$ and $n_\sigma$ are the particle numbers and densities of the two spin components. The 1D equivalent of the FFLO state is extremely stable due to perfect nesting: the Fermi surfaces in 1D are just points, see the discussion in section~\ref{MFstudies}.
On the other hand, no long range order exists in 1D, and the polynomially decaying FFLO correlations are highly vulnerable to thermal fluctuations.
In higher dimensions, the parameter regime (polarization, interaction, chemical potentials, etc.) for a stable FFLO state is extremely small, although lattice geometries provide a bigger parameter window, as explained in section~\ref{MFstudies}.
Thermal fluctuations, in contrast, are known to have less effect on quantum states the higher the dimension is, and true long range order is possible in 3D.
This has inspired the hypothesis that an intermediate dimensionality between 1D and 3D might optimally combine the large parameter regime of FFLO in 1D with the stabilization towards finite temperature typical for higher dimensions. Mean-field~\cite{parish_quasi-one-dimensional_2007,sun_pair_2013} and effective field theory~\cite{zhao_theory_2008} calculations for coupled 1D tubes have suggested that this could be possible.
The question has also been considered by DMRG calculations on a Hubbard-ladder~\cite{feiguin_pair_2009}.

The behaviour of the FFLO state in 1D-3D crossover in lattices was considered with the DMFT method by Kim and T\"orm\"a~\cite{kim_fulde-ferrell-larkin-ovchinnikov_2012} and Heikkinen {\it et al.}~\cite{heikkinen_finite-temperature_2013,heikkinen_nonlocal_2014}. In~\cite{kim_fulde-ferrell-larkin-ovchinnikov_2012}, the attractive Hubbard model combined with a trap in one direction was approached by DMFT with an exact diagonalization impurity solver, allowing access to zero and low temperatures. The interpolation between 1D and 3D was done by keeping tunneling parallel to one fixed direction (the direction of the trap potential), $t_\parallel$, while tuning the tunneling in the perpendicular direction, $t_\perp$. 
The interaction strength $U$ was chosen for each value of the transverse hopping $t_\perp$
such that it corresponded to the lattice equivalent of the unitarity limit~\cite{Burovski2006}.
For small $t_\perp \le 0.3 t_\parallel$, it was found that the shell structure within the trap was of the 1D type: the FFLO state in the middle of the trap surrounded by balanced superfluid areas.
For larger coupling between the chains, the 3D-like behaviour of balanced superfluid in the central trap regions surrounded by FFLO areas was found, in contrast.
This 1D-3D crossover behaviour occurred for small polarizations, while for larger polarizations, the FFLO state exists everywhere in the trap, and was replaced by the normal state for polarizations beyond the critical value.
A clear difference to the previous mean-field predictions was found: the DMFT results predicted the FFLO state to be stable for a wide range of $t_\perp$ values throughout the crossover, extending well in to the 3D-like regime.
This contrasts with having an optimal region near the transition from 1D to 3D, as suggested in articles presenting mean field~\cite{parish_quasi-one-dimensional_2007} and effective field theory~\cite{zhao_theory_2008} studies. There can be several reasons for this difference, first that the work~\cite{kim_fulde-ferrell-larkin-ovchinnikov_2012} includes beyond mean-field fluctuations. Second it considers coupled 1D lattice systems while the earlier mean-field works studied coupled 1D tubes (continuum systems), thus nesting effects are present in~\cite{kim_fulde-ferrell-larkin-ovchinnikov_2012} in all dimensions. 

\begin{figure}
  \centerline{
    \includegraphics[width=0.95\columnwidth,trim=20mm 0 20mm 0]{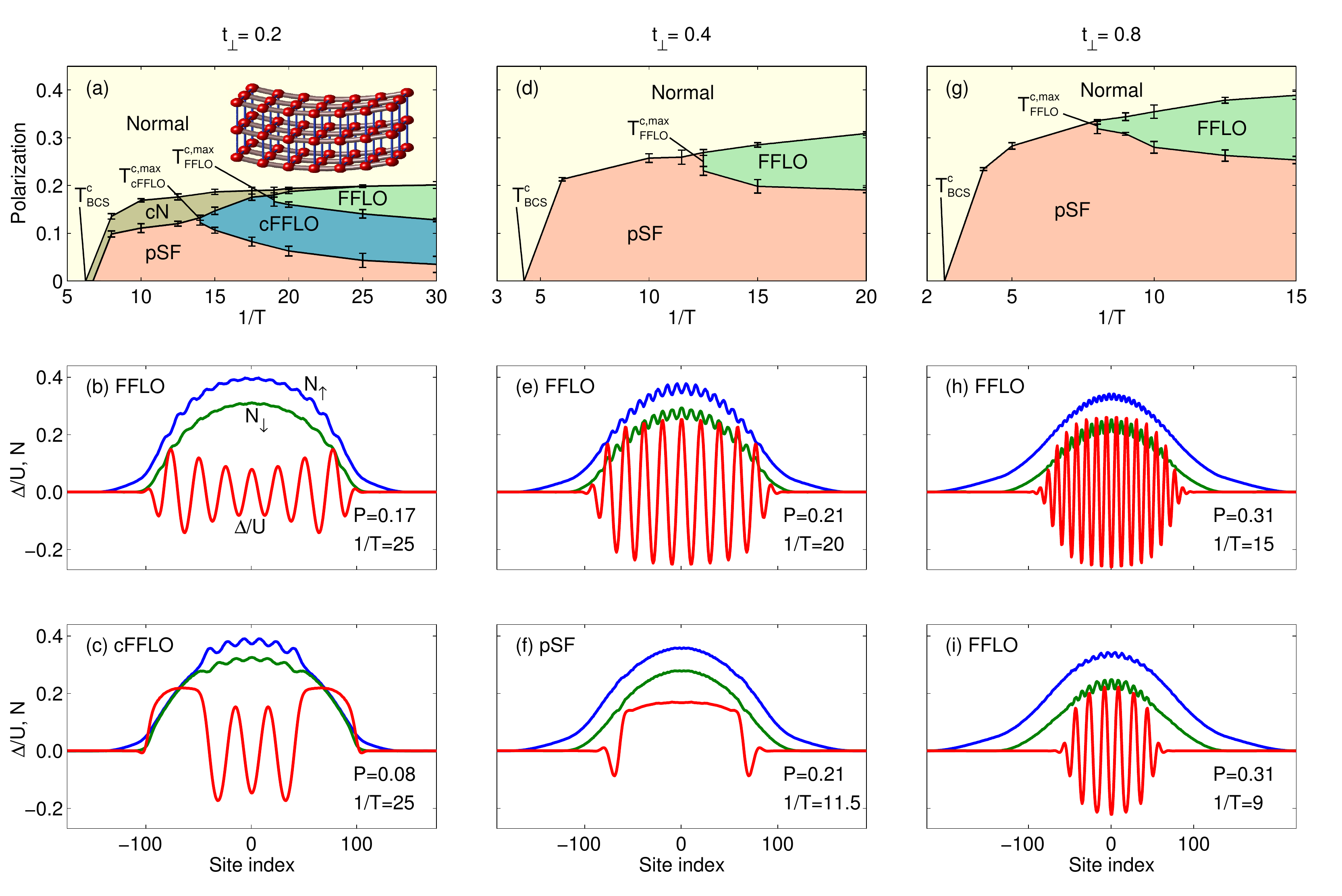}
  }
  \caption{ 
    The phase diagram of a trapped, spin-polarized Fermi gas in a 1D-3D dimensional crossover
    predicted by real-space DMFT. The inset of panel (a) shows the geometry of the system. Labels: pSF is a polarized superfluid phase, cN is a shell structure with a normal state
    at the center of the trap and a superfluid state towards the edges, cFFLO means FFLO in the middle and superfluid at the edges.
    Panel (a) shows the phase diagram for a quasi-1D lattice ($t_\perp=0.2$).
    Panels (b) and (c) give examples of the FFLO and the cFFLO cases.
    Panel (d) displays the phase diagram for $t_\perp=0.4$ which is quite in the middle of the crossover; (e) and (f)
    compare the FFLO and the polarized superfluid phases at constant polarization.
    The phase diagram in a quasi-3D geometry with $t_\perp=0.8$ is given in (g), with examples of the effect of increasing temperature given in (h) and (i). Energies and temperatures are in units of $t_\parallel = 1$. 
    Reproduced with permission from~\cite{heikkinen_finite-temperature_2013}.
    \label{fig:rdmft_results}
  }
\end{figure}

Heikkinen {\it et al.}~\cite{heikkinen_finite-temperature_2013} extended these studies to finite temperatures by using a QMC solver in the DMFT impurity problem. The configuration and the parameters were similar to reference~\cite{kim_fulde-ferrell-larkin-ovchinnikov_2012} described above. The total particle number was held approximately constant
while the polarization was varied. The main results are summarized in figure~\ref{fig:rdmft_results}.
Again the FFLO state is found to be stable over a large range of parameters throughout the crossover. The critical temperature for the FFLO state is found to be about one-third of the critical temperature of the balanced case, see figure~\ref{fig:rdmft_results} (a), (d), (g).
 
\begin{figure}
	\centerline{
		\includegraphics[width=0.95\columnwidth]{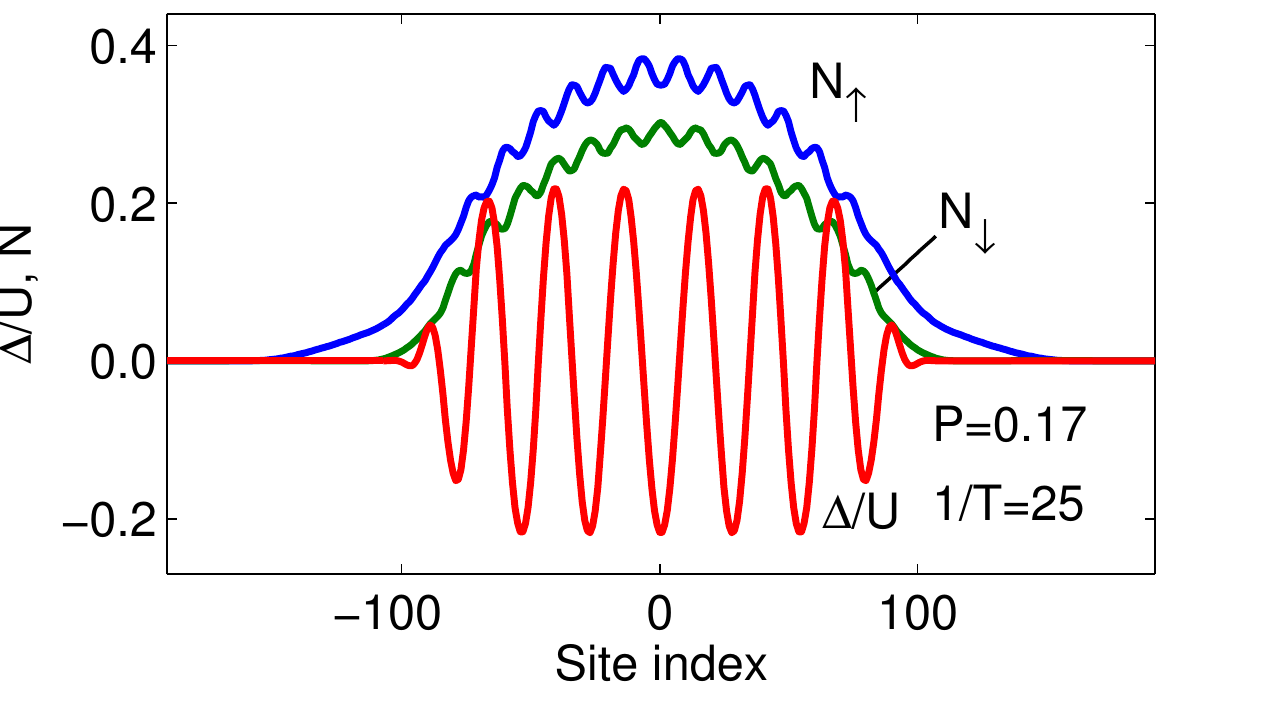}
	}
	\caption{ 
		The FFLO state exhibits an especially uniform order parameter profile, despite the overall harmonic trap, in the crossover point between 1D and 3D, near $t_\perp=0.3$. Here $U=-3.75$, $P=0.17$ and $T=0.04$.  
		Reproduced with permission from~\cite{heikkinen_finite-temperature_2013}.
		\label{fig:uniformDelta}
	}
\end{figure}

Both references~\cite{kim_fulde-ferrell-larkin-ovchinnikov_2012} and~\cite{heikkinen_finite-temperature_2013} showed that 
at the turning point between 1D-like and 3D-like behavior,
at $t_\perp=0.3 t_\parallel$, the FFLO order parameter demonstrates an intriguing behavior: it has a remarkably constant oscillation amplitude despite the overall trapping potential, see figure~\ref{fig:uniformDelta}. This is because in the 1D limit, the FFLO oscillations occur at the edges, and in the 3D limit at the center of the trap. Thus in the intermediate regime both tendencies balance to produce a flat profile~\cite{parish_quasi-one-dimensional_2007}. There is thus a sweet spot in the intermediate regime where the FFLO order parameter in a trapped system resembles the one expected for a homogeneous system. Moreover many suggested experimental probes of the FFLO state give more clear signatures in case of homogeneous FFLO oscillations compared to a typical trapped system FFLO order parameter profile.

\begin{figure}
  \centerline{
    \includegraphics[width=0.95\columnwidth]{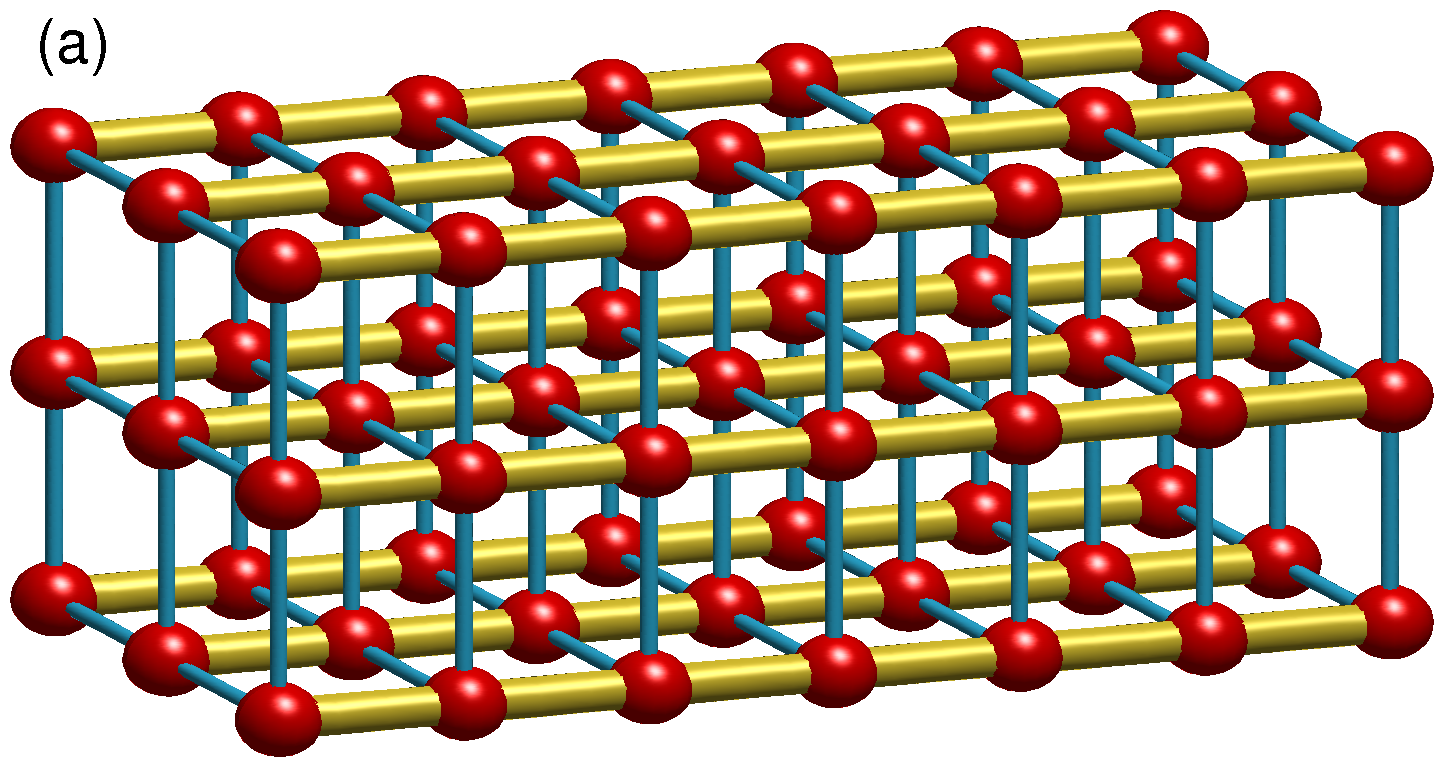}
  }
  \caption{Schematic of the system considered. A chain in 1D forms the cluster used in the DMFT method, the hopping energy within the chain is set to unity, $t=1$ and all other energies are in these units. Periodic boundary conditions are applied to the cluster. The chains are connected in the perpendicular directions by a hopping $t_\perp$ which is varied to drive the 1D -- 3D crossover. Reproduced with permission from~\cite{heikkinen_nonlocal_2014}.
    \label{fig:ClusterResultsMiikka1}
  }
\end{figure}

Since fluctuations at the level of single sites (as captured by DMFT) already change the picture in comparison to mean-field results, it is essential to ask what might non-local fluctuations do to the stability of the FFLO state. This question was considered in~\cite{heikkinen_nonlocal_2014} using a cluster version of DMFT. The choice of the cluster is illustrated in figure~\ref{fig:ClusterResultsMiikka1}: clusters consisting of 1D chains of 36-42 sites were solved exactly, and the chains were connected in two perpendicular directions, and usual DMFT was applied to the colletion of the chains. The self-energy is of the form
\begin{eqnarray}
	\bm{\Sigma}_{ii';ll'}(i\omega_n)=\delta_{l,l'}\bm{\Sigma}_{i i';l}(i\omega_n), 
\end{eqnarray}
that is, it is block diagonal in the interchain index $l$,
while all the indices $i$ and $i'$ run through all the self-energy terms,
both local and non-local, within the 1D chain.
Periodic boundary conditions were applied in the 1D cluster but otherwise arbitrary spatial dependence of the order parameter was allowed. This approach is different from the DCA version of DMFT where translational invariance is imposed, but it also deviates from the standard CDMFT with fixed cluster size. In this way, a spontaneous breaking of the translational invariance as expected for the FFLO state was allowed, but it was not imposed by construction as in CDMFT. 

\begin{figure}
  \centerline{
    \includegraphics[width=0.9\columnwidth]{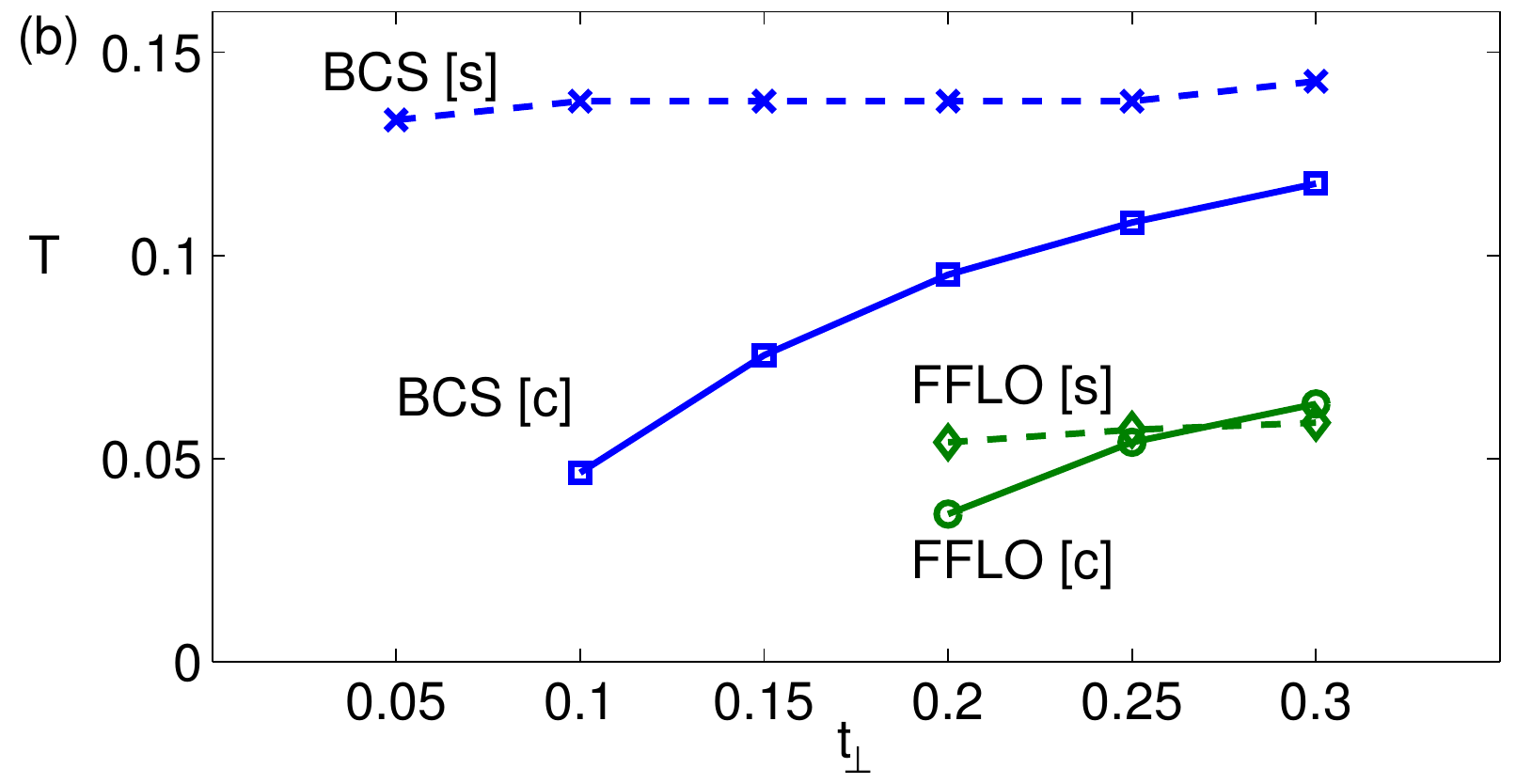}
  }
  \caption{Critical temperatures for the BCS and FFLO states as a function of the interchain hopping $t_\perp$ showing how the temperatures go down in approaching the 1D limit when non-local fluctuations are taken into by the cluster version (label [c]). In contrast, the single site (label [s]) DMFT which approximates the self-energy to be local gives critical temperatures essentially independent of dimensionality. All critical temperatures are smaller than the BCS mean-field $T_{c,MF} = 0.52$ and the mean-field prediction for the FFLO critical temperature of $0.5 T_{c,MF}$.     
    The calculations are for total density at half filling, but in the FFLO case the spin-polarization varies. Reproduced with permission from~\cite{heikkinen_nonlocal_2014}.
    \label{fig:ClusterResultsMiikka2}
  }
\end{figure}

As the main result, the FFLO state was found to be stable even in the presence of non-local quantum fluctuations. The critical temperatures for both the FFLO and BCS phases were found to be smaller than those predicted by mean-field theory, see figure~\ref{fig:ClusterResultsMiikka2}. There was a clear difference between the cluster and single-site, real-space DMFT results: the cluster version captured the increasing fluctuations when the 1D limit was approached, while the critical temperature in the single-site version remained rather constant when the interchain tunneling was decreased to realize quasi-1D systems. Notably, even the BCS critical temperature was strongly affected by non-local quantum fluctuations at the quasi-1D limit. This indicates that the cluster DMFT approach used was capable of describing the decreasing robustness of long-range coherence when going towards the 1D limit and the related drop in critical temperature.  

Hulet, Mueller and coworkers~\cite{liao_spin-imbalance_2010} have experimentally studied an imbalanced Fermi gas in a continuum 1D system and found consistency with the FFLO state, although no smoking gun signature.
We do not discuss this important experiment here because it belongs to the topic of a 1D FFLO phase, which was excluded from this review.
The same experimental group recently studied the 1D-3D crossover~\cite{revelle_1d-to-3d_2016}, note that the difference to the crossover studied in this section is that their 1D system is a continuum tube, not a 1D lattice system as here. The crossover was driven by tuning the tunneling strength between the 1D tubes.
They observed that indeed there is a critical tunneling energy separating the 1D and 3D regimes: in the 1D regime, the polarized superfluid appeared in the middle of the tube, while in the 3D case it was found at the edges, consistent with our discussion of what happens in a corresponding lattice 1D-3D crossover. Direct signatures of the FFLO state were not observed.

\subsection{Quantum Monte Carlo studies of the FFLO state in 2D square lattices}
\label{QMCstudies}

As discussed in section~\ref{MFstudies}, in 2D the nesting effects near van Hove singularities are predicted to be particularly strong.
However, quasi-long range order should be possible in 2D, in contrast to 1D, via BKT type mechanisms. This makes 2D lattices a promising setting for the FFLO state but also imposes demands on the theoretical treatment since fluctuations should be important. In 2012, Wolak {\it et al.}~\cite{Wolak2012} used determinant QMC to tackle this question and found for low fillings that the pair correlator indeed had a feature at finite momentum, suggesting instability towards the FFLO phase.
Simultaneously they calculated normalized double occupancy that describes pairing. At higher temperatures, the peak in the pair correlator reduces to zero instead of the finite value, while the normalized double occupancy stays constant. This was interpreted as a polarized paired phase (PPP), without the broken translational symmetry typical for the FFLO state. However, when the authors calculated the pairing susceptibility, they found no $s$-wave superfluidity in the polarized case for the parameter regime they were able to study. The fermion sign problem limited their study to low densities and temperatures at $0.1-0.2 t$, where $t$ is the hopping, or higher. Thus the existence of an FFLO superfluid remained to be confirmed. 

\begin{figure}
  \centerline{
    \includegraphics[width=0.9\columnwidth]{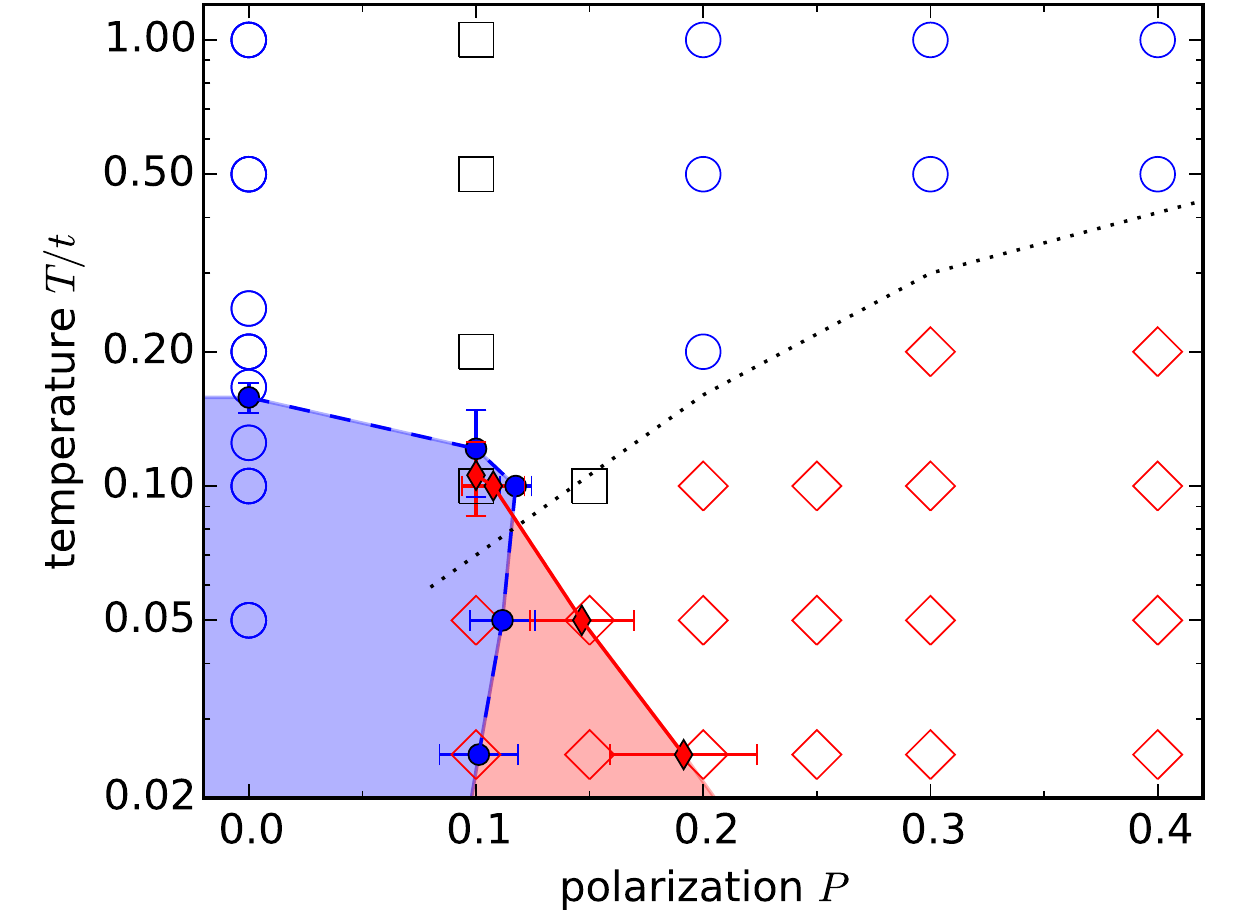}
  }
  \caption{ 
    The phase diagram from~\cite{gukelberger_fulde-ferrell-larkin-ovchinnikov_2015}. Here quarter filling $f=n/2=(\langle n_\uparrow + n_\downarrow \rangle)/2 = 0.25$ is considered and the interaction is chosen to be $U/t = −4$. The white region Fermi
    liquid is unstable
    towards conventional (Q = 0) pairing in the blue shaded region. In the red shaded
    region an exclusive FFLO instability with finite pair
    momentum occurs. Open symbols indicate whether zero- (blue
    circles) or finite-momentum pairing (red diamonds) is dominant
    (black squares: no significant difference). Instability of the Fermi liquid towards the FFLO/BCS state is determined from the divergence of the pair susceptibility. 
  }
  \label{fig:QMCcomparison1}
\end{figure}

Gukelberger {\it et al.}~\cite{gukelberger_fulde-ferrell-larkin-ovchinnikov_2015} used unbiased diagrammatic QMC simulations and determined the FFLO superfluid phase boundaries from the divergence of the pair susceptibility.
For attractive interactions of strength of about half the band width, they found a reasonably large parameter area in the polarization - temperature phase diagram where the FFLO state is predicted. A zero momentum pairing phase is found, and the high temperature - high polarization phase is identified as a Fermi liquid, see figure~\ref{fig:QMCcomparison1}. The critical temperatures for the FFLO phase were found to be optimally about half of the critical temperature of the balanced case, in accordance with the DMFT studies~\cite{heikkinen_finite-temperature_2013,heikkinen_nonlocal_2014}. For very large polarizations, a triplet superfluid was found, but at an exponentially small temperature.  

\begin{figure}
  \centerline{
    \includegraphics[width=0.9\columnwidth]{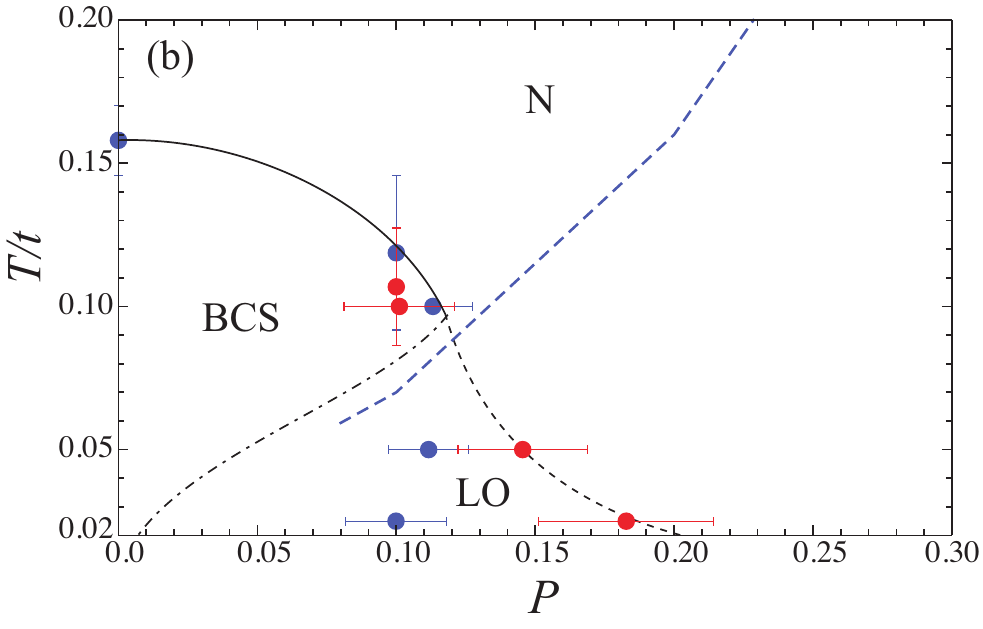}       
  }
  \caption{ 
  Mean-field phase diagram of a 2D square lattice from~\cite{baarsma_larkin-ovchinnikov_2016} at fixed filling f = 0.25 as function of temperature $T$, in units of the hopping $t$, and
    scaled polarization $P$. The results shown by the lines are scaled by $T'$ and $P'$ as explained in the text. Normal (N) phase as well as BCS and LO superfluid phases are found. The three phases meet at the Lifshitz
    point. All transition lines are continuous phase transitions. The diagrammatic Monte Carlo results from~\cite{gukelberger_fulde-ferrell-larkin-ovchinnikov_2015} are shown for comparison. The mean-field results match the QMC ones with the scaling. The blue dots mark the phase boundaries of the BCS
    superfluid phase and the red dots the transitions from the normal to an FFLO state. The BCS
    pairing and the FFLO pairing dominated regions are separated by the dashed blue line.
  }
  \label{fig:QMCcomparison2}
\end{figure}

Interestingly the phase diagram matches the mean-field result with simple scaling of the temperatures and polarizations, as was found by Baarsma and T\"orm\"a~\cite{baarsma_larkin-ovchinnikov_2016}. In that work, the two-component Fermi gas in a 2D square lattice was considered using the mean-field ansatz, which describes the LO state, see section~\ref{MFstudies}. As expected, there is a clear quantitative difference between the mean-field and the QMC results, the latter producing smaller critical temperatures and polarizations by a factor of about four. However, when the temperatures are rescaled by the scaling factor $T'$ between the two critical temperatures at $P=0$, and the polarizations similarly by the ratio of polarizations at a low temperature $P'$, the two phase diagrams match remarkably well for all values of $T$ and $P$, see figure~\ref{fig:QMCcomparison2}. 
	
In~\cite{Karmakar2016} a hybrid of mean-field theory and the QMC approach that aims to capture thermal fluctuations of the order parameter was used. Critical temperatures were found to be about 4-5 times smaller than predicted by mean field theory, similar to the difference in the QMC and mean-field calculations in~\cite{baarsma_larkin-ovchinnikov_2016} and~\cite{gukelberger_fulde-ferrell-larkin-ovchinnikov_2015} discussed above. However the temperature-magnetization (magnetization divided by the density equals polarization) phase diagram obtained by their hybrid method predicts the LO region to be much smaller and in a different place than the exact QMC results of~\cite{gukelberger_fulde-ferrell-larkin-ovchinnikov_2015}, although direct comparison is not possible due to the near half-filling and quarter filling densities used in~\cite{Karmakar2016} and~\cite{gukelberger_fulde-ferrell-larkin-ovchinnikov_2015}, respectively. Similarly to~\cite{Wolak2012}, the authors of~\cite{Karmakar2016} find a finite momentum signature in the pairing structure factor well above $T_c$. 

Note that since the focus of this review is on the FFLO state in lattices, we do not discuss here beyond mean-field theory studies treating continuum imbalanced Fermi gases. Likewise, we will not give a comprehensive review of beyond mean-field treatments of trapped imbalanced gases, but a few examples of such studies are discussed in section~\ref{section_traps}. 

\section{FFLO in trapped gases and phase separation}

\label{section_traps}

The effect of trapping potential and its relation with FFLO has already been reviewed in~\cite{gubbels_imbalanced_2013}, so here we give only a brief description and then concentrate on topics not covered by their review.
Particularly we consider here the effect of trap elongation, beyond mean-field theories, and we give a more detailed analysis of the Bogoliubov-de Gennes (BdG) method.
Most of the discussion in this section involves gases without optical lattices, although a lattice can be used for producing highly elongated potentials, and the BdG method can be used also in lattices~\cite{chen_exploring_2009}.

Ultracold atom gas experiments are always conducted in some kind of external trapping potential.
This has profound effects on the behaviour of the gas as the translational symmetry is broken and atom gas density becomes position dependent.
Furthermore it has particularly dramatic effects on spin-imbalanced gases, as the position dependence of the trap naturally drives a phase separation in which, in 3D, the edges of the trap become fully polarized while the center of the trap hosts both atomic species.

Fortunately, the external trapping potential can be incorporated in the mean-field theory.
The external trapping potential in the Hamiltonian~\eqref{eq:a} is described by the term
\begin{eqnarray}
   \hat V_\mathrm{ext} = \sum_{\sigma=\uparrow,\downarrow} V({\bf r}) \hat \psi_\sigma^\dagger({\bf r}) \hat \psi_\sigma ({\bf r}).
\end{eqnarray}
In practice the trapping potential is often assumed to be spherically symmetric $V({\bf r}) = V(r)$, or at least cylindrically symmetric. 
The mean-field approximation proceeds as outlined in section~\ref{BasicFFLOTheoryLattice}, with the main differences being that the order parameter becomes position dependent $\Delta({\bf r})$ with a priori unknown position dependence, and that the Hartree energy shift, which follows the shape of the density profile $n_\sigma({\bf r})$, can no longer be simply included in the chemical potential. The mean-field Hamiltonian in a trap becomes
\begin{eqnarray}
  \hat H = &\int d{\bf r} \, \sum_{\sigma=\uparrow,\downarrow} \hat \psi_\sigma^\dagger ({\bf r}) K_\sigma({\bf r}) \hat \psi_\sigma ({\bf r}) + \\
  &\int d{\bf r} \left(\Delta({\bf r}) \hat \psi_\uparrow^\dagger({\bf r}) \hat \psi_\downarrow^\dagger ({\bf r}) + H.c. \right),
\label{eq:trappedBCS}
\end{eqnarray}
where the local single-particle Hamiltonian is defined as
\begin{equation}
  K_\sigma({\bf r}) = -\frac{\hbar^2 \nabla^2}{2m} - \mu_\sigma + V_0 n_{-\sigma}({\bf r})+ V_\mathrm{ext}({\bf r}).
\label{eq:singlepartH}
\end{equation}
In practice the Hartree shift, the $V_0 n_{-\sigma}$-term in the single-particle Hamiltonian, is usually completely neglected when studying trapped gases as it is ill-behaved in the strongly interacting limit. 
For strong attractive interactions $|k_\mathrm{F}a| > \frac{4}{3\pi}$, the Hartree shift predicts an unphysical collapse of the gas.
This instability is driven by the uncontrolled increase of the Hartree shift when the density of the gas increases.
This critical bound is easily obtained by considering the total energy cost of adding a particle: the kinetic energy equals the Fermi energy $E_\mathrm{F} = \frac{\hbar^2 k_\mathrm{F}^2}{2m}$ and the (Hartree) interaction energy gain is $\frac{4\pi\hbar^2 a}{m} n$, where $n=k_\mathrm{F}^3/(6\pi^2)$ is the density of one of the spin components. 
Together these energies yield
\begin{equation}
   E = \frac{\hbar^2 k_\mathrm{F}^2}{2m} + \frac{4\pi\hbar^2 a}{m} \frac{k_\mathrm{F}^3}{6\pi^2} = \frac{\hbar^2 k_\mathrm{F}^2}{2m} \left[1 + \frac{4}{3\pi} k_\mathrm{F}a\right].
\end{equation}
As seen from the equation, the interaction energy increases faster than the kinetic energy cost as function of the Fermi momentum $k_\mathrm{F}$, and hence the density $n$.
In particular when the factor $1 + \frac{4}{3\pi} k_\mathrm{F}a$ becomes negative, the total energy cost of adding a particle becomes negative. 
It is thus energetically favorable for the gas to collapse when attractive interaction becomes strong: $\left|k_\mathrm{F}a\right| > \frac{3\pi}{4}$.
Such collapse is completely unphysical but it plagues mean-field theories and the BdG method in particular.
The problem can be solved by using correct (i.e., determined by Monte Carlo methods or experimentally) energy functional for the Hartree energy shift~\cite{bulgac_zero-temperature_2007}, but often the Hartree energy shift is simply neglected in mean-field studies. 

The mean-field Hamiltonian equation~\eqref{eq:trappedBCS} can be diagonalized formally in the same way as in the homogeneous case, leading eventually to the BdG equations:
\begin{equation}
\left(\begin{array}{cc} 
 K_\uparrow({\bf r}) & \Delta ({\bf r}) \\
 \Delta({\bf r}) & -K_\downarrow({\bf r}) \end{array}\right) \left(\begin{array}{c} 
   u_i({\bf r}) \\
   v_i({\bf r}) \end{array} \right) = E_i \left(\begin{array}{c} 
   u_i({\bf r}) \\
   v_i({\bf r}) \end{array} \right),
\label{eq:bdg}
\end{equation}
where $i$ is the eigenstate index.
The physical interpretation of each of the elements in these BdG equations are the same as in the homogeneous case discussed in section~\ref{subsec:bogoliubov_transformation}.
However, the apparent simplicity of these equations is misleading. The position dependence of both the order parameter $\Delta({\bf r})$ and the differential operators in the single-particle Hamiltonian $K_\sigma({\bf r})$ require one to solve either coupled differential equations or turn each of the elements into matrices.
Notice that this is also the case of a homogeneous LO-state, which does not allow separation into different momentum states ${\bf k}$, but instead involves a coupling between all states, see section~\ref{subsec:bogoliubov_transformation}.

Before analyzing the BdG equations in more detail, we first briefly review the use of the local density approximation for solving equation~\eqref{eq:bdg}.
We also shortly review other more advanced theories used for describing FFLO physics in trapped systems, not discussed in earlier reviews.

\subsection{Local density approximation}

In the local density approximation (LDA), the gas is assumed to be locally homogeneous, with the external potential $V_\mathrm{ext}({\bf r})$ providing only a position-dependent energy shift.
This assumption is expected to be valid for sufficiently large systems in which all physical quantities are only slowly varying with position.
With such an assumption, the position coordinate ${\bf r}$ can be neglected in BdG equations~\eqref{eq:bdg}, becoming only a parameter that determines the local chemical potential
\begin{equation}
  \tilde \mu_{{\bf r},\sigma} = \mu_\sigma + V_\mathrm{ext}({\bf r}),
\end{equation}
and the BdG equations become identical to the ones in the homogeneous case
\begin{equation}
\left(\begin{array}{cc} 
 \epsilon_k - \tilde \mu_\uparrow & \Delta \\
 \Delta & -\epsilon_k + \tilde \mu_{\downarrow} \end{array}\right) \left(\begin{array}{c} 
   u_{i,{\bf k}} \\
   v_{i,{\bf k}} \end{array} \right) = E_{i,{\bf k}} \left(\begin{array}{c} 
   u_{i,{\bf k}} \\
   v_{i,{\bf k}} \end{array} \right),
\label{eq:ldabdgmom}
\end{equation}
where $\epsilon_k= \frac{\hbar^2 {\bf k}^2}{2m}$.
However this needs to be solved separately for each different value of the (local) chemical potential $\tilde \mu_\sigma$.
Thus the LDA to the BdG equations means solving the homogeneous mean-field theory locally for each spatial point ${\bf r}$, with chemical potentials that are shifted by the value of the external potential $V_\mathrm{ext}({\bf r})$.

\begin{figure}
\includegraphics[width=0.95\columnwidth]{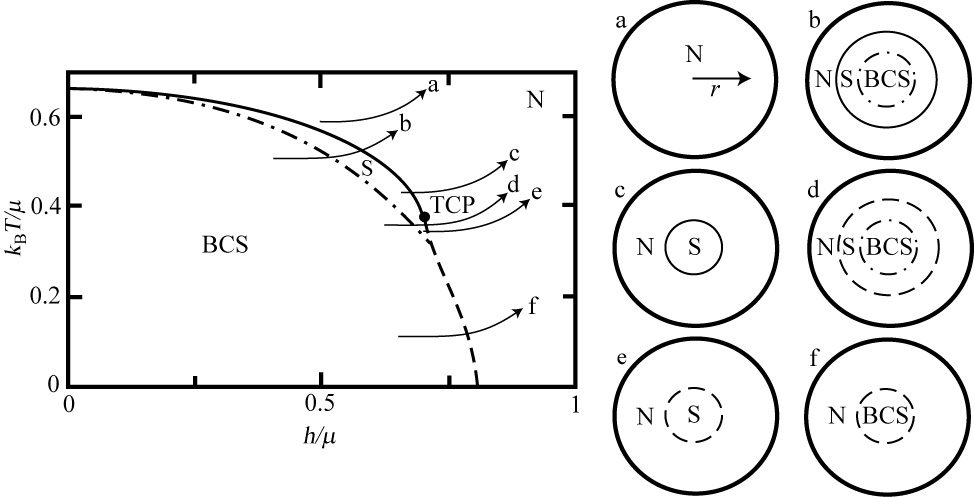}
\centering
\caption{A schematic picture of how LDA in a trap involves a range of parameters in homogeneous phase diagram. The tails of the arrows correspond to the center of the trap. Moving towards the edge of the cloud, the average chemical potential $\mu$ decreases, and the relative chemical potential difference $\delta \mu/\mu=h/\mu$ and scaled temperature $k_\mathrm{B}T/\mu$ increase. Crossing a phase boundary in the phase diagram corresponds to an interface between shells of different phases in the trapped gas. Different polarizations, temperatures and atom numbers yield the different cases a-f. Here BCS and N phases correspond to balanced superfluid and normal fluid phases. The $S$-phase corresponds to a gapless superfluid Sarma phase that is stabilized by finite temperature. The FFLO phase was not considered in the diagram. Reproduced with permission from~\cite{gubbels_imbalanced_2013}.}
\label{fig:phasediagramtolda}
\end{figure}
Assuming that the two spin components feel the same trapping potential, the chemical potential difference $\delta \mu = \tilde \mu_\uparrow - \tilde \mu_\downarrow$ does not depend on the position, whereas the average chemical potential $\mu_\mathrm{avg}({\bf r}) = (\tilde \mu_{{\bf r},\uparrow} + \tilde \mu_{{\bf r},\downarrow})/2$ does. 
To solve the mean-field BCS theory for a trapped gas in the LDA, it is thus sufficient to solve the homogeneous order parameter $\Delta$ and the densities $n_\sigma$ for a range of average chemical potentials.
Once one has the functions $\Delta(\mu_\mathrm{avg})$, $n_\sigma(\mu_\mathrm{avg})$ mapped for a fixed chemical potential difference $\delta \mu$ and temperature $T$, the local densities $n_{\sigma,{\bf r}}$ and order parameter $\Delta_{\bf r}$ can immediately be determined as 
\begin{equation}
   \Delta_{\bf r}^\mathrm{LDA} = \Delta \left(\frac{\mu_\uparrow + \mu_\downarrow}{2} - V_\mathrm{ext}({\bf r})\right)
\end{equation}
and
\begin{equation}
   n_{{\bf r},\sigma}^{\mathrm{LDA}} = n_\sigma\left(\frac{\mu_\uparrow + \mu_\downarrow}{2} - V_\mathrm{ext}({\bf r})\right).
\end{equation}
The trapped gas thus effectively spans a range of parameters in the homogeneous gas phase diagram, and from the phase diagram one can immediately read the corresponding layered phase structure of the gas, see figure~\ref{fig:phasediagramtolda}.
In practice, one needs to iterate the chemical potential difference $\delta \mu$ if one wants to solve the model for a specific number of atoms.

As described by the LDA, the inhomogeneous potential drives the system to phase separation~\cite{sheehy_becbcs_2007,zhang_finite-temperature_2007}. 
In 3D, at the edge one has a single-component gas in a normal state, whereas the center of the trap involves an equal density BCS-superfluid. 
This general property is seen in experiments~\cite{zwierlein_fermionic_2006,zwierlein_direct_2006}, but also in practically all theories using LDA, both mean-field~\cite{gubbels_imbalanced_2013} and beyond mean-field theories~\cite{pieri_trapped_2006,chien_finite_2006}.

Indeed LDA is very versatile, as it is not limited to mean-field theories, but it can be applied to any theory for homogeneous gases.
The homogeneous mean-field results suggest that the FFLO phase involves only a sliver of parameter space between phase separated and normal states in the homogeneous gas phase diagram~\cite{sheehy_bec-bcs_2006}.
This translates into a narrow shell of FFLO state in a trapped gas for specific parameters for atom numbers, interaction strength and temperature.
More involved beyond mean-field theories have similarly been applied to trapped gases using the LDA~\cite{pieri_trapped_2006,chien_finite_2006}. 
These works considered the role played by pseudogap physics at temperatures above the critical temperature, but they did not consider the FFLO state.

\begin{figure}
\centering
\includegraphics[width=0.95\columnwidth]{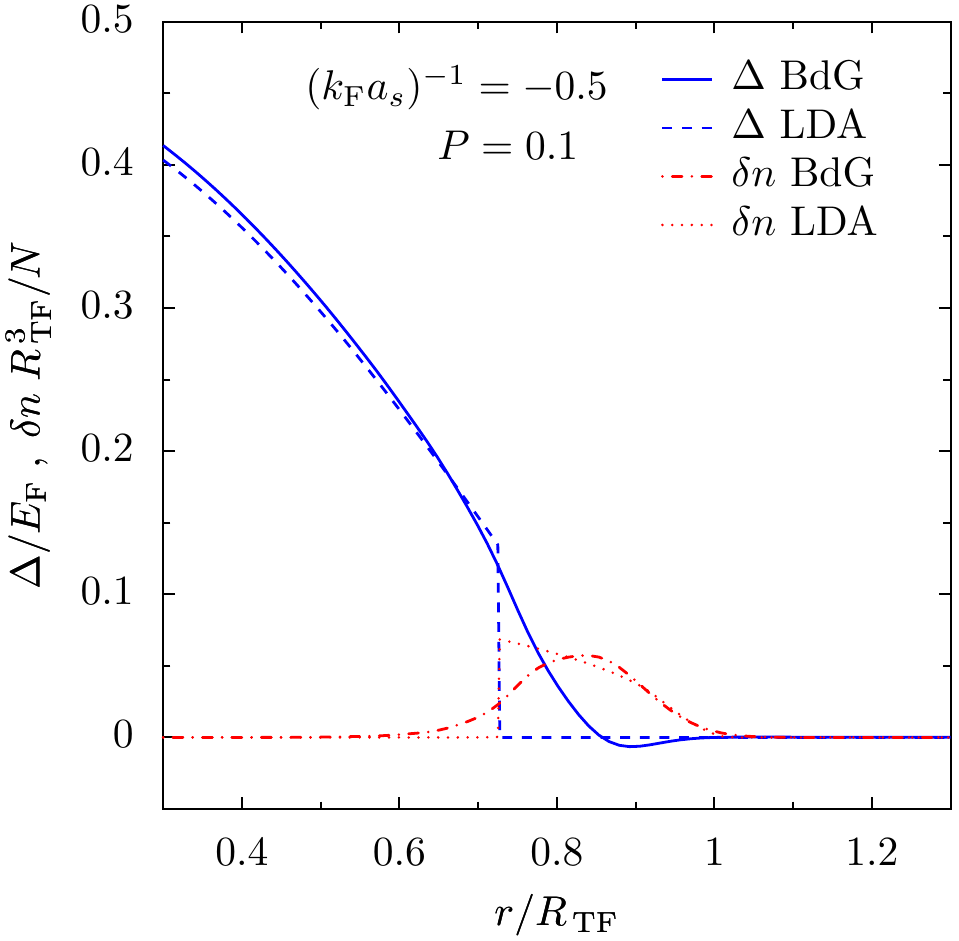}
\caption{Mean-field theory together with LDA produces sharp order parameter $\Delta(r)$ and density (here density difference $\delta n(r)$) changes at phase boundaries. Surface tension provides an energy cost for such rapid variations, leading to smoother functional dependencies of densities and superfluid order parameter as seen in profiles obtained with BdG equations. Reproduced with permission from~\cite{jensen_non-bcs_2007}.}
\label{fig:densityprofile_bdg_lda}
\end{figure}
LDA involves sharp phase boundaries, and the zero-temperature mean-field theory in particular produces sharp density changes at the superfluid-normal state interfaces as seen in figure~\ref{fig:densityprofile_bdg_lda}.
Such rapid variations violate the assumptions of slowly varying potentials in LDA.
Improvements to LDA involve gradient corrections that describe the surface tension or domain wall energies~\cite{de_silva_surface_2006, imambekov_breakdown_2006, caldas_surface_2007, baur_theory_2009,gubbels_imbalanced_2013}.
When an energy cost is associated with the superfluid-normal interfaces, rapid variations of the order parameter $\Delta({\bf r})$ and densities $n_\sigma({\bf r})$ are suppressed.  
Mean-field theories showed that this should lead to deformation of the superfluid core to minimize the size of the superfluid-normal interfaces.
These effects were hoped to solve discrepancies observed in the experiments by Rice~\cite{partridge_pairing_2006} and MIT groups~\cite{zwierlein_fermionic_2006,zwierlein_direct_2006}.
The surface tension was too weak~\cite{baur_theory_2009} to provide an explanation.
Instead it was suggested~\cite{parish_evaporative_2009} and later experimentally confirmed~\cite{liao_metastability_2011} that the features in the Rice experiment were due to the evaporative cooling from highly elongated trap, resulting in a depolarized nonequilibrium state at the center of the trap. 
However the concept of surface tension is useful for understanding deformations of the density profiles predicted by BdG equations in elongated traps, which will be discussed below.

\subsection{Bogoliubov-de Gennes equations in spherical traps}

The BdG equations~\eqref{eq:bdg} can be solved also without the LDA. 
However this requires either the solution of coupled differential equations or expressing the different elements in matrix form.
The first approach using differential equations was reviewed by Gubbels and Stoof~\cite{gubbels_imbalanced_2013}, so here we review the matrix formalism, which is the approach used by a number of authors.
We focus on the case of spin-imbalanced Fermi gas in a spherically symmetric harmonic trap~\cite{castorina_nonstandard_2005,kinnunen_strongly_2006,machida_generic_2006}, but the formalism itself also works with minor modifications in non-spherically symmetric traps~\cite{kim_exotic_2011,baksmaty_concomitant_2011,baksmaty_bogoliubovgennes_2011}.
Also, the early works on FFLO by Mizushima {\it et al.} involved BdG equations in cylindrically symmetric system with periodic boundary conditions~\cite{mizushima_direct_2005}.

The standard approach is to expand single-particle operators in the eigenbasis of the trapping potential.
This results in a diagonal form for the single-particle part of the mean-field Hamiltonian. 
Another possible choice is to use an evolving quasiparticle basis, in which the eigenbasis of the previous iteration of the gap equation $\left(u_i({\bf r}), v_i({\bf r}) \right)^T$ is used as the basis for the next iteration.

In the case of a spherically symmetric harmonic trap, $V(r) = \frac{1}{2} m\omega^2 r^2$, where $\omega$ is the trapping frequency, the single-particle creation and annihilation operators are expressed in the basis of 3D harmonic oscillator eigenstates
\begin{equation}
\hat{\psi}_{\sigma}(\vec{r})=\sum_{nlm}R_n^l(r)Y_{lm}(\hat{r})\hat{c}_{nlm\sigma},
\end{equation}
where the quantum numbers $n,l,m$ are: the radial quantum number $n$ counting the nodes in the radial function, the orbital angular momentum $l$, and the projected angular momentum $m$. 
The angular part of the eigenstate $Y_{lm}(\hat{r})$ is the spherical harmonic with the radial unit vector $\hat{r}=(\theta,\varphi)$.
The radial part of the wavefunction is 
\begin{equation}
R_n^l(r)=\frac{\sqrt{2}}{r_\mathrm{osc}^{3/2}} \sqrt{\frac{n!}{\left(n+l+\frac12\right)!}}e^{-\bar{r}^{2}/2}\bar{r}^{l}L_{n}^{l+1/2}\left(\bar{r}^2\right),
\end{equation}
where $\bar{r}=r/r_{\mathrm{osc}}$, the harmonic oscillator length $r_{\mathrm{osc}}=[\hbar/(m\omega)]^{1/2},$ and $L_{n}^{l+1/2}(\bar{r}^{2})$ is the associated Laguerre polynomial. 
The corresponding single-particle eigenenergies are $\varepsilon_{nl} = \hbar \omega \left(2n+l+3/2\right)$, with different $m$-states being degenerate.
While such analytical formulas for the single-particle eigenstates are useful, the BdG method is also quite applicable with numerically solved eigenstates.

The spherical symmetry, which is assumed to also hold for the order parameter $\Delta(r)$, allows performing the angular integrations.
This makes the mean-field Hamiltonian in equation~\eqref{eq:trappedBCS} diagonal in $m$-quantum numbers and the resulting radial 1D Hamiltonian $(2l+1)$-fold degenerate. 
Dropping the redundant $m$-quantum number, the Hamiltonian becomes
\begin{eqnarray}
H_{\mathrm{MF}} & = & \sum_{n,l,\sigma}(2l+1)\left(\varepsilon_{nl}-\mu_{\sigma}\right)\hat c_{nl\sigma}^{\dagger} \hat c_{nl\sigma}\nonumber \\
 & + & \sum_{n,n^{\prime},l,\sigma}J_{nn^{\prime}\bar{\sigma}}^{l} \hat c_{nl\sigma}^{\dagger} \hat c_{n^{\prime}l\sigma}\nonumber \\
 & + & \sum_{n,n^{\prime},l}F_{nn^{\prime}}^{l} \hat c_{nl\uparrow}^{\dagger} \hat c_{n^{\prime}l\downarrow}^{\dagger}+\mathrm{H.c}.
\end{eqnarray}
Here the Hartree interaction is described by the elements 
\begin{equation}
J_{nn^{\prime}\sigma}^{l}=\int_{0}^{\infty}dr\ r^{2}R_n^l(r)Un_{\sigma}(r)R_{n^\prime}^l(r),
\end{equation}
and the pairing field is described by 
\beq
F_{nn^{\prime}}^{l}=\int_{0}^{\infty}dr\ r^{2}R_n^l(r)\Delta(r)R_{n^\prime}^l(r).
\enq
We note that the $\sigma$ dependence of the Hartree term is due to the population imbalance which implies that the corrections are different for the two components. 
The density of $\sigma$ atoms is 
\begin{equation} 
n_{\sigma}(r)=\sum_{n,n^{\prime},l}\frac{2l+1}{4\pi}R_n^l(r)R_{n^\prime}^l(r)\langle \hat c_{nl\sigma}^{\dagger}\hat c_{n^{\prime}l\sigma}\rangle,
\end{equation}
and the order parameter is 
\begin{equation}
\Delta(r)=U\sum_{n,n^{\prime},l}\frac{2l+1}{4\pi}R_n^l(r)R_{n^\prime}^l(r)\langle \hat c_{nl\uparrow}^{\dagger} \hat c_{n^{\prime}l\downarrow}^{\dagger}\rangle.
\end{equation}
The additional factor $(2l+1)$ comes from the degeneracy of the $m$-states.

The nature of the four matrix entries in the BdG equations~\eqref{eq:bdg} can now be explained.
The upper left corner element, $\hat K_\uparrow({\bf r})-\mu_\uparrow$ becomes a matrix with diagonal elements
$\epsilon_{nl} + J_{nn\downarrow}^l- \mu_\uparrow$ describing single-particle energies and Hartree shifts, and off-diagonal elements $J_{nn' \downarrow}^l$ provide transitions between the $n$-levels due to a position dependent Hartree shift.
Similarly, the upper right corner element, $\Delta({\bf r})$, becomes a matrix with elements $F_{nn'}^l$.
The diagonal elements of these blocks, $F_{nn}^l$, describe intrashell pairing in a manner analogous to the zero center-of-mass momentum ${\bf k,-k}$ pairing in a homogeneous system. Similarly, the off-diagonal elements of the block, $F_{nn'}^l$ describe pairing between different harmonic trap levels $n$ and $n'$ (the $l$-quantum number being the same for both).
Solving the BdG equations involves diagonalizing the resulting matrix, yielding the eigenenergies $E_i$ and the eigenstates $(u_i,v_i)$. 
Obviously one needs to impose a cutoff to keep the dimensionality of the matrices finite, but even more importantly since the gap equation is ultraviolet divergent.
The regularization of the gap equation, and the associated renormalization of the bare interaction $U$ is generally done using LDA.
However it is not clear how renormalization should be done when studying 1D-3D crossover, as LDA based schemes are either for 1D (in which case regularization/renormalization is not needed) or for 3D.
For the present purposes, it is enough to simply assume some large energy cutoff $\epsilon_\mathrm{c}$.
With the energy spectrum of the 3D harmonic oscillator, the energy cutoff implies an angular momentum $l$-dependent cutoff for the radial quantum number $n < N_l = \epsilon_\mathrm{c}/(2\hbar\omega)- l/2$.

In spherically symmetric systems, the matrix equation has a block structure, since the Hamiltonian can be expressed as a sum of mutually commuting terms for different $l$-quantum numbers.
One can thus express it as $H_\mathrm{MF} = \sum_l H_\mathrm{MF}^l$, and write a separate set of BdG equations for each $H_\mathrm{MF}^l$ block.
Diagonalizing these yields eigenenergies $E_{jl}$ and the corresponding eigenstates $(u_j^l,v_j^l)^T$,
where the $u_j^l$ and $v_j^l$ are now vectors in the initial single-particle basis and $j$-index enumerates the different eigenstates of the sub-Hamiltonian $H_\mathrm{MF}^l$. 
Notice that the $j$-index also includes the $+/-$-quasiparticle branches introduced in section~\ref{subsec:bogoliubov_transformation}, and consequently the sign convention for the quasiparticle energies $E_{jl}$ is altered from the one in equation~\eqref{eq:qp_energies}.
Here half of the eigenstates have negative energies, and the zero-temperature ground state corresponds to the state in which all quasiparticle states with negative energy states are occupied and others empty, whereas in section~\ref{subsec:bogoliubov_transformation}, the corresponding quasiparticle states describe particle and hole excitations.

The atom densities can now be written as
\begin{eqnarray}
  n_{\uparrow}(r) =& \\
\nonumber  &\sum_{jl}\frac{2l+1}{4\pi}\sum_{n,n^{\prime}=0}^{N_l}R_n^l(r)R_{n^\prime}^l(r) u_{jn}^l u_{jn'}^l n_{\mathrm{F}}(E_{jl}),
\end{eqnarray}
where the Fermi distribution $n_{\mathrm{F}}(E)=1/(1+e^{E/k_{\mathrm{B}}T})$ and $u_{jn}^l$-factors describe the weight of the spin-$\uparrow$ component in each state $nl$.
Likewise, the density of atoms in $\downarrow$ state is 
\begin{eqnarray}
  n_{\downarrow}(r)=&\\
  \nonumber &\sum_{jl}\frac{2l+1}{4\pi}\sum_{n,n^{\prime}=0}^{N_l}R_n^l(r)R_{n^\prime}^l(r) v_{jn}^{l}v_{jn^{\prime}}^{l}n_{\mathrm{F}}(-E_{jl}),
\end{eqnarray}
where the sign in the Fermi function is changed because the $\downarrow$ component of the eigenstates correspond to holes as compared to the particles in the  $\uparrow$ components.
The order parameter is given by
\begin{eqnarray}
  \label{eq:finalgap}
  \Delta(r) = & \\
  \nonumber &U\sum_{jl}\frac{2l+1}{4\pi}\sum_{n,n^{\prime}=0}^{N_l}R_n^l(r)R_{n^\prime}^l(r)u_{jn}^{l}v_{jn^{\prime}}^{l} n_{\mathrm{F}}(E_{jl}).
\end{eqnarray}

BdG equations are ideally suited for describing FFLO-type physics, since they do not in principle restrict the spatial dependence of the order parameter.
However in practice, most of the studies have assumed that the order parameter is real and spherically symmetric $\Delta({\bf r}) = \Delta(r)$.
These assumptions neglect the possibility of FF-type superfluids in which the phase of the order parameter is changing, and the angular FFLO-state in which the order parameter varies not only radially but also around the trap.
However, later BdG studies for gases in elongated (non-spherical) traps seem to confirm these early assumptions.
On the other hand, angular FFLO has been studied in toroidal traps~\cite{yanase_angular_2009} and in those systems even the complex FF-type states can be stabilized when the system is rotated~\cite{yoshida_rotating_2011}. Angular FFLO seems to also be present in 2D lattices at high filling~\cite{chen_exploring_2009}.

\subsection{The FFLO state and proximity effects in trapped 3D gases}

\begin{figure}
\centering
\includegraphics[width=0.95\columnwidth]{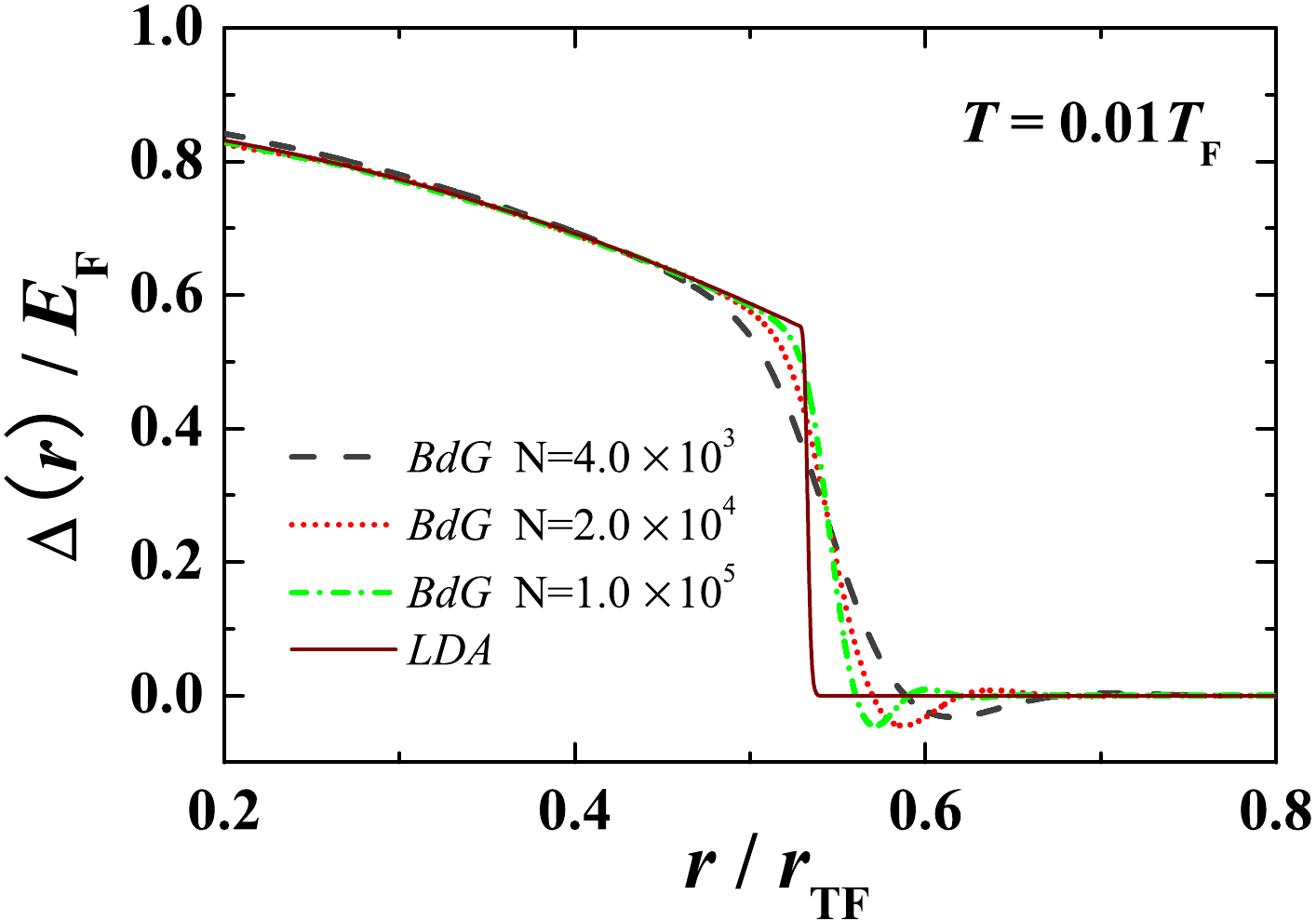}
\caption{Increasing the atom numbers in BdG theory produces profiles that are increasingly like the corresponding LDA mean-field results. Reproduced with permission from~\cite{liu_mean-field_2007}.} 
\label{fig:ndependence}
\end{figure}

The BdG theory yields density and order parameter profiles that are in good qualitative agreement over a wide range of parameters with experimental results but also with the LDA~\cite{zwierlein_fermionic_2006,zwierlein_direct_2006,kinnunen_strongly_2006,machida_generic_2006,liu_mean-field_2007,jensen_non-bcs_2007,gubbels_imbalanced_2013}.
Figure~\ref{fig:ndependence} shows how increasing the system size, or atom numbers, produces profiles that are increasingly similar to the LDA results.
In particular, BdG equations provide the correct general features of a balanced BCS-type superfluid in the center of the trap and fully polarized edges.

However between the balanced superfluid in the center and unpaired normal fluid at the edge, BdG equations provide a narrow region with an oscillating order parameter that has been associated with the FFLO-state.
Despite the similarity with the usual FFLO state, the oscillations are present whenever one has a superfluid-polarized normal state interface~\cite{machida_generic_2006}.
On the other hand, this interface region becomes relatively narrower when the size of the system is increased.
The interface region and the associated density and order parameter modulations can be understood as a proximity effect~\cite{liu_mean-field_2007,jensen_non-bcs_2007, gubbels_imbalanced_2013}.
The proximity effect is caused at the interface of the two phases by the correlations inside the superfluid penetrating the partially polarized normal gas.
The effect has been studied in interfaces of superconductors and ferromagnets in solid state systems~\cite{buzdin_proximity_2005}.
Indeed as the center of the trap is a balanced BCS-type superfluid, and the edge is a polarized normal gas, one does expect superfluid correlations to reach into the normal gas.
This manifests as oscillations of the order parameter, with the period of the oscillations determined by the difference in the local Fermi momenta of the two atomic species, in this sense similar to the FFLO state.
Similar order parameter oscillations have also been predicted for dipolar gases, in which the $p$-wave interaction channel itself provides explicit intershell coupling~\cite{Baranov_superfluidity_2004,Baranov_bcs_pairing_2004}.

The penetration depth has been shown to be independent of the system size~\cite{jensen_non-bcs_2007}.
On one hand, this implies that the region with order parameter oscillations cannot be considered to form a separate shell of an FFLO-type superfluid phase, as the relative size of the region shrinks when the number of atoms is increased.
On the other hand, it is not a finite size effect or an artifact of the BdG method, but rather an interface effect that can be understood through the surface tension discussed above.

\subsection{FFLO in elongated traps and dimensional crossover}

In the case of elongated traps, as is the case of actual experiments, the external trapping potential becomes
\begin{equation}
   V(\rho,z) = \frac{1}{2}m\omega_\rho^2 \rho^2 + \frac{1}{2} m \omega_z^2 z^2, 
\end{equation}
where $\omega_\rho$ and $\omega_z$ are the trapping frequencies in the radial and longitudinal directions, respectively.
For $\omega_z < \omega_\rho$, the trap is elongated in a cigar-like fashion, with the trap aspect ratio defined as $\lambda = \omega_\rho/\omega_z$.

The elongation has trivial effects when using LDA, since the local value of the potential $V_\mathrm{ext}({\bf r})$ uniquely determines the state in point ${\bf r}$ and the state is the same along an equipotential surface.
This also implies that all phase boundaries and density profiles in the trap follow the shape of the trap, and particularly, they have the same aspect ratios as the trap itself.
As discussed above, gradient corrections to the LDA change the picture by introducing energy costs to domain walls.
These surface tension effects are most naturally included in the BdG method and, as will be seen, the BdG method predicts significant deformation for the minority component atoms for highly elongated traps.
Interestingly more advanced theories, such as asymmetric superfluid local density approximation (ASLDA) density functional theory study by Pei {\it et al.}~\cite{pei_competition_2010} (see also~\cite{bulgac_zero-temperature_2007}) and dynamical mean-field theory (DMFT) in~\cite{kim_exotic_2011}, did not find similar elongation of the minority component density.
With these caveats, we can analyze the effect of elongation on BdG mean-field theory.

\begin{figure}
\includegraphics[width=0.95\columnwidth]{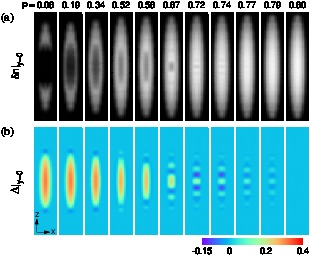}
\centering
\caption{Density difference $\delta n$ and order parameter $\Delta$ in an elongated system calculated using dynamical mean-field theory for a lattice gas. The FFLO-type order parameter oscillations align themselves in the longitudinal direction, and they encompass the whole gas for sufficiently large polarization and trap elongation. Reproduced with permission from~\cite{kim_exotic_2011}.}
\label{fig:dmft}
\end{figure}

Unfortunately breaking the spherical symmetry increases the numerical complexity of the BdG method, as the block diagonal form of the Hamiltonian is lost.
Actually the early works on the FFLO-state~\cite{mizushima_direct_2005,mizushima_fuldeferrelllarkinovchinnikov_2005} involved BdG equations applied to traps with cylindrical symmetry but with periodic boundary conditions in the longitudinal direction.
Similarly the case of extreme elongation, in which the system is predominantly 1D, has been considered in the review by Guan, Batchelor and Lee~\cite{guan_fermi_2013}.
In that limit, the spin-imbalance results in density difference at the center of the trap instead of depletion of the minority component at the edges~\cite{feiguin_pairing_2007}. 
Consequently the FFLO-type state is strongly enhanced in 1D traps, the case which has also been studied experimentally~\cite{liao_spin-imbalance_2010}.
However already moderate elongation of the trap has important effects on the density and order parameter profiles.
In particular, a DMFT study involving a trapped gas in a lattice in~\cite{kim_exotic_2011} predicts a polarization window for elongated traps in which the FFLO state can be stabilized, see figure~\ref{fig:dmft}.

\begin{figure}
\includegraphics[width=0.95\columnwidth]{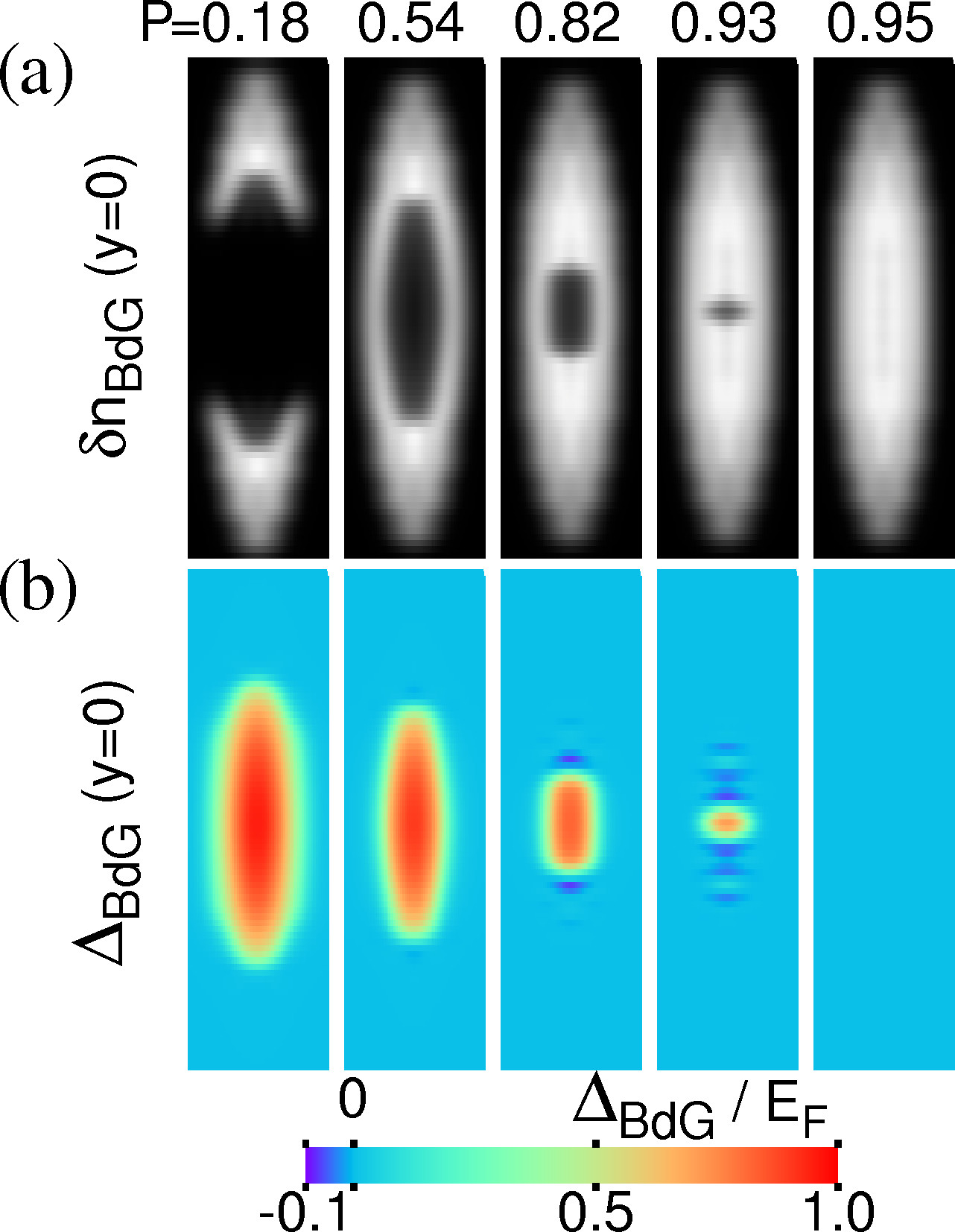}
\centering
\caption{Density difference $\delta n$ and order parameter $\Delta$ in an elongated system calculated using BdG mean-field theory. Comparing with figure~\ref{fig:dmft}, BdG equations produce qualitatively similar oscillations of the order parameter, but in addition the minority component densities are deformed and critical polarization is overestimated. Reproduced with permission from~\cite{kim_exotic_2011}.}
\label{fig:bdgdeform}
\end{figure}

The BdG method applied to elongated traps yields significant deformation in the minority component density profile, evident in figure~\ref{fig:bdgdeform}.
Baksmaty {\it et al.}~\cite{baksmaty_bogoliubovgennes_2011} studied the effect in detail, showing that it is enhanced in highly elongated systems.
At the highest elongation they studied, the radial density distributions of the minority and majority components were equal, reflecting that the minority component was, at the largest polarizations, more 2D rather than cigar-like as the majority component.
The cause for the deformation is in the surface tension at the superfluid-normal gas interface, which prefers the radial density profiles of the two components to match at the expense of longitudinal direction.
BdG studies for elongated traps also predict order parameter oscillations, like in spherically symmetric traps. 
With increasing elongation, these oscillations arranged themselves along the longitudinal axis, at the same time becoming more prominent~\cite{baksmaty_bogoliubovgennes_2011}.
While these interface effects are expected to also be present in larger systems, the role they play in the actual aspect ratios of the atom gases must decrease when the system size is increased.

\begin{figure}
\includegraphics[width=0.95\columnwidth]{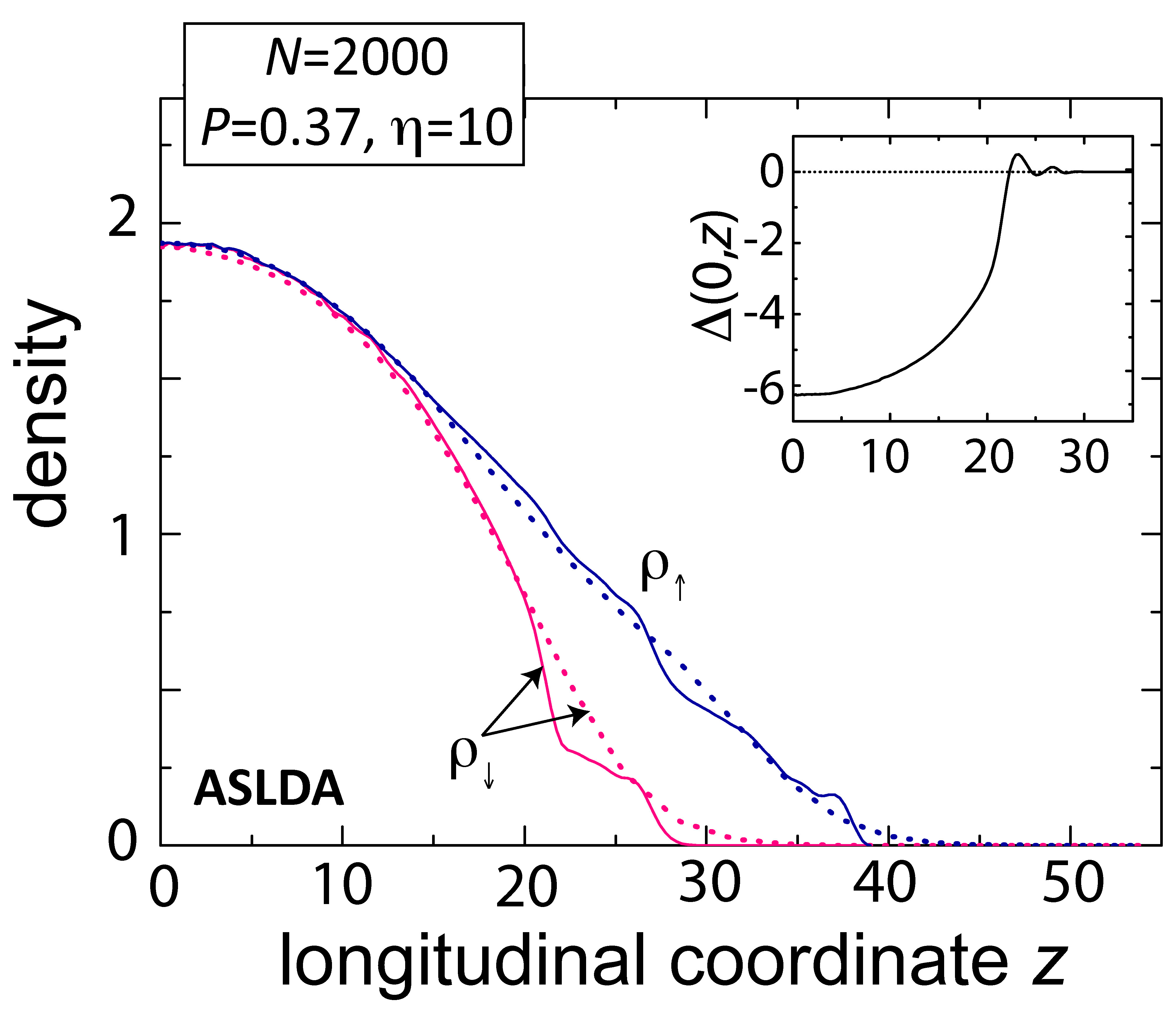}
\centering
\caption{Density and order parameter (inset) profiles of a unitary spin-imbalanced gas obtained using density functional ASLDA theory. Order parameter $\Delta(0,z)$ has minor modulations at the phase boundary, in agreement with mean-field BdG theory. However, density profiles (solid lines for longitudinal direction and dashed lines for transverse direction) show both spin components to exhibit the same aspect ratio as the trap ($\eta = 10$). Reproduced with permission from~\cite{pei_competition_2010}.}
\label{fig:asldaprofiles}
\end{figure}
More advanced theories, such as the ASLDA study by Pei {\it et al.}~\cite{pei_competition_2010} and the DMFT study in~\cite{kim_exotic_2011} did not exhibit similar deformation of the minority component density profile, see Figs.~\ref{fig:dmft} and~\ref{fig:asldaprofiles}.
This would suggest that surface tension may not play a significant role after all.
The DMFT work was done for 210 atoms, but the ASLDA study involved significantly larger systems.
These findings underline the importance of normal state correlations and spin-dependent Hartree energy shifts.
The ASLDA theory particularly is, despite the name, essentially a BdG theory but with proper Monte Carlo-based variational coefficients for the Hartree and superfluid self-energies.
The lack of a proper normal state description in the BdG method is also believed to be the main reason for the anomalously high critical polarization $P_\mathrm{c}$ predicted by the theory.

\subsection{BdG in practice}

\begin{figure}
\centering
\includegraphics[width=0.95\columnwidth]{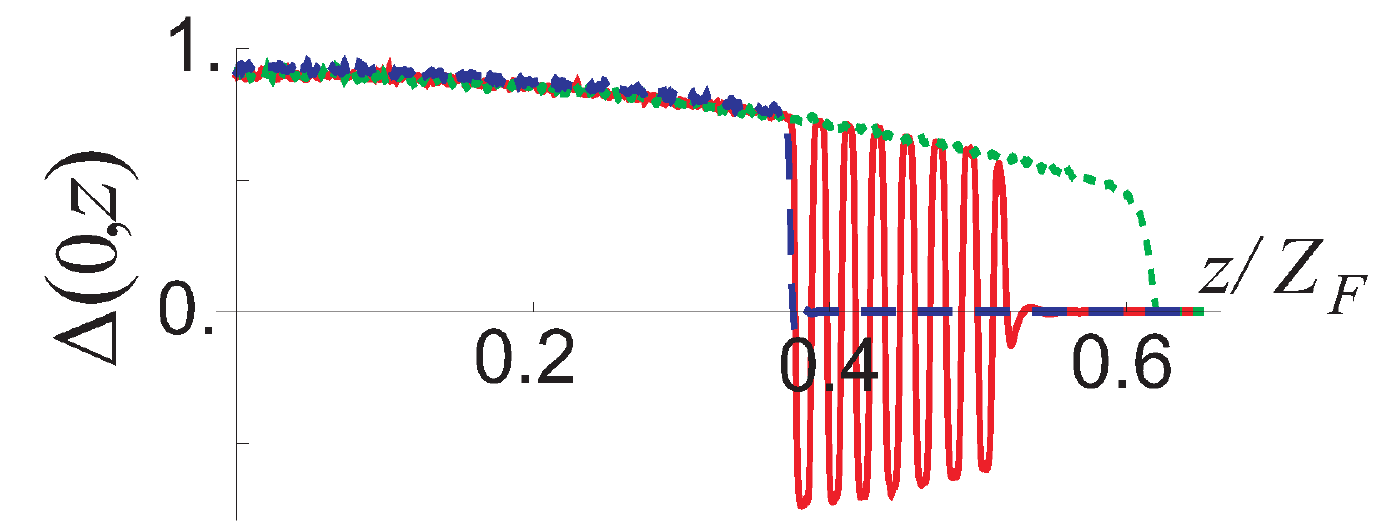}
\caption{Different initial order parameter ansatzes result in different converged states (here shown different order parameter profiles along the longitudinal $z$-axis). The failure of the BdG theory to find the ground state when starting from different initial states may be a consequence of metastable states. Reproduced with permission from~\cite{baksmaty_concomitant_2011}.} 
\label{fig:baksmaty}
\end{figure}

Spin-imbalanced gases are particularly difficult to solve with BdG equations.
The main reason is that the nodes or zeroes of the order parameter $\Delta({\bf r})$ propagate very slowly in an iterative process.
This can be easily understood from the LDA point of view: since zero order parameter $\Delta = 0$ is always a solution to the gap equation, any zeroes in the order parameter profiles remains.
Since BdG involves non-localized single-particle eigenstates, these nodes are still somewhat mobile.
However the larger the number of atoms, the more localized the high energy eigenstates are, and the bigger the problem.
Compared with balanced gases, solving spin-imbalanced BdG profiles can take up to a hundred times more iterations for convergence, and even then one has no guarantee of finding the ground state.
More advanced root solving methods, such as the Broyden method, do improve the convergence and have been used by some groups~\cite{kim_exotic_2011,baksmaty_bogoliubovgennes_2011}.
Also the initial state for the gap profile is a good approximation of the final solution as big changes in the profiles are often associated with appearance of zeroes in the gap profile during the iteration.
However the rigidity of the order parameter zeroes in BdG equations has also been used for explaining the observations of the Rice experiment~\cite{partridge_pairing_2007,liao_metastability_2011}.
In references~\cite{pei_competition_2010,baksmaty_concomitant_2011}, initial states with different oscillating order parameter ansatzes were considered, and given sufficiently large atom numbers, the iteration was found to converge to different solutions, see figure~\ref{fig:baksmaty}.
Ground state solutions were the ones with the lowest number of order parameter zeroes, but the excited states obtained through this procedure were argued to describe the findings of the Rice experiment as a signature of a metastable state.
Notice that the work in~\cite{pei_competition_2010} involved the ASLDA theory and thus goes beyond the usual mean-field BdG method.

\section{FFLO and spin-orbit coupling}\label{FFLOSOC}

\subsection{Spin-orbit coupling in the context of cold atomic gases}
Recently there has been rapid progress both on the theory and experiments
of spin-orbit coupled ultracold quantum gases (UQG).
Review articles by Galitski and Spielman~\cite{galitski_spin-orbit_2013} and
Dalibard {\it et al.}~\cite{dalibard_textitcolloquium_2011}
summarize the status well, but let us also review
the most salient features of these system here. After this quick
introduction, we proceed to review the growing body of literature
on spin-orbit coupled fermions in the context of FFLO states.

Spin-orbit coupling (SOC) is of great interest since it plays a crucial role in many condensed matter
phenomena of current interest such as Majorana fermions~\cite{elliott_textitcolloquium_2015} and topological insulators~\cite{hasan_textitcolloquium_2010}.
SOC is often the key ingredient giving rise to a non-zero topological index for insulators.
This in turn implies closing the gap and hence conducting states on the surface of the insulator where
two topologically different insulating states are in contact. Somewhat analoguous phenomena can appear in (gapped)
superconductors, where an energy gap exists for fermionic excitations. For a topologically non-trivial superconductor, this gap can close on the surface.

Since the early days of quantum mechanics, in solid state systems, spin-orbit coupling has been a distinct possibility. If we consider an electron moving in an electric field and 
move to the electrons rest frame, the associated Lorentz transformation will generate a magnetic field.
This magnetic field can then couple with electron spin giving rise to a spin-orbit coupling that will
depend on the electron velocity. A detailed form of this coupling will 
be related to the directions of electron
velocity and electric field, but for the special case of a motion in the xy-plane $\hat{{\bf e}}_x$ and electric field 
$E_0\hat{{\bf e}}_z$ along the z-axis, we find magnetic field
\beq
{\bf B}_{SO}=\frac{E_0\hbar}{mc^2}(k_x\hat{\bf e}_y-k_y\hat{\bf e}_x),
\enq
where $m$ is the particle mass and $c$ is the speed of light.  This magnetic field then interacts with the magnetic moment $\hat{{\bf \mu}}=\gamma \hat{{\bf \sigma}}$, where $\gamma$ is the gyromagnetic ratio
and $\hat{{\bf \sigma}}$ is the spin operator. As a result a spin-orbit coupling
\beq
H_{SO,R}=-\hat{{\bf \mu}}\cdot {\bf B}_{SO}=\alpha (k_y\hat{\sigma}_x-k_x\hat{\sigma}_y)
\enq
appears, where $\alpha$ is a constant determining the strength of the coupling. Sometimes the coupling coefficient 
$\alpha$ is expressed in terms of the characteristic velocity $v_R=\alpha/\hbar$.
The above coupling is called
Rashba-coupling, and the system with it lacks mirror symmetry. 
In the absence of inversion symmetry, the
spin-orbit coupling is of Dresselhaus type, namely
\beq
H_{SO,D}=\alpha (k_y\hat{\sigma}_x+k_x\hat{\sigma}_y).
\enq

In atomic physics, the $\sim \hat{L}\cdot \hat{S}$ term (with $\hat{L}$
the orbital angular momentum and $\hat{S}$ the spin angular momentum) in the Schr\"{o}dinger equation has its origin in the non-relativistic limit of the  Dirac equation.
Momentum and spin dependent
contributions to the system energy  are called spin-orbit couplings for this reason.
Notice that unlike the Zeeman interaction for example, the above examples of SOC preserve time reversal symmetry since under time reversal both spin and momentum change sign. Consequently the Kramers theorem ensures that each eigenstate is at least doubly degenerate, and one expects a degenerate ground state in a system with spin-orbit coupling (in the absence of Zeeman terms). 

Electric fields inside solid state systems can be very large $\sim 10^{12}\, V/m$ (unit is volts per meter) and comparable to
fields inside atoms. Such large fields can give rise to substantial spin-orbit couplings, which are clearly
unfeasible in a laboratory setting. Fortunately SOCs do not have to be created using electric fields and real spins, but can also be generated using synthetic fields~\cite{dalibard_textitcolloquium_2011} and pseudo-spins of atomic systems.
Indeed spin-orbit coupling has been recently demonstrated in UQG
both for bosons~\cite{lin_spin-orbit-coupled_2011} as well as for fermions~\cite{wang_spin-orbit_2012,cheuk_spin-injection_2012}.
In these experiments, demonstrations relied on synthetic spin-orbit coupling generated with laser beams coupling 
to multilevel atoms in specific ways. These were based on earlier theoretical 
proposals~\cite{ruseckas_non-abelian_2005,osterloh_cold_2005}.
The basic idea is that different atomic internal states are coupled via
a two-photon
Raman transition. The atom absorbs a photon from one laser beam, and then via stimulated emission,
emits it to the second beam. To conserve momentum, this process can impart momentum 
on the atom, and its strength depends on the Doppler shifts between the atoms and laser beams, which creates the characteristic momentum dependence of spin-orbit coupling~\cite{higbie_periodically_2002}.

An example of a spin-orbit coupling that has been realized is an equal mixture of Rashba and Dresselhaus couplings. In other words, the couplings have been of the type
\beq
H_{SO}=2\alpha k_x\hat{\sigma}_y
\enq
so that the full non-interacting Hamiltonian in the experiment by Lin {\it et al.}~\cite{lin_spin-orbit-coupled_2011} is
\beq
\hat{H}=\frac{\hbar^2k^2}{2m} \hat{I}+\frac{\Omega_z}{2}\hat{\sigma}_z+\frac{\delta}{2}\hat{\sigma}_y
+2\alpha k_x\hat{\sigma}_y,
\label{eq:Spielman_H}
\enq
where $\hat{I}$ is the identity matrix and $\hat{\sigma}$ are the Pauli matrices. The parameters $\Omega_z=-g\mu_B B_z$ and $\delta=-g\mu_B B_y$ 
of the Hamiltonian are related to experimentally controlled magnetic field components $B_z$
and $B_y$. The strength of the SOC $\alpha=E_L/k_L$ can be expressed in terms of the recoil energy $E_L=\hbar^2k_L^2/2m$ of the laser beam photons, where $k_L=\sqrt{2}\pi/\lambda$ and $\lambda$
is the wavelength of the Raman beam photons.
This $2\times 2$ Hamiltonian can be easily diagonalized to find the eigenstates and eigenenergies.
These states form the so-called helicity basis~\cite{wang_spin-orbit_2012}.
In the above Hamiltonian, the term with $\sigma_z$ corresponds
to out-of-plane Zeeman field while the (constant) term with $\sigma_y$
is an in-plane Zeeman field. 

\begin{figure} 
\centering
\includegraphics[width=0.45\columnwidth]{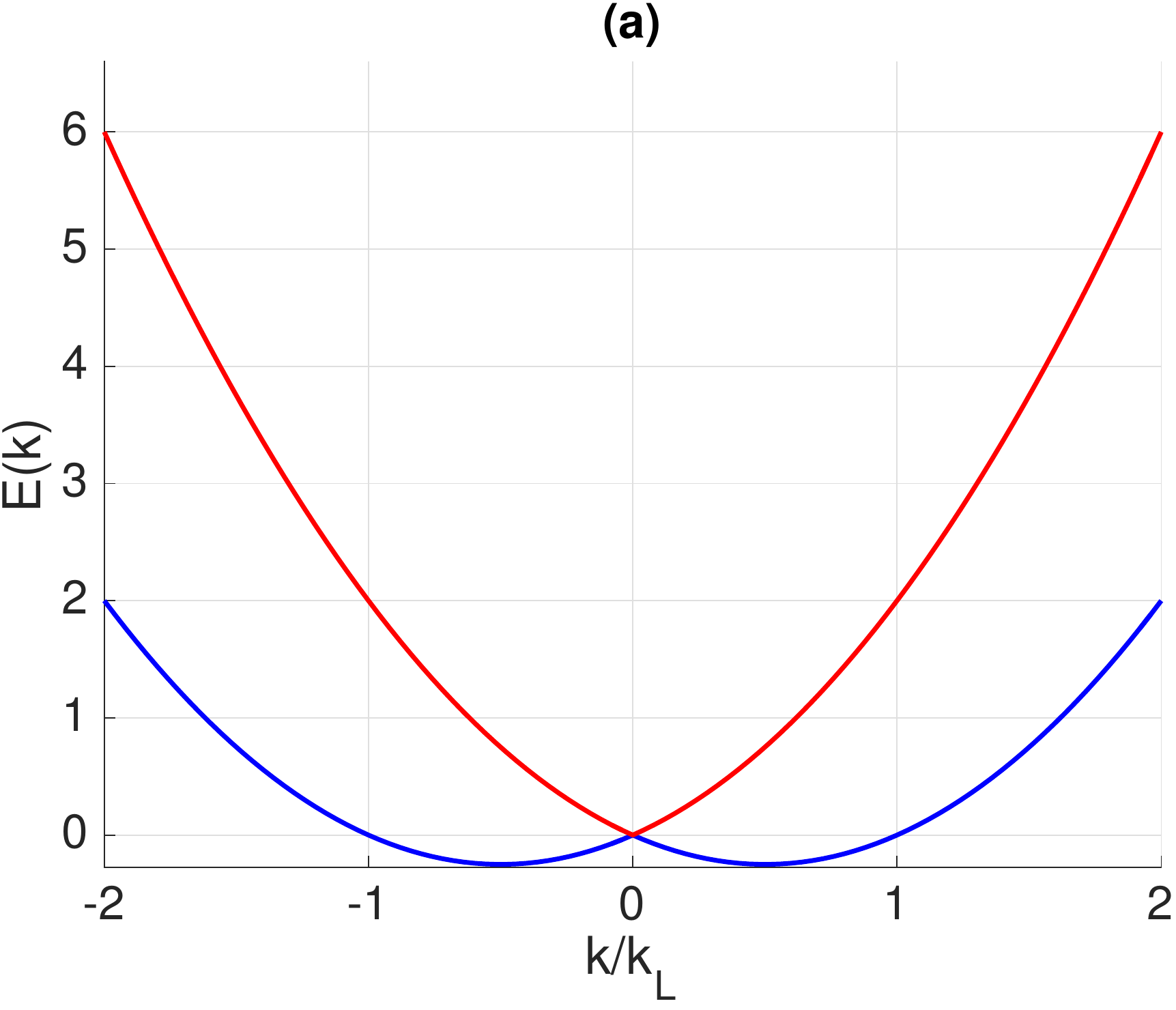}
\includegraphics[width=0.45\columnwidth]{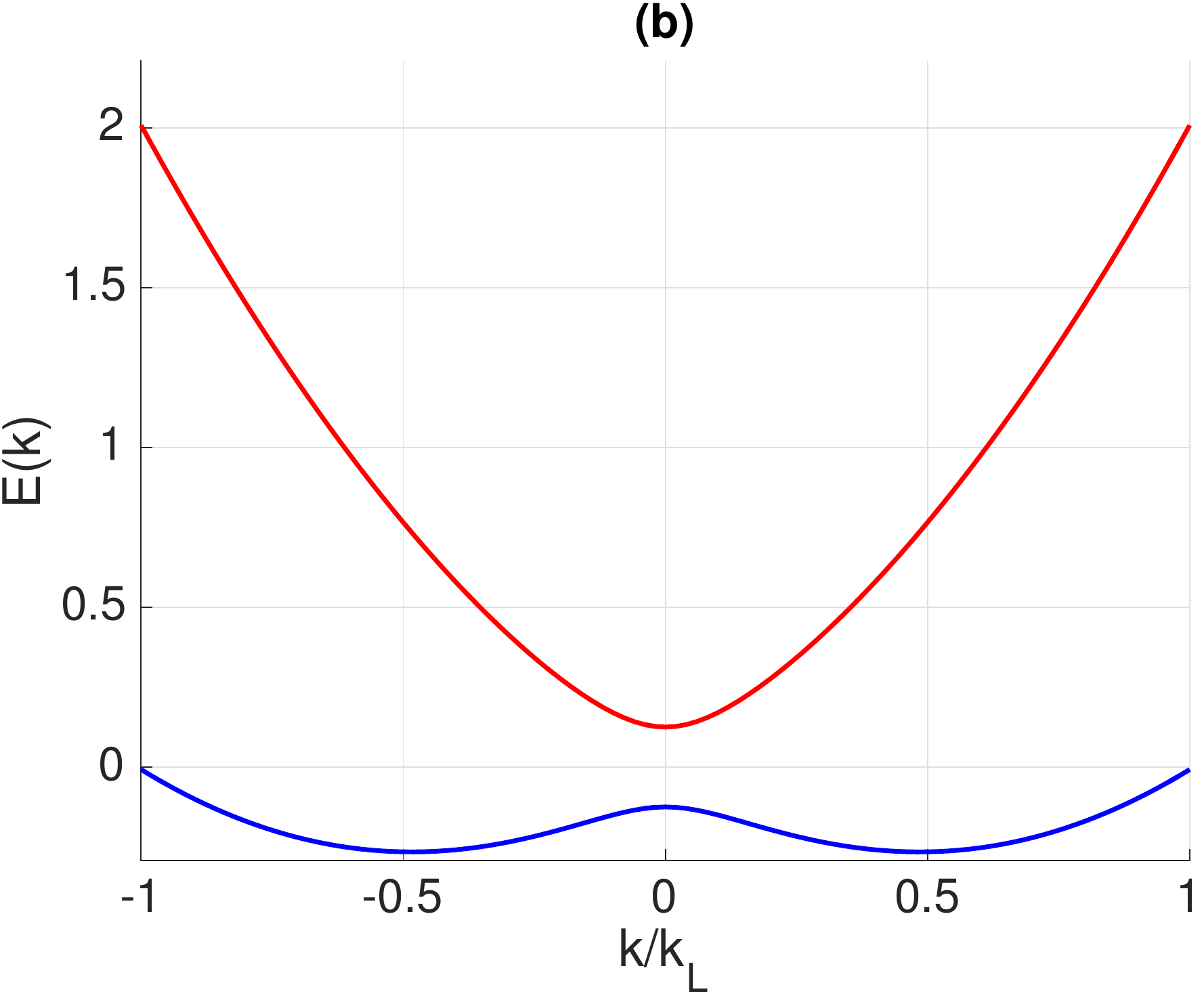}\\
\includegraphics[width=0.45\columnwidth]{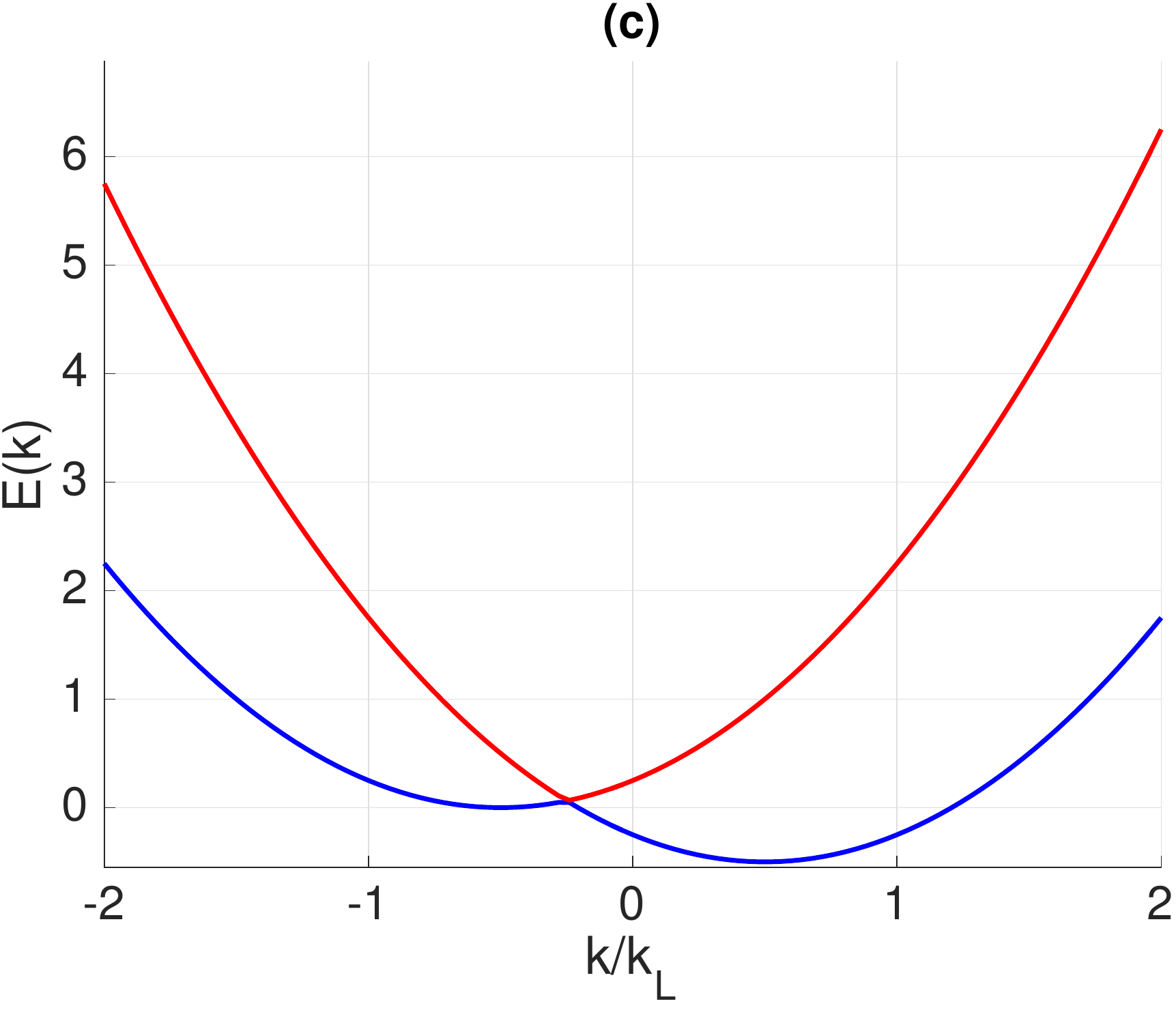}
\includegraphics[width=0.45\columnwidth]{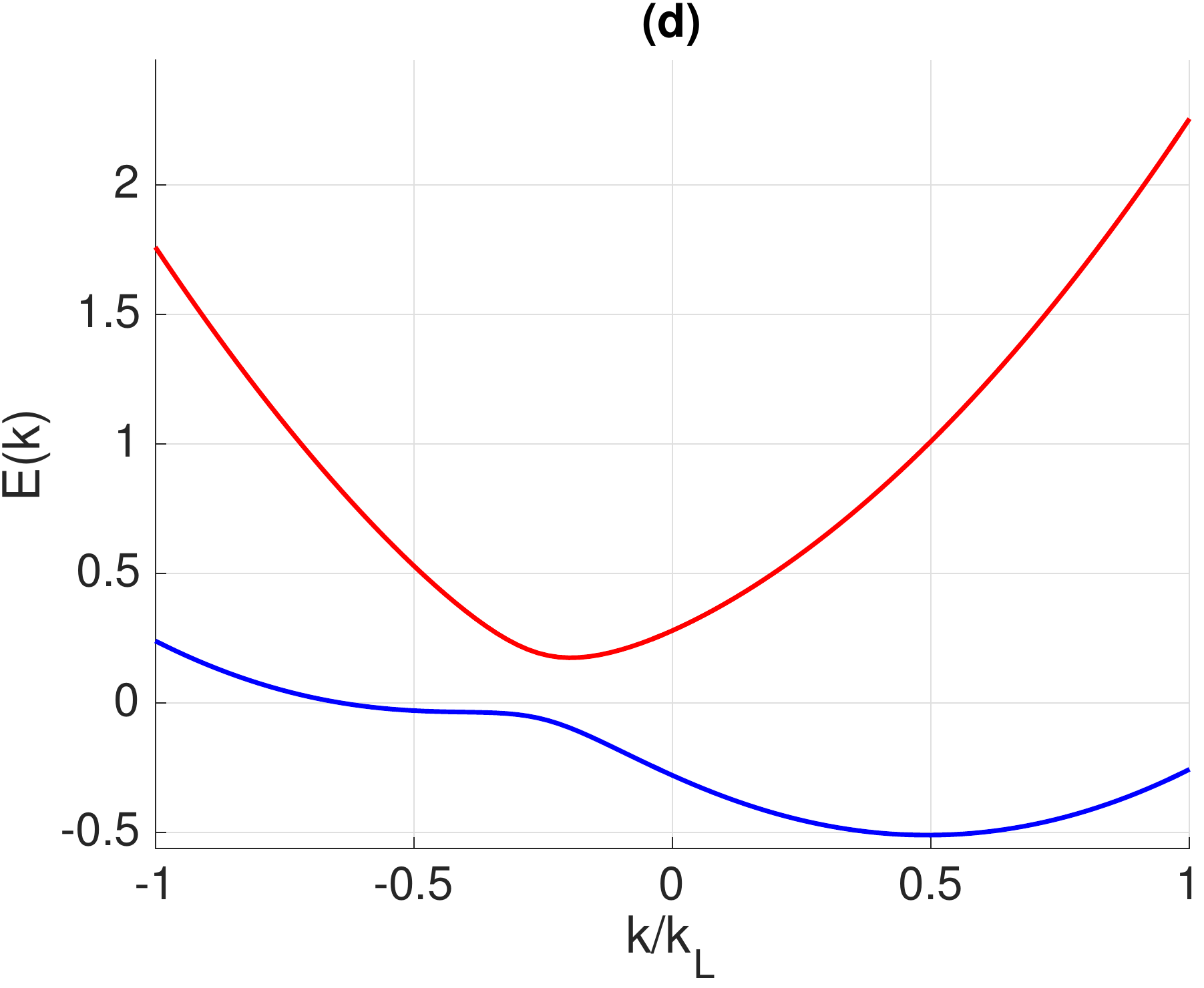}
\caption[Fig1]{Band structure for the Hamiltonian in equation~\eqref{eq:Spielman_H}. We choose $E_L$ as the unit of energy.
 In a) $\alpha=0.5$, $\Omega_z=0$, and $\delta=0$, in b)
 $\alpha=0.5$, $\Omega_z=0.25$, and $\delta=0$ in
 c)  $\alpha=0.5$, $\Omega_z=0$, and $\delta=0.5$, and
 in d)  $\alpha=0.5$, $\Omega_z=0.25$, and $\delta=0.5$.
}
\label{fig:dispersions_ownfig}
\end{figure}

Figure~\ref{fig:dispersions_ownfig} demonstrates the eigenstates
and the role of different terms in the Hamiltonian~(\ref{eq:Spielman_H}) along the x-axis.
With Zeeman fields and SOC coupling zero, we simply have two degenerate parabolic dispersions with their minimum at $k_x=0$.
As we turn on the spin-orbit coupling, the two 
parabolic dispersions move in opposite directions so that their
(degenerate) 
minima occur at non-zero momentum. The dispersions cross at 
$k=0$. As the out-of-plane Zeeman field $\Omega$ is turned on, the
band crossing disappears and a gap opens between bands 
at $k=0$. However the ground state remains degenerate.
Finally if we turn on the in-plane Zeeman field $\delta$, the ground
state degeneracy is lifted. When all terms are non-zero, 
we have an avoided crossing between bands with a non-degenerate ground
state at $k\neq 0$. 

The above 
Hamiltonian is somewhat special but does contain 
many features that reoccur in various spin-orbit coupled systems.
A different SOC was experimentally realized by
Huang {\it et al.}~\cite{huang_experimental_2016} in a gas of fermionic 
potassium atoms. In their case, the spin-orbit coupling was 
of Dresselhaus type $H_{SO}=\lambda_x k_x\sigma_x+\lambda_y k_y\sigma_y$. Another experiment by the same group~\cite{meng_experimental_2015}
studied a system with spin-orbit coupling
$H_{SO}=-\alpha k_y\sigma_x+(\beta_xk_x-\beta_yk_y)\sigma_y$.
Finally very recently, Wu {\it et al.}~\cite{wu_realization_2016} experimentally realized a 2D SOC of type
$H_{SO}=h(q_x^2+q_z^2)\sigma_z+\lambda_{so}(q_x\sigma_y+q_z\sigma_x)$ in a cloud of bosonic 
$^{87}Rb$ atoms.
For a more thorough theoretical discussion on how different spin-orbit couplings might be realized experimentally, see references~\cite{ruseckas_non-abelian_2005,osterloh_cold_2005,campbell_realistic_2011,dalibard_textitcolloquium_2011}.

The SOC in equation~\eqref{eq:Spielman_H} only depends on the velocity along x-axis. If we instead assume
a Rashba SOC, the single particle Hamiltonian is
\beq
\hat{H}=\frac{\hbar^2k^2}{2m} \hat{I}+\alpha (k_y\hat{\sigma}_x-k_x\hat{\sigma}_y).
\enq
For (pseudo) spin-$\frac12$ particle, this can be expressed as a matrix 
\beq
\hat{H}=\left(\begin{array}{cc}
\frac{\hbar^2k^2}{2m}  & \alpha (k_y+ik_x)\\
\alpha (k_y-ik_x) &  \frac{\hbar^2k^2}{2m}
\end{array}\right),
\enq
which we can diagonalize to find the eigenenergies
\beq
E_{\pm}(k_x,k_y)=\frac{\hbar^2k^2}{2m}\pm \alpha \sqrt{k_x^2+k_y^2}.
\enq
In this case, it is clear that the ground state has a large degeneracy since all states on the ring defined
by $\sqrt{k_x^2+k_y^2}=\alpha m/\hbar^2$ have the same energy. In the presence of interactions, it has been suggested that this degeneracy might lead to interesting effects such as quantum many body Schr\"odinger cat-states~\cite{stanescu_spin-orbit_2008}.  

Creating such SOCs as Rashba and Dresselhaus couplings analoguous with those appearing
in solid state systems is useful since it allows us to explore the associated physics in a very clean and tunable environment. However the SOCs that have been experimentally realized until now only capture a small segment of theoretical proposals~\cite{juzeliunas_generalized_2010,anderson_synthetic_2012}, and it is expected that synthetic SOCs that have no counterpart in solid state systems will be demonstrated in ultracold atomic systems. This will
open a way to probe novel physics with unexpected properties.

\subsection{Spin-orbit coupling in fermionic systems and FFLO}
For fermions, spin-orbit coupling provides many interesting possibilities. By changing the structure of Fermi surfaces, SOC can reduce the
effective dimensionality of the system
and thus enhance pairing, and for example, make superconductivity possible 
at higher temperatures. By enabling topologically non-trivial
band structures, it could imply topologically non-trivial superconductors as well. For identical fermions, $s$-wave interactions, which usually dominate physics in ultracold quantum gases, vanish.
For a mixture with two different fermionic species, $s$-wave interactions
are relevant and usually dominate over
$p$- and $d$-wave interactions which could be of interest regarding topological states. 
When $s$-wave interactions are small, attempts to experimentally realize some interesting quantum states in, for example, a $p$-wave interacting  fermionic gas are hampered by large losses in the region where $p$-wave interactions would be large~\cite{regal_tuning_2003}.
Spin-orbit coupling provides an interesting way around these problems: a system with usual $s$-wave interactions can give rise to $p$-wave interactions in the basis of pseudo-spin eigenstates~\cite{zhang_p_x+ip_y_2008}. 

\begin{figure}
\includegraphics[width=0.95\columnwidth]{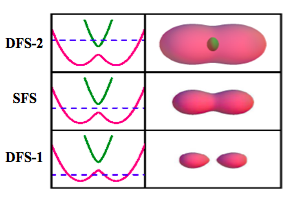}
\caption[Fig2]{Illustration of Lifshitz transitions and changes in the structure of the
Fermi surfaces as chemical potential (or density) of the fermions is increased.  For small atom numbers there are two disconnected
Fermi surfaces. As the atom number is increased the surfaces merge and
eventually a second surface appears inside the other one as the 
higher band starts to become occupied. This example assumed equal Rashba and Dresselhaus SOCs.
(Here DFS means double Fermi surface and SFS single Fermi surface.) 
Reproduced with permission from~\cite{wang_spin-orbit_2012}. 
}
\label{fig:FermiTopology}
\end{figure}

In addition to these possibilities, SOC in a Fermi gas can 
change the topology of the Fermi surfaces and give rise to Lifshitz transitions~\cite{seo_emergence_2012}
already for non-interacting systems. For low densities, fermions
accumulate to the lower helicity branch and start filling the states around each minima. The Fermi surfaces
are then disconnected. For higher densities, the surfaces eventually merge, and for even higher densities, a separate closed Fermi surface appears inside the larger one since fermions start to fill the
higher helicity branch as well. We demonstrate this in figure~\ref{fig:FermiTopology} reproduced from the article by Wang {\it et al.}~\cite{wang_spin-orbit_2012} who studied spin-orbit interactions in a harmonically trapped Fermi gas.
Changes in the Fermi surfaces can affect phase diagrams dramatically since they can, for example, influence nesting properties and densities of states in an important way.

Let us explore how FFLO states in systems with spin-orbit coupling are approached theoretically.  Following the
notation of Zhang and Yi~\cite{zhang_topological_2013} 
for a system of two-component
fermions interacting via contact interaction with strength $U$, the Hamiltonian is given by
\begin{eqnarray}
  \label{eq:HamiltonianSO}
\hat{H}=&\sum_{k,\sigma=\{\uparrow,\downarrow\}}\xi_{k,\sigma}
\hat{c}_{{\bf k},\sigma}^\dagger \hat{c}_{{\bf k},\sigma}+\hat{H}_{SOC}+\hat{H}_\mathrm{Z}+\\
\nonumber &U\sum_{{\bf k},{\bf k'},{\bf q}}
\hat{c}_{{\bf k+q},\uparrow}^\dagger \hat{c}_{{\bf -k+q},\downarrow}^\dagger \hat{c}_{{\bf -k'+q},\downarrow}
\hat{c}_{{\bf k'+q},\uparrow},
\end{eqnarray}
where for attractive interactions $U<0$. 
Futhermore, $\xi_{k,\sigma}=\epsilon_{k,\sigma}-\mu$, $\epsilon_{k,\sigma}=\hbar k^2/2m_\sigma$,
and $\mu$ is the chemical potential. If the system is in a lattice, this single particle dispersion
in the absence of SOC would be changed to lattice dispersion.
The Hamiltonian $\hat{H}_{SOC}$ describes the spin-orbit couplings, and
in a 2D system with Rashba coupling, this term would be
\beq
\hat{H}_{SOC}=\alpha \sum_{{\bf k}}  (\hat{c}_{{\bf k},\uparrow}^\dagger  \hat{c}_{{\bf k},\downarrow}^\dagger) 
(k_y\hat{\sigma}_x-k_x\hat{\sigma}_y)
\left(\begin{array}{c}
\hat{c}_{{\bf k},\uparrow}\\
\hat{c}_{{\bf k},\downarrow}
\end{array}\right).
\enq
The last term $\hat H_\mathrm{Z}$ is due to Zeeman interactions and has a form
\beq
\hat{H}_\mathrm{Z}=-\sum_{{\bf k}, \beta=\{x,y,z\}} h_\beta (\hat{c}_{{\bf k},\uparrow}^\dagger  \hat{c}_{{\bf k},\downarrow}^\dagger) 
\hat{\sigma}_\beta
\left(\begin{array}{c}
\hat{c}_{{\bf k},\uparrow}\\
\hat{c}_{{\bf k},\downarrow}
\end{array}\right).
\enq
Often simplifications are assumed
for the Zeeman fields $h_\beta$ such that, for example, $h_y=0$ and only the out-of-plane
Zeeman field $h_z$ and in-plane Zeeman field $h_x$ contribute.

The Hamiltonian in~\eqref{eq:HamiltonianSO} is
one typical example, and variations exist.
For example, the spin-orbit coupling could
be expressed on a different basis so that an equivalent Hamiltonian might superficially look
different~\cite{hu_fuldeferrell_2013}. Also one might consider SOC that is a mixture of
Rashba and Dresselhaus couplings with equal Rashba and Dresselhaus (ERD) spin-orbit 
couplings~\cite{iskin_topological_2013,wu_unconventional_2013a,wu_unconventional_2013}
being especially relevant to most experiments done so far. 
In lattices the momentum summations would be restricted to the
first Brillouin zone and renormalization of contact interaction is unnecessary due to natural cut-off provided by the lattice.

Dimensionality can also be varied
from 1D to 3D and sometimes SOC might be 2D even if the system
itself is 3D~\cite{zheng_route_2013}.
The type of SOC is essential for symmetry breaking properties of the potential FFLO state: some forms of SOC explicitly break the rotational symmetry, for instance, which therefore cannot be broken spontaneously. However even in the presence of SOC, the density is typically uniform in the normal state, and the FFLO state can break the translational symmetry. The existence of a preferred momentum as given by SOC, may also influence the collective mode spectrum of the FFLO state.

Without interactions, the Hamiltonian~\eqref{eq:HamiltonianSO}
could be diagonalized into a helicity basis straight away, but
interactions necessitate additional approximations. There are two dominant approaches.
In the first approach, one assumes a Fulde-Ferrell type order parameter (note: we mostly follow the notation
from reference~\cite{zhang_topological_2013} and therefore notation here deviates slightly from~\eqref{eq:order_parameter_FF}
\beq
\Delta=\Delta_{\bf Q}e^{i{\bf Q}\cdot {\bf r}}=Ue^{i{\bf Q}\cdot {\bf r}}
\sum_{{\bf k}} \langle \hat{c}_{{\bf Q-k},\downarrow}\hat{c}_{{\bf k},\uparrow}\rangle
\enq
and then derives a usual BCS type mean-field theory around it.
This implies a mean-field Hamiltonian~\cite{zhang_topological_2013}
\begin{eqnarray}
H_{MF}=\sum_{{\bf k}} \left[\xi_{{\bf Q-k}}-\frac{|\Delta_{{\bf Q}}|^2}{U}\right] \mathcal{I} +
\frac{1}{2}\sum_{{\bf k}} \mathcal{M},
\end{eqnarray}
where
\begin{eqnarray}
  \mathcal{M} = \\
\nonumber  \left(\begin{array}{cccc}
\xi_{{\bf k}}-h_z & \Delta_{{\bf Q}}& 0& \Lambda_{{\bf k}}\\
\Delta_{{\bf Q}}^* & -\xi_{{\bf Q-k}}-h_z& -\Lambda_{{\bf Q-k}} & 0\\
0 & -\Lambda_{{\bf Q-k}}^*& -\xi_{{\bf Q-k}}+h_z& -\Delta_{{\bf Q}}^*\\
 \Lambda_{{\bf k}}^* & 0& -\Delta_{{\bf Q}}& \xi_{{\bf k}}+h_z
\end{array}\right),
\end{eqnarray}
and $\mathcal{I}$ is a 4x4 unit matrix.
Here a basis of $(a_{{\bf k},\uparrow}, a_{{\bf Q-k},\uparrow}^\dagger, a_{{\bf Q-k},\downarrow}, a_{{\bf k},\downarrow}^\dagger)^T$ was used and $\Lambda_{\bf k}=\alpha (k_x+ik_y)-h_x$.
From this mean-field Hamiltonian, the grand potential
\beq
\Omega=-\frac{1}{\beta} \ln Tr \left[e^{-\beta H_{MF}}\right]
\enq
can be computed and then minimized to find the ground state for a set of parameters, see section~\ref{EnergiesAll}.

In the second related but more general and numerically more demanding approach, one allows arbitrary spatial variation
of the order parameter $\Delta({\bf r})$ and then solves the BdG equations
for the quasiparticle amplitudes of the mean-field theory. Since we have already
explained the BdG equations in section~\ref{section_traps}, here we simply guide the reader to references~\cite{iskin_trapped_2012,iskin_spin-orbit-coupling-induced_2013,seo_topological_2013,xu_competing_2014} for descriptions of the SOC case.

\begin{figure}
\includegraphics[width=0.95\columnwidth]{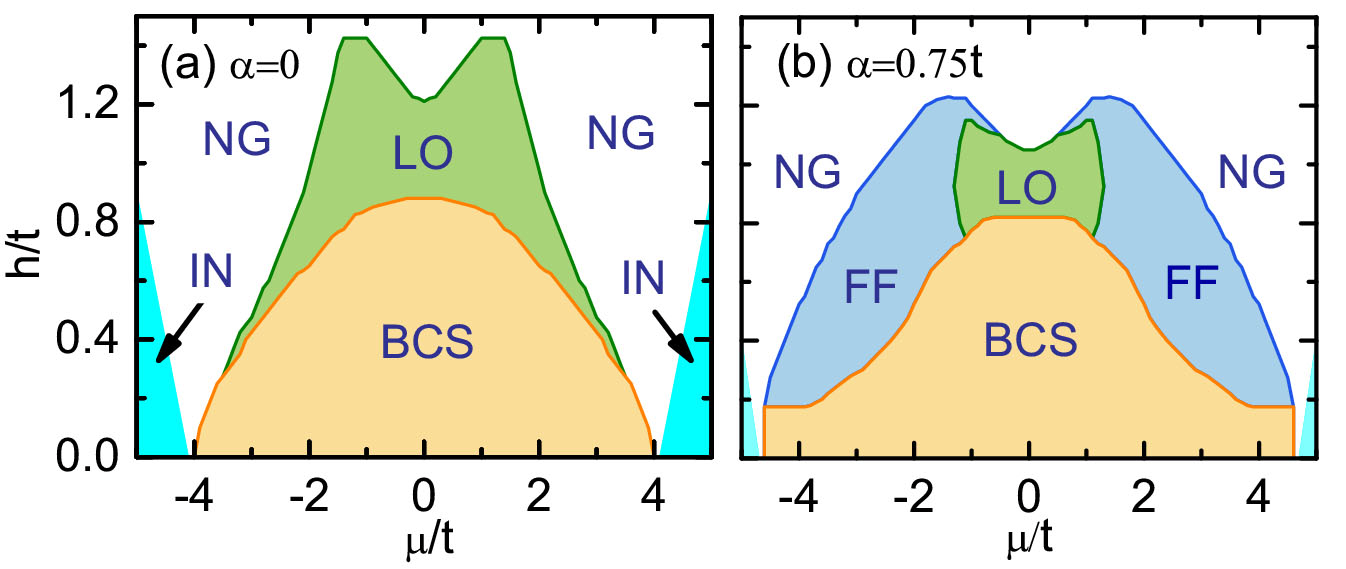}
\caption[Fig3]{Illustration of the competition between FF and LO phases in a lattice. As the strength $\alpha$ of the spin-orbit coupling increases the FF phase starts to dominate over the LO phase.
  Reproduced with permission from~\cite{xu_competing_2014}.
}
\label{fig:XuCompDiagram}
\end{figure}

In the absence of SOC, limitation to the FF-ansatz can sometimes be too restrictive since LO states typically have lower energy. However in the presence of SOC, the assumption of the FF-ansatz is often better justified. As explained previously, SOC can create ground state degeneracy while (in-plane) Zeeman fields break this degeneracy, and one minimum away from ${\bf k}=0$ becomes energetically favorable.
This makes it more natural
to assume a single plane wave ansatz for the order parameter as well.
This has been studied by Xu {\it et al.}~\cite{xu_competing_2014} who explored the  
competition of LO and FF states in a lattice in 2D optical lattices and found that FF states dominate once relatively large spin-orbit coupling is present. We
demonstrate this with figure~\ref{fig:XuCompDiagram}.

For polarized Fermi gases without SOC, it is known that population imbalance tends
to drive the system towards phase separation. When such considerations were extended
to systems with SOC, it was found by Iskin and Suba\c{s}i~\cite{iskin_stability_2011} 
and by Yi and Guo~\cite{yi_phase_2011} in continuum that
SOC can counteract the tendency towards phase separation and stabilize uniform superfluid
phases. Furthermore in the presence of SOC, nodes of the quasiparticle dispersions
might behave in a complex way.
However these studies did not consider FFLO phases.
A later article by Iskin~\cite{iskin_trapped_2012}  
solved BdG equations in the presence of Rashba SOC in a 2D
system in a continuum and found that SOC implies polarized superfluid states replacing superfluids
with oscillating order parameters. However the chosen polar coordinate system imposed rotational symmetry  on the solutions, and for example, FF type solutions were then excluded by construction.
 
As explained in earlier sections, usually in the absence of SOC one can conclude that FFLO states occupy only a narrow region of the parameter space in free space; although a considerable one in lattices. When researchers started to explore
FFLO in systems with SOC, they noticed that this conclusion was no longer valid.
Zheng {\it et al.}~\cite{zheng_route_2013} studied 3D systems
with Rashba SOC and an in-plane Zeeman field in continuum. They assumed an FF ansatz for the order parameter and 
found a large region of the parameter space where the FFLO state might appear.
Since the system was 3D, this phase was expected to have long-range superfluid order as well.
Zheng {\it et al.} explained the sudden dominance of the FFLO state by observing that 
the Zeeman field with SOC distort the Fermi surfaces and shift them in such
a way that they are no  longer centered around the origin. This suppresses the usual BCS state
and gives rise to FFLO states.

\begin{figure}
\includegraphics[width=0.95\columnwidth]{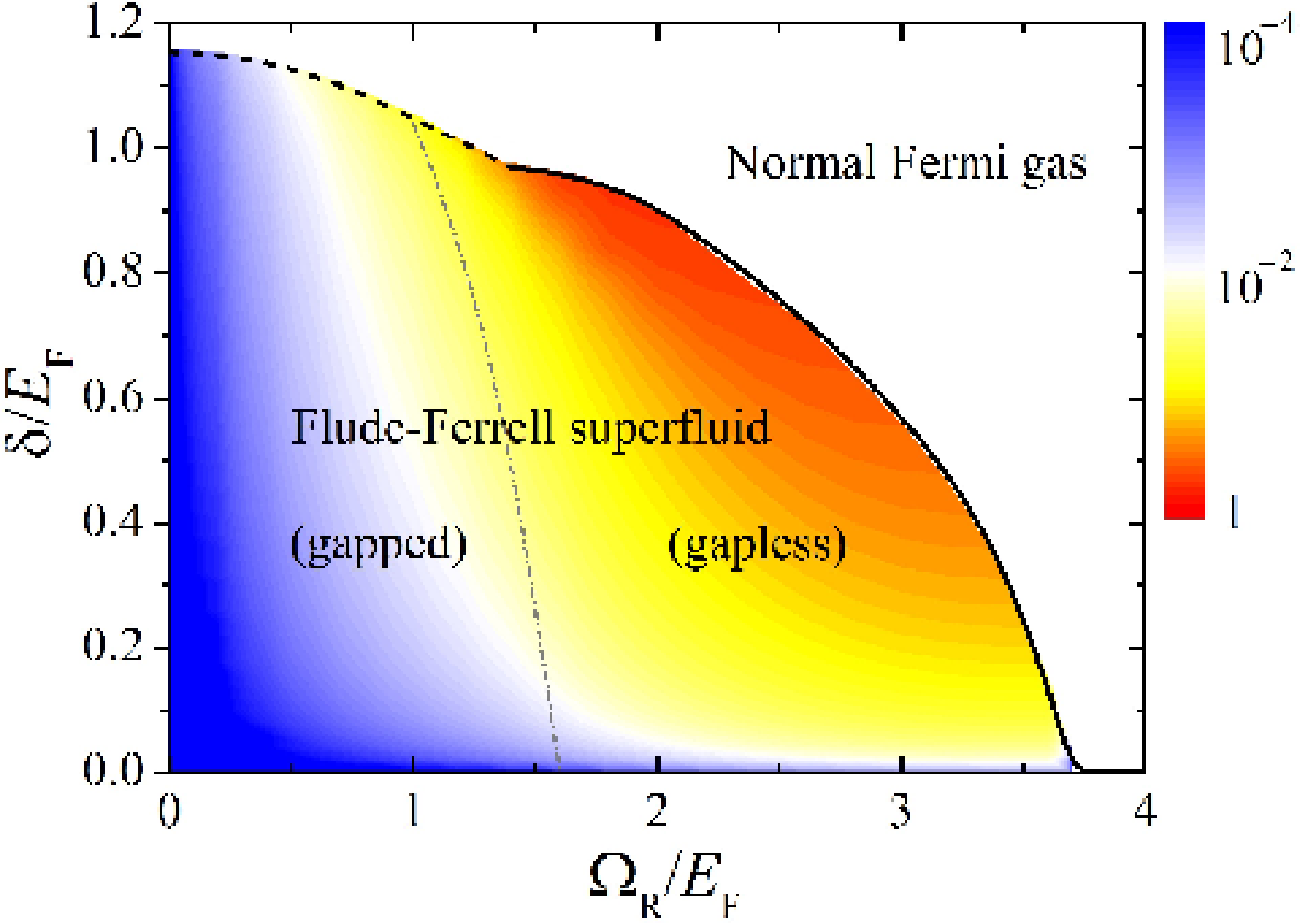}
\caption[Fig4]{A phase diagram in a 3D spin-orbit coupled Fermi gas.
Here $\delta$ corresponds to the in-plane Zeeman
field and $\Omega_R$ to the out-of-plane Zeeman field. The color shows the magnitude of the
center of mass momentum of Cooper pairs, $q/k_F$. Reproduced with permission from~\cite{liu_inhomogeneous_2013}.}
\label{fig:LiuHuFFLOSchematic}
\end{figure}

Hu and Liu studied a closely related system in reference~\cite{hu_fuldeferrell_2013} and outlined
a finite temperature phase diagram at a broad Feshbach resonance. They predicted that for sufficiently
large SOC, the superfluid state is always an FF state. They also estimated 
a promisingly high critical temperature of  $T_c\sim 0.2\, T_F$ for the transition, where $T_F$ is the Fermi temperature. In 
reference~\cite{liu_inhomogeneous_2013}, Liu and Hu extended their model to also
include an out-of-plane Zeeman field and again found large regions of the parameter space
occupied by the FFLO state. We show their predicted phase diagram
in figure~\ref{fig:LiuHuFFLOSchematic}.  They again predicted that the critical temperature could be a
substantial fraction of the Fermi temperature.

The instability of the normal state to FFLO with Rashba SOC was then elaborated by Liu~\cite{liu_fulde-ferrell_2013}, who used the Thouless criterion and beyond mean-field corrections to compute
(among other things) the critical temperature. With Rashba SOC coupling $\alpha k_\text{F}/E_\text{F}=1$ (where $E_\text{F}$ is the Fermi energy), 
in-plane Zeeman field $h_x=0.5 E_F$, and in the unitarity limit the critical temperature for the FF
superfluid was found to be quite high, $T_c=0.2\, T_F$.

Around the same time, similar conclusions were found by Dong {\it et al.}~\cite{dong_fuldeferrell_2013}
and Zhou {\it et al.}~\cite{zhou_exotic_2013}.
In these studies, the model was somewhat different in that they focused on the symmetric 3D
spin-orbit coupling in continuum $v_xk_x\hat{\sigma}_x+v_yk_y\hat{\sigma}_y+v_zk_z\hat{\sigma}_z$.
Despite this difference, by assuming FF type order parameter  Dong {\it et al.}~\cite{dong_fuldeferrell_2013} found that the FF state dominated the
phase diagram both at zero as well as at non-zero temperatures. In addition 
to solving the mean-field phase diagram, Dong {\it et al.} also solved the two-body problem
for the dimer bound state and found dimers with finite momentum. 
The Zhou {\it et al.}~\cite{zhou_exotic_2013} study complements these results by
exploring the instability towards the FFLO state using a small $Q$ expansion of the grand potential.
Furthermore it classifies FFLO states according to the presence of a gap and
 structure of the nodal surfaces  in momentum space.

\begin{figure}
\includegraphics[width=0.95\columnwidth]{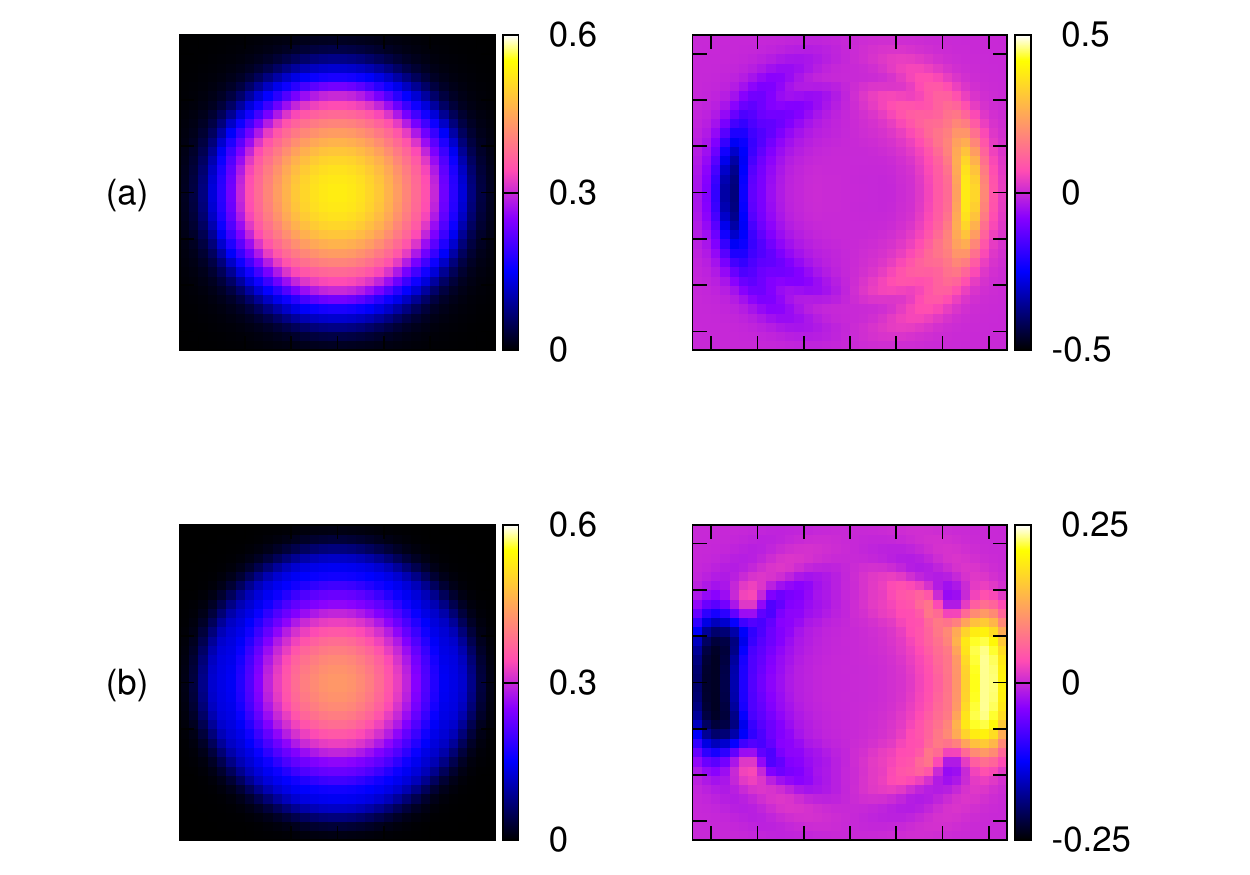}
\caption[Fig5]{Amplitude and phase of the order parameter $\Delta$ as a function 
of position in the lattice when there was an in-plane Zeeman field. 
Variation in the amplitude is due to harmonic trapping potential while
the phase behaves in a way expected from FF-type state. 
Upper row shows results with Rashba SOC while the lower
one is based on equal Rashba and Dresselhaus SOC.  
Figure from Iskin~\cite{iskin_spin-orbit-coupling-induced_2013}.
}
\label{fig:IskinTrappedFF}
\end{figure}

In an experiment, a trapping potential would be present. 
A harmonically trapped system of two-component fermions in an optical lattice with SOC
was studied by Iskin~\cite{iskin_spin-orbit-coupling-induced_2013}. 
In this study, BdG equations were solved in 
a 2D lattice model for $150$ fermions. Iskin found that with Rashba SOC and
in-plane Zeeman field, the order parameter was of the FF type, without
a nodal structure in the order parameter expected from the LO state.
We show an example of this in figure~\ref{fig:IskinTrappedFF}.

\subsection{Topological states in systems with spin-orbit coupling}

Topological states of matter have become an intense research area in recent years~\cite{bernevig_topological_2013,hasan_textitcolloquium_2010}.
In the context of spin-orbit coupling, Tewari {\it et al.}~\cite{tewari_topologically_2011} 
pointed out that SOC with Zeeman fields in a fermionic system can be used to
drive a topological phase transition into a topologically non-trivial superfluid. This happens
when the Zeeman splitting becomes so large that the chemical potential is in the 
gap (opened by the Zeeman splitting) of the different dispersion branches.
The question then arose whether or not the FFLO phases 
that appear in systems with spin-orbit couplings 
might also have non-trivial topological properties.
This was answered affirmatively
in several recent studies, that were published close to each other in time~\cite{zhang_topological_2013,qu_topological_2013,cao_gapless_2014,xu_anisotropic_2014}.
The key feature making this possible was the presence of Zeeman fields in addition to spin-orbit coupling.

Let us first clarify what is meant by the term ``topological'' here.
Earlier we pointed out that in the presence of the spin-orbit coupling, the 
topology of the Fermi surface can change for non-interacting particles.
Similar arguments
can be extended to the intertacting case by considering the locations on momentum space
where quasi-particle dispersions change signs.
Such an approach has been used to classify
different phases as being topologically distinct~\cite{iskin_stability_2011,yi_phase_2011,iskin_trapped_2012,wu_unconventional_2013a}.
Historically transitions between phases with different Fermi surface topologies have sometimes
been called topological transitions, but due to the present-day use of the term topological, it is better to refer to them as Lifshitz transitions.

Changes in the topological structure of the nodes of the dispersions should not be 
confused with classifying states according to their topological properties
by computing their topological invariants, such as the Chern number. 
For each k-vector, diagonalization of the BdG equation gives rise to four eigenenergies, and
we can label these eigenenergies in terms of quasi-particle and quasi-hole
index $\eta=\{+,-\}$ and helicity index $\nu=\{1,2\}$
as $E_{{\bf k},\nu}^\eta$. If we then denote the associated wavefunctions with the shorthand notation $|n\rangle$
$n=(\eta,\nu)$, we can calculate the Berry curvature, see for instance~\cite{zhang_topological_2013}
\begin{eqnarray}
  \Gamma^\eta_\nu({\bf k})=\\
\nonumber  i\sum_{n\neq n'}\frac{\langle n|\partial_{k_x} H_{MF}|n'\rangle \langle n'|\partial_{k_y} H_{MF}|n\rangle-(k_x\leftrightarrow k_y)}{(E_{{\bf k},\nu}^\eta-E_{{\bf k},\nu'}^{\eta'})^2}.
\end{eqnarray}
We then get the Chern number for each helicity band of quasi-particles and holes by integrating over
the Berry curvature over the Brillouin zone
\beq
\gamma_\nu^\eta=\frac{1}{2\pi}\int d{\bf k} \Gamma^\eta_\nu({\bf k}).
\enq
The Chern number for the superfuid is then the sum of the Chern numbers of the
quasi-hole (i.e. $\eta=-$) bands $C=\gamma_1^{-}+\gamma_2^{-}$. For topologically trivial 
phases, the Chern number vanishes when it is non-zero for topological phases.

Note that here we compute the Chern number for the interacting state. However one could also
compute the Chern number for single particle bands and find that some bands are topologically
non-trivial.
This adds an additional layer of potential confusion, since having a topologically non-trivial single particle band does not imply that the realized state would also be topologically non-trivial.
However it may lead to other interesting consequences, such as 
superfluidity in a flat band~\cite{peotta_superfluidity_2015,julku_geometric_2016,tovmasyan_effective_2016,liang_band_2017,liang_semiclass_2017}.

Zhang {\it et al.}~\cite{zhang_topological_2013} and Qu {\it et al.}~\cite{qu_topological_2013} studied a 2D Fermi gas with Rashba SOC.
They identified a gapped FF state with a center of mass momentum along
the x-axis as well as a gapless nodal FF state (with two disconnected gapless contours in momentum space), and finally a topological FF state.
A topologically trivial
gapped FF state appears for large SOC or, alternatively a small out-of-plane Zeeman field.
At the phase boundary between nFF and gapped FF state, quasi-hole and quasiparticle spectra
touch the Fermi surface at different points. This is in contrast with the boundary between topological FF
state and the gapped FF state where the gaps close at the same point.
By solving the BdG equations of a tight-binding model in a 2D strip geometry,
Qu {\it et al.}~\cite{qu_topological_2013} also confirmed the existence of chiral edge states of topological
FF states using a square lattice model.

\begin{figure}
  \includegraphics[width=0.95\columnwidth]{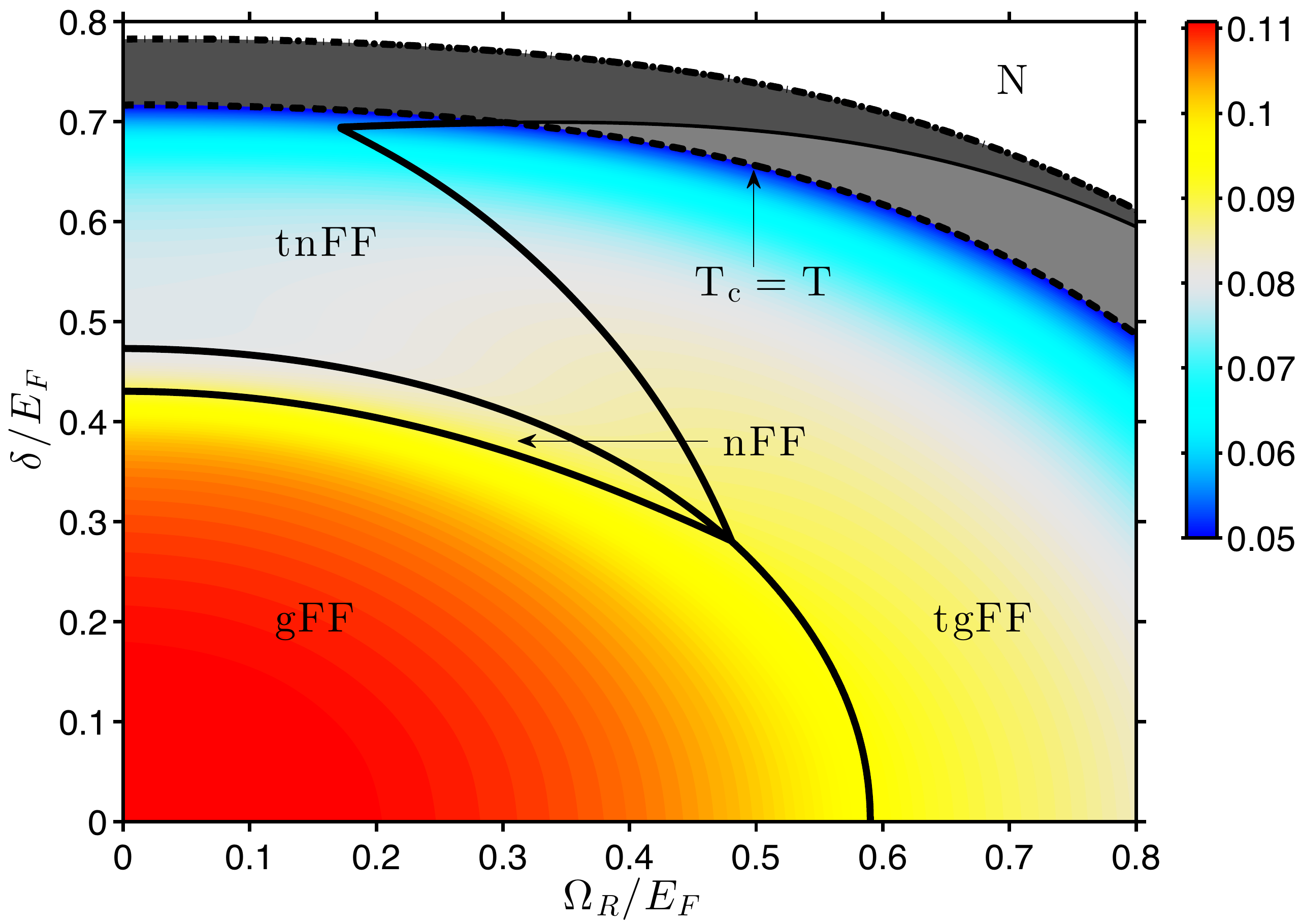}
\caption[Fig6]{
Phase diagram of a two component 2D
Fermi gas with Rashba SOC. The horizontal axis gives the strength of the out-of-plane Zeeman field while the
vertical axis indicates in-plane Zeeman field. Possible phases are gapped
Fulde-Ferrell (gFF), nodal Fulde-Ferrell (nFF) with nodes in the quasiparticle dispersions,
gapless topological Fulde-Ferrell (tnFF), gapped
topological Fulde-Ferrell (tgFF), and normal phase (N).
Strength of the Rashba coupling was chosen as $\lambda=E_F/k_F$. Temperature was set at 
$T=0.05\, T_F$ and the 
binding energy $E_b=0.2\, E_F$ of the two-body bound state of the 2D system without SOC fixed the interaction strength. Dashed line is a prediction for
the BKT transition temperature.
Adapted with permission from~\cite{cao_gapless_2014}.
}
\label{fig:Cao_PhaseDiagramtFF}
\end{figure}

Cao {\it et al.}~\cite{cao_gapless_2014} also found the possibility of gapped and gapless
topological FF superfluid in a 2D Fermi gas with SOC in continuum. For an example of their full phase diagram,
see the figure~\ref{fig:Cao_PhaseDiagramtFF}.
By adding 
a disorder potential along the y-axis, they confirmed the robustness of the Majorana edge modes. In conventional systems without SOC, transverse superfluid stiffness of the FF states vanishes~\cite{radzihovsky_fluctuations_2011,yin_fulde-ferrell_2014}.
Interestingly Cao {\it et al.}~\cite{cao_gapless_2014} found a positive definite
 superfluid density tensor when SOC was present.
Furthermore they also extended the study beyond Rashba coupling by considering
SOC of the type
\beq
H_{SOC}=\lambda \cos\psi k_x\hat{\sigma}_x+ \lambda\sin\psi k_y\hat{\sigma}_y
\enq
and varying the angle parameter $\psi$. If $\psi=\pi/4$, SOC is of Rashba type
while $\psi=\pi/2$ corresponds to the experimentally realized equal Rasha and Dresselhaus
SOC. Importantly Cao {\it et al.} confirmed that while topological states 
disappear in the limit of equal Rashba and Dresselhaus (ERD) coupling, 
the existence of topological states does not require pure Rashba
type SOC.

Closely related to these studies is the paper by Xu {\it et al.}~\cite{xu_anisotropic_2014}
who also identified the possibility of realizing (massless chiral) Weyl fermions
in these systems. Dirac fermions are massive and described by four (or more) components, but 
in the zero mass limit, the Dirac equation allows solutions with two components known
as Weyl fermions. 
 By increasing the out-of-plane Zeeman field $h_z$, one can move from
a gapped FF phase to a topological FF phase which is gapped except at Weyl nodes
around which the dispersions are linear and can be described by the chiral Weyl equation.
For relatively large values of $h_x$, a phase where 
excitations are gapless also outside Weyl nodes was found and designated as a gapless topological FF state.
A quasiparticle excitation gap closes at the transition
and identifies the points where Weyl fermions might appear. Xu {\it et al.}~\cite{xu_anisotropic_2014}
noted that properties of these Weyl fermions could be controlled with  Zeeman fields
and suggested they could be observed by studying the changes in the speed of sound
around the Weyl nodes.

SOC with the experimentally realized equal Rashba and Dresselhaus SOC was studied
by Wu {\it et al.}~\cite{wu_unconventional_2013a,wu_unconventional_2013} in 2D
systems
and by Iskin and Suba\c{s}i~\cite{iskin_topological_2013} in a 3D system.
Although Cao {\it et al.}~\cite{cao_gapless_2014} did not find topologically non-trivial states
(meaning non-zero Chern number) for ERD, Wu {\it et al.}~\cite{wu_unconventional_2013a,wu_unconventional_2013} pointed out highly non-trivial nodal structures
in momentum space that could be used to further classify, for example, FFLO states.
Similar conclusions were found by Iskin and Suba\c{s}i~\cite{iskin_topological_2013}.
Since SOC can generate effective $p$-wave interactions, the symmetry of the realized paired
states is of interest. This question was studied by Chan and Gong~\cite{chan_pairing_2014}
who found, for example, triplet pairing in the limit with vanishing in-plane Zeeman fields.

Dimensionality influences the appearance of topologically non-trivial states.
Liu and Hu~\cite{liu_topological_2012} solved BdG equations in a 1D system.
In the absence of SOC, one expects either a BCS superfluid or an FFLO superfluid (or their 1D
equivalents without long range order), but
SOC with Zeeman field could drive the system into a superfluid 
with zero energy Majorana fermions~\cite{elliott_textitcolloquium_2015}
at the edge. This is called a topological superfluid since a zero energy edge mode in real space is taken
as an indicator of an interface with two topologically distinct regions with the surrounding
vacuum being obviously topologically trivial. Physically the appearance of Majorana fermions
can be understood from the analogy with chiral $p$-wave superfluids. As explained earlier,
SOC mixes states in such a way as to create effective $p$-wave interactions in a system
with underlying $s$-wave interactions. A Zeeman field can then open a gap between
two branches so that, for sufficiently large Zeeman fields, all atoms occupy the lowest branch, thus creating a possibility for $p$-wave pairing.

Liu and Hu~\cite{liu_topological_2012} found that the FFLO state appears
for small SOC while topological superfluid appears at large values of SOC.
Chen~\cite{chen_inhomogeneous_2013} also studied the mean-field theory with different SOCs including ERD, 
using a 1D lattice model. In the system of~\cite{chen_inhomogeneous_2013}, instead of the
earlier mentioned Chern number, one could compute the $\mathbb{Z}_2$ number which is 
$1$ for topologically trivial systems and $-1$ for topologically non-trivial ones.
With ERD, Chen found the possibility of topologically non-trivial FFLO state 
with $\mathbb{Z}_2=-1$
by tuning the band filling.
These states could be further classified based on whether or not they were also gapped.
Chen also solved the BdG equations in a harmonic trap and intepreted that 
the topologically non-trivial
state would manifest itself there as zero energy Majorana fermions at the edge of the system.

Dimensional crossover and the competition between FFLO and Majorana states
was studied further by Seo {\it et al.}~\cite{seo_topological_2013} by tuning the dispersions
from multi-dimensional to 1D using a lattice potential in the $xy$-plane. In agreement with Li and Hu
~\cite{liu_topological_2012}, they found FFLO states for smaller values of the (Rashba)
SOC and (topological) uniform superfluid state with Majorana fermions at the edge for higher values
of SOC. It was also found that lowering the dimensionality from 3D to 1D enables a crossover
from the 3D uniform superfluid phase into a FFLO phase or to the topological
uniform superfuid state.
In a bilayer system with SOC, FFLO can appear without any population or chemical potential imbalance, and can transit to topological superconductivity as the interaction increases~\cite{wang_fulde-ferrell_2017}.

Qu {\it et al.}~\cite{qu_fulde-ferrell-larkin-ovchinnikov_2014} used a 1D lattice model
with SOC and a large out-of-plane Zeeman field
to study the competition between FFLO superfluids and superfluids with Majorana fermions.
They computed the phase diagram for the model and, among other things, found that
with increasing strength of SOC superfluids with Majorana fermions
occupy ever larger areas of the parameter space. This result is consistent with references~\cite{liu_topological_2012,seo_topological_2013} which did not consider a lattice model.

As dimensionality is reduced, fluctuations become more pronounced 
and might destroy the ordered states predicted by simple mean-field theories.
In particular in a 2D superfluid, Berezinsky - Kosterlitz - Thouless (BKT)
transition is expected to appear. As temperature is lowered, exponentially
decreasing correlations of the disordered state above the critical temperature
are replaced with just algebraicly decaying correlations below the BKT transition temperature. This phase with quasi long-range order is a superfluid, and there is a universal jump in the superfluid density at the BKT critical temperature.
BKT physics has recently been studied in the context of spin-orbit
coupled Fermi gases for FF states~\cite{xu_berezinskii-kosterlitz-thouless_2015,cao_superfluid_2015}. In the absence of SOC, the transverse superfluid density
for the FF state vanishes, and zero critical temperature is expected in that case~\cite{yin_fulde-ferrell_2014,jakubczyk_renormalization_2017}.
When SOC with in- and out-of-plane Zeeman fields are introduced, Xu {\it et al.} and Cao {\it et al.}  found non-zero BKT transition temperature
for both gapped as well as gapless FF superfluids
since transverse superfluid density could be non-zero. This suggests that
observations of FFLO phases in 2D systems with spin-orbit couplings
might be possible at non-zero temperatures.

All articles on SOC and FFLO that we have considered so-far rely on mean-field descriptions.
An interesting deviation from this pattern is the work by 
Zvyagin and Schlottmann~\cite{zvyagin_effects_2013} who considered
an exactly solvable 1D fermionic model with spin-orbit interactions and strongly attractive onsite interactions. Using Bethe ansatz  they obtained 
the critical exponents for superfluidity and density waves and found that the exponent
for the superfluid was the smallest. This suggested 
 an instability to a FFLO phase  in the presence of weak interchain coupling.

 \subsection{Summary of the status of FFLO in the presence of spin-orbit coupling}
 
As is clear the literature on FFLO states in the presence of SOC has expanded rapidly.
All studies so far have been mean-field theories and mostly focus on Rashba or equal Rashba and Dresselhaus spin-orbit couplings. Studies have revealed that compared to
systems without SOC, FFLO states become favorable 
in the presence of SOC. Furthermore, in contrast with systems without SOC, FF states
are expected to dominate over the LO ones. In 1D systems, it has also been pointed out that
states with Majorana fermions at the edge of the system compete with the FFLO states.
As SOC strength increases, eventually FFLO gives way to Majorana states.
In higher dimensions, Majorana states can also appear if there are vortices
in superfluids with spin-orbit couplings~\cite{Read_paired_states_2000,chan_non-abelian_2017,chan_generic_2017}.
%In higher dimensions, states can be classified further, but the most remarkable insight
But the most remarkable insight in higher dimensions is the prediction of topological FFLO states in systems with Rashba SOC and both in- and out-of-plane Zeeman 
fields. In such topologically non-trivial states, it is possible that the gap closes even in the bulk.

It will be interesting to see how accurate the predictions of mean-field theories are. Beyond mean-field effects might very well lead to pronounced 
changes in, for example, lower dimensional systems as well as in 
optical lattices with relatively large filling fractions.
 Orbital physics in optical lattices can also
enable new possibilities. A recent proposal to realize FFLO states in unpolarized Fermi gas by Zheng {\it et al.}~\cite{zheng_fulde-ferrell_2016} relied on moving optical lattices which couples the $s$- and $p$-orbital bands
with each other.
This gives rise to a single particle Hamiltonian equivalent to the
presence of spin-orbit coupling, although in a lattice model, the off-diagonals
also contain terms that are higher than lowest order in $k$.

\section{Detection of FFLO state}

\label{section_detection}

The challenge in realizing the FFLO state is accompanied by the difficulty of actually detecting it. 
The FFLO state is predicted to occupy only a rather small part of the homogeneous phase diagram in 3D.
Thus, in spatially inhomogeneous trapped gases, the state will be present in only a small part of the gas, except possibly in 1D and spin-orbit coupled systems where the FFLO state involves a larger parameter range. 
The effect of trapping potential has been considered in section~\ref{section_traps}.
In particular, the size of the FFLO phase domain in the cloud should be larger than the FFLO wavelength $1/q_\mathrm{FF}$.
Atomic gas experiments with more box-like trapping potentials~\cite{meyrath_bose-einstein_2005,es_box_2010,gaunt_bose-einstein_2013,mukherjee_homogeneous_2016} are believed to improve the situation~\cite{roscher_-medium_2016}, as the size of the FFLO domain in the cloud can then be increased and the experimental signature of the state improved.

Here we provide a comprehensive review of the various schemes that have been proposed for detection of the FFLO state.
Indeed, most of the theoretical works on the topic propose some way of probing the state, although only a few analyze the problem of detection in detail.
We roughly divide them into six different categories: ramps and quenches, density profiles, interferometric methods, collective modes, noise correlations, and spectroscopies.
Many of the methods have been proposed and analyzed in 1D systems, but we will consider them all on equal footing since many of the methods are applicable in any dimension.
Likewise we consider methods proposed both for continuum and lattice cases, and in spin-orbit coupled systems; many of the methods are applicable for all systems.

\subsection{Ramps and quenches}
\label{sec:ramps}

\begin{figure}
\label{fig:riegger}
\centering
\includegraphics[width=0.95\columnwidth]{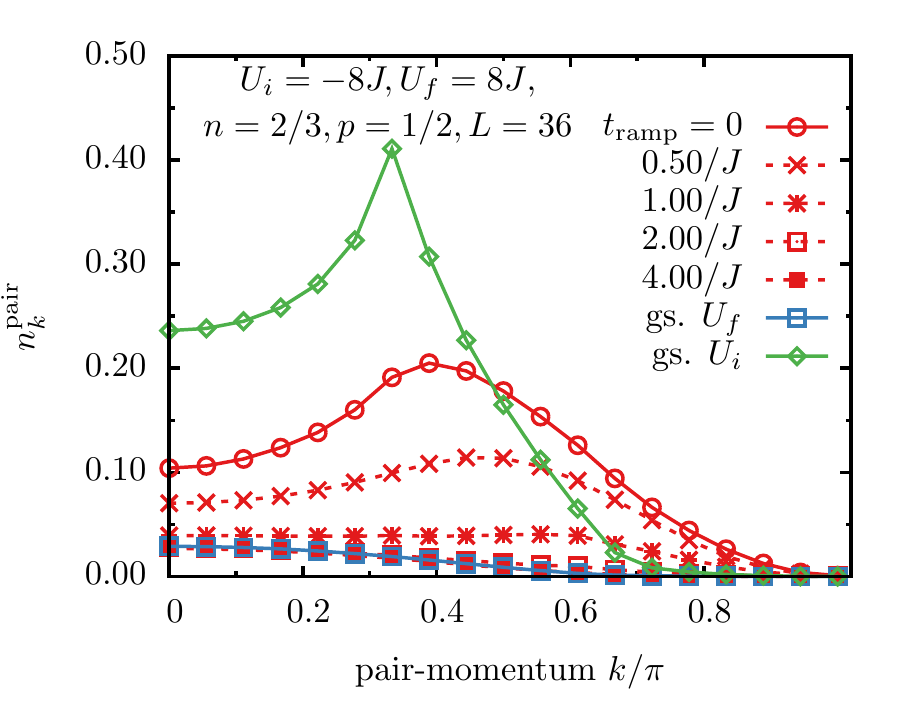}
\caption{The pair-momentum distribution of repulsive bound pairs reveal pre-quench FFLO-state correlations. However, correlations are rapidly lost for finite interacting strength sweep rates. Here $U_i=-8\,J$ is the initial interaction strength and $U_f=8\,J$ is the interaction strength after the quench. Shown are also the pair correlations in ground states for interaction strengths $U=U_i$ and $U_f$. Reproduced with permission from~\cite{riegger_interaction_2015}.}
\end{figure}

The now standard methods for detecting balanced BCS-type pairing in ultracold gases were already proposed early on also for detecting FFLO correlations~\cite{yang_realization_2005}.
One method is an interaction strength quench, in which an external magnetic field is swept across the BCS-BEC crossover. 
This process projects the Cooper pairs in the BCS side onto bound molecules in the BEC side, and the momentum distribution of the molecules is then probed by releasing the gas from the trap and following the free expansion of the molecular cloud~\cite{regal_observation_2004} or by using momentum-resolved Raman spectroscopy~\cite{fu_momentum-resolved_2012}.
Finding a molecular condensate on the BEC side is a signature of an initial condensate of Cooper pairs on the BCS side.
In the present context, since the Cooper pairs in the FFLO state will have a finite center-of-mass momentum, the momentum distribution of the corresponding molecules would show the same center-of-mass motion~\cite{sheehy_becbcs_2007,casula_quantum_2008,tezuka_density-matrix_2008, wang_quantum_2009,wolak_finite-temperature_2010, baur_fulde-ferrell-larkin-ovchinnikov_2010,zhang_modulated_2010,lee_asymptotic_2011, cai_stable_2011,dalmonte_dimer_2012,liang_unconventional_2015}.
While the projection to Feshbach molecules can be used for probing the FFLO state in lattices also~\cite{zhang_modulated_2010}, the presence of an optical lattice allows for other schemes.
Indeed quenches and the expansions can be done in various ways, but the common denominator when considering detection of the FFLO state is that the pairs in the FFLO state are projected onto some more stable configuration (bound molecules, repulsively bound pairs, doublons) and the momentum distribution of these is then probed, see Fig.~\ref{fig:riegger}.
Futhermore an interesting variant was proposed in~\cite{zapata_triplet_2012} in which triplet correlations between atoms residing in the nodes of the oscillating LO-order parameter are mapped onto $p$-wave molecules.
In 3D systems, the pair formation dynamics in the unitary regime were found to be very rapid~\cite{sanner_correlations_2012}. 
However it is not clear whether this imposes constraints for seeing FFLO correlations in molecular momentum distributions, since formation of molecules is actually a desired property.
However molecules are strongly interacting in the unitary regime, and a quantitative understanding of how FFLO correlations survive the BCS-BEC ramp in 3D systems is missing.

\begin{figure}
\label{fig:kajala}
\centering
\includegraphics[width=0.8\columnwidth]{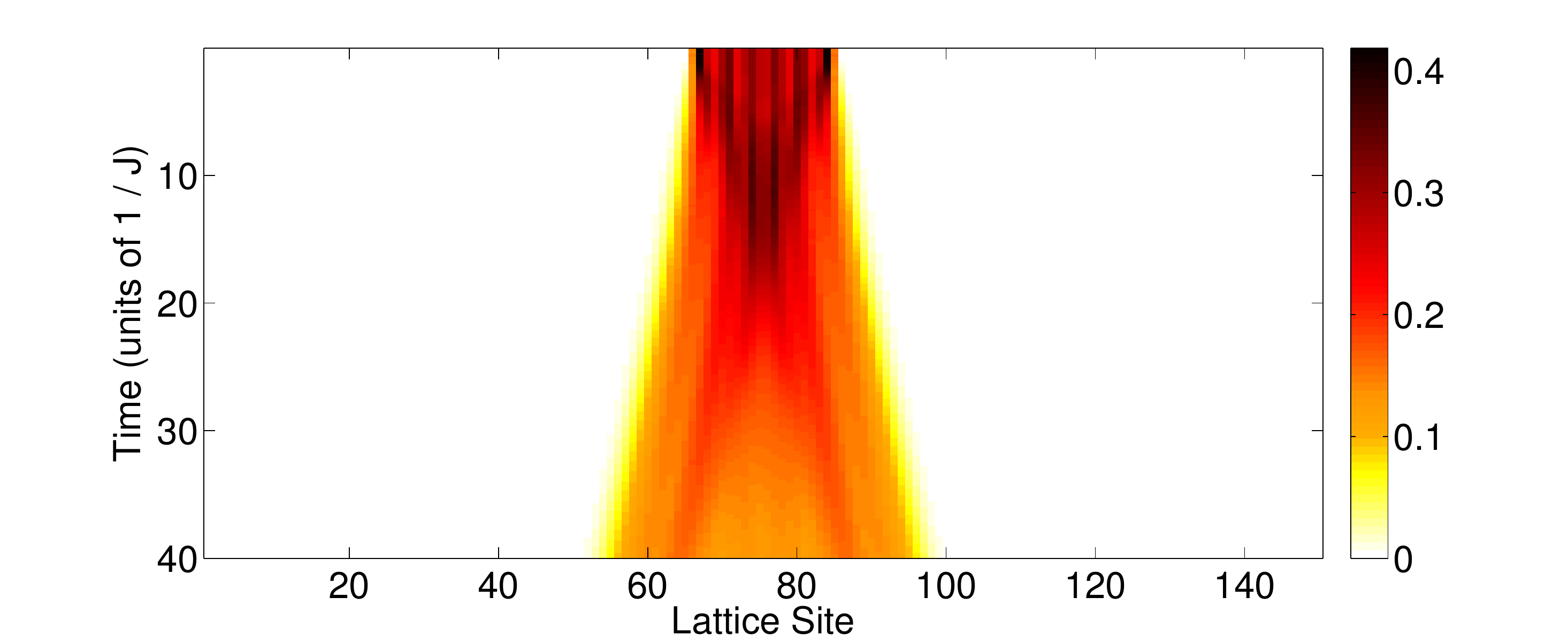}\\
\includegraphics[width=0.8\columnwidth]{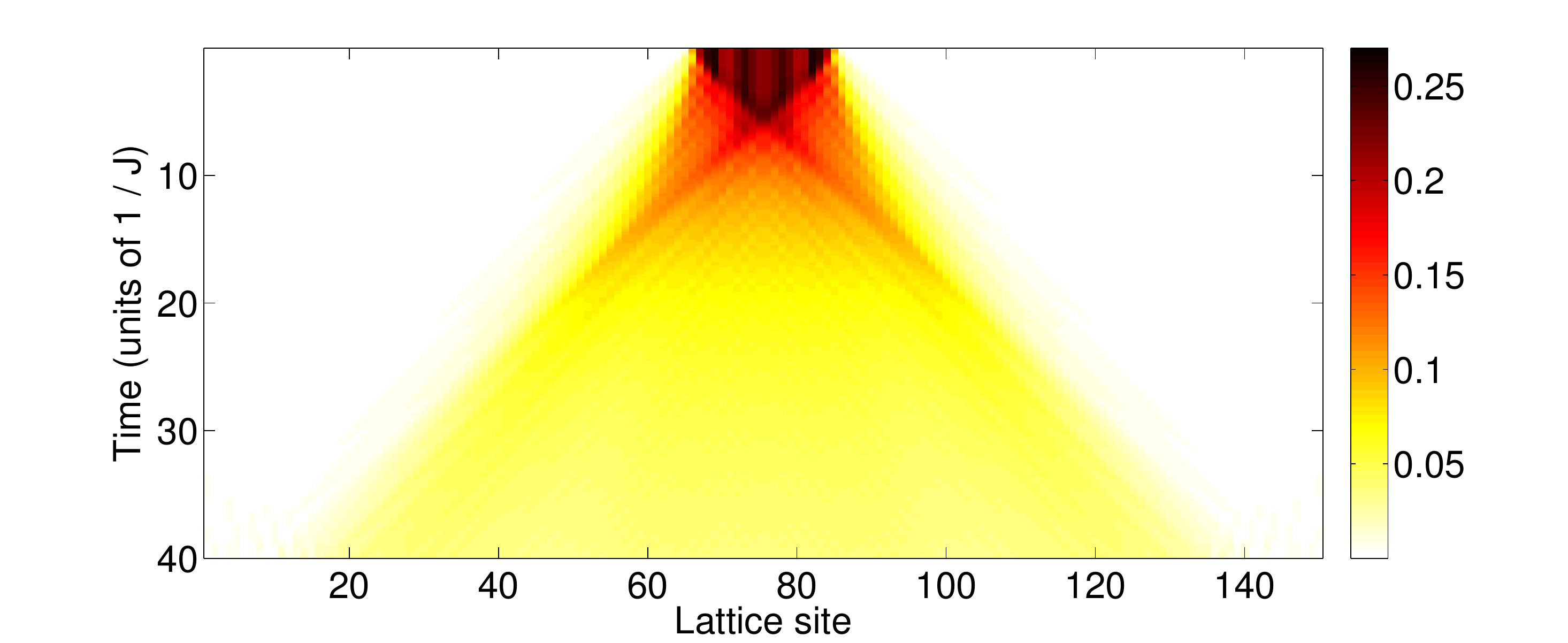}\\
\caption{Releasing the spin-imbalanced gas from a trap (here shown for a 1D lattice and a box trap), the clouds of doublons and unpaired atoms expand. The center-of-mass momenta of the pairs can be obtained from the expansion speeds. Reproduced with permission from~\cite{kajala_expansion_2011}.}
\end{figure}

The stability of the correlations has been studied in 1D systems. 
Ref.~\cite{riegger_interaction_2015} considered a 1D system in which interactions are quenched from the attractive to the repulsive side, and the initial FFLO correlations are imprinted on repulsively bound pairs.
However signatures of FFLO correlations were found to vanish rapidly in the interaction strength sweep, showing the importance of a sufficiently rapid sweep.
Similar decay of the FFLO correlations in the pair momentum distribution were found also in Refs.~\cite{bolech_long-time_2012,lu_expansion_2012,yin_quench_2016}. 
However also interestingly the work suggests that features in the spin-density are more robust and offer a possible way for detecting the FFLO state in the expansion of a 1D gas.
Since the loss of FFLO correlations was caused by the unavoidable collisions when expanding along 1D lattice, the problem may be less prominent in higher dimensions.
On the other hand, Ref.~\cite{kajala_expansion_2011} considered similar 1D setting, and the separation of the doublons and single atoms was beneficial, as the expansion velocity of unpaired atoms was found to be connected with the FFLO momentum $q$. 
In this case, the atoms remained interacting during the expansion, that is, only the trap potential was quenched, see Fig.~\ref{fig:kajala}.
Consequently the FFLO momentum $q$ was directly reflected in the expansion velocities of the unpaired fermions because in 1D they form a Fermi sea of non-interacting quasiparticles with Fermi momentum $k_{\mathrm{F}\uparrow} - k_{\mathrm{F}\downarrow}$.
The use of time-of-flight expansion for measuring momentum distributions and detecting the FFLO state have also been considered in spin-orbit coupled systems~\cite{iskin_topological_2013,zheng_route_2013,zheng_fflo_2014}.

Alternatively one could quench to a non-interacting regime, in which case the release of the gas will allow the measurement of the momentum distributions of the two atomic components separately.
While this will not yield direct access to pair-distribution function, many works suggest that FFLO state will have observable effects in single-particle momentum distributions as well~\cite{koponen_fermion_2006,koponen_finite-temperature_2007,batrouni_exact_2008,wang_quantum_2009,loh_detecting_2010}.

\subsection{Density profiles}
\label{sec:density}

The FFLO state, particularly the LO state, has also been shown to influence the density distributions of the two atomic components~\cite{mora_transition_2005,tezuka_density-matrix_2008,bulgac_unitary_2008,kakashvili_paired_2009,loh_detecting_2010,heidrich-meisner_bcs-bec_2010,guan_quantum_2011,cai_stable_2011,chiesa_phases_2013,heikkinen_nonlocal_2014,baarsma_larkin-ovchinnikov_2016}.
While the total density profile is largely unaffected by the FFLO state, the local polarization appears to be a viable option for the detection~\cite{loh_detecting_2010,baarsma_larkin-ovchinnikov_2016}. 
The state might thus be observable in species selective absorption imaging~\cite{mizushima_direct_2005}, or in phase contrast images~\cite{shin_observation_2006}, that directly probe the density difference between the two species.
Unfortunately in 3D, even if the FFLO phase domain would be large enough to exhibit actual density modulations, any such off-centered density modulations are easily lost in the columnar absorption image, which probes the integrated density through the cloud (along the direction of the imaging laser).
\begin{figure}
\centering
\includegraphics[width=0.8\columnwidth]{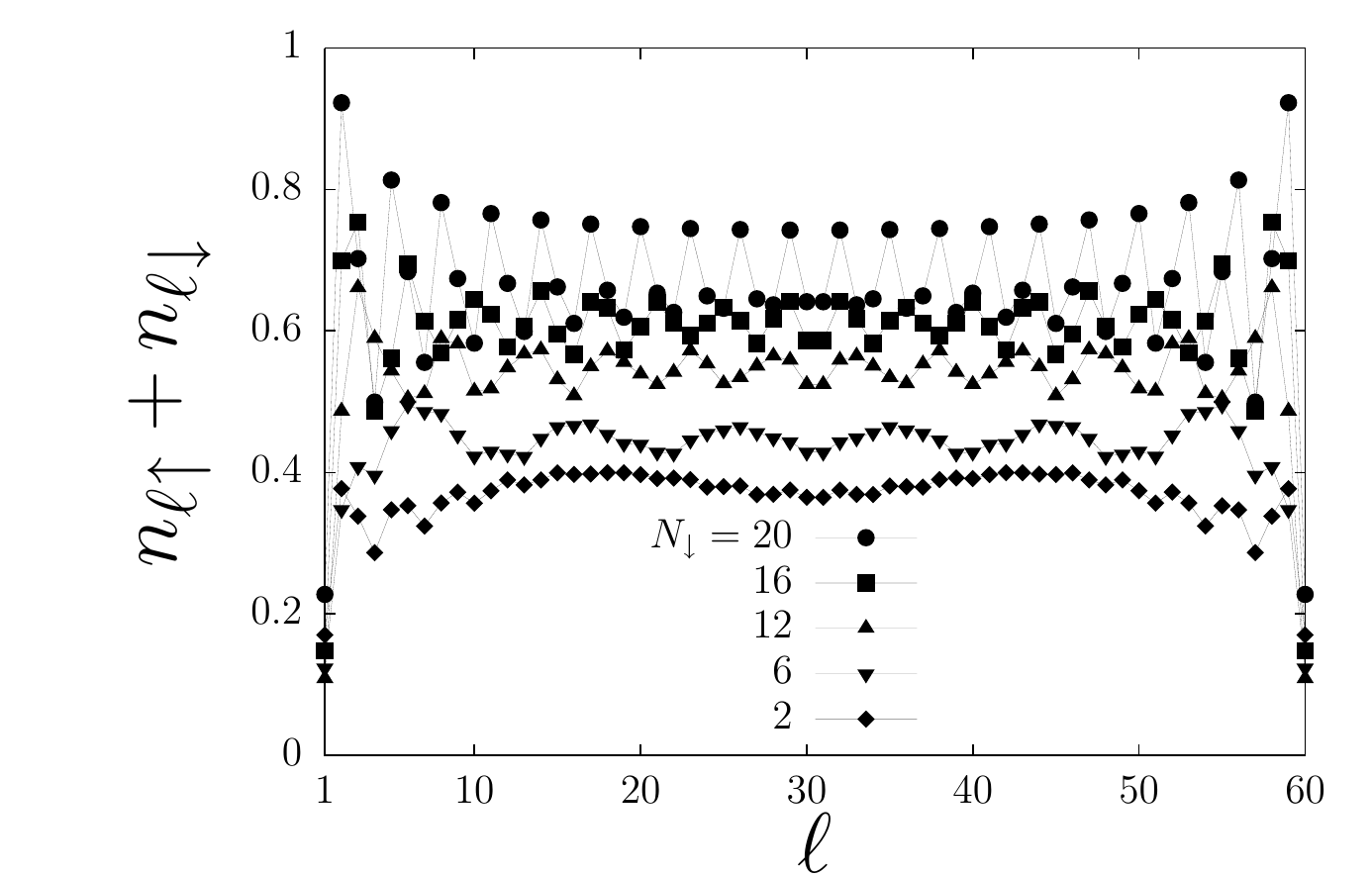}\\
\includegraphics[width=0.8\columnwidth]{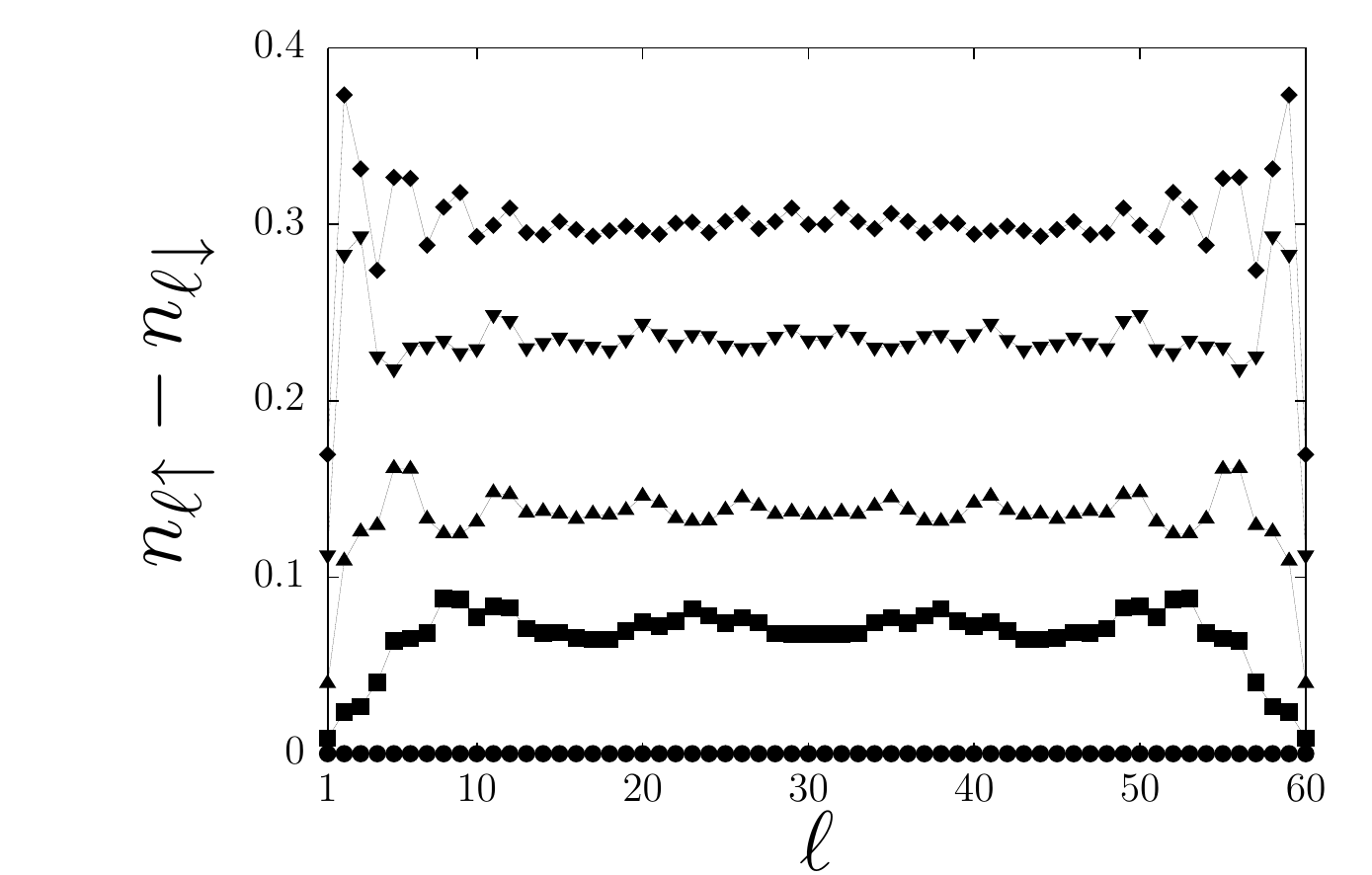}\\
\caption{FFLO-type pairing in 1D lattice results in minor modulations in the density difference between the two atomic components. Reproduced with permission from~\cite{rizzi_fulde-ferrell-larkin-ovchinnikov_2008}.}
\label{fig:rizzi}
\end{figure}
 
Since the FFLO momentum tends to align itself in the longitudinal direction in elongated traps, one would expect density modulations, and FFLO correlations, to be most easily observed in quasi 1D-systems~\cite{parish_quasi-one-dimensional_2007}.
Also, quantum gas microscope setups for fermions~\cite{edge_imaging_2015,omran_microscopic_2015,cheuk_quantum-gas_2015,haller_single-atom_2015,parsons_site-resolved_2015,Mitra2017} with 2D lattices can avoid the problems with columnar integration altogether.
At sufficiently high densities, 2D systems are predicted to exhibit a special angular FFLO state~\cite{chen_exploring_2009} in which the modulations are positioned in an angular fashion instead of radially.
Still the density variations predicted even in 1D systems are quite minor, see for example Fig.~\ref{fig:rizzi} and Refs.~\cite{rizzi_fulde-ferrell-larkin-ovchinnikov_2008,kim_exotic_2011}. 
Minor modulations in the densities may be better observed through spectroscopic means that specifically couple to density modulations of a given periodicity, such as Bragg spectroscopy~\cite{edge_signature_2009,baarsma_inhomogeneous_2013,devreese_controlling_2011}.

A quite different approach would be to utilize the connection between the trapping potential, local density and pressure for determination of the equation of state~\cite{horikoshi_measurement_2010,nascimbene_exploring_2010,nascimbene_equation_2010}. 
Equation of state, or particularly the heat capacity~\cite{kinast_heat_2005}, is influenced by the FFLO state, and one would expect jumps in the specific heat at various phase transitions~\cite{sheehy_becbcs_2007,konschelle_anomalous_2007,dutta_lifshitz_2013}, although precise thermometry of the gas is still one of the outstanding experimental problems.
From local densities one can calculate also local entanglement entropy, which was studied in Ref.~\cite{Franca_entanglement_2017} and shown to be a viable candidate for detecting the normal state to FFLO state transition.

\subsection{Interferometric methods}

\begin{figure*}
\includegraphics[width=1.8\columnwidth]{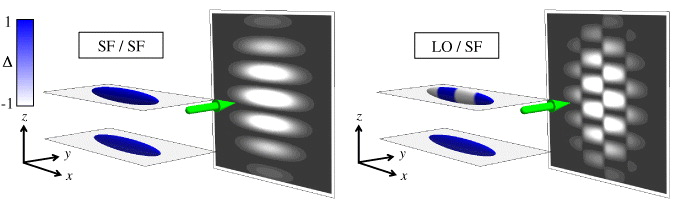}
\caption{Two interfering balanced superfluids will result in interference fringes (left), but if one of the clouds is in FFLO state, the resulting interference fringes will become staggered. Re-used under Creative Commons Attribution CC BY licence 3.0 from~\cite{swanson_proposal_2012}.}
\label{fig:swanson}
\end{figure*}
Interference patterns between two atomic clouds with FFLO-correlations have also been considered by a number of groups~\cite{gritsev_interferometric_2008,rizzi_fulde-ferrell-larkin-ovchinnikov_2008,swanson_proposal_2012}.
Fig.~\ref{fig:swanson} shows one such setup, in which two atomic clouds are allowed to expand, and upon overlapping an interference pattern is seen. 
The appearing interference fringes probe the phase difference between the sources, and in the case of an FFLO state with an oscillating order parameter, the fringes become staggered~\cite{swanson_proposal_2012}.
To be able to probe the interference between pairing fields in the first place, the method needs to be combined with the magnetic field sweep into the BEC side, in a similar fashion as considered in section~\ref{sec:ramps}.

An interesting alternative is the suggestion of using a Josephson current for detecting an FFLO state~\cite{hu_josephson_2011,xu_competing_2014}.  
The Josephson current is proportional to the local order parameter, and a local probe can thus identify position dependent order parameter fields. This can be realized with two clouds coupled via a small dimple potential channel~\cite{hu_josephson_2011}.
Since the scheme couples directly to fermionic pair correlations, no BCS-BEC ramps are needed.

For interference studies, one needs a scheme for producing two (or more) clouds. 
One possibility is to prepare these in a double well potential or in an optical lattice.
Alternatively a single atomic cloud could be split in two by lasers, or Bragg beams could be used for extracting parts of the cloud for the interference protocol~\cite{carusotto_atom_2005,rizzi_fulde-ferrell-larkin-ovchinnikov_2008}.

\subsection{Collective modes}

Superfluidity is a collective phenomenon, and it has profound effects on collective excitations. 
A natural question is then whether collective modes or excitations can be used for identifying the superfluid FFLO state.

Most studies of the collective modes in FFLO state consider uniform lattice or continuum systems and apply mean-field theories with the random-phase approximation~\cite{edge_signature_2009,edge_collective_2010, samokhin_goldstone_2010, heikkinen_collective_2011, mendoza_superfluidity_2013,mendoza_collective_2014, koinov_collective_2015}, although 1D systems have also been studied with exact numerical methods~\cite{hu_phase_2007,zheng_floquet_2015}.
The breaking of the rotational symmetry in the FFLO state has been shown to result in anisotropic speed of sound~\cite{heikkinen_collective_2011}. 
This is discussed in detail in section~\ref{CollectiveSection}.
Probing the anisotropic speed of sound has also been considered in the context of spin-orbit coupled systems as a possible way of detecting the FFLO state~\cite{xu_anisotropic_2014,xu_berezinskii-kosterlitz-thouless_2015,zheng_fulde-ferrell_2016}.
Furthermore as one might expect, the FFLO state influences the nature of vortices in the superfluid state~\cite{radzihovsky_quantum_2009}.
These may offer a way for detecting the FFLO state~\cite{samokhalov_fulde-ferrell-larking-ovchinnikov_2010,shang_single_2010}, although the low visibility of vortices in the BCS and unitary regimes can prove challenging~\cite{Zwierlein2005}.

However more research would be needed for understanding the effect of FFLO state in other collective modes, such as the often used trap modes (breathing modes, dipole modes, etc.), and the interplay between the second sound and unpaired atoms in the FFLO state would deserve more analysis.

\subsection{Noise correlations}

Simple density probes suffer from rather weak signals (small density perturbations) and strong noise (large background density).
Therefore higher order correlations, such as density-density and current-current correlations, have been considered as potential probes for the FFLO state.
The use of shot noise correlations, in which the atom gas in the FFLO state is first quenched to a weakly interacting state and allowed to expand, and the correlated appearance of fermions at momentum states $-k$ and $k+q$ for variable $k$ but fixed $q$ was already proposed by Yang~\cite{yang_realization_2005}.

\begin{figure}
\includegraphics[width=0.95\columnwidth]{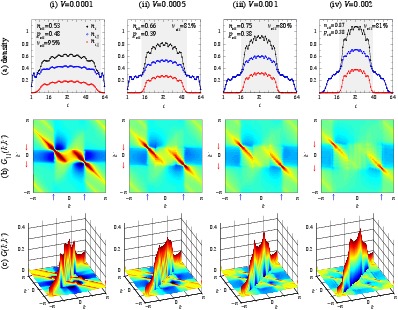}
\caption{Density profiles (top row) and spin-spin correlators (middle row) reveal the FFLO state as oscillations in density or shifted correlator peaks. In particular, the spin-spin correlator has peaks shifted by the FFLO momentum ${\bf q}$. The bottom row is the total shot noise correlator and the different columns correspond to different harmonic trapping potential strengths $V$. Reproduced with permission from~\cite{luscher_fulde-ferrell-larkin-ovchinnikov_2008}.}
\label{fig:luscher}
\end{figure}

The use of noise correlations as a probing scheme in general was reviewed in Refs.~\cite{bloch_many-body_2008,torma_quantum_2014-2}, but here we briefly review the ideas that have been presented for the detection of the FFLO state itself.
Ref.~\cite{luscher_fulde-ferrell-larkin-ovchinnikov_2008} shows a detailed analysis of the various density and spin correlations in a 1D lattice and their applicability for detecting the FFLO state. 
While practically all higher order correlators show signatures of the FFLO state~\cite{rizzi_fulde-ferrell-larkin-ovchinnikov_2008,batrouni_exact_2008}, the $\langle n_\uparrow n_\downarrow \rangle$ correlator appears to provide particularly clear signatures, with peaks separated by the FFLO momentum, shown in Fig.~\ref{fig:luscher}. 
These results seem to also hold in higher dimensions~\cite{dey_current_2011,dey_effect_2011}. 
The use of noise correlations in spin-orbit coupled systems has also been proposed~\cite{xu_anisotropic_2014,zheng_fflo_2014,zheng_fulde-ferrell_2016,yin_quench_2016}.

In practice, the noise correlations can be studied either in-situ or in expanding clouds.
In a freely expanding cloud, the observed spatial correlations yield information regarding the initial momentum correlations~\cite{paananen_noise_2008,koponen_fflo_2008}. 
Such time-of-flight probes need to be combined with interaction quenches to molecular states~\cite{wang_quantum_2009} or noninteracting states. 
Still one would expect the expansion dynamics to be problematic for the conservation of the correlations.  
Alternatively, the various density-density correlations can also be studied in-situ.
A spatially resolved quantum polarization spectroscopy protocol was suggested~\cite{roscilde_quantum_2009,chen_pure_2012} as a way to probe spin-spin correlations, and density-density correlations might be used to probe using quantum gas microscopes~\cite{edge_imaging_2015,omran_microscopic_2015,cheuk_quantum-gas_2015,haller_single-atom_2015,parsons_site-resolved_2015}.
Furthermore although the ground state FFLO state does not have a total mass current (see section~\ref{subsec:bloch_theorem}), it does contain mutually cancelling superfluid and quasiparticle currents in the ground state. Thus current-current correlations, if one could come up with an idea on how to probe them, could provide a good signature of the FFLO state.

\subsection{Spectroscopies}

\begin{figure}
\includegraphics[width=0.95\columnwidth]{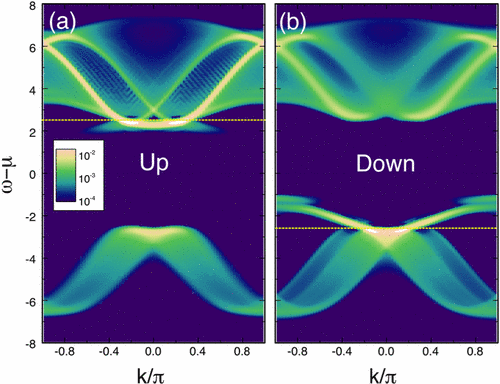}
\caption{Spectral function for the quarter-filled spin-polarized Hubbard chain, with $U=-8t$, $N_\uparrow = 24$, $N_\downarrow = 16$. The upper (lower) parts of the spectra describe the response for adding (removing) a fermion of corresponding spin-component. The upper part could be probed with usual radio-frequency spectroscopy, while the lower part could be accessed with inverse radio-frequency spectroscopy. Chemical potential of each component is shown with horizontal lines. The unpaired majority atoms show up as the in-gap states at frequencies below the chemical potential, and the in-gap states in the minority component spectral function reflect the possibility of forming pairs in the inverse spectroscopy. Reproduced with permission from~\cite{feiguin_spectral_2009}.}
\label{fig:feiguin}
\end{figure}

Since the FFLO state involves a spatially varying order parameter, an obvious way for detecting the state would be to use probes that have been used for probing the order parameter and excitation gap $\Delta$ in spin-balanced systems~\cite{torma_physics_2016}.
The spectral function can be probed by numerous methods~\cite{Torma2014_ch10,Torma2014_ch11}: radio-frequency spectroscopy, Raman spectroscopy, Bragg spectroscopy, and lattice modulation spectroscopy, to name just a few commonly used methods.
Radio-frequency spectroscopy is already a standard method for probing the single-particle excitation spectrum, although possible interactions in the final state make interpretation of results complicated. 
Bragg spectroscopy can be used for probing elementary particle-hole excitations, but in the case of long-wavelength Bragg spectroscopy, also collective excitations. 
A particularly appealing aspect of Bragg spectroscopy is the periodic coupling to atomic density fields, and thus a possibility of probing the weak density modulations in the FFLO state~\cite{edge_signature_2009,baarsma_inhomogeneous_2013,devreese_controlling_2011}. 
Lattice modulation spectroscopy for observing the FFLO state was considered in Ref.~\cite{korolyuk_probing_2010}.
In this method, a periodic variation of the depth of the lattice potential excites the system, creating doublon excitations.
The width of the double occupancy spectrum was shown to provide a measure of the FFLO momentum.
In similar fashion, other spectroscopies probe different parts of the system's spectral function, see for example Fig.~\ref{fig:feiguin}. 

Inhomogeneous order parameter results in Andreev mid-gap states~\cite{an_phase_2009,loh_detecting_2010} and unpaired atoms yield gapless excitations~\cite{lutchyn_spectroscopy_2011,heikkinen_finite-temperature_2013}.
These provide single-particle excitations at energies below the superfluid gap or even at negative energies~\cite{bakhtiari_spectral_2008}.
However a major problem in deducing the presence of the FFLO state from spectroscopic probes is that the inhomogeneity due to the trapping potential already results in low-energy excitations in the spectral function~\cite{kinnunen_strongly_2006}, as discussed in section~\ref{section_traps}.
To avoid these problems, local probes such as methods utilizing lasers (Bragg spectroscopy, Raman spectroscopy) may be most useful, and they have indeed been used for probing the essentially homogeneous order parameter $\Delta$~\cite{hoinka_goldstone_2017}.
Also methods acting on the whole cloud, such as radio-frequency spectroscopy, can be used as long as the response is considered only in a limited region of the gas, such as in the experiments done in JILA~\cite{sagi_measurement_2012}.
Again box traps will also help make the spectroscopic signature clearer.

An interesting addition to usual spectroscopies is the use of impurities. 
Spin-selective scattering potentials~\cite{sheehy_spin-selective_2011} and impurities~\cite{li_single_2012,jiang_single_2011,li_interaction-induced_2012} are influenced by the local pairing gap, allowing a very precise spatially resolved spectroscopy of the FFLO order parameter.

\begin{figure}
\includegraphics[width=0.95\columnwidth]{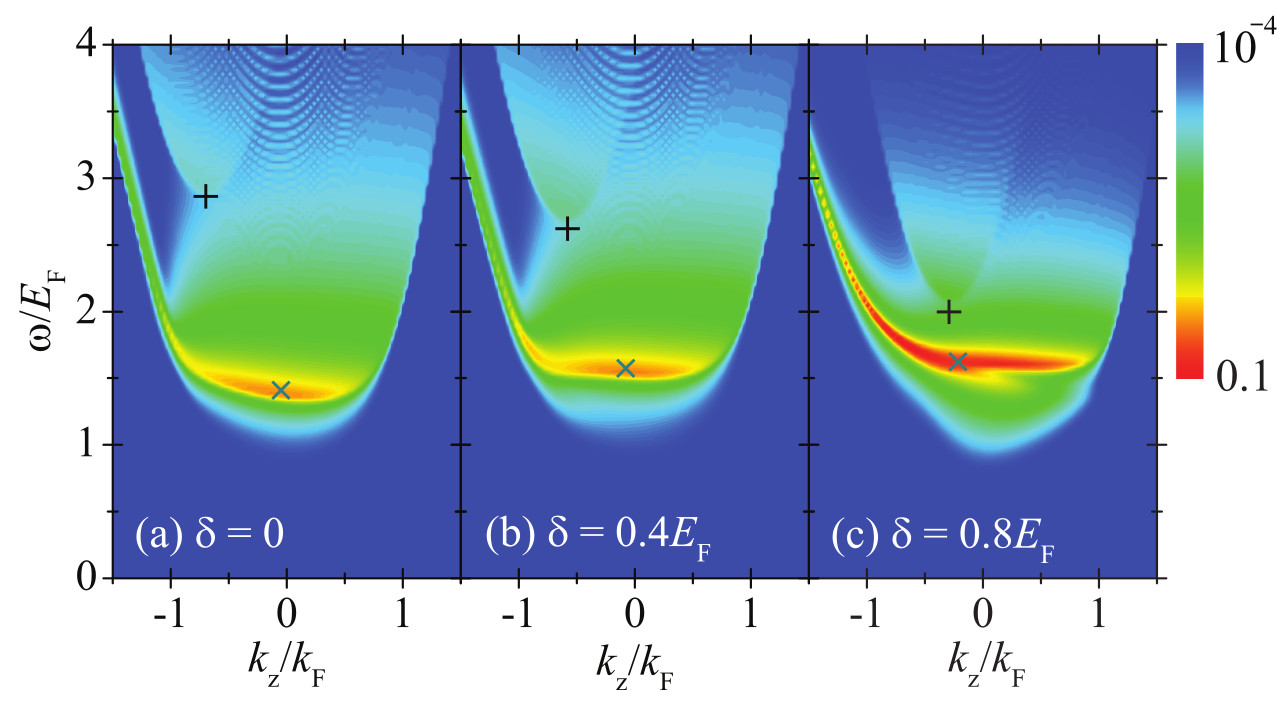}
\caption{Logarithmic contour plot showing momentum resolved rf-spectra in a spin-orbit coupled system. The FFLO phase in b) and c) reveals itself in the spectral peaks (marked by crosses 'x' and '+') as a shift in the momentum $k_z$. Reproduced with permission from~\cite{liu_inhomogeneous_2013}.} 
\label{fig:spinorbitmomres}
\end{figure}

Use of spatially- and momentum-resolved spectroscopies have also been considered in spin-orbit coupled systems~\cite{qu_topological_2013,zhang_topological_2013,zhou_exotic_2013,xu_anisotropic_2014,liu_topological_2012,wu_unconventional_2013,xu_anisotropic_2014}, and were studied in more detail in~\cite{liu_inhomogeneous_2013}.
In spin-orbit coupled systems, the connection between spin-imbalance and the FFLO state is not clear, since the spin-orbit coupling itself is sufficient for producing a finite FFLO momentum. 
Spin-orbit coupled systems are analyzed in more detail in section~\ref{FFLOSOC}. 
Fig.~\ref{fig:spinorbitmomres} shows momentum-resolved radio-frequency spectrum of a spin-orbit coupled system. 
A particularly interesting proposal is to probe the edge modes in a topological FFLO phase using spatially resolved spectroscopy~\cite{chen_inhomogeneous_2013}. In the case of both spatially- and momentum-resolved spectroscopy, the topological FFLO state is expected to manifest as gapless excitation at finite momentum $k$~\cite{chan_pairing_2014}.

\section{Conclusions and outlook}
\label{conclusions}

In this review, we have presented the theory of the FFLO state in lattice and trap geometries, and reviewed literature on theory and simulations of the lattice FFLO state in the context of ultracold quantum gases (UQG). A review of recent literature on the FFLO state in the presence of spin-orbit coupling was also given, as well as a comprehensive description of possible methods to observe this elusive state of matter in quantum gas experiments. 

As section~\ref{section_detection} shows, there are plenty of options for experimental probes that may reveal salient features of the FFLO state. Promising recent developments that may be crucial for unambiguous observation of the FFLO state are the quantum gas microscopes~\cite{bakr_probing_2010,sherson_single-atom-resolved_2010,torma_situ_2014}, in particular their fermionic versions~\cite{edge_imaging_2015,omran_microscopic_2015,cheuk_quantum-gas_2015,haller_single-atom_2015,parsons_site-resolved_2015,Mitra2017}. In those systems, one has unprecedented access to small spatial variations in densities as well as spin and density correlations that all carry specific signatures of the FFLO state. The development has been similarly impressive in realizing spin-orbit coupled systems in UQG~\cite{lin_spin-orbit-coupled_2011,dalibard_textitcolloquium_2011}, and it will be interesting to extend such studies to the case of imbalanced fermion systems.
One more promising development is the realization of box potentials and homogeneous gases~\cite{meyrath_bose-einstein_2005,gaunt_bose-einstein_2013,Schmidutz_quamtum_joule2014,mukherjee_homogeneous_2016,hueck_two-dimensional_2017,mazurenko_cold-atom_2017} instead of the usual harmonically trapped gases. This can make possible FFLO signatures more clear, counteract the tendencies to phase separation, and realize the fixed particle number situation instead of fixed chemical potential, which may lead even to the stability of the Sarma state~\cite{gubankova_breached_2003,koponen_fermion_2006}. 
In the future, there will probably also be many artificial solid state systems where the orbital magnetism and Pauli paramagnetic effects can be tailored in a way amenable to the FFLO state. In general, examples of such {\it designer matter} range from microscopic superconducting structures realizing artificial atoms~\cite{you_atomic_2011} to nanostructured~\cite{moriarty_nanostructured_2001} and thin-film materials~\cite{martin_thin-film_2016} to atomic scale artificial lattices~\cite{schulz_many-body_2015}.

The FFLO state in lattices poses some major open questions to be approached by future theory work, furthermore, several conceptually new directions of research can be envisioned. To connect with near future UQG experiments, further work on finite temperature properties is needed. Interestingly, some beyond-mean-field calculations \cite{Wolak2012,Karmakar2016} find a finite momentum signature in the pairing structure factor well above the critical temperature for superfluidity $T_c$. It would be important to study in more detail whether and when such precursors of the FFLO state exist, since they may be observable in the UQG experiments even at presently available temperatures. This relates to the fundamental but almost
unstudied question of the nature of the normal state right above the critical temperature systems susceptible to FFLO correlations, especially for strong interactions. One can also search for precursors of the FFLO state at the limits of imbalance: polaron systems in which there are only individual atoms of the minority spin state, and the case of vanishing polarization where the FFLO state is connected with pinned solitons and soliton trains that can arise either in the ground or excited state~\cite{lutchyn_spectroscopy_2011}.

Another key question is the stability of the FFLO state at temperatures finite but well below $T_c$. In continuum systems, the stability of the FFLO state towards fluctuations of the superfluid phase is a subtle issue. Already the original works by Fulde and Ferrell \cite{FF}, and Larkin and Ovchinnikov \cite{LOeng}, point out the fundamental difference of the supercurrents in the directions parallel and transverse to the FFLO wave vector $\mathbf{q}$. The question of stability was studied in~\cite{Shimahara1998} where the Ginzburg-Landau theory in the context of type II superconductors predicted that any spatial dependence of the order parameter that varies in one dimension only (such as the (double)planewave FF(LO) ansatz) is unstable, however if the order parameter structure is essentially 2D such as triangular, stability is established. A more elaborate analysis based on mean-field approximation with quantum fluctuations, and taking into account anisotroptic superfluid density \cite{yin_fulde-ferrell_2014}, showed that the FF state in a 2D continuum system is unstable due to vanishing superfluid density in the direction transverse to the FF wavevector. However the fate of the LO state remained an open question since the microscopic BCS-based approach does not easily accommodate the LO ansatz (c.f.\ the discussion in section~\ref{subsec:bogoliubov_transformation}). Radzihovsky and coworkers have extensively studied the stability of the continuum FFLO state and its collective modes in various dimensions using a Ginzburg-Landau type approach~\cite{radzihovsky_quantum_2009,radzihovsky_imbalanced_2010,radzihovsky_fluctuations_2011,radzihovsky_quantum_2012}. They showed that the spontaneously broken rotational symmetry leads to a soft Goldstone mode, which limits the superfluid density in the transverse (to the FFLO momentum) direction. Indeed in 3D, the FF state was argued to have vanishing transverse superfluid density, and even in the LO state it was suppressed. Still, FF and LO states were argued to be energetically stable at zero temperature for dimensions $d > 1$.

All the above theory knowledge on FFLO stability concerns continuum systems: much less is known about the effect of fluctuations in the lattice case. The Mermin-Wagner theorem, denying true long-range order in 1D and 2D systems is valid there as well, as is the possibility of the Berezinskii-Kosterlitz-Thouless (BKT) mechanism to produce quasi-long-range order and superfluidity in 2D. However due to the breaking of the rotational symmetry and continuous translation invariance as well as nesting effects in lattices, the role of fluctuations may differ from the continuum case. The existing beyond-mean-field studies of the density-imbalanced Fermi gases in lattices indicate stability of the FFLO state against fluctuations: a stable FFLO state was found by DMFT calculations that take into account local fluctuations~\cite{kim_fulde-ferrell-larkin-ovchinnikov_2012,heikkinen_finite-temperature_2013}, and by cluster DMFT which also includes non-local fluctuations~\cite{heikkinen_nonlocal_2014}, as well as by QMC methods ~\cite{gukelberger_fulde-ferrell-larkin-ovchinnikov_2015}. Furthermore, recent cluster DMFT~\cite{Vanhala2017} and other numerical~\cite{Zheng2017} studies found the particle-hole equivalent of the FFLO state~\cite{moreo_cold_2007}, namely the striped phase, to persist in the doped repulsive Hubbard model. There is thus mounting theoretical and computational evidence for the general existence of the FFLO state in lattices, but which form of the order parameter is the stable one, and at exactly what temperatures, can be still regarded as essentially open questions. This calls for the evaluation of the superfluid density, even in the mean-field case, for estimating the BKT temperature of superfluidity. Furthermore, beyond-mean-field studies of unexplored parameter regimes and with less approximative treatment of fluctuations are needed. 

Understanding FFLO superfluidity beyond the mere ground-state phase diagrams also demands studies of collective modes. In spin-orbit coupled systems, it would be interesting to study which symmetries are spontaneously broken by trying to identify the corresponding Goldstone (Anderson-Bogoliubov) modes. For instance the equal Rashba and Dresselhaus SOC breaks the rotational symmetry already at the Hamiltonian level and the dispersion has minima at finite momenta, which could affect the collective mode spectrum of the FFLO state. Further the Higgs mode related to order parameter amplitude fluctuations, and the Leggett mode appearing in multiband systems, are to our knowledge completely unexplored in the FFLO context. 
Considering the extended (nearest-neighbor interacting) Hubbard model~\cite{dhar_population_2016}, $p$-wave interactions~\cite{massignan_creating_2010}, dipolar interactions~\cite{baranov_superfluid_2002,Baranov_theoretical_2008,Lee2017}, or dark-state control of Feshbach resonances~\cite{he_realizing_2017} can also bring new features to FFLO physics.

Observation of the FFLO state, after half-a-century of theory work and experimental effort, would be a major breakthrough. It would profoundly affect the way we understand both superconductivity and magnetism. Realization of the FFLO state with quantum gases in optical lattices would also be a landmark in determining the phase diagram of the Hubbard model, the paradigm model of strongly correlated electrons~\cite{micnas_superconductivity_1990}. Observation of the FFLO phase would suggest, assuming that the system is particle-hole symmetric (see section~\ref{U-Umapping}), that in the repulsive Hubbard model the striped phase is stable in the corresponding parameter regimes; this would give information about the possible existence of the d-wave superfluid and thereby the role of the Hubbard model in explaining high temperature superconductivity. 

Observation and understanding of the classic FFLO state is a grand goal in itself, however, even more is expected when extending the basic ideas of the FFLO physics in simple square lattices to new contexts. In~\cite{kim_topological_2013,sarjonen_topological_2015} the idea of {\it mixed geometry pairing} was introduced: the $\uparrow$ and $\downarrow$ fermionic species were residing in different lattice geometries, which resulted in stable Sarma/BP states, among other things. Rich new physics is expected from multiband systems, especially those that can host flat, that is, dispersionless bands. In a flat band, interaction effects dominate over kinetic energy, which can lead to predictions of room temperature superconductivity~\cite{kopnin_high-temperature_2011,Khodel_1990}, for example. Remarkably even when the dispersion is zero, supercurrent can still flow: it was recently found~\cite{peotta_superfluidity_2015,julku_geometric_2016,tovmasyan_effective_2016,liang_band_2017,liang_semiclass_2017}
that superfluidity in flat bands is guaranteed whenever the band has non-zero Berry curvature, and a direct relation between flat band superfluidity and the quantum metric was discoverd. The quantum metric~\cite{provost_riemannian_1980} describes the distance between quantum states and is the real part of the quantum geometric tensor whose imaginary part is the Berry curvature. This makes one obviously curious about the fate of the FFLO state in a lattice supporting flat bands. Flat bands can be achieved, for instance, by lattice geometry, as in case of the Lieb and kagome lattices, or by (effective) magnetic fluxes; both approaches have been already experimentally demonstrated in UQG~\cite{aidelsburger_realization_2013,miyake_realizing_2013,jotzu_creating_2015,taie_coherent_2015}. 
Another avenue for rich new physics would be to consider topological superconductivity~\cite{qi_topological_2011,bernevig_topological_2013} for spin-density imbalance, a topic already touched by some of the SOC FFLO studies. Ultimately one may find a new density-imbalanced superfluid whose excitation spectra, broken symmetries and collective modes, and topological invariants differ so much from the FFLO and Sarma state paradigms that it is essentially a new state of matter.

\ack
This work was supported by the Academy of Finland through its
Centres of Excellence Programme (Projects No. 284621, 13272490 and 303351) and by the European Research Council (ERC-2013-AdG-340748- CODE). We thank Leo Radzihovsky, Dong-Hee Kim, Aleksi Julku, Pramod Kumar, Long Liang, Marek Tylutki, and Tuomas Vanhala for careful reading of the manuscript and their valuable comments.

\section*{References}
\bibliographystyle{unsrt}

\bibliography{bibli_all_wurl_fixed}

\end{document}